\documentclass[amsmath, amssymb, superscriptaddress, nofootinbib, 12pt, singlespace, lot, lof, oneside]{UAthesis}
\usepackage{graphics}		
\usepackage{latexsym}		
\usepackage{amsmath}
\usepackage{amssymb}		
\usepackage{citesort}		
\usepackage{palatino}
\usepackage{makeidx}

\newcommand{\be}{\begin{equation}}
\newcommand{\ee}{\end{equation}}
\newcommand{\ba}{\begin{eqnarray}}
\newcommand{\ea}{\end{eqnarray}}
\newcommand{\eq}[1]{(\ref{#1})}
\newcommand{\non}{\nonumber}
\newcommand{\n}[1]{\label{#1}}

\newcommand{\tens}[1]{{\boldsymbol{#1}}}
\newcommand{\eps}{\varepsilon}
\newcommand{\grad}{{\tens{d}}}
\newcommand{\pa}{{\tens{\partial}}}
\newcommand{\covd}{{\tens{\nabla}}}

\newcommand{\cKY}{h}

\newcommand{\PKY}{{\tens{\check h}}} 
\newcommand{\PKV}{{\tens{\xi}}}   

\newcommand{\muv}[1]{{\tens{m}_{\hat #1}}}                  
                
\newcommand{\mbv}[1]{{\bar{\tens{m}}_{\hat #1}}}            
  
\newcommand{\Du}[1]{{D_{\hat #1}}}                
\newcommand{\Db}[1]{{\bar{D}_{\hat #1}}}     

\newcommand{\env}[1]{{\tens{e}_{\hat #1}}}                 
                 
\newcommand{\ehv}[1]{{{\tens{\tilde e}}_{\hat #1}}}

\newcommand{\cxv}[1]{{\tens{\epsilon}_{#1}}}       
\newcommand{\cxf}[1]{{\tens{\epsilon}^{ #1}}}

\newcommand{\chv}[1]{{{\tens{\tilde \epsilon}}_{ #1}}}     
\newcommand{\chf}[1]{{{\tens{\tilde \epsilon}}^{#1}}}

\newcommand{\ix}[1]{{\raisebox{0pt}[0pt][0pt]{$\scriptstyle #1$}}}

\newcommand{\cp}{C}
\newcommand{\pder}{{\tens{\partial}}}

\newcommand{\cv}{{\tens{\partial}}}

\newcommand{\phsp}{{\boldsymbol{\Gamma}}}
\newcommand{\coTB}{{\mathbf{T}^{*}M}}

\DeclareMathOperator{\sign}{\mathrm{sign}}

\newcommand{\lied}{\pounds}

\newcommand{\psic}{{\Psi}}

\newcommand{\Ric}{{\mathbf{Ric}}}

\newcommand{\hh}{\, ,\hspace{0.5cm}}
\newcommand{\hhh}{\, ,\hspace{0.2cm}}

\newcommand{\hook}{\raisebox{-0.35ex}{\makebox[0.6em][r]
{\scriptsize $-$}}\hspace{-0.15em}\raisebox{0.25ex}{\makebox[0.4em][l]{\tiny $|$}}}

\newcommand{\bss}[1]{{\boldsymbol{#1}}}

\newcommand{\bs}[1]{{\boldsymbol{#1}}}

\newcommand{\nbs}[1]{{{#1}}}
\newcommand{\nbsi}[2]{{}^{#1}\!{{#2}}}
\newcommand{\bsi}[2]{{}^{\tiny #1}\!{\boldsymbol{#2}}}

\newcommand{\fb}[3]{{}_{\tiny #1}\!{\boldsymbol{#2}}^{\hat{#3}}}
\newcommand{\vb}[3]{{}^{#1}\!{\boldsymbol{#2}}_{\hat{#3}}}
\newcommand{\vnb}[3]{{}^{#1}\!{{#2}}_{\hat{#3}}}

\begin{document} 

%
%

\title{Hidden Symmetries of Higher-Dimensional Rotating Black Holes}
\author{David Kubiz\v n\'ak}
\UAdegree{Doctor of Philosophy}
\UAdepartment{Physics}
\UAconvocation{Fall}
\UAsubmitdate{September 15, 2008}
\UAaddress{Department of Physics \\ 
   University of Alberta \\
   Edmonton, Alberta\\
   Canada  T6G 2G7 
}
\UAyear{2008}

\UAabstract{
In this thesis we study higher-dimensional rotating black holes. 
Such black holes are widely discussed in string theory and brane-world models
at present.
We demonstrate that even the most general known Kerr-NUT-(A)dS spacetime, describing 
the general rotating higher-dimensional asymptotically (anti) de Sitter black hole 
with  NUT parameters, is in many aspects similar to its
four-dimensional counterpart.
Namely, we show that it admits a fundamental hidden
symmetry associated with the principal conformal Killing--Yano tensor.
Such a tensor generates towers of hidden and explicit symmetries.
The tower of Killing tensors is responsible for the existence of irreducible, quadratic in momenta, conserved integrals of geodesic motion. 
These integrals, together with the integrals corresponding to the tower of explicit symmetries, make geodesic equations in the Kerr-NUT-(A)dS spacetime 
completely integrable. 
We further demonstrate that in this spacetime 
the Hamilton--Jacobi, Klein--Gordon, and stationary string  
equations allow complete separation of variables and 
the problem of finding parallel-propagated frames reduces to the set of the first order ordinary differential equations.  
Moreover, we show that the Kerr-NUT-(A)dS spacetime is the most general Einstein space which possesses all these properties. We also explicitly derive the most general (off-shell) canonical metric admitting the principal conformal Killing--Yano tensor and demonstrate 
that such a metric is necessarily of the special algebraic type D of the higher-dimensional algebraic classification.
The results presented in this thesis describe the new and complete picture of 
the relationship of hidden symmetries and rotating black holes in higher dimensions. 

}

\UApreface{
When in 1963 Kerr discovered an astrophysically relevant but relatively complicated metric describing the gravitational field of a  rotating black hole, it seemed that no analytical predictions were possible even for the simplest particle geodesic motion.
However, a `miracle' happened, and it turned out that not only the geodesic motion can be analytically solved, but also the equations describing 
various perturbations  of this background can be `drastically' simplified. 
This opened a way for studying astrophysical processes, such as 
the plasma accretion around black holes, the radiation produced by infalling matter,
the origin of jets, the production and propagation of waves produced in the vicinity 
of black holes, and it even led to estimates of the gravitational wave production in 
star collisions and galaxy merges. It also facilitated the study of more theoretical problems, such as the problem of stability of the Kerr solution, 
the calculation of the quasinormal modes,
or the study of the Hawking radiation.  
A hidden symmetry responsible for this miracle can be mathematically described by a simple antisymmetric object, called the Killing--Yano tensor.

Recently, higher-dimensional rotating black hole spacetimes have become of high interest due to various developments in gravity and high energy physics.
One of the reasons is the popularity of the scenarios with large extra dimensions.
 In these so-called brane world models, our Universe is represented by a four-dimensional brane floating in the higher-dimensional bulk space---where only gravity can propagate. 
One of the main predictions of these models is the possibility to produce higher-dimensional mini black holes in particle colliders. 
Such black holes may provide a window into higher dimensions as well as into 
non-perturbative gravitational physics which may already appear on the TeV scale.
Naturally, one wants to study the properties of these black holes which are the higher-dimensional generalizations of the Kerr geometry.

Another motivation for studying higher-dimensional black hole spacetimes comes from string theory. When quantizing a string action, 
a conformal anomaly appears unless the spacetime dimension is $D=26$ for 
bosonic strings or $D=10$ for superstrings. 
Black holes in string theory are widely discussed 
in connection with the problem of microscopical explanation of the black hole entropy. 
The study of black holes in an anti-de Sitter background can improve the understanding 
of the AdS/CFT correspondence. 
For this reason, it is useful to understand the particle and field propagation in these backgrounds. For example, recently the structure of the black hole singularity was probed using geodesics and correlators on the dual CFT on the boundary.

This thesis is devoted to the study of hidden symmetries of higher-\break dimensional rotating black holes (with spherical horizon topology).
We shall demonstrate that, despite their complexity, the higher-dimensional rotating black holes admit a similar hidden symmetry as their four-dimensional `cousins'. In consequence,
these black holes possess the following properties: the geodesic motion 
in these backgrounds is completely integrable, the Hamilton--Jacobi, Klein--Gordon, 
and Dirac equations allow the separation of variables, the 
metric element is of the algebraic type D, it allows a generalized Kerr--Schild form, and it is uniquely determined by the presence of the hidden symmetry.
In this context, four dimensions are not exceptional, and four-dimensional miraculous properties of rotating black holes are, in some sense, even more miraculous 
in higher dimensions.

In {\bf Part I} of the thesis we focus on hidden symmetries. The next chapter 
introduces the concept of hidden symmetries and outline their importance for   understanding  the properties of four-dimensional rotating black holes. 
It also overviews higher-dimensional black hole spacetimes and recapitulates
the results on hidden symmetries which 
laid the groundwork for new developments described in this thesis.
The basic definitions and properties of the Killing and Killing--Yano tensors, 
the objects responsible for hidden symmetries, are summarized in Chapter 2.

In Chapter 3, we introduce the central notion of this thesis, the notion of the {\em principal conformal Killing--Yano} (PCKY) tensor. We demonstrate how, based on a simple property of this object, one can generate a whole tower of Killing--Yano tensors and a corresponding tower 
of rank-2 Killing tensors. We also discuss another, in some sense more physical, method for generating various Killing tensors. 
This method is based on the construction of integrals of motion for geodesic motion which are 
of higher order in geodesic momenta and therefore associated with the corresponding Killing tensors.
Finally, we demonstrate that the 
PCKY tensor also generates a tower of Killing vector fields.
One of these Killing vectors plays an exceptional role and we reserve for it the notion 
{\em primary}.

In {\bf Part II},  applying the previous results, we study the remarkable properties of higher-dimensional rotating black holes. In Chapter 4,
 we demonstrate that the general Kerr--NUT--(A)dS spacetime,
describing the higher-dimensional arbitrarily rotating black hole with NUT parameters and the cosmological constant, possesses the PCKY tensor. Moreover, this tensor determines uniquely preferred (canonical) coordinates 
for this metric and hence its canonical form. This form is especially useful
for the subsequent calculations.  We also learn that it is possible to consider
a broader class of the (off-shell) metrics which possess the (same) PCKY tensor.
It will be shown later, that this class describes the most general metric element admitting the PCKY tensor. We call it the {\em canonical metric} element. 

In Chapter 5, we demonstrate the {\em complete integrability} of geodesic motion in the 
canonical background. Namely, we prove 
that the constants of geodesic motion corresponding to the extended tower of Killing tensors and the constants of geodesic motion corresponding to the tower of Killing vectors are all functionally independent and that they all mutually Poisson commute of one another.
The latter property is closely related to the fact that the 
corresponding Killing tensors and/or Killing vectors  
Schouten--Nijenhuis commute.
We use the opportunity to briefly review the theory of the Schouten--Nijenhuis brackets and to remind that with respect to these brackets Killing tensors form a Lie algebra. 

The {\em separability} of the Hamilton--Jacobi and Klein--Gordon 
equations in the background of the canonical metric is demonstrated in Chapter 6.
Such a separability provides an independent proof of complete integrability of geodesic motion. It also allows to study the contribution of scalar field to the Hawking evaporation of these black holes. Several related results, directly connected with these developments, are mentioned. Namely, we recapitulate the theory of the
{separability structures} and describe a recent achievement on the symmetry operators
which underly the separability of the
Hamilton--Jacobi and Klein--Gordon 
equations. Some open questions, primarily connected with the separability problem for  higher spin equations in this background, are also briefly discussed.

In Chapter 7, we address the question of {\em uniqueness} and generality of these  developments. 
This leads us to the study of 
metric elements admitting the PCKY tensor. 
In particular, we demonstrate the following two important results:
First, we establish that the Kerr-NUT-AdS spacetime is the most general solution of the vacuum Einstein equations with the cosmological constant which possesses the  PCKY tensor.
Second, without imposing the Einstein equations, we explicitly derive the 
most general metric admitting such a tensor and show that it coincides with the canonical metric element. These results naturally generalize 
the results obtained earlier in four dimensions.

{\bf Part III} of the thesis is devoted to further developments connected with the PCKY tensor. In Chapter 8, we demonstrate the separability of the Nambu--Goto equations for a stationary string in the background of the canonical spacetime. 
Such a string is generated by a 1-parameter family of Killing trajectories and the problem of finding its configuration reduces to a problem of finding a
geodesic line in a (one dimension lower) effective background. 
The resulting integrability of this geodesic problem is connected with the existence of hidden symmetries which are inherited from the black hole background.
More generally, we introduce the concept of $\xi$-branes, that is 
more dimensional objects with the worldvolume aligned along the set of Killing 
vector fields, and discuss their integrability in the Kerr-NUT-(A)dS spacetime.

In Chapter 9, we study the equations describing the parallel
transport of orthonormal frames along timelike geodesics in the spacetime
admitting the PCKY tensor.
It is demonstrated how, in the presence of this tensor, these  equations
can be reduced to a set of the first order ordinary differential
equations.  Concrete examples of parallel-propagated frames in $D=3, 4, 5$ 
canonical spacetimes are constructed and it is shown that the obtained 
set of equations can be solved by the separation of variables.
In the last chapter we summarize the overall picture of the obtained results, 
link these results to the related achievements, 
and discuss possible future directions.

To keep the main text concise and fluent, we have moved the complementary material of various character to the {\bf appendices}.
In Appendix A some four-dimensional aspects of hidden symmetries are discussed.
The first section plays the role of an introduction for newcomers to the problematic of hidden symmetries. On a simple four-dimensional example we describe the
main ideas of the, much more complicated, higher-dimensional theory.
In the second section we study hidden symmetries of the Pleba\'nski--Demia\'nski
class of solutions. Some physically important subcases are discussed in more detail. 
An account of historical developments leading to the discovery of the PCKY tensor in the general Kerr-NUT-(A)dS spacetimes is recorded in Appendix B.
Miscellaneous results are gathered in Appendix C. 

The results presented in this {\bf doctoral thesis} were obtained during the course of the author's Ph.D. program at the University of Alberta between years 2005 and 2008.
The thesis is based on the following published papers in peer reviewed journals:
\cite{FrolovKubiznak:2007}, \cite{KubiznakFrolov:2007}, \cite{PageEtal:2007},
\cite{FrolovEtal:2007}, \cite{KrtousEtal:2007prd}, \cite{KrtousEtal:2007jhep}, 
\cite{KubiznakKrtous:2007},
\cite{KubiznakFrolov:2008}, \cite{FrolovKubiznak:2008}, \cite{ConnellEtal:2008},
\cite{KrtousEtal:2008}.

\subsection*{Notations and conventions}

Throughout the thesis, we use a mixture of invariant, tensorial, and matrix notations.
The tensors are typed in boldface and their components (with due indices) in normal letters. We consider a $D$-dimensional manifold $M^D$ equipped with a metric $\tens{g}$. 
Except Chapter 9 and the appendices the  metric is symbolically of the Euclidean signature;  
it is related to the physical metric by a simple Wick rotation (see Chapter 4). The coordinate indices are denoted by the Latin letters from the 
beginning of the alphabet, ${a,b,c,\dots,=1,\dots,D}$;
we use the Einstein summation convention for them. These 
indices may be also understood as abstract, in the sense of 
\cite{Wald:book1984}.
A dot above tensors denotes the differentiation along the vector field; $\tens{\dot T}\equiv \nabla_u \tens{T}\equiv u^a\nabla_a \tens{T}$.
In Chapter 7, we use the symbol $\,\check{\  }\,$ to distinguish an operator from the corresponding 2-form. For example, having a 2-form  
$\tens{F}$ with the (abstract) indices  $F_{ab}$,  $\tens{\check F}$ denotes the  operator $F^a{}_b$.
Where it cannot lead to confusion, 
a dot between tensors indicates contraction, e.g.,
${\tens{a}\cdot\tens{b}\equiv a^c b_c}$. 
Similarly, $\PKY\cdot \PKY\cdot\tens{v}$ 
denotes a vector with components ${h^{a}{}_{b}\,h^{b}{}_{c}\,v^c}$.
The symbols $\tens{d}x^a$, ${\cv_{x^a}}$, denote the coordinate 1-form, vector, 
associated with the coordinate~${x^a}$. 
In several place we also use matrix notations. Matrices are typed in normal letters and the standard matrix notations are used for them. 
For example, having a rank-2 tensor ${\tens{A}}$, 
the symbol ${A}$ stands for the matrix 
of its components ${A^{a}_{\ b}}\,$, $A^T$ denotes the transposed matrix, and ${\rm Tr}{A}$ stands for its trace.

The central object of the thesis is a {\em principal conformal Killing--Yano tensor}, that is a non-degenerate, closed, conformal Killing--Yano 2-form. 
We reserve for it an abbreviation PCKY and (except Chapter 8) the symbol $\tens{h}$.
The associated primary (Killing) vector is denoted by $\tens{\xi}$; $\,\xi^b\equiv 1/(D-1)\nabla_a h^{ab}$.
The (orthonormal) Darboux bases of 1-forms and vectors determined by the PCKY tensor are 
called the {\em canonical} bases and denoted by $\{\tens{\omega}\}$ and $\{\tens{e}\}$
(see Chapter 3).
To distinguish the basis indices from the coordinate indices we use $\hat{\ }$\,\,.
For example, $u^{\hat a}$ denotes the basis components of the velocity $\tens{u}$.
The same symbol is also used to denote (differential) operators, for example, $\hat{\xi}\equiv i \xi^a\nabla_a$.
Further conventions are introduced later in the text (see, e.g., Chapter 2).

}

%
%


%
%

\maketitle


\part{Hidden Symmetries}

\chapter{Introduction}

\section{Symmetries}

In modern theoretical physics one can hardly overestimate the  role
of symmetries. They comprise the most fundamental laws of nature, 
they allow us to classify solutions, in their presence  complicated
physical problems become tractable.  The value of symmetries is
especially high in nonlinear theories, such as general relativity. 

In curved spacetime continuous symmetries (isometries)  are generated
by Killing vector fields. Such symmetries have clear geometrical
meaning. Let us assume that in a given manifold we have a
$1$-parameter family of  diffeomorphisms generated by a vector field
$\bss{\xi}$. Such a vector field determines the dragging  of
tensors by the diffeomorphism transformation. If a tensor field $\bss{T}$ is
invariant with respect to this dragging, that is, its Lie derivative
along $\bss{\xi}$ vanishes, ${\cal L}_{\xi}\bss{T}=0$, we have a
symmetry. A  vector field which generates transformations
preserving the metric is called a {\em Killing vector field}, and the
corresponding diffeomorphism---an {\em isometry}. 
According to the first Noether theorem continuous symmetries of the theory
imply the existence of conserved quantities.  For a covariant theory in an external
gravitational field for each of the Killing
vectors there exists a conserved quantity. For example, for a
particle geodesic motion this conserved quantity is a projection of
the particle momentum on the Killing vector. 

Besides isometries the spacetime may also possess {\em
hidden symmetries}, generated by either symmetric or antisymmetric tensor
fields.  Such symmetries are not directly related to the metric invariance
under diffeomorphisms. They
represent the genuine symmetries of the phase space rather than the
configuration space. For example, the symmetric Killing tensors  
give rise to the conserved quantities of higher order in particle
momenta, and underline the separability of the scalar field
equations.  Less known but even more fundamental are the antisymmetric
Killing--Yano tensors which are related to the separability of field
equations with spin,  the existence of `quantum' symmetry operators, and
the presence of  conserved charges.

\section{Miraculous properties of the Kerr geometry}

To illustrate the role of hidden symmetries in general relativity let
us  recapitulate the ``miraculous'' properties \cite{Chandrasekhar:1983} of the
Kerr geometry. This astrophysical important solution was obtained in
1963 by Kerr \cite{Kerr:1963}. 
The metric is
stationary and axially symmetric; it possesses two Killing vectors, $\pa_t$ and $\pa_\phi$,
generating the time translation and the rotation. 
The Kerr solution is of the special algebraic type {D} of Petrov's classification
\cite{Petrov:1954}, \cite{Petrov:1969}, it belongs to 
the special class of solutions which can be presented
in the Kerr--Schild form \cite{KerrSchild:1965}, \cite{KerrSchild:1969}, \cite{DebneyEtal:1969},
\be\n{KS}
g_{ab}=\eta_{ab}+2H l_{a}l_{b}\, .
\ee
Here, $\tens{\eta}$ is a flat metric and $\tens{l}$ is a null vector,
in both metrics $\tens{g}$ and $\tens{\eta}$.\footnote{If the ansatz \eq{KS} is inserted into
the Einstein equations, one effectively reduces the problem to a
linear one (see, e.g., \cite{GursesGursey:1975}). This gives a powerful tool for the study
of special solutions of the Einstein equations. This method works in higher 
dimensions as well. For example, 
the Kerr--Schild ansatz was used by  Myers and Perry
to obtain their higher-dimensional black hole solutions \cite{MyersPerry:1986}.}

Although the Killing vector fields $\pa_t$ and $\pa_{\phi}$ are not
enough to provide a sufficient number of integrals of
motion\footnote{For example, for a particle motion these isometries
generate the conserved energy and azimuthal component of the angular
momentum, which, together with the conservation of $p^2$, gives only three
integrals of motion. For separability of the Hamilton--Jacobi
equation in the Kerr spacetime the fourth integral of motion is
required.} in 1968 Carter \cite{Carter:1968pr}, \cite{Carter:1968cmp}
demonstrated that both---the Hamilton--Jacobi and  scalar field
equations---can be separated. This proved, apart from other things,
that there  exists an additional integral of motion, `mysterious'
Carter's constant, which makes the particle geodesic motion
completely integrable. In 1970, Walker and Penrose \cite{WalkerPenrose:1970} pointed
out that Carter's constant is quadratic in particle momenta and its
existence is directly connected with the symmetric Killing tensor
\cite{Stackel:1895}
\be\label{KT_4D}
K_{ab}=K_{(ab)}\,,\quad \nabla_{\!(c}K_{ab)}=0\,.
\ee
During the several following years it was discovered that it is not
only the Klein--Gordon equation which allows the separation of
variables in the Kerr geometry. In 1972, Teukolsky decoupled the
equations for electromagnetic and gravitational perturbations, and
separated variables in the resulting master equations \cite{Teukolsky:1972}.
One year later the massless neutrino equation by Teukolsky and Unruh
\cite{Teukolsky:1973}, \cite{Unruh:1973}, and in 1976 the massive Dirac equation by
Chandrasekhar and Page \cite{Chandrasekhar:1976}, \cite{Page:1976} were separated.

Meanwhile a new breakthrough was achieved in the field of hidden
symmetries when in 1973 Penrose and Floyd \cite{Penrose:1973}, \cite{Floyd:1973} discovered
that the Killing tensor for the Kerr metric can be written in the
form 
\begin{equation}\label{square}
K_{ab}=f_{ac}f_b{}^c\,,
\end{equation}
where the antisymmetric tensor
$\tens{f}$ is the Killing--Yano (KY) tensor
\cite{Yano:1952} 
\begin{equation}\label{KY4}
f_{ab}=f_{[a b]}\,,\quad \nabla_{\!(c}f_{a) b}=0\,.
\end{equation}
A Killing--Yano tensor is in many aspects more fundamental than a
Killing tensor. In particular, 
having a Killing--Yano tensor one can always construct the corresponding 
Killing tensor using Eq. \eqref{square}. 
On the other hand, not every Killing tensor can be decomposed 
in terms of a Killing--Yano tensor (for necessary conditions see \cite{Collinson:1976}, \cite{FerrandoSaez:2002}).

Many of the remarkable properties of the Kerr spacetime are consequences
of the existence of  the Killing--Yano
tensor. In particular, in 1974 Collinson
demonstrated that the integrability conditions for a non-degenerate
Killing--Yano tensor imply that the spacetime is necessary of the
Petrov type {D} \cite{Collinson:1974}.\footnote{All the vacuum type {D} solutions were
obtained by Kinnersley \cite{Kinnersley:1969}. Demia\'nski and Francaviglia showed
that in the absence  of acceleration these solutions admit
Killing and Killing--Yano tensors \cite{DemianskiFrancaviglia:1980}. 
A general (off-shell) metric element admitting a Killing--Yano tensor 
in four dimensions was  obtained  by Dietz and R\"udiger
\cite{DietzRudiger:1981}, \cite{DietzRudiger:1982}, see also \cite{Taxiarchis:1985}.
}
In 1975, Hughston and
Sommers showed that in the Kerr geometry the Killing--Yano
tensor $\tens{f}$ generates both of its isometries 
\cite{HughstonSommers:1973}. Namely, the Killing vectors $\pa_t$ and $\pa_\phi,$ can be written as follows:
\begin{equation}\label{kvv}
\xi^a\equiv \frac{1}{3}\,\nabla_{\!b}(*f)^{b a}=(\partial_t)^{a}\,,\quad
\eta^{a}\equiv -K^{a}_{\ b}\xi^{b}=(\partial_\phi)^{b}\,.
\end{equation}  
This means that, in fact, all the symmetries necessary for complete 
integrability of geodesic motion in the Kerr spacetime are `derivable' from the existence
of a single Killing--Yano tensor.

In 1977, Carter demonstrated
\cite{Carter:1977} that given an isometry $\tens{\xi}$ and/or a Killing tensor
$\tens{K}$ one can construct the operators
\begin{equation}\label{opK}
\hat \xi \equiv i\xi^a \nabla_a\,,\quad
\hat K \equiv \nabla_a K^{a b}\nabla_b\,,
\end{equation}
which commute with the scalar Laplacian\footnote{In fact, the operator
$\hat K$ defined by \eqref{opK} commutes with $\Box$ provided that the background metric satisfies the vacuum Einstein  or source-free Einstein--Maxwell equations. In more general spacetimes, however, a quantum anomaly proportional to a contraction of  $\tens{K}$ with the
Ricci tensor may appear.
Such anomaly is not present if 
an additional condition \eqref{square} is satisfied  \cite{Cariglia:2004}.
} 
\begin{equation}
\Bigl[\Box\,, \hat\xi \,\Bigr]=0=\left[\Box\,, \hat K\right]\,,\quad
\Box\equiv \nabla_a g^{a b}\nabla_b\,.
\end{equation}
Moreover, in the Kerr geometry these operators 
commute also between themselves and provide therefore good `quantum' numbers
for scalar fields. In 1979 Carter and McLenaghan found that an operator
\begin{equation}
\hat f \equiv i\gamma_5\gamma^a\Bigl(f_{a}{}^{b}\nabla_b
-\frac{1}{6}\,\gamma^b\gamma^c \nabla_{\!c}f_{a b}\Bigr)
\end{equation}
commutes with the Dirac operator $\gamma^a\nabla_a$ 
\cite{CarterMcLenaghan:1979}. This
gives a new quantum number for the spinor wavefunction and explains 
why  separation of the Dirac equation can be achieved.
Similar symmetry operators for other equations with spin, including
electromagnetic and gravitational perturbations, were constructed
later \cite{KamranMcLenaghan:1984}, \cite{Kamran:1985}, \cite{TorresdelCastillo:1988}, \cite{KalninsMiller:1989},\cite{KalninsEtal:1996}. 

In 1983, Marck solved equations for the parallel transport of an
orthonormal frame along geodesics in the Kerr spacetime 
\cite{Marck:1983kerr}, \cite{Marck:1983null} 
and used this
result for the study of tidal forces. For this construction
he used a simple fact that the vector 
\begin{equation}\label{Lc}
L_a\equiv f_{ab}p^b\,,\quad L_a p^a=0\,,
\end{equation}
is parallel-propagated along a geodesic $\tens{p}$.  

In 1987, Carter \cite{Carter:1987} pointed out that 
the Killing--Yano tensor itself is derivable from a
$1$-form $\tens{b}\,,$ 
\begin{equation}\label{C87}
\tens{f}=\tens{*d b}\,.
\end{equation} 
We call such a form $\tens{b}$ a (KY) {\em potential}.
It satisfies the Maxwell equations and can be interpreted as a
4-potential of an electromagnetic field with the source current 
proportional to the {\em primary} Killing vector field $\pa_t\,,$ cf. Eq.
\eqref{kvv}. 
In 1989, Frolov {\em et al.} 
\cite{FrolovEtal:1989}, \cite{CarterFrolov:1989}, \cite{CarterEtal:1991},  
separated equations for an
equilibrium configuration of a cosmic  string near the Kerr black
hole. In 1993, Gibbons {\em et al.} demonstrated  that
due to the presence of Killing--Yano tensor the classical spinning
particles in this background possess enhanced worldline supersymmetry
\cite{GibbonsEtal:1993}. Conserved quantities in the Kerr geometry generated by
$\tens{f}$ were discussed in 2006 by Jezierski and \L ukasik
\cite{JezierskiLukasik:2006}. 

To conclude this section we  mention that many of the above
statements and results, which we have formulated for the Kerr
geometry, are in fact more general. Their validity can be extended to
more general spacetimes,  or even to an arbitrary number of spacetime
dimensions. For example, the whole Carter's class of solutions
\cite{Carter:1968pl}, \cite{Carter:1968cmp} (see also \cite{Debever:1971}, \cite{Plebanski:1975}) 
admits a KY tensor and possesses many of the discussed properties (see Appendix A). General results on Killing--Yano tensors  and algebraic properties were gathered by Hall \cite{Hall:1987}.  
A  relationship among  the existence of Killing tensors and separability
structures for the
Hamilton--Jacobi equation in an arbitrary number of spacetime
dimensions  was discussed in \cite{Woodhouse:1975}, \cite{BenentiFrancaviglia:1979}, 
\cite{KalninsMiller:1981} (see also Section 6.3.1). We refer to Appendix A for 
further details on hidden symmetries in 4D.

\section{Higher-dimensional black holes}

Higher-dimensional black hole solutions have been studied for a long time. 
Already in 1963, Tangherlini \cite{Tangherlini:1963} obtained a higher-dimensional
generalization of the Schwarzschild metric \cite{Schwarzschild:1916}. 
The charged version of the Tangherlini metric was found  in 1986 by
Myers and Perry \cite{MyersPerry:1986}.  In the same paper a general vacuum
rotating black hole in higher dimensions was obtained. This solution,
often called the Myers--Perry (MP) metric,
generalizes the four-dimensional Kerr solution.\footnote{%
To be more precise,
it is well known that physics of higher-dimensional black holes can be substantially different, and much richer, than in four dimensions. 
Whereas in four dimensions only the black holes with spherical horizon topology are allowed, it is expected that non-spherical horizon topologies are a generic feature
of higher-dimensional gravity. Besides the class of rotating black holes solutions with spherical horizon, such as the Myers--Perry metrics and their generalizations, there exist other rotating `black objects', for example black rings and their generalizations. 
In this thesis we concentrate only on the class of rotating black holes with spherical horizon topology. Such black holes can be considered as natural higher-dimensional generalizations of the Kerr metric.
} 
Main new feature of the MP metrics
in $D$ dimensions  is that, instead of one rotation parameter, they have
$m\equiv [(D-1)/2]$ rotation parameters, corresponding to $m$
independent $2$-planes of rotation.
Later, in 1998,  Hawking, Hunter, and Taylor-Robinson \cite{HawkingEtal:1999}
found a $5$D generalization of the 4D rotating black hole in
asymptotically (anti) de Sitter space (Kerr-(A)dS metric 
\cite{Carter:1968cmp}).  In 2004 Gibbons, L\"u, Page, and Pope
\cite{GibbonsEtal:2004}, \cite{GibbonsEtal:2005} discovered the general Kerr-(A)dS metrics in 
arbitrary number of  dimensions.    After several
attempts to include  NUT \cite{NewmanEtal:1963} parameters \cite{ChongEtal:2005}, \cite{ChenEtal:2007},
in  2006 Chen, L\"u, and Pope \cite{ChenEtal:2006cqg}  found a general
Kerr-NUT-(A)dS solution of the Einstein equations for all $D$. 
These metrics were obtained in special coordinates which are
the natural higher-dimensional generalization of 
the Carter's $4$D {\em canonical coordinates} \cite{Carter:1968cmp}, \cite{Debever:1971}, \cite{Plebanski:1975}. So far,
they remain the most general  black-hole-type solutions of the Einstein 
equations with the cosmological constant (and spherical horizon topology)
which are  known
analytically.\footnote{ Besides the brane-world scenarios, these black
holes find their applications for the ADS/CFT correspondence. In the
BPS limit the odd-dimensional metrics lead to the Sasaki--Einstein
metrics \cite{HashimotoEtal:2004}, \cite{CveticEtal:2005a}, \cite{CveticEtal:2005b} whereas the even-dimensional metrics lead to the
Calabi--Yau spaces \cite{OotaYasui:2006}, \cite{LuPope:2007}.
There have been also several attempts to generalize these solutions.
For example, to find a similar solution of the Einstein--Maxwell equations
either in an analytical form \cite{AlievFrolov:2004}, \cite{Aliev:2006a}, \cite{Aliev:2006b},\cite{Aliev:2007}, \cite{KunzEtal:2006b}, \cite{ChenLu:2008}, \cite{Krtous:2007} or numerically \cite{KunzEtal:2005},\cite{KunzEtal:2006a},\cite{KunzEtal:2007}, \cite{BrihayeDelsate:2007}, \cite{KleihausEtal:2008}. 
See also \cite{CharmousisGregory:2004}, \cite{PodolskyOrtaggio:2006}, \cite{PravdaEtal:2007},\cite{OrtaggioEtal:2008}, or \cite{HouriEtal:2008b}, \cite{LuEtal:2008a}.} 
For a recent extended review on higher-dimensional black holes see, e.g., 
\cite{EmparanReall:2008}.

In connection with these black holes the following natural questions arise: To
what extent are the remarkable properties described for the four-dimensional black
holes innate to four dimensions? Do some of them transfer 
to higher dimensions as well?
And in particular, do some of the higher-dimensional black holes possess hidden symmetries?

\section{Hidden symmetries in higher dimensions}

The hidden symmetries of higher-dimensional rotating black holes were
first discovered  for the $5$D Myers--Perry metrics \cite{FrolovStojkovic:2003a} ,\cite{FrolovStojkovic:2003b}. It
was demonstrated that both, the Hamilton--Jacobi and  scalar field
equations,  allow the separation of variables and the corresponding
Killing tensor was obtained. 
Later it was shown that 5D
results  can be extended to an arbitrary number of dimensions,
provided that rotation parameters of the MP metric can be
divided into two classes, and within each of the classes these
parameters are equal of one another \cite{VasudevanEtal:2005b}. Similar results were found in the presence of the cosmological constant and NUT parameters
\cite{LopezOrtega:2003}, \cite{VasudevanEtal:2005a}, \cite{KunduriLucietti:2005a},  \cite{VasudevanStevens:2005}, \cite{Vasudevan:2006}, \cite{ChenEtal:2006jhep}, \cite{Davis:2006b}, \cite{Vasudevan:PhD}. It was also demonstrated that a stationary string configuration in the $5$D Myers--Perry spacetime is  
completely integrable  \cite{FrolovStevens:2004}.

Were these results the most general possible? Or, were there somewhere 
hidden other symmetries which would allow a further progress? In particular,
do the higher-dimensional rotating black holes admit the fundamental symmetry of a 
Killing--Yano tensor? These were the questions which stimulated 
further research.
 
The outcome of this research can be briefly summarized as follows:\footnote{Let us emphasize that not all 
of the results were obtained by the author of this thesis and/or his collaborators.
What we summarize here is the overall progress which has been recently achieved in this direction.}
Even the most general known higher-dimensional
Kerr-NUT-(A)dS black holes possess many of the remarkable 
properties of their four-dimensional `cousins' .  
Namely, the geodesic motion in these backgrounds is completely integrable, 
the Hamilton--Jacobi, Klein--Gordon, Dirac, and stationary string equations are completely separable. The metrics are of the type D of higher-dimensional algebraic classification and allow a generalized Kerr--Schild form.
Many of these properties follow directly from 
the existence of a fundamental hidden symmetry associated with the 
{\em principal conformal Killing--Yano} (PCKY) tensor.

The PCKY tensor was first discovered for 
the Myers--Perry metrics \cite{FrolovKubiznak:2007}, and soon after that for the completely general Kerr-NUT-(A)dS spacetimes \cite{KubiznakFrolov:2007}. 
Starting with this tensor, one can generate the
whole {\em tower} of hidden symmetries \cite{KrtousEtal:2007jhep}
which are responsible for complete integrability of geodesic
motion in these spacetimes \cite{PageEtal:2007}, \cite{KrtousEtal:2007prd}. 
Such an integrability was
independently proved by separating the Hamilton--Jacobi equation \cite{FrolovEtal:2007}.
Due to the presence of hidden symmetries, also the Klein--Gordon \cite{FrolovEtal:2007} and Dirac \cite{OotaYasui:2008} equations allow the separation of variables in these backgrounds. Recently, extending the work of \cite{HouriEtal:2007}, \cite{HouriEtal:2008a}, the uniqueness of these results was demonstrated \cite{KrtousEtal:2008}. In particular, it was proved that, similar to the 4D case, the most general Einstein space admitting the PCKY tensor is the Kerr-NUT-(A)dS spacetime.
Meanwhile, it was shown that the Kerr-NUT-(A)dS spacetime is of the algebraic type D of higher-dimensional algebraic classification \cite{HamamotoEtal:2007} and that it allows a generalized Kerr--Schild form \cite{ChenLu:2008}.
By relaxing some of the requirements imposed on the PCKY tensor
more general spacetimes were recently discovered \cite{HouriEtal:2008b}, \cite{HouriEtal:2008c}. 
Directly related to the PCKY tensor is also a 
proved complete integrability of a stationary  
string configuration in the vicinity of the Kerr-NUT-(A)dS black hole \cite{KubiznakFrolov:2008}
and the possibility of constructing a parallel propagated frame in such a background
\cite{ConnellEtal:2008}. (For other related works see Chapter 10.)

\chapter{Killing--Yano and Killing tensors}
In this chapter we review the basic objects responsible for symmetries of the spacetime and briefly discuss their properties.
In particular, we introduce the Killing and Killing--Yano tensors, and their conformal generalizations. We also exploit the 
opportunity to establish some notations used later in the text.

\section{Definitions}
\subsubsection{Killing vectors}
Let us consider a $D$-dimensional spacetime with a metric $\bss{g}$.
A spacetime  possesses an {\em isometry} generated by the
{\em Killing vector} field $\tens{\xi}$ if this vector obeys the {
Killing equation} 
\be 
\nabla_{(a}\xi_{b)}=0\, . 
\ee 
For a geodesic motion of a particle in such a curved spacetime the quantity
$p^{a}\xi_{a}$, where $\tens{p}$ is the momentum of the particle, remains constant
along the particle's trajectory.
Similarly, for a null geodesic,
$p^{a}\xi_{a}$ is conserved provided that $\tens{\xi}$ is a {\em conformal
Killing vector} obeying the equation
\be\label{CKV_d1}
\nabla_{(a}\xi_{b)}=\tilde{\xi}g_{ab}\,,\quad \tilde{\xi}=
D^{-1}\nabla_b\xi^{b}\, .
\ee
An equivalent defining equation for the conformal Killing vector is 
\be\label{CKV_d2}
\nabla_a\xi_b=\nabla_{[a}\xi_{b]}+g_{ab}\tilde \xi\,.
\ee
That is, the conformal Killing vector is an object, the covariant derivative of which  splits into the `{\em exterior}' and `{\em divergence}' parts.\footnote{%
For a general vector an additional term, the `{\em harmonic}' part, is present.
It is the lack of this term what makes conformal Killing vectors `special'.}
When the first term vanishes the conformal Killing vector is {\em closed}.
The vanishing of the second term means that we are dealing with the Killing vector. 
In the case when both parts are zero we have a {\em covariantly constant} (Killing) vector. 

There exist two natural (symmetric and antisymmetric) generalizations
of a (conformal) Killing vector. (For a `hybrid' proposal, see, e.g., \cite{CollinsonHowarth:2000}.)

\subsubsection{Killing tensors}
A symmetric (rank-$p$) {\em conformal Killing tensor} \cite{WalkerPenrose:1970}
$\tens{Q}$ obeys the equations
\be\n{CK}
Q_{a_1a_2 \ldots a_p}=
Q_{(a_1a_2 \ldots a_p)}\,,\quad 
\nabla_{(b}Q_{a_1a_2 \ldots a_p)}=g_{(b a_1}\tilde{Q}_{a_2 \ldots
a_p)}
\, .
\ee
As in the case of a conformal Killing vector, the tensor
$\tilde{\tens{Q}}$ is  determined by tracing both sides of Eq.
\eq{CK}. In particular, a rank-2 conformal Killing tensor obeys the equations
\be
\nabla_{(a} Q_{bc)}=g_{(ab}\tilde Q_{c)}\;,\quad
\tilde Q_{a}=\frac{1}{D+2}\,(2\nabla_{c}Q^{c}_{\ a}+\nabla_{a}Q^{c}_{\ c})\,.
\ee

If $\tilde{\tens{Q}}$ vanishes, the tensor $\tens{Q}$ is called
a {\em Killing tensor} \cite{Stackel:1895} and it is usually denoted by $\tens{K}$.
So we have
\be\n{KT_def}
K_{a_1a_2 \ldots a_p}=
K_{(a_1a_2 \ldots a_p)}\,,\quad 
\nabla_{(b}K_{a_1a_2 \ldots a_p)}=0\, .
\ee 
Obviously, the metric is a (trivial) rank-2 Killing tensor.
In the presence of the Killing
tensor $\tens{K}$ the conserved quantity for a geodesic motion is 
\begin{equation}
K=K_{a_1a_2 \ldots a_p}p^{a_1}p^{a_2}\ldots p^{a_p}\,.
\end{equation}
For null geodesics this quantity is conserved not only for a Killing
tensor, but also for a conformal Killing tensor.
Let us finally mention that a symmetrized tensor product of Killing vectors
is a (reducible) Killing tensor. More generally, we call the Killing tensor 
{\em reducible} if it is a linear combination of the products of  
Killing tensors of a lower rank.

\subsubsection{Killing--Yano tensors}
The {\em conformal Killing--Yano} (CKY) tensors were first proposed in 1968 by Kashiwada and Tachibana \cite{Kashiwada:1968}, \cite{Tachibana:1969} as a generalization  of the  Killing--Yano (KY) tensors introduced by Yano in 1952 \cite{Yano:1952}.

One of the simplest approaches to the CKY tensors is based on a natural generalization of the definition \eqref{CKV_d2} of conformal Killing vectors. 
The CKY tensor $\tens{k}$ of rank-$p$ is a $p$-form the covariant derivative of which 
has vanishing harmonic part, that is it splits into the exterior and divergence parts as follows:
\ba
\nabla_{a}k_{b_1\dots b_p}\!\!&=&\!\!
\nabla_{\![a}k_{b_1\dots b_p]}+p\,g_{a[b_1\!}{\tilde k}_{b_2\dots b_p]}\,,
\label{CKY_d1}\\
{\tilde k}_{b_2\dots b_{p}}\!\!&=&\!\!\frac{1}{D-p+1}\,\nabla_{c}k^{c}{}_{b_2\dots b_p}\,.
\label{CKY_ktilde}
\ea
[The tensor $\tens{\tilde k}$, \eqref{CKY_ktilde}, is determined by tracing both sides of the first equation.]
The defining equation \eqref{CKY_d1} is invariant under the Hodge duality;
the exterior part transforms into the divergence part and vice versa.
This implies that the dual $*\tens{k}$ is a CKY tensor whenever $\tens{k}$ is a CKY tensor
(see also the next section).

Three special subclasses of CKY tensors are of particular interest:
(a) {\em Killing--Yano} tensors with 
zero divergence part in \eqref{CKY_d1} (b) {\em closed CKY} tensors with vanishing exterior part in \eqref{CKY_d1} and (c) {\em covariantly constant} (KY) tensors with  both parts vanishing. The subclasses (a) and (b) transform into each other under the Hodge duality.

In particular, a rank-2 CKY tensor $\tens{k}$ obeys the equations
\begin{equation}\label{CKY_4rank2}
\nabla_{a}k_{bc}= \nabla_{[a}k_{bc]}+ {2}\, g_{a[b} \xi_{c]}\;,\quad
\xi_{a}=\frac{1}{D-1}\,\nabla_{c}k^{c}{}_{a}\;.
\end{equation}
The vector $\tens{\xi}$, defined by the last equation, is called {\em primary}.
It satisfies \cite{Tachibana:1969} (see also 
Appendix C.2)
\be\label{xi_cond}
\nabla_{(a}\xi_{b)}=\frac{1}{D-2}R_{c(a}k_{b)}{}^c\,.
\ee
Thus, in an Einstein space, that is when
$R_{ab}=\Lambda g_{ab}$, $\tens{\xi}$ is the Killing vector.

An alternative (equivalent) definition of a rank-$p$ CKY tensor naturally generalizes
the definition \eqref{CKV_d1} \cite{Tachibana:1969}, \cite{Jezierski:1997}, \cite{Cariglia:2004}. It reads
\be\label{CKY_d2}
\nabla_{(a}k_{b_1)b_2\ldots b_{p}}=
g_{a b_1}\tilde{k}_{b_2 \ldots b_{p}}-
(p-1)\,g_{[b_2(a}\tilde{k}_{b_1) \ldots b_{p}]}\,,
\ee
where $\tens{\tilde k}$ (obtained again by tracing procedure) is given by \eqref{CKY_ktilde}.

Let us finally mention two additional important properties of KY tensors. 
Having a KY tensor $\tens{f}$:
\begin{enumerate}
\item The tensor 
\begin{equation}\label{Lprop}
L_{a_1a_2 \ldots a_{p-1}}\equiv f_{a_1a_2 \ldots a_p}p^{a_p}\,
\end{equation}
is parallel-propagated along the geodesic $\tens{p}$.
\item
The object 
\begin{equation}\label{CKT} 
K_{ab}=\frac{c_p}{(p-1)!}\,f_{a a_2 \ldots
a_p}f_{b}^{\ \,a_2 \ldots
a_p}
\end{equation}
is an associated Killing tensor. Here $c_p$ is an arbitrary constant,
which is often taken to be one. For a different convenient choice
see Section~3.2. 
\end{enumerate}
Similarly, for a rank-2 CKY tensor $\tens{k}$ the object 
\begin{equation}\label{cKT}
Q_{ab}=k_{ac}k_{b}{}^{c}\;,
\end{equation} 
is an associated conformal Killing tensor.

Let us also mention that the exterior product of Killing vectors 
does not generally produce a KY tensor.\footnote{%
A trivial example when this works is , for example, the case of maximally symmetric spacetimes (see also Footnote 1 in Appendix A). 
}
 However, in Section 3.2.1 we shall prove the important
fact that the exterior product of two closed CKY tensors is again a closed CKY tensor.

\section{CKY tensors as differential forms}
The CKY tensors are forms and operations with them are greatly
simplified if one uses the `invariant' language of differential forms. 
In this section we establish some of these notations. We also 
recast the CKY equation \eqref{CKY_d1} into this language and prove its invariance under the Hodge duality. 

If $\tens{\alpha}_p$ and $\tens{\beta}_q$ are $p$- and $q$-forms,
respectively, the external derivative $\tens{d}$ of their exterior (wedge) product
$\wedge$ obeys a relation
\be\n{dd}
\tens{d}(\bss{\alpha}_p\wedge \bss{\beta}_q)=
\tens{d}\bss{\alpha}_p\wedge \bss{\beta}_q+
(-1)^p\bss{\alpha}_p\wedge \tens{d}\bss{\beta}_q\, .
\ee
For an arbitrary form $\tens{\alpha}$ we denote  
\be
\bs{\alpha}^{\wedge m}\equiv \underbrace{\bs{\alpha}\wedge \ldots \wedge
\bs{\alpha}}_{\mbox{\tiny{total of m factors}}}\, .
\ee
A Hodge dual $*\tens{\alpha}_p$ of a $p$-form $\tens{\alpha}_p$ is
a $(D-p)$-form defined as 
\be\label{Hodge_dual}
(* {\alpha}_p)_{a_1 \ldots \,a_{D-p}}= {1\over p!}\, \alpha^{b_1\ldots\,
b_p}e_{b_1\ldots\, b_p a_1 \ldots\, a_{D-p}}\, ,
\ee
where $e_{a_1 \ldots\, a_D}$ is a totally antisymmetric tensor.
The {\em co-derivative}
$\tens{\delta} $ is defined as follows:
\be\label{coderivative}
\tens{\delta \alpha}_p=(-1)^p\epsilon_p \,\tens{*d\!*\!\alpha}_p\,,\quad
\epsilon_p=(-1)^{p(D-p)}\frac{\det(g)}{|\det(g)|}\, .
\ee
One also has $\tens{*\!*\!\alpha}_p=\epsilon_p\tens{\alpha}_p$.

If $\{ \tens{e}_{\hat a}\}$ is a basis of vectors, then the dual basis of 1-forms
$\{\tens{\omega}^{\hat a}\}$ is defined by the relations
$\tens{\omega}^{\hat a}(\tens{e}_{\hat b})=\delta^a_b$. We denote
$g_{\hat a \hat b}=\tens{g}(\tens{e}_{\hat a},\tens{e}_{\hat b})$ 
and by $g^{\hat a \hat b}$ the inverse
matrix. The operations with the indices enumerating the basic
vectors and forms are performed by using these matrices. In
particular, $\tens{e}^{\hat a}=g^{\hat a \hat b}\tens{e}_{\hat b}$, and so on. 
We denote a covariant derivative along the vector $\tens{e}_{\hat a}$ by
${\nabla}_{\!\hat a}$\,;\,${\nabla}_{\!\hat a}\equiv {\nabla}_{{e}_{\hat a}}$.
 One has
\be\label{2.15}
\tens{d}=\tens{\omega}^{\hat a}\wedge {\nabla}_{\!\hat a}\,,\quad
\tens{\delta}=-\tens{e}^{\hat a}\hook {\nabla}_{\!\hat a}\, .
\ee
In tensor notations the `hook' operator (inner derivative) 
along a vector $\tens{X}$, applied to a $p$-form
$\tens{\alpha}_p\,,$ $\tens{X}\hook \tens{\alpha}_p$\,, corresponds to a contraction 
\be\label{hook_op}
({X}\hook {\alpha}_p)_{a_2 \ldots a_p}= X^{a_1} (\alpha_p)_{a_1 a_2
\ldots a_p}\, .
\ee
It satisfies the properties
\begin{equation}\label{hook2}
\tens{e}^{\hat a} \hook (\tens{\alpha}_p\wedge \tens{\beta}_q)=
(\tens{e}^{\hat a} \hook \tens{\alpha}_p)\wedge \tens{\beta}_q +(-1)^p
\tens{\alpha}_p\wedge (\tens{e}^{\hat a} \hook \tens{\beta}_q)\,,
\end{equation}
\be\label{hook1}
\tens{e}^{\hat a}\hook \tens{\omega}_{\hat a}=D\,,\quad 
\tens{\omega}_{\hat a}\wedge (\tens{e}^{\hat a}\hook \tens{\alpha}_{p})=p\,\tens{\alpha}_{p}\,.
\ee

For a given vector $\bss{X}$ one  defines $\bss{X}^{\flat}$   as a
corresponding 1-form with the components
$(X^{\flat})_{a}=g_{ab}X^{b}$. 
In particular, one has $(\tens{e}_{\hat a})^{\flat}=g_{\hat a \hat b}\tens{\omega}^{\hat b}$.
An inverse to $\flat$ operation is
denoted by $\sharp$. Namely if $\tens{\alpha}$ is a 1-form  then
$\tens{\alpha}^{\sharp}$ denotes a  vector with components 
$({\alpha}^{\sharp})^a=g^{a b}\alpha_{b}$. 
We refer to \cite{Sternberg:1964}, \cite{Kress:phd} where  these and many other useful
relations can be found.

The definition \eq{CKY_d1} of the (rank-$p$) CKY tensor $\tens{k}$ 
reads \cite{BennEtal:1997}, \cite{BennCharlton:1997}, \cite{Kress:phd}:
\be\n{CKY_def3}
{\nabla}_{X} \tens{k}={1\over p+1} \tens{X}\hook \tens{d}\tens{k}
-{1\over D-p+1}\tens{X}^{\flat}\wedge
\tens{\delta} \tens{k}\, .
\ee 
Here, the first term on the right-hand-side denotes the exterior part, the second term 
denotes the divergence part, and $\tens{X}$ is an arbitrary vector.

Using the relation
\be\label{prop}
\tens{X}\hook \tens{*\omega}=\tens{*}(\tens{\omega}\wedge \tens{X}^{\flat})\,, 
\ee
it is easy to show that under the Hodge duality the exterior part 
transforms into the divergence part and vice versa.
Indeed, we find
\be\label{closedCKY_KY}
\tens{*}(\tens{X}\hook \tens{d}\tens{k})=
-\tens{X}^{\flat}\wedge \tens{\delta}(*\tens{k})\,,\quad
-\,\tens{*}(\tens{X}^\flat\wedge \tens{\delta k})=
\tens{X}\hook \tens{d}(\tens{*k})\,,
\ee
where we have used the definition \eqref{coderivative}.
In particular, \eqref{CKY_def3} implies
\be
{\nabla}_{X}(\tens{* k})={1\over p_*+1} \tens{X}\hook \tens{d}(\tens{*k})
-{1\over D-p_*+1}\tens{X}^{\flat}\wedge
\tens{\delta}(\tens{* k})\,,\quad p_*=D-p\, .
\ee
That is, the Hodge dual $\tens{*k}$ of a CKY tensor $\tens{k}$ is again
a CKY tensor. 
Moreover, the Hodge dual of a closed CKY tensor is a KY tensor and vice versa.

For a (rank-$p$) closed CKY tensor $\tens{h}$,  
characterized by vanishing of the exterior part, $\tens{dh}=0$, 
there exists locally a (KY) {\em potential} $\tens{b}$,
which is a $(p-1)$-form, such that 
\begin{equation}\label{db}
\tens{h}=\tens{db}\,.
\end{equation}
The Hodge dual of such a tensor $\tens{h}\,,$
\begin{equation}\label{fdb}
\tens{f}=\tens{*h}=\tens{*db}\, ,
\end{equation}
is a Killing--Yano tensor ($\tens{\delta f}=0$). 


\chapter[PCKY tensor and towers of hidden symmetries]{Principal conformal Killing--Yano tensor and towers of hidden symmetries}
\label{ch3}
\chaptermark{Principal Conformal Killing--Yano tensor}

In this chapter we introduce a notion of a {\em principal conformal Killing--Yano}
(PCKY) tensor---the central object of this thesis.
Starting with the PCKY tensor and the metric in any
$D$-dimensional spacetime we show how to generate 
a tower of $n-1=[D/2]-1$ Killing--Yano tensors, of rank $D-2j$ for all
$1\leq j \leq n-1$, and an extended tower of $n$ rank-2 Killing tensors, giving $n$ quadratic in momenta constants of geodesic motion that are in involution.
We also discuss another, more physical, method for generating Killing tensors and 
outline a construction of $D-n$ vectors which turn out to be the independent 
commuting Killing vectors.
Based on these results,  we shall prove in Part II
many of the remarkable properties of 
higher-dimensional rotating black hole spacetimes. 
This chapter is based on \cite{PageEtal:2007}, \cite{KrtousEtal:2007jhep}, \cite{KrtousEtal:2007prd}, and \cite{FrolovKubiznak:2008}.

\section{Principal conformal Killing--Yano tensor}

\subsection{Definition}
In what follows we consider a $D$-dimensional spacetime $M^D$,
equipped with the metric 
\be\label{1}
\tens{g}=g_{ab}\tens{d}x^{a}\tens{d}x^{b}\, .
\ee
To  treat both cases of even and odd  dimensions simultaneously we denote 
\be
D=2n+\eps\,,
\ee
where $\varepsilon=0$  and
$\varepsilon=1$ for  even and odd number of dimensions, respectively. 

{{\bf Definition} (\cite{KrtousEtal:2007jhep}).} {\em A principal conformal Killing--Yano tensor $\tens{h}$  is a 
closed non-degenerate CKY 2-form, ${\tens h}={1\over 2}\,h_{ab}\, \tens{d}x^a\!\wedge \tens{d}x^b,$ obeying the following equation:
\be\label{PCKY}
\nabla_{X}\tens{h}=\tens{X}^{\flat}\wedge \tens{\xi}^\flat\, ,
\ee 
where $\tens{X}$ is an arbitrary vector field. 
}

The condition of {\em non-degeneracy} means that the skew symmetric  matrix $h_{ab}$ 
has the (matrix) rank $2n$ and that the eigenvalues of $\tens{h}$ are 
functionally independent in some spacetime domain. So, we exclude the possibility that 
$\tens{h}$ possesses the constant eigenvalues, and in particular, that it is 
covariantly constant; $\tens{\xi}\neq 0$.
The equation \eqref{PCKY} implies
\be\label{xi}
\tens{dh}=0\,,\quad \tens{\xi}^\flat=-\frac{1}{D-1}\tens{\delta h}\,.
\ee
In particular, this means that there exists a 1-form (KY) potential
$\tens{b}$ such that 
\be\label{hdb}
\tens{h}=\tens{db}\,.
\ee
The dual tensor 
\be
\tens{f}=\tens{*h}
\ee
is a principal Killing--Yano tensor [$(D-2)$-form]. 
In tensor notations the definition \eq{PCKY} of the PCKY 
tensor $\tens{h}$ reads
\be\label{PCKY_coords}
\nabla_{c} h_{ab}=2g_{c[a}\xi_{b]},\quad
\xi_b=\frac{1}{D-1}\nabla_dh^{d}_{\ b}\,.
\ee

\subsection{Canonical basis and canonical coordinates}
\label{canonical}
Let us consider an eigenvalue
problem for a conformal Killing tensor $\tens{Q}$ associated with $\tens{h}$ [cf. Eq. \eqref{cKT}],
\be \label{Q_def}
Q_{ab}\equiv h_{ac}h_b^{\ \,c}\,.
\ee
It is easy to show that in the Euclidean domain
its eigenvalues $x^2$, 
\be\n{problem_Q}
Q^a_{\ b}v^{b}=x^2 v^a\, ,
\ee 
are real and non-negative. Using a modified Gram--Schmidt procedure it
is possible to show that there exists such an orthonormal basis in
which the operator $\tens{h}$ has the following structure:
\be\n{dar}
\mbox{diag}(0,\ldots,0,\Lambda_1,\ldots,\Lambda_p)\, ,
\ee
where $\Lambda_i$ are matrices of the form
\be
\Lambda_i=\left(   
\begin{array}{cc}
0 & -x_i I_i\\
x_i I_i& 0
\end{array}
\right)\, ,
\ee
and $I_i$ are unit matrices. We call such a basis an orthonormal {\em
Darboux basis} (see also Section 9.2). Its elements are unit
eigenvectors of the problem \eq{problem_Q}.

For a non-degenerate 2-form $\tens{h}$ the number of zeros
in the Darboux decomposition \eq{dar} coincides with $\eps$. Since all the
eigenvalues $x$ in \eq{problem_Q} are different (we denote them
$x_{\mu}$, $\mu=1,\ldots,n$), the matrices
$\Lambda_i$ are 2-dimensional. We denote the vectors of the Darboux
basis by $\tens{e}_{\hat \mu}$ and $\tens{\tilde e}_{\hat \mu}\equiv \tens{e}_{\hat n+\hat \mu}$, where $\mu=1,\ldots,n$, and 
enumerate them so that the orthonormal vectors $\tens{e}_{\hat \mu}$ and $\tens{\tilde e}_{\hat \mu}$ span a 2-dimensional plane of eigenvectors of \eq{problem_Q} with the same eigenvalue $x_{\mu}$. 
In an odd-dimensional spacetime we also have
an additional basis vector $\tens{e}_{\hat 0}$ (the eigenvector of
\eq{problem_Q} with $x=0$).  We further denote by
$\tens{\omega}^{\hat \mu}$ and $\tens{\tilde \omega}^{\hat \mu}\equiv \tens{\omega}^{\hat n+\hat \mu}$ (and
$\tens{\omega}^{\hat 0}$ if $\eps=1$) the dual basis of 1-forms.
The metric $\tens{g}$ and the PCKY tensor $\tens{h}$ in this basis take
the form
\ba
\tens{g}\!\!&=&\!\!\delta_{a b}\tens{\omega}^{\hat a}\tens{\omega}^{\hat b}=\sum_{\mu=1}^n (\tens{\omega}^{\hat \mu}\tens{\omega}^{\hat \mu}+
\tens{\tilde \omega}^{\hat \mu}\tens{\tilde \omega}^{\hat \mu})+\eps \tens{\omega}^{\hat 0}\tens{\omega}^{\hat 0}\, ,\n{gab}\\
\tens{h}\!\!&=&\!\!\sum_{\mu=1}^n x_{\mu} \tens{\omega}^{\hat \mu}\wedge \tens{\tilde \omega}^{\hat \mu}\,
.\n{hab}
\ea
In what follows we shall refer to bases $\{\tens{\omega}\}$ and $\{\tens{e}\}$ as
the {\em cononical bases} of 1-forms and vectors associated with the PCKY tensor. 
These bases are fixed uniquely by the PCKY tensor up to a 2D rotation
in each of the (KY) 2-planes $\tens{\omega}^{\hat \mu}\wedge \tens{\tilde \omega}^{\hat \mu}$.  

Since the `eigenvalues' $x_\mu$ are functionally independent in some spacetime domain, 
we may use them as `natural' coordinates. In fact, we shall demonstrate in Chapter 7  
that these $n$ coordinates  can be `upgraded' by adding $n+\eps$ new coordinates $\psi_i$, determined completely by the PCKY tensor. Therefore, the PCKY tensor `determines'
in $D$ dimensions $D$ preferred coordinates.  
We call such preferred coordinates $(x_\mu, \psi_i)$ the {\em canonical coordinates}. 
The most general (off-shell) {\em canonical metric} element admitting the PCKY tensor
is derived in Chapter 7. When it is written in the canonical coordinates, many of the coefficients of rotation vanish. We call the corresponding (special) canonical basis, the {\em principal canonical basis}. Such a basis is fixed uniquely; the 
freedom of 2D rotations was already exploited. (For more details see Chapter 7.)
 
To summarize, the PCKY tensor determines uniquely the class of canonical spacetimes,
together with the preferred canonical coordinates and the preferred principal canonical basis, in which these spacetimes take a `simple form'.

\section{Towers of hidden symmetries}
In this section we present a simple way how, from a single PCKY tensor, one can generate the whole towers of hidden symmetries. Our approach is based on the lemma
of the following subsection. We also derive an explicit form of the tower of Killing tensors in the canonical basis.

\subsection{Important property of closed CKY tensors}
{{\bf Lemma} (\cite{KrtousEtal:2007jhep}).} {\em Let $\tens{k}^{(1)}$ and $\tens{k}^{(2)}$ be two closed CKY tensors. Then their exterior product 
$\tens{k}\equiv\tens{k}^{(1)}\wedge\tens{k}^{(2)}$ is also a closed CKY tensor.}

We shall prove this lemma in two steps. The fact that $\tens{k}$ is closed is trivial, it follows from Eq. \eq{dd}. Let us first show that for a $p$-form $\tens{\alpha}_p$ obeying the equation
\be
\nabla_X\tens{\alpha}_p=\tens{X}^{\flat}\wedge \tens{\gamma}_{p-1}\,,
\ee
one has
\be\n{eq0}
\tens{\gamma}_{p-1}=-\frac{1}{D-p+1}\,\tens{\delta}\tens{\alpha}_p\, .
\ee 
Indeed, using Eqs. \eqref{2.15}, and relations 
\eqref{hook2}, \eqref{hook1}, we find
\ba\n{eq2}
-\tens{\delta}\tens{\alpha}_p&=&\tens{e}^{\hat a} \hook \nabla_{\hat a} \tens{\alpha}_p=
\tens{e}^{\hat a} \hook (\tens{\omega}_{\hat a} \wedge \tens{\gamma}_{p-1})\nonumber\\
&=&(\tens{e}^{\hat a}\hook \tens{\omega}_{\hat a})\tens{\gamma}_{p-1}
-\tens{\omega}_{\hat a}\wedge(\tens{e}^{\hat a}\hook
\tens{\gamma}_{p-1})=(D-p+1)\tens{\gamma}_{p-1}\, .\nonumber
\ea

The second step in the proof of the lemma is to show that if
$\tens{\alpha}_p$ and $\tens{\beta}_q$ are two closed CKY tensors then
\be
\nabla_X(\tens{\alpha}_p\wedge
\tens{\beta}_q)=\tens{X}^{\flat}\wedge\tens{\gamma}_{p+q-1}\, .
\ee
Really, one has
\be\n{eq3}
\begin{split}
\nabla_X(\tens{\alpha}_p\wedge
\tens{\beta}_q)=&\ 
\nabla_X\tens{\alpha}_p\wedge
\tens{\beta}_q+
\tens{\alpha}_p\wedge
\nabla_X\tens{\beta}_q\\
=&\, -\frac{1}{D-p+1}(\tens{X}^{\flat}\!\wedge \tens{\delta} \tens{\alpha}_p)\wedge
\tens{\beta}_q
-\frac{1}{D-q+1}
\tens{\alpha}_p\wedge (\tens{X}^{\flat}\!\wedge \tens{\delta}
\tens{\beta}_q)\\
=&\ \tens{X}^{\flat}\wedge \tens{\gamma}_{p+q-1}\,,
\end{split}
\ee
where
\be
\tens{\gamma}_{p+q-1}=-\frac{1}{D-p+1} \tens{\delta} \tens{\alpha}_p\wedge
\tens{\beta}_q
-\frac{(-1)^p}{D-q+1}
\tens{\alpha}_p\wedge \tens{\delta}
\tens{\beta}_q\, .\nonumber
\ee
Combining \eq{eq3} with \eq{eq0} we arrive at the statement of the lemma.$\quad \heartsuit$

\subsection{Towers of hidden symmetries}
\label{Towers}
According to the lemma of the previous subsection, the PCKY tensor
generates a set (tower) of new closed CKY tensors
\be\label{hj}
\tens{h}^{(j)}\equiv \tens{h}^{\wedge j}=\underbrace{\tens{h}\wedge \ldots \wedge
\tens{h}}_{\mbox{\tiny{total of $j$ factors}}}\, .
\ee 
$\tens{h}^{(j)}$ is a $(2j)$-form, and in particular $\tens{h}^{(1)}=\tens{h}$.
Since $\tens{h}$ is non-degenerate, one has a
set of $n$ non-vanishing closed CKY tensors. In an even dimensional
spacetime  $\tens{h}^{(n)}$ is proportional to the totally antisymmetric
tensor, whereas it is dual to a Killing vector in odd dimensions. In both cases 
such a CKY tensor is trivial and can be excluded from the tower of hidden symmetries. 
Therefore we take $j=1,\dots, n-1$.
The CKY tensors \eqref{hj} can be generated from the potentials $\tens{b}^{(j)}$
[cf. Eq. \eqref{db}]
\be\label{bj}
\tens{b}^{(j)}\equiv\tens{b}\wedge \tens{h}^{\wedge (j-1)}\,,\quad
\tens{h}^{(j)}=\tens{d}\tens{b}^{(j)}\,.
\ee 
Each $(2j)$-form $\tens{h}^{(j)}$ 
determines a $(D-2j)$-form of the Killing--Yano tensor [cf. Eq. \eq{fdb}]
\be\label{fj}
\tens{f}^{(j)}\equiv\tens{*}\tens{h}^{(j)}\, .
\ee
In their turn, these tensors give rise to the Killing tensors
$\bss{K}^{(j)}$
\be\n{Kj}
K^{(j)}_{ab}\equiv{1\over (D-2j-1)!(j!)^2} f^{(j)}_{\, \, \, \, \, a c_1\ldots c_{D-2j-1}}
f_{\, \, \, \, \, b}^{(j) \, \,  c_1\ldots c_{D-2j-1}}\, .
\ee
A choice of the coefficient in the definition \eq{Kj} is adjusted 
to the canonical basis (see the next subsection).
It is also convenient
to include the metric $\bss{g}$, which
is a trivial Killing tensor, as an element $\bss{K}^{(0)}$ of the
tower of the Killing tensors. The total number of irreducible 
elements of this {\em extended
tower} is $n$.

\subsection{Explicit form of Killing tensors}
Let us now explicitly evaluate the Killing tensors \eqref{Kj}
in the canonical basis. 
Using identities 
\be\label{asid1}
\eps^{a_1\dots a_rc_{r+1}\dots c_D} \eps_{b_1\dots b_rc_{r+1}\dots c_D}
 =r! (D-r)! \delta^{[a_1}_{b_1}\dots\delta^{a_r]}_{b_r}\;,
\ee
\be \label{asid2}
  (r+1)\,\delta^{[a}_{[b}\delta^{a_1}_{b_1}\dots\delta^{a_r]}_{b_r]}
= \delta^{a}_{b}\delta^{[a_1}_{[b_1}\dots\delta^{a_r]}_{b_r]}
  -r\,\delta^{a}_{[b_1}\delta^{[a_1}_{|b|}\dots\delta^{a_r]}_{b_r]}\,,
\ee
we can rewrite \eqref{Kj} as 
\ba\label{pom}
K^{(j)\,a}{}_b\!\!\!&=&\!\!\!
\frac{(2j+1)!}{(2^j j!)^2}\; \delta{}^{[a}_{[b} 
    h^{a_1b_1}_{}\dots h^{a_jb_j]}_{}
    h^{}_{a_1b_2}\dots h^{}_{a_jb_j]}\nonumber\\
\!\!\!&=&\!\!\!\frac{(2j)!}{(2^{j}j!)^2}
\Bigl(\delta^a_b h^{[a_1b_1}\!\dots h^{a_jb_j]}h_{[a_1b_1}\!\dots h_{a_jb_j]}
-2j\,h^{a[b_1}\!\dots h^{a_jb_j]}h_{b[b_1}\!\dots h_{a_jb_j]}\Bigr)\nonumber\\
\!\!\!&=&\!\!\! A^{(j)}\delta^a_b-{\tilde K}^{(j)\,a}{}_b\,.
\ea
Here we have introduced  
\ba
A^{(j)}\!\!\!&\equiv&\!\!\frac{(2j)!}{(2^{j}j!)^2} h^{[a_1b_1}\!\dots h^{a_jb_j]}h_{[a_1b_1}\!\dots h_{a_jb_j]}\,,\nonumber\\ 
{\tilde K}^{(j)\,a}{}_b
\!\!\!&\equiv&\!\!\frac{2j(2j)!}{(2^{j}j!)^2} h^{a[b_1}\!\dots h^{a_jb_j]}h_{b[b_1}\!\dots h_{a_jb_j]}\,.\nonumber
\ea
In the canonical basis, using \eqref{gab} and \eqref{hab}, we find
\ba
\tens{h}^{(j)}\!\!&=&\!\! j!\!\!\!\sum_{\nu_1<\dots<\nu_j} x_{\nu_1}\dots x_{\nu_j} 
\tens{\omega}^{\hat \nu_1}\wedge \tens{\tilde \omega}^{\hat \nu_1}
\wedge\dots\wedge\tens{\omega}^{\hat \nu_j}\wedge \tens{\tilde \omega}^{\hat \nu_j}\,.\\
\tens{\tilde K}^{(j)}\!\!&=&\!\!
\sum_{\mu=1}^n x_\mu^2A_\mu^{(j-1)}(\tens{\omega}^{\hat \mu}\tens{\omega}^{\hat \mu}+
\tens{\tilde \omega}^{\hat \mu}\tens{\tilde \omega}^{\hat \mu})\,,\\
A^{(j)}\!\!&=&\!\!\sum_{\nu_1<\dots<\nu_j}\! x_{\nu_1}^2\dots x_{\nu_j}^2\;,\quad
A^{(j)}_\mu=\!\! \sum_{\substack{\nu_1<\dots<\nu_j\\\nu_i\ne\mu}}\!
   x_{\nu_1}^2\dots x_{\nu_j}^2\;,\label{AA}
\ea
From the obvious fact that quantities \eqref{AA} obey, $A^{(j)}=A_\mu^{(j)}+x_\mu^2A_\mu^{(j-1)}$, we obtain the following form of  
the Killing tensors in the canonical basis:
\begin{equation}\label{KTdef2}
\tens{K}^{(j)} =\sum_{\mu=1}^n A^{(j)}_\mu(\tens{\omega}^{\hat \mu}\tens{\omega}^{\hat \mu}+ \tens{\tilde \omega}^{\hat \mu}\tens{\tilde \omega}^{\hat \mu})+\eps A^{(j)}\tens{\omega}^{\hat 0}\tens{\omega}^{\hat 0}\,.
\end{equation}   
Let us finally mention that it was shown in \cite{HouriEtal:2008a} that 
$\tilde K^{(j)\,a}{}_b=Q^a{}_cK^{(j-1)\,c}{}_b$. Here $\tens{Q}$ is the conformal Killing tensor introduced in Section \ref{canonical}. Eq. \eqref{pom} 
therefore gives the following recursive relation for $\tens{K}^{(j)}$:
\be\label{recursive}
\tens{K}^{(j)}=A^{(j)}\tens{g}-\tens{Q}\cdot \tens{K}^{(j-1)}\,,\quad
\tens{K}^{(0)}=\tens{g}\,. 
\ee 

\section{Other method for generating Killing tensors}
\label{other_method}
In this section we describe, in some sense a more physical method, how to generate 
from the PCKY tensor various towers of Killing tensors. 
This method is based on the fact that Killing tensors are in one to one correspondence with conserved quantities for a geodesic motion which are of higher order 
in geodesic momenta \cite{WalkerPenrose:1970}. In particular, we extract these constants as 
invariants of the parallel-transported 2-form $\tens{F}$---obtained as a projection of the PCKY tensor to a subspace orthogonal to the velocity
of a geodesic motion \cite{PageEtal:2007}.
Depending on how these invariants are extracted one obtains different (related) sets of 
constants of motion and corresponding towers of Killing tensors.
For example, the traces of powers of the operator $\tens{F}^2$ lead to 
the set of Killing tensors of increasing rank \cite{PageEtal:2007}. 
The advantage of this approach is that one can generate constants of geodesic motion with the help of {\em generating functions}. This gives a powerful tool how to study  properties of these constants, such as their independence or Poisson commutativity (see Chapter 5). In particular, we introduce two generating functions: the first one 
generates constants given by the traces of powers of the operator $\tens{F}^2$, 
the second one leads to the earlier described tower of Killing-tensors \cite{KrtousEtal:2007jhep}.

\subsection{2-form $\tens{F}$}

Let $\gamma$ be a geodesic affine parametrized by $\tau$, $u^a=dx^a/d\tau$ be its `velocity' tangent vector, and $w\equiv u^au_a$ be its norm. We denote the covariant derivative of a tensor $\tens{T}$ along $\gamma$ by 
\begin{equation}
\tens{\dot T} \equiv \nabla_u \tens{T}=u^a \nabla_a \tens{T}\,.
\end{equation}
In particular $\tens{\dot u}=0$.
Let us now consider the following 2-form $\tens{F}$ \cite{PageEtal:2007}:
\begin{equation}\label{formF}
\tens{F}\equiv \tens{u}\hook(\tens{u}^\flat\wedge \tens{h})=
w\tens{h}-\tens{u}^\flat\!\wedge \tens{s}\,,\quad 
\tens{s}\equiv \tens{u}\hook \tens{h}\,.
\end{equation}
From the construction, such a form is automatically parallel-transported 
along $\gamma$. Indeed, we have
\be
\nabla_u (\tens{u}^\flat\wedge \tens{h})=\tens{u}^\flat\wedge\nabla_u\tens{h}=
\tens{u}^\flat \wedge \tens{u}^\flat \wedge \tens{\xi}^\flat=0\,.
\ee
So, already $\tens{u}^\flat\wedge \tens{h}\propto \tens{u}\hook\tens{f}$ is parallel-transported [cf. Eq. \eqref{Lprop}], and the last contraction 
with $\tens{u}$ in \eqref{formF} is just to obtain a 2-form with which it is easier to work.
Since $\tens{F}$ is parallel-propagated along $\gamma$, any
object constructed from $\tens{F}$ and the metric $\tens{g}$ is also
parallel-propagated. In particular, this is true for the invariants constructed from  $\tens{F}$, such as its eigenvalues. These are therefore constants of motion.\footnote{Actually, from the fact that $\tens{F}$ is parallel-transported along $\gamma$ one can obtain slightly more, see Chapter 9.}

Let us notice that $\tens{F}$ can be also written as 
\begin{equation}\label{Fdef}
  F_{ab}=w P^c_a h_{cd}P^d_b\;.
\end{equation}
Here, ${P_a^b=\delta_a^b-w^{-1}u^bu_a}$ is the projector to the $(D-1)$-dimensional
subspace $V$ orthogonal to the 1-dimensional space $U$ generated by $\tens{u}$. $P_{ab}=g_{ab}-w^{-1}u_a u_b$ can be also  understood as a metric in $V$ induced by its embedding into the tangent space $T$; $T=U\oplus V\,.$
This means that $\tens{F}$ has a clear geometrical meaning: {\em it is the 
projection of the PCKY tensor $\tens{h}$ along the tangent vector $\tens{u}$ of geodesic ${\gamma}$. }
$F_{ab}$ and $F^a{}_b=g^{ac}F_{cb}$ can be considered  as a 2-form and an
operator, respectively,  either in the subspace $V$ or in the
complete tangent space $T$. Since $F^a{}_b u^b=0$, the vector $\tens{u}\in T$
is an eigenvector of $\tens{F}$ with a zero eigenvalue. One also immediately has
\be
\dot F_{ab}=w P^c_a \dot h_{cd} P^d_b=
w P^c_a u_{[c}\xi_{d]}P^d_b=0\,,
\ee
where we used the defining equation \eqref{PCKY_coords}.

\subsection{Killing tensors of increasing rank}

One of the convenient ways \cite{PageEtal:2007} how to extract the invariants of $\tens{F}$ is to 
consider the traces of powers of the operator $\tens{F}^2$:
\be
C_{j} \equiv w^{-j}{\rm Tr}[(-\tens{F}^2)^j]\,.
\ee 
(The traces of odd powers of $\tens{F}$ are zero, because of the antisymmetry of
$\tens{F}$.) 

In what follows we shall use matrix notation in which $F$ is the antisymmetric matrix with orthonormal components $F^{\hat a}{}_{\hat b}\,$, $H$ is the antisymmetric matrix with components $h^{\hat a}{}_{\hat b}\,$, $Q \equiv -H^2$ is the symmetric matrix with components $Q^{\hat a}{}_{\hat b} = - h^{\hat a}{}_{\hat c}
h^{\hat c}{}_{\hat b}\,$, $W$ is the symmetric matrix with components $u^{\hat a} u_{\hat b}\,$, $w \equiv
{\rm Tr}(W) = u^{\hat c} u_{\hat c}\,$, $P \equiv I-p$ is the projection onto the hyperplane $V$
orthogonal to the velocity, and $p=W/w$. These matrices have the
properties that $P^2=P$ and $WH^{2j+1}W = 0$ for all nonnegative integers $j$. 
The component Eq. (\ref{Fdef}) becomes the matrix equation
\begin{equation}\label{F}
F = wPHP 
\end{equation}
whose square is the symmetric matrix $F^2 =w^2 P(HP)^2.$
So, we get the constants of motion 
\be\label{Cj2}
C_{j} = (-w)^{j}{\rm Tr}[(HP)^{2j}]\,. 
\ee
The trace of the matrix product can be viewed diagrammatically as a loop
formed by joined vertices (each with two `legs') corresponding to matrices 
in the product. In our case the loop is formed by alternating $H$ and $P$
vertices. Substituting $P=I-p$ we get a sum over all possible loops
in which $P$ is replaced either by $I$ or by $-p$. In the 
case of the identity $I$ the corresponding vertex is effectively
eliminated, in the case of one dimensional projector 
$p$ the loop splits into disconnected pieces. Namely, we can use identity
\begin{equation}
{\rm Tr}\bigl(H^{k_1}\!p\,H^{k_2}\!p\,\cdots H^{k_c}\!p\bigr)=
{\rm Tr}\bigl(H^{k_1}\! p\bigr){\rm Tr}\bigl(H^{k_2}\! p\bigr)\cdots
{\rm Tr}\bigl(H^{k_c}\! p\bigr)\;.
\end{equation}
The trace in \eqref{Cj2} thus leads to
\be\label{trhPj}
{\rm Tr}\bigl[(HP)^{2j}\bigr]={\rm Tr}\bigl(H^{2j}\bigr)
  +\sum_{c=1}^{2j}\sum_{\substack{k_1\le \dots\le k_c\\k_1+\dots+k_c=2j}}
  \!\!\!(-1)^c N^{2j}_{k_1\dots k_j} \prod_{i=1}^c {\rm Tr}\bigl(H^{k_i}\!p\bigr)\;.
\ee
The sum over ${c}$ is the sum over number of `splits' of the loop,
the indices ${k_i}$ are the `lengths' of the splitted pieces,
and the combinatorial factor ${N^{2j}_{k_1\dots k_c}}$ 
gives a number of ways in which the loop of the length ${2j}$
can be split to ${c}$ pieces of lengths ${k_1,\dots,k_c}$.
From the antisymmetry of $H$ it follows that traces of odd powers of $H$ 
(optionally multiplied by a projector) are zero. Setting ${k_i=2l_i}$ 
and introducing $Q$ as earlier we have
\be\label{trhPjinQ}
C_j=w^j{\rm Tr}\bigl(Q^j\bigr)
  +\sum_{c=1}^j\sum_{\substack{l_1\le \dots\le l_c\\l_1+\dots+l_c=j}}
  \!\!\!(-1)^c\, 2\,N^j_{l_1\dots l_j} w^j\prod_{i=1}^c {\rm Tr}\bigl(Q^{l_i}\!p\bigr)\;,
\ee
where we have used ${N^{2j}_{2l_1\dots2l_c}=2N^j_{l_1\dots l_c}}$ 
which follows from the definition of ${N}$'s.
Let us define the following quantities:
\begin{equation}\label{wjdef}
  w_j \equiv w {\rm Tr}(Q^j p)={\rm Tr}(Q^j W)\equiv Q_{ab}^{(j)}u^au^b\,.
\end{equation}
Here,  $Q_{ab}^{(j)}$ denote covariant components that form the tensor
$\tens{Q}^{(j)}$ corresponding to the $j$-th power of the matrix $Q$. 
We also denote $Q^{(j)}\equiv {\rm Tr}\bigl(Q^j\bigr)$.
For example, $Q^{\ix{(1)}}\!=\!Q_c^{\ c}$, $Q_{ab}^{(1)}\!=\!Q_{ab}$,
$Q^{\ix{(2)}}\!=\!Q_c^{\ d} Q_d^{\ c}$, $Q_{ab}^{(2)}\!=\!Q_a^{\ c} Q^{}_{cb}$,
$Q^{\ix{(3)}}\!=\!Q_c^{\ d} Q_d^{\ e} Q_e^{\ c}$,
and $Q_{ab}^{(3)}\!=\!Q_a^{\ c} Q_c^{\ d} Q^{}_{db}$.
Then we finally obtain
\be\label{trhPjinw}
C_j=w^j Q^{(j)}-2jw^{j-1}w_j
  +2\sum_{c=2}^j\sum_{\substack{l_1\le \dots\le l_c\\l_1+\dots+l_c=j}}
  \!\!\!(-1)^c \,N^j_{l_1\dots l_j} w^{j-c}\prod_{i=1}^c w_{l_i}\;.
\ee
We can easily see that the ${C_j}$'s have the form
\begin{equation}\label{cisKus}
  C_j = K_{a_1\dots a_{2j}}u^{a_1}\dots u^{a_{2j}}\,,
\end{equation}
where ${K_{a_1\dots a_{2j}}}$, formed from combinations of
the metric $g_{ab}$, ${\rm Tr}(Q^j)$, and the $Q^{(i)}_{ab}$'s for $i\leq j$,  
are Killing tensors \cite{WalkerPenrose:1970} in the sense of Eq.~\eqref{KT_def}.

In particular, we get the first four constants of motion
\ba\label{c1-4}
C_1 \!\!&=&\!\! w Q^{(1)} - 2 w_1\;,\non\\
C_2 \!\!&=&\!\!w^2 Q^{(2)} - 4 w\, w_2 + 2 w_1^2\;,\non\\ 
C_3 \!\!&=&\!\! w^3 Q^{(3)} - 6 w^2 w_3 + 6 w\, w_1 w_2 - 2 w_1^2\;,\non\\
C_4 \!\!&=&\!\! w^4 Q^{(4)} - 8 w^3 w_4 + 4w^2w_2^2+8w^2w_1w_3
        -8 w\, w_1^2w_2+2w_1^4\;.
\ea
Comparing with \eqref{cisKus} we obtain the corresponding tower of (reducible) Killing tensors of increasing rank:
\ba\label{KT1-4}
K_{ab} \!\!\!&=&\!\! g_{ab} Q^{(1)}- 2 Q_{ab}^{(1)}\;,\non\\
K_{abcd} \!\!\!&=&\!\! g_{(ab}g_{cd)} Q^{(2)}- 4 g_{(ab}Q_{cd)}^{(2)} 
+ 2 Q_{(ab}^{(1)}Q_{cd)}^{(1)}\;,\non\\ 
K_{abcdef}\!\!\!&=&\!\! g_{(ab}g_{cd}g_{ef)} Q^{(3)} 
- 6 g_{(ab}g_{cd} Q_{ef)}^{(3)} + 6 g_{(ab} Q_{cd}^{(1)} Q_{ef)}^{(2)}-
2 Q_{(ab}^{(1)} Q_{cd}^{(1)} Q_{ef)}^{(1)} \;,\non\\
K_{abcdefgh}\!\!\!&=&\!\! g_{(ab}g_{cd}g_{ef}g_{gh)}  Q^{(4)} - 
8 g_{(ab}g_{cd}g_{ef} Q_{gh)}^{(4)}
+ 4g_{(ab}g_{cd}Q_{ef}^{(2)} Q_{gh)}^{(2)} \non\\
\!\!\!&+&\!\! 8g_{(ab}g_{cd}Q_{ef}^{(1)} Q_{gh)}^{(3)}
-8 g_{(ab}Q_{cd}^{(1)} Q_{ef}^{(1)}Q_{gh)}^{(2)} 
+2Q_{(ab}^{(1)} Q_{cd}^{(1)} Q_{ef}^{(1)} Q_{gh)}^{(1)} \;.
\ea

To write an explicit form of the constants of motion or the Killing tensors obtained 
we can use the canonical basis. There  
we have 
\begin{equation}\label{tensorQ}
\tens{Q}^{(j)} = Q_{\hat a \hat b}^{(j)} \tens{\omega}^{\hat a} \tens{\omega}^{\hat b}= \sum_{\mu=1}^n x_\mu^{2j}
 (\tens{\omega}^{\hat \mu} \tens{\omega}^{\hat \mu}  + \tens{\tilde \omega}^{\hat \mu} \tens{\tilde \omega}^{
\hat \mu} )\,,
\end{equation}
and also
\begin{equation}\label{scalarQ}
Q^{(j)} = 2\sum_{\mu=1}^n x_\mu^{2j}\,,\qquad
w_j=\sum_{\mu=1}^n x_\mu^{2j}
 (u_{\hat \mu}^2+{\tilde u}_{\hat \mu}^2)\,,
\end{equation}
where $u_{\hat a}$ denotes the basis components of the velocity
\begin{equation}
\tens{u}^\flat=\sum_{\mu=1}^n \bigl( u_{\hat \mu} \tens{\omega}^{\hat \mu} + {\tilde u}_{\hat \mu} \tens{\tilde \omega}^{\hat \mu}\bigr)
+\eps\,u_{\hat 0} \tens{\omega}^{\hat 0}\,.
\end{equation}

\subsection{Generating functions}
\label{sc:genfc}
Although formula \eqref{trhPjinw} gives the constants of motion explicitly, the presence of combinatoric factors 
makes calculations difficult in practice. In this subsection we introduce 
generating functions \cite{KrtousEtal:2007jhep} which allow to, aside from other things, write down a more useful formula how to evaluate these constants.

We introduce the {\em generating function} ${W(\beta)}$, 
\begin{equation}\label{Wdef}
  W(\beta)\equiv \det\bigl(I+\sqrt{\beta}w^{-1}F\bigr)\;.
\end{equation}
Due to the antisymmetry of $F$ and properties of the determinant,
$W(\beta)$ can be rewritten as a function of ${\beta}$ instead of
$\sqrt{\beta}$, and in terms of $H$ and $P$ instead of $F$,
\begin{equation}\label{Wdef2}
  W(\beta)=\det{}^{\!1/2}\bigl(I-\beta w^{-2}F^2\bigr)
    =\det\bigl(I-\sqrt{\beta}\, H P\bigr)\;.
\end{equation}
Because it is constructed only in terms of 
covariantly conserved quantities $F$ and $w$, 
the generating function is conserved along $\gamma$,
and the same is true for its derivatives with respect to ${\beta}$.
We can thus define constants of motion $\kappa_j$ as
the coefficients in the $\beta$-expansion of $W(\beta)$:
\begin{equation}\label{cdef}
  W(\beta) \equiv \frac{1}{w} \,\sum_{j=0}^\infty \kappa_j\, \beta^j\;.
\end{equation}
It turns out that all terms with $j>n$ are zero.
To evaluate the observables $\kappa_j$, we can split $W(\beta)$
in the following way:
\begin{equation}\label{Wsep}
  W(\beta) =W_0(\beta)\;\Sigma(\beta)\;,
\end{equation}
where
\begin{equation}\label{WSigmadef}
\begin{aligned}
  W_0(\beta) &= \det\bigl(I-\sqrt{\beta}H\bigr)= 
\det{}^{\!1/2}\bigl(I+\beta Q\bigr)\;,\\
  \Sigma(\beta) &= \det\Bigl(I\!+\!\frac{\sqrt{\beta}H}{I\!-\!\sqrt{\beta}H}\,p\Bigr) 
  ={\rm Tr}\Bigl[\sum_{j=0}^\infty \bigl(\sqrt{\beta}H\bigr)^j p\Bigr]
=\sum_{j=0}^\infty {\rm Tr}(H^{2j} p)\beta^j\\
&=
\sum_{j=0}^\infty(-1)^j{\rm Tr}(Q^jp)\beta^j
={\rm Tr}\bigl[(I+\beta Q)^{-1}p\bigr]=\frac{1}{w}\sum_{j=0}^\infty (-1)^jw_j\beta^j\;.
\end{aligned}
\end{equation}
Here we have used the fact that the matrix in the determinant  in the
expression for ${\Sigma(\beta)}$ differs from ${I}$ only  in the
one-dimensional subspace U given by ${\tens{u}}$. 
The generating function $W(\beta)$ thus
splits into a part $W_0(\beta)$ independent of ${\tens{u}}$ and a part
$\Sigma(\beta)$ linear in $p$.
Such generating function therefore leads to the tower of 2nd-rank Killing tensors;
these will be described in the next section.

Let us now consider a different generating function $Z(\beta)$, 
\be\label{Z}
Z(\beta)\equiv\log W(\beta)\,.
\ee
Using the relation
\be
\log \Bigl[\det\bigl(I-A\bigr)\Bigr]=-\sum_{n=1}^\infty \frac{1}{n}\,{\rm Tr}(A^n)\,,
\ee
we find
\begin{equation}\label{Zbeta}
\begin{split}
Z(\beta)&=\log W_0(\beta)+\log \Sigma(\beta)=
\log\Bigl[\det{}\bigl(I\!-\!\sqrt\beta\,H P\bigr)\Bigr]\\
&=-\sum_{j=1}^\infty \frac{1}{2j}{\rm Tr}\bigl[(H P)^{2j}\bigr]\beta^j=
\sum_{j=1}^\infty \frac{(-1)^{j+1}}{2j}\frac{C_j}{w^j}\beta^j\,.
\end{split}
\end{equation}
Constants $C_j$, given by \eqref{trhPjinw}, therefore  correspond to terms proportional to ${\beta^j}$ in the power expansion of ${Z(\beta)}$.
The first term of \eqref{trhPjinw} is obtained from ${\log W_0(\beta)}$,
and the sum over all possible splittings of the loop corresponds 
to the ${\beta^j}$ term of ${\log\Sigma(\beta)}$. Clearly, the $j$-th derivative 
of $\log\Sigma(\beta)$ (evaluated at ${\beta=0}$) contains 
the sum over all possible products of ${l}$-th derivatives ${\Sigma^{(l)}(0)}$
which are proportional to ${w_l}$ defined in \eqref{wjdef}. 
The factors ${N^j_{l_1\dots l_2}}$ can thus be obtained by the explicit
computation of the derivatives of the generating function ${\log\Sigma(\beta)}$:
\be
C_j=w^j Q^{(j)}-\frac{2(-w)^{j}}{(j-1)!}\frac{d^j}{d\beta^j}\left[\log
\bigl(1+\sum_{k=1}^j(-1)^k\frac{w_k}{w}\beta^k\bigr)\right]_{\beta=0}.
\ee
With the help of this formula and a software for algebraic manipulation 
one can easily generate constants of motion $C_j$.

The relation between ${W(\beta)}$ and ${Z(\beta)}$ implies that
constants $C_j$ and constants $\kappa_j$ [introduced in \eqref{cdef}]
are related as follows:
\begin{equation}\label{Ckck}
C_j=-\frac{2(-w)^{j}}{(j-1)!}\frac{d^j}{d\beta^j}\left[\log\bigl(w+
\sum_{k=1}^j\kappa_k\beta^k\bigr)\right]_{\beta=0}.
\end{equation}
So, $C_j$'s are polynomial combinations of $\kappa_j$'s and $w$ 
with constant coefficients. In particular, we get
\ba\label{cCrel}
C_1&=&2\kappa_1\;,\non\\
C_2&=&-4w\kappa_2+2\kappa_1^2\;,\non\\
C_3&=&6w^2\kappa_3-6w\kappa_1\kappa_2+2\kappa_1^3\;,\non\\
C_4&=&-8w^3\kappa_4+8w^2\kappa_1\kappa_3+4w^2\kappa_2^2
-8w\kappa_1^2\kappa_2+2\kappa_1^4\;.
\ea

\subsection{Rank-2 Killing tensors}
\label{2ndrank}
Using the canonical basis, let us now explicitly evaluate the 2nd-rank 
Killing tensors generated by function $W(\beta)$. 
We again introduce the quantities 
\begin{equation}\label{Aj}
  A^{(j)}=\!\!\sum_{\nu_1<\dots<\nu_j}\! x_{\nu_1}^2\dots x_{\nu_j}^2\;,\quad
  A^{(j)}_\mu=\!\!
  \sum_{\substack{\nu_1<\dots<\nu_j\\\nu_i\ne\mu}}\!
   x_{\nu_1}^2\dots x_{\nu_j}^2\;.
\end{equation} 
Then, with the help of relations \eqref{tensorQ} and \eqref{scalarQ}, we find
\ba\label{Woinx}
W_0(\beta)\!\!&=&\!\!\!\det{}^{\!1/2}\bigl(I+\beta Q\bigr)=\prod_{\mu=1}^n(1+\beta x_\mu^2)= \sum_{j=0}^n A^{(j)} \beta^j\;,\\
\Sigma(\beta)\!\!&=&\!\!\!\frac{1}{w}\sum_{j=0}^\infty (-1)^jw_j\beta^j=
\frac{1}{w}\Bigl(\eps u_{\hat 0}^2+
     \sum_{j=0}^{\infty}
      (-1)^j\beta^j\sum_{\mu=1}^n(u_{\hat \mu}^2+{\tilde u}_{\hat \mu}^2)\, x_\mu^{2j}\Bigr).\ \ 
\ea
The original generating function \eqref{Wsep} reads
\begin{equation}\label{Winw}
W(\beta) = \frac1w \sum_{j=0}^n 
  \Bigl(\sum_{l=0}^j (-1)^l A^{(j-l)} w_l \Bigr)\beta^j
=\frac{1}{w}\sum_{j=0}^n\Bigl[\eps A^{(j)} u_{\hat 0}^2 +
       \sum_{\mu=1}^nA^{(j)}_\mu \bigl(u_{\hat \mu}^2+{\tilde u}_{\hat \mu}^2\bigr)
        \Bigr]\beta^j\;.
\end{equation}
Comparing Eq.~\eqref{Winw} with Eq.~\eqref{cdef}, we
can identify $n+\eps$ conserved quantities $\kappa_j$
(constants of geodesic motion,  $j=0,\dots,n+\eps-1$),
\begin{equation}\label{const}
\kappa_j = \sum_{l=0}^j (-1)^l A^{(j-l)} w_l 
        = \eps A^{(j)} u_{\hat 0}^2 + \sum_{\mu=1}^nA^{(j)}_\mu
   \bigl(u_{\hat \mu}^2+{\tilde u}_{\hat \mu}^2\bigr)\;,
\end{equation}
which are quadratic in velocities. 
They are generated \cite{WalkerPenrose:1970} by the 2nd-rank Killing tensors  $\tens{K}^{(j)}$
\begin{equation}\label{constfromKT}
  \kappa_j=K^{(j)}_{ab} u^a u^b\;,\quad 
\nabla_{(a}K^{(j)}_{bc)}=0\;,
\end{equation}
where 
\begin{equation}\label{KTdef}
\tens{K}^{(j)} =\sum_{l=0}^j (-1)^l A^{(j-l)} \tens{Q}^{(l)}
    =\sum_{\mu=1}^n A^{(j)}_\mu(\tens{\omega}^{\hat \mu}\tens{\omega}^{\hat \mu}+
\tens{\tilde \omega}^{\hat \mu}\tens{\tilde \omega}^{\hat \mu})+\eps A^{(j)}\tens{\omega}^{\hat 0}\tens{\omega}^{\hat 0}\,.
\end{equation} 
The Killing tensor $\tens{K}^{(n)}$ is present only in an odd number of spacetime dimensions and it is reducible. Similar to Section \ref{Towers} we exclude it from the
set. The remaining tensors  \eq{KTdef} coincide with those of the extended tower 
introduced in Section \ref{Towers}.

The first expression for $\tens{K}^{(j)}$ in \eqref{KTdef} can be easily derived from the recursive relation \eqref{recursive} \cite{HouriEtal:2008a}. It immediately implies that 
\be
\tens{K}^{(i)}\cdot \tens{K}^{(j)}=\tens{K}^{(j)} \cdot \tens{K}^{(i)}\,,
\ee
and therefore $\tens{K}^{(j)}$'s have common eigenvectors (see also Section 
\ref{separability structures}).

We shall prove in Chapter 5, that 
observables $\kappa_j$ are in {\em involution}, that is, they mutually Poisson commute:
\begin{equation}\label{Poisson}
  \{\kappa_i,\kappa_j\}=0\;.
\end{equation} 
This is equivalent to vanishing of the Schouten--Nijenhuis (SN) brackets 
\cite{Schouten:1940}, \cite{Schouten:1954}, \cite{Nijenhuis:1955}
for the corresponding Killing tensors (see Section \ref{Lie}):
\begin{equation}\label{PoissonK}
  \bigl[K^{(j)}, K^{(l)}\bigl]_{abc}\equiv K^{(j)}_{e(a}\, \nabla^{e} K^{(l)}_{bc)}- 
  K^{(l)}_{e(a}\, \nabla^{e} K^{(j)}_{bc)}=0\;.
\end{equation}
Once \eqref{Poisson} is proved for $\kappa_j$'s, 
the  relation \eqref{Ckck} shows that also observables ${C_j}$ 
are in involution, and vice versa. An independent proof of mutual Poisson commutativity of observables $\kappa_j$ 
was later demonstrated in \cite{HouriEtal:2008a}, 
using the method of generating functions.

\section{Tower of Killing vectors}
In the previous two sections we have seen that the PCKY tensor 
determines the whole set of hidden symmetries.
In this section we demonstrate that it also naturally generates $n+\eps$ vectors $\tens{\xi}^{(k)}$ $(k=0,\dots,n-1+\eps)$ which turn out to be the independent commuting Killing vector fields \cite{KrtousEtal:2007jhep}, \cite{KrtousEtal:2008}.
 
The {\em primary} (Killing) vector $\tens{\xi}^{(0)}\equiv\tens{\xi}$ is defined by \eqref{xi}, 
\be\label{Primary}
\xi_b^{(0)}\equiv\xi_b=\frac{1}{D-1}\nabla_dh^{d}{}_b\,.
\ee
The {\em secondary} (Killing) vectors $\tens{\xi}^{(j)}\equiv\tens{\eta}^{(j)}$ ($j=1,\dots,n-1$) can be constructed as
\begin{equation}\label{etaj}
\xi^{(j)\,a}\equiv\eta^{(j)\,a}\equiv K^{(j)}\,\!^a_{\ b}\xi^b\,. 
\end{equation}
In odd dimensions the last Killing vector is given by
the ${n}$-th Killing--Yano tensor (see Section \ref{Towers})
\be\label{xin}
\tens{\xi}^{(n)}\equiv\tens{f}^{(n)}\,.
\ee

The proof that all these vectors are the mutually commuting Killing fields 
which also (Schauten-Nijenhuis) commute with the Killing tensors constructed in Section \ref{2ndrank},
\be\label{commutation}
\bigl[\tens{\xi}^{(i)},\tens{K}^{(j)}\bigr]=0\,,\quad
\bigl[\tens{\xi}^{(i)},\tens{\xi}^{(j)}\bigr]=0\,,
\ee  
is demonstrated in Chapter 7.\footnote{%
Historically, this fact was first proved 
\cite{HouriEtal:2008a} under the additional conditions 
\begin{equation}\label{Liexihg}
  \lied_\xi \tens{g} =0\,,\quad \lied_\xi \tens{h}  =0\,.
\end{equation}
The first condition requires that $\tens{\xi}$ is a Killing vector. [This condition is trivially satisfied in any Einstein space, cf. Eq. \eqref{xi_cond}.]
It is easy to see, that from the second condition it follows that also  $\tens{\eta}^{(j)}$'s are the Killing vectors. Indeed, from \eqref{PCKY} we have $\nabla_\xi \tens{h}=0$. Using \eqref{Liexihg}, we find
\be
\lied_\xi \tens{K}^{(j)}=0\,,\quad \nabla_{\xi} \tens{K}^{(j)}=0\,,
\ee
and therefore 
\be
\nabla_{(a}\eta_{b)}^{(j)}=\frac{1}{2}\lied_\xi {K}^{(j)}_{ab}-
\nabla_{\xi} {K}^{(j)}_{ab}=0\,.
\ee
It is shown in Chapter 7 that both conditions \eqref{Liexihg} follow from the existence 
of the PCKY tensor. 
}


\part[Higher-Dimensional Rotating Black Holes]{Remarkable Properties of Higher-Dimensional Rotating Black Holes}


\chapter[PCKY tensor in Kerr-NUT-(A)dS spacetimes]
{PCKY tensor in the Kerr-NUT-(A)dS spacetimes}
\label{ch5}
\chaptermark{PCKY tensor in Kerr-NUT-(A)dS spacetimes}

In this chapter, based on \cite{KubiznakFrolov:2007}, we demonstrate that the general Kerr--NUT--(A)dS spacetime,
describing the higher-dimensional arbitrarily rotating black hole with NUT parameters and the cosmological constant, possesses the PCKY tensor. We write the Kerr--NUT--(A)dS metric in canonical coordinates, completely determined by the PCKY tensor.
In this (canonical) form, the metric can be considered as a natural higher-dimensional generalization of the 
Carter's canonical form for the 4D Kerr-NUT-(A)dS solution. 
The invariant (geometrical) definition of canonical coordinates makes the canonical form 
convenient for calculations. For example, it is these coordinates in which the Hamilton--Jacobi equation separates (see Chapter 6). We also introduce, a more general, (off-shell) canonical metric and its principal canonical basis.

\section{Overview of the Kerr-NUT-(A)dS metrics}

The most general known higher-dimensional ($D>2$) solution describing rotating black
holes with NUT parameters in an asymptotically (Anti) de Sitter
spacetime (Kerr-NUT-(A)dS metric) was found by Chen, L\"u, and Pope \cite{ChenEtal:2006cqg}. We write it in the following symmetric (analytically continued) form: 
\begin{equation}\label{metric_coordinates}
\tens{g}=\sum_{\mu=1}^n\left[\frac{\tens{d}x_{\mu}^2}{Q_{\mu}}
  +Q_{\mu}\!\Bigl(\sum_{j=0}^{n-1} A_{\mu}^{(j)}\tens{d}\psi_j\!\Bigr)^{\!2}\right]
  -\frac{\eps c}{A^{(n)}}\Bigl(\sum_{j=0}^n A^{(j)}\tens{d}\psi_j\!\Bigr)^{\!2} \!.
\end{equation}
Here, functions $A^{(j)}$, $A^{(j)}_\mu$ are given by \eqref{Aj}, and    
\be\label{Q}
Q_{\mu}=\frac{X_{\mu}}{U_{\mu}}\,,\quad 
U_{\mu}=\prod_{\substack{\nu=1\\\nu\ne\mu}}^{n}(x_{\nu}^2-x_{\mu}^2)\,.
\ee
{\em Metric functions} $X_{\mu}$ are functions of $x_{\mu}$ only, 
and for the Kerr-NUT-(A)dS solution take the form 
\be\n{X}
X_{\mu}=\sum\limits_{k=\varepsilon}^{n}c_kx_{\mu}^{2k}-2b_{\mu}x_{\mu}^{1-\varepsilon}+\frac{\varepsilon c}{x_{\mu}^2}\,.
\ee 
Time is denoted by $\psi_0$, azimuthal coordinates by $\psi_j$,
${j=1,\dots,m\equiv D-n-1}$, $x_n$ is an analytical continuation of 
the Boyer--Lindquist type radial coordinate,
and ${x_\mu}$, ${\mu=1,\dots, n-1}$, stand for latitude
coordinates.\footnote{%
Similar to the 4D
case (see Appendix A.1.3), the signature of the symmetric form of the metric depends on
the domain of $x_\mu$'s and signs of $X_\mu$'s.
The physical Kerr-NUT-(A)dS spacetime is recovered when standard radial coordinate $r=-ix_n$, and new parameter $M=(-i)^{1+\epsilon}b_n$, are introduced, that is, by a simple Wick rotation. See also Section 9.4.1.
} 
The parameter $c_n$ is proportional to the cosmological
constant \cite{HamamotoEtal:2007} 
\begin{equation} 
{R_{ab}=(-1)^{n}(D-1)c_n\,g_{ab}}\,, 
\end{equation}
and the remaining constants $c_k$, $c>0$, and
$b_{\mu}$ are related to rotation parameters, mass, and NUT
parameters. One of these constants may be eliminated due to the scaling symmetry. The metric therefore constitutes the $(D-1-\varepsilon)$--parametric Einstein space
(see \cite{ChenEtal:2006cqg} for more details).
The limit of flat spacetime is recovered when $c_n=0$ and all of the parameters $b_{\mu}$ are zero (equal of one another) in the even (odd) dimensional case.

The Kerr-NUT-(A)dS spacetime \eqref{metric_coordinates}--\eqref{X}  
may be understood as a higher-dimensional
generalization of the four-dimensional Kerr-NUT-(A)dS solution obtained by
Carter \cite{Carter:1968pl}, \cite{Carter:1968cmp}.
Moreover, the coordinates $(x_\mu,\psi_j)$ used in the metric
are the higher-dimensional analogue of the canonical coordinates 
\cite{Carter:1968cmp}, \cite{Carter:1968pl}, \cite{Debever:1971}, \cite{Plebanski:1975}. As discussed in the next section, they have a well defined geometrical meaning. 
More generally, it is possible to consider a broader class of metrics 
\eqref{metric_coordinates} where $X_\mu$'s are arbitrary functions of one variable; $X_\mu=X_\mu(x_\mu)$.
To stress that such metrics do not necessarily satisfy the Einstein equations we call them {\em off-shell} metrics.
It will be shown in Chapter 7, that the most general metric element admitting the PCKY tensor, the {\em canonical metric} element, coincides with the off-shell spacetime 
\eqref{metric_coordinates}. Therefore, from now on we refer to the off-shell spacetime 
\eqref{metric_coordinates}, without imposing \eqref{X}, as to the canonical metric.
The canonical metric is of the special algebraic type D 
\cite{HamamotoEtal:2007} of the higher-dimensional algebraic classification 
\cite{MilsonEtal:2005}, \cite{ColeyEtal:2004a}, 
\cite{Coley:2008}. 
Let us finally remark that 
formulas \eqref{metric_coordinates}--\eqref{X} are applicable also in $D=3$
where  one recovers the $2$--parametric BTZ black hole \cite{BanadosEtal:1992}.

In what follows we shall also use the orthonormal form of the metric
\ba
\tens{g}\!\!&=&\!\!\delta_{a b}\tens{\omega}^{\hat a}\tens{\omega}^{\hat b}=\sum_{\mu=1}^n (\tens{\omega}^{\hat \mu}\tens{\omega}^{\hat \mu}+
\tens{\tilde \omega}^{\hat \mu}\tens{\tilde \omega}^{\hat \mu})+\eps \tens{\omega}^{\hat 0}\tens{\omega}^{\hat 0}\,,\label{metric_vielbein}\\
\tens{\omega}^{\hat \mu} \!\!&=&\!\! \frac{\tens{d}x_{\mu}}{\sqrt{Q_{\mu}}}\,,\quad 
\tens{\tilde  \omega}^{\hat \mu} = \sqrt{Q_{\mu}}
 \sum_{j=0}^{n-1}A_{\mu}^{(j)}\tens{d}\psi_j\;,\quad 
\tens{\omega}^{\hat 0} =\sqrt{\frac{-c}{A^{(n)}}}
\sum_{j=0}^nA^{(j)}\tens{d}\psi_j\;.\label{omega}
\ea
The inverse metric reads
\be\label{inv_coords}
\tens{g}^{-1}=\sum_{\mu=1}^n\Bigl[ Q_\mu(\pa_{x_\mu})^2
  +\frac{1}{Q_\mu U_\mu^2}\Bigl(\sum_{k=0}^m(-x_\mu^2)^{n\!-\!1\!-\!k}\pa_{\psi_k}\Bigr)^{\!2}\Bigr]-\frac{\eps}{cA^{(n)}}(\pa_{\psi_n})^2 \;,
\ee
or, in the orthonormal form
\ba
\tens{g}^{-1}\!\!&=&\!\!\delta^{ab} \tens{e}_{\hat a} \tens{e}_{\hat b}
= \sum_{\mu=1}^n (\tens{e}_{\hat \mu} \tens{e}_{\hat \mu} + \tens{\tilde e}_{\hat\mu} \tens{\tilde e}_{\hat \mu})
       + \eps \tens{e}_{\hat 0}\tens{e}_{\hat 0},\label{inv_vielbein}\\
\tens{e}_{\hat \mu} \!\!&=&\!\! \sqrt{Q_{\mu}}\pa_{x_{\mu}}\,, \ \ 
\tens{\tilde e}_{\hat \mu} =\frac{1}{\sqrt{Q_{\mu}}U_{\mu}}
\sum_{j=0}^{m}(-x_\mu^2)^{n-1-j} \pa_{\psi_{j}}\,,\ \ 
\tens{e}_{\hat 0}=\frac{ \pa_{\psi_n}}{\sqrt{-cA^{(n)}}}\,.\label{e_vielben}
\ea
The inverse relations to \eqref{e_vielben} are
\be\label{psi_vielbein}
\pa_{x_\mu}=\frac{\tens{e}_{\hat \mu}}{\sqrt{Q_{\mu}}}\,, \ \ 
\pa_{\psi_{j}}\!\!=\!\sum_{\mu=1}^{n}\!\sqrt{Q_\mu}A_\mu^{(j)}\tens{\tilde e}_{\hat \mu}
\!+\!\eps A^{(j)}\sqrt{\frac{-c}{A^{(n)}}}\tens{e}_{\hat 0}\,,\ \ 
\pa_{\psi_n}\!\!=\!\sqrt{\!-\!cA^{(n)}}\tens{e}_{\hat 0}\,.
\ee
The determinant of the metric $\tens{g}$ reads
\be\label{detg}
g=\det(\tens{g})=\bigl(-c A^{(n)}\bigr)^\eps\, U^2\,,\quad
U\equiv \det\bigl[A_{\mu}^{(j)}\bigr]=\prod_{\substack{\mu, \nu=1\\\mu<\nu}}^n(x_\mu^2-x_\nu^2)\;.
\ee
In the last expression, $A_{\mu}^{(j)}$, given by \eqref{Aj}, is understood 
as the $n\times n$ matrix. Some algebraic identities regarding these functions or other 
properties of the canonical metric are gathered in Appendix C.4.

\section{Principal conformal Killing--Yano tensor}
The general canonical metric \eqref{metric_coordinates} described in the previous section,
and in particular the Kerr-NUT-(A)dS spacetime \eqref{metric_coordinates}--\eqref{X},
possesses a PCKY tensor \cite{KubiznakFrolov:2007}.\footnote{%
In fact, the PCKY tensor in higher dimensions was first discovered for the Myers--Perry metrics \cite{FrolovKubiznak:2007}, and only after that 
for the general Kerr-NUT-(A)dS spacetimes. For an account of these historical developments we refer the reader to Appendix B.
}   
The corresponding  1-form (KY) potential $\tens{b}$ reads
\be\n{bb}
\tens{b}=\frac{1}{2}\sum_{j=0}^{n-1} A^{(j+1)} \tens{d}\psi_j\, .
\ee
The PCKY tensor, $\tens{h}=\tens{db}$, takes the following forms:
\be\label{PCKY_metric}
\tens{h}=\frac{1}{2}\sum_{j=0}^{n-1}\tens{d}A^{(j+1)}\!\wedge \tens{d}\psi_j=
\frac{1}{2}\sum_{\mu=1}^n\Bigl[\tens{d}x_\mu^2\wedge \sum_{j=0}^{n-1}A_{\mu}^{(j)}
\tens{d}\psi_j\Bigr]
= \sum_{\mu=1}^{n} x_\mu \tens{\omega}^{\hat \mu}
\wedge \tens{\tilde \omega}^{\hat \mu}\,.
\ee
The last expression shows that the basis $\{\tens{\omega}\}$, introduced in \eqref{omega},
is a canonical basis associated with the PCKY tensor $\tens{h}$ 
(see Section \ref{canonical}). 
In fact, this canonical basis has an additional nice property that many of the
Ricci coefficients of rotation vanish \cite{HamamotoEtal:2007}, \cite{KrtousEtal:2008};
it is a {\em principal canonical basis} (see also Chapter 7).
  
Having a PCKY tensor and its canonical basis, we may employ the machinery of Chapter 3.
In particular, we obtain the following extended tower of the 2nd-rank irreducible Killing tensors ($j=0,\dots, n-1$): 
\begin{equation}\label{KT_metric}
\tens{K}^{(j)} =
\sum_{\mu=1}^n A^{(j)}_\mu(\tens{\omega}^{\hat \mu}\tens{\omega}^{\hat \mu}+
\tens{\tilde \omega}^{\hat \mu}\tens{\tilde \omega}^{\hat \mu})+\eps A^{(j)}\tens{\omega}^{\hat 0}\tens{\omega}^{\hat 0}\,.
\end{equation} 
The Killing fields \eqref{Primary}--\eqref{xin} become ($i=1,\dots, n-1$) 
\be\label{xik_metric}
\tens{\xi}^{(0)}=\pa_{\psi_0}\,,\quad 
\tens{\xi}^{(i)}=\pa_{\psi_i}\,,\quad 
\tens{\xi}^{(n)}=\pa_{\psi_n}\,.
\ee
This means that coordinates $(x_\mu, \psi_j)$ are {\em canonical coordinates}. 
All of them are completely determined by the PCKY tensor: `essential' coordinates $x_{\mu}$ are connected with its eigenvalues (see Section \ref{canonical}), Killing coordinates $\psi_{j}$ $(j=0,\dots, m)$ are defined by the tower of Killing vectors generated from this tensor. It is this invariant definition of coordinates what makes the form \eqref{metric_coordinates} of the canonical metric so convenient for calculations.
For example, we shall see in Chapter 6 that these coordinates are the normal separable coordinates for which the Hamilton--Jacobi and Klein--Gordon equations allow the separation of variables.  


\chapter[Complete  integrability of geodesic motion]
{Complete integrability of geodesic motion}
\label{ch6}
\chaptermark{Complete  integrability of geodesic motion}

In this chapter we demonstrate that in the canonical spacetime \eqref{metric_coordinates}, the $n$ constants of geodesic motion 
corresponding to the extended tower of Killing tensors and the $D-n$
constants of geodesic motion corresponding to the tower of Killing vectors are  
functionally independent of one another, making a
total of $D$ independent constants of motion in all dimensions $D$.
The Poisson brackets of all pairs of these $D$ constants are zero, so, the geodesic
motion in these spacetimes is completely integrable \cite{PageEtal:2007}, \cite{KrtousEtal:2007prd}.

\section{Constants of motion}
In the previous chapter we have seen that the (off-shell) canonical spacetime \eqref{metric_coordinates} admits the PCKY tensor $\tens{h}$, \eqref{PCKY_metric}, which in its
turn generates the extended tower of $n$ Killing tensors $\tens{K}^{(j)}$, \eqref{KT_metric}, and the tower of $D-n$ Killing vectors $\pa_{\psi_k}$, \eqref{xik_metric}.
Together, these objects give $D$ constants of geodesic motion,\footnote{%
Instead of constants $\kappa_j\,,$ one can consider a different set of $n$ constants
corresponding to various invariants of the form $\tens{F}$, \eqref{formF}. For example, we may consider [cf. Eq. \eqref{Cj2}]
\be\label{Cj_tilde}
\tilde C_0\equiv w=\kappa_0=\tens{u}\cdot\tens{u}\,,\quad
\tilde C_j\equiv{\rm Tr}\bigl[(H\tilde P)^{2j}\bigr]=(-w)^j C_j\,,\quad \tilde P\equiv wP=wI-W\,.
\ee
In Section \ref{sc:pb},  we shall use this choice to prove the Poisson commutativity
of $\kappa_j$'s. 
} 
\be\label{psikappa}
\Psi_k=\xi^{(k)}_{\ a} u^a=\tens{u}\cdot\cv_{\psi_k}\,,\quad
\kappa_j=K^{(j)}_{\ \, ab}u^au^b=\tens{u}\cdot \tens{K}^{(j)}\cdot \tens{u}\,.
\ee
Here, we have denoted the momentum of the geodesic motion ${\tens{u}}$, $u^a=dx^a/d\tau$, and we understand all mentioned quantities as observables
(i.e. functions) on the phase space ${\phsp\equiv \coTB}$. (For a review of the 
canonical mechanics on the phase space ${\phsp}$ see, e.g., Appendix of 
\cite{KrtousEtal:2007prd}.)

Let us now explicitly evaluate constants \eqref{psikappa} in the orthonormal basis 
\eqref{metric_vielbein}. There we have 
\begin{equation}\label{uu}
\tens{u}^\flat=\sum_{\mu=1}^n \bigl( u_{\hat \mu} \tens{\omega}^{\hat \mu} + {\tilde u}_{\hat \mu} \tens{\tilde \omega}^{\hat \mu}\bigr)
+\eps\,u_{\hat 0} \tens{\omega}^{\hat 0}\,.
\end{equation}
Using \eqref{psi_vielbein} and \eqref{const} we find 
\ba
\Psi_{k}\!\!&=&\!\!\sum_{\mu=1}^{n}\!\sqrt{Q_\mu}A_\mu^{(k)} {\tilde u}_{\hat \mu}
+\eps A^{(k)}\sqrt{\frac{-c}{A^{(n)}}}\,u_{\hat 0}\,,\quad 
\Psi_n=\sqrt{-cA^{(n)}}\,u_{\hat 0}\,,\label{constants_psij}\\
\kappa_j\!\!&=&\!\! \sum_{l=0}^j (-1)^l A^{(j-l)} w_l =
\sum_{\mu=1}^nA^{(j)}_\mu
   \bigl(u_{\hat \mu}^2+{\tilde u}_{\hat \mu}^2\bigr)+\eps A^{(j)} u_{\hat 0}^2\;.
\label{constants_kappaj}
\ea
These formulas may be easily inverted using relation \eqref{psi_vielbein} and identities 
\begin{equation}
\sum_{j=0}^{n-1}\frac{(-x_\nu^2)^{n\!-\!1\!-\!j}}{U_\nu}A_\mu^{(j)}=\delta_\nu^\mu\;,\quad
\sum_{\mu=1}^n \frac{A_{\mu}^{(k)}}{x_\mu^2 U_\mu}
=\frac{A^{(k)}}{A^{(n)}}\,,
\end{equation}
proved in Appendix C.4. The result is given by formula \eqref{basisvectors} below.

\section{Complete integrability}
{\bf Definition.} {A motion in $M^D$ is {\em completely integrable} if there exist $D$ 
functionally independent integrals of motion which are in {\em involution}, that is, 
they mutually Poisson commute of one another \cite{Arnold:book}, \cite{Kozlov:1983}.}

{\bf Proposition.} {\em The geodesic motion in the
canonical spacetime \eqref{metric_coordinates} is completely integrable.
The geodesic momentum $\tens{u}$ can be written in the form \eqref{uu}, where 
the basis components (expressed 
in terms of integrals of motion $\Psi_k$ and $\kappa_j$) are:
\be\label{basisvectors}
u_{\hat \mu} =\frac{\sigma_\mu}{(X_\mu U_\mu)^{1/2}}\,\bigl(
X_\mu V_\mu-W_\mu^2 \bigr)^{1/2}\,,\ \ 
{\tilde u}_{\hat\mu}=\frac{1}{\sqrt{Q_\mu}}\frac{W_\mu}{U_\mu}\,,\ \ 
u_{\hat 0} =\frac{\Psi_n}{A^{(n)}}\sqrt{\frac{A^{(n)}}{-c}}\,. 
\ee 
Constants $\sigma_\mu=\pm 1$ are independent 
of one another, and 
\begin{equation}\label{VW}
V_\mu\equiv\sum_{j=0}^{m}(-x_{\mu}^2)^{n-1-j}\kappa_j\,,\quad  
W_\mu\equiv\sum_{k=0}^{m}(-x_{\mu}^2)^{n-1-k}\Psi_{\!k}\,,\quad
\kappa_n\equiv -\frac{\Psi_n^2}{c}\,.
\end{equation}
}
In order to prove this proposition, we need to establish  
the functional independence and Poisson commutativity of integrals of motion 
$\kappa_j$ and $\Psi_k$. This is done in the following two sections.   

The coordinate components of the velocity are
\be\label{xpsi_dot}
\dot x_\mu=\frac{\sigma_\mu}{|U_\mu|}\,\bigl(X_\mu V_\mu-W_\mu^2\bigr)^{1/2}\,,\quad 
\dot \psi_k=\sum_{\mu=1}^n \frac{(-x_\mu^2)^{n-1-k}}{U_\mu X_\mu}\,W_\mu
-\eps\frac{\Psi_n}{cA^{(n)}}\,.
\ee
To obtain these expressions we have used \eqref{omega}, and the explicit form of the inverse metric \eqref{inv_coords}.
Using formula \eqref{separ_CC} proved in Appendix C.5 we can symbolically integrate 
equations for $\psi_k\,$:
\be
\psi_k=\sum_{\mu=1}^n\int\frac{\sigma_\mu{\rm sign}(U_\mu)f_\mu^{(k)}dx_\mu}{\sqrt{X_\mu V_\mu-W_\mu^2}}\,,\quad
f_\mu^{(k)}\equiv \frac{W_\mu}{X_\mu}(-x_\mu^2)^{n-1-k}-\eps\frac{\Psi_n}{c x_\mu^2}\,\,.
\ee 
Similarly, we can express the affine parameter $\tau$ as [cf. Eq. \eqref{tau_CC}]
\be
\tau=\sum_{\mu=1}^n\int\frac{\sigma_\mu{\rm sign}(U_\mu)(-x_\mu^2)^{n-1}dx_\mu}{\sqrt{X_\mu V_\mu-W_\mu^2}}\,.
\ee

\section{Independence of constants of motion}
In this section we want to demonstrate that quantities $\kappa_j$ and ${\Psi_k}$ are independent 
at a generic point of the phase space ${\phsp=\coTB}$. This means that 
their gradients on the phase space are linearly independent. 
To prove that it is sufficient to show that these gradients
are independent in the vertical direction of the cotangent bundle ${\coTB}$,
i.e., that the derivatives of these quantities with respect to the momentum ${\tens{u}}$,
are linearly independent. To achieve this we will study the wedge product of
the `vertical' derivatives. We denote the vertical derivative by $\pder$. For observable $f$, $\,{\pder f}\equiv \pder f/\pder \tens{u}$ denotes a {\em vector} field on the manifold $M^D$, with components ${\partial{f}/\partial{u_a}}$. 

Let us, instead of $\kappa_j$ consider the equivalent set of observables
\be\label{pom44}
2\tilde \kappa_j\equiv (-1)^j\kappa_j=w_j+\dots\,.
\ee
Here we have used the first relation \eqref{constants_kappaj}. `Dots' denote terms which contain $w_k$ with $k<j$. 
We are interested in the quantity\footnote
{
The wedge product is, strictly speaking, defined for (antisymmetric) forms. 
However, we can easily define the wedge product also for the vectors or
lower the vector indices with the help of the metric to get forms.
} 
\begin{equation}\label{jac}
  \tens{J} = \pder\tilde{\kappa}_0\wedge\dots\wedge\pder\tilde{\kappa}_{n\!-\!1}\wedge
      \pder\Psi_0\wedge\dots\wedge\pder\Psi_m\;.
\end{equation}
Due to \eqref{psikappa}, \eqref{pom44}, and the definition of $w_j$, \eqref{wjdef}, we have 
\begin{equation}\label{pderofcj}
 \pder\Psi_j=\pa_{\psi_j}\,,\quad 
  \pder\tilde\kappa_j=\tens{Q}^j\cdot\tens{u}+
\dots\,,
\end{equation}
where `dots' denote linear combinations of ${\tens{Q}^k\cdot\tens{u}}$ with ${k<j}\,;$ 
${\tens{Q}^l\cdot\tens{u}}$ represents the vector with components
${Q^a_{a_1}Q^{a_1}_{a_2}\cdots Q^{a_{l\!-\!1}}_{a_j}u^{a_j}}$.
From the antisymmetry of the wedge product it follows that
\begin{equation}\label{jacinu}
  \tens{J}=\tens{u}\wedge(\tens{Q}\cdot\tens{u})\wedge\dots\wedge(\tens{Q}^{n\!-\!1}\cdot\tens{u})\wedge
  \cv_{\psi_0}\wedge\dots\wedge\cv_{\psi_m}\;.
\end{equation}
Let us now use the explicit form of ${\tens{Q}^j}$ [cf. Eq. \eqref{tensorQ}] 
\begin{equation}\label{powQ}
  \tens{Q}^j = \sum_{\mu=1}^n x_\mu^{2j}\tens{e}_{\hat \mu}\,\tens{\omega}^{\hat \mu}+
\sum_{\mu=1}^n x_\mu^{2j}\tens{\tilde e}_{\hat\mu}\,\tens{\tilde \omega}^{\hat\mu}\;.
\end{equation}
The second term acts on the subspace of vectors spanned on ${\cv_{\psi_j}}$.
Thus, thanks to the term ${\cv_{\psi_0}\wedge\dots\wedge\cv_{\psi_m}}$ in the wedge product, this part can be ignored in \eqref{jacinu}. 
Moreover, taking into  account that $\tens{e}_{\hat \mu}\,\tens{\omega}^{\hat \mu}=
\pa_{x_\mu}\grad x_\mu\,,$ and $u^\mu=\grad x_\mu\cdot \tens{u}$,
the substitution of \eqref{powQ} into \eqref{jacinu} leads to
\begin{equation}\label{jacfin}
  J = u^1 \dots u^n\, U\, \cv_{x_1}\wedge\dots\wedge\cv_{x_n}\wedge\cv_{\psi_0}\wedge\dots\wedge\cv_{\psi_{D\!-\!n\!-\!1}}\;,
\end{equation}
where 
\begin{equation}\label{Udef}
  U = \mspace{-10mu}\sum_{\substack{\text{permutations}\\\text{${\sigma}$ of ${1\dots n}$}}}\mspace{-15mu}
  \sign\sigma\; x_1^{2\sigma_1}\dots x_n^{2\sigma_n}
    = \mspace{-10mu}\prod_{\substack{\mu,\nu=1\dots n\\\mu<\nu}}\mspace{-8mu} (x_\mu^2-x_\nu^2)\;.
\end{equation}
In a generic point of the phase space we have ${u^\mu\ne0}$ and ${x_\mu^2\ne x_\nu^2}$ (for ${\mu\ne\nu}$) and
therefore ${\tens{J}\ne0}$ there. Thus we have shown that the constants of motion  are independent.

\section{Poisson brackets}
\label{sc:pb}

Finally, we need to show that observables $\kappa_j$ and $\Psi_j$
Poisson commute. 
The Poisson bracket of two functions on the phase space $\phsp$ can be written as
\begin{equation}\label{PBdef}
  \{A,B\}=\covd A \cdot \pder B - \pder A \cdot \covd B\;,
\end{equation}
where ${\covd F}$ represents an arbitrary (torsion-free) covariant derivative 
which ignores the dependence of ${F}$ on the momentum ${\tens{u}}$,
and ${\pder B}$ is the derivative of ${B}$ with respect to the momentum ${\tens{u}}$.
${\covd F}$ and ${\pder F}$ is a 1-form and a vector field on 
the spacetime ${M}^D$, respectively; the dot indicates a contraction between them.
Naturally, we use the covariant derivative ${\covd}$ generated by the metric connection on ${M}^D$.

Clearly, the commutation of any observable with
the Hamiltonian 
\be\label{Hamiltonian}
H=\frac{1}{2}w=\frac{1}{2}\tens{u}\cdot \tens{u}=\frac{1}{2}\kappa_0 
\ee
of the geodesic motion is equivalent to 
the conservation of the observable. So we have
\begin{equation}\label{PBw}
  \{\kappa_0,\kappa_j\}=0\;,\quad\{\kappa_0,\Psi_j\}=0\;.
\end{equation}
The Poisson bracket between observables 
${\Psi_j=\tens{u}\cdot\cv_{\psi_j}}$ reduces to the Lie bracket of the 
Killing vector fields ${\cv_{\psi_j}}\,,$ which vanishes because 
${\cv_{\psi_j}}$ are coordinate vector fields:
\be\label{PBclcl}
\{\Psi_i,\Psi_j\} 
    = \cv_{\psi_j}\cdot(\covd\cv_{\psi_i})\cdot\tens{u}
     -\cv_{\psi_i}\cdot(\covd\cv_{\psi_j})\cdot\tens{u}
    = \bigl[\cv_{\psi_j},\cv_{\psi_i}\bigr]\cdot\tens{u}=0\,.
\ee
The Poisson bracket of $\kappa_i$ with the observable ${\Psi_j=\tens{\pa_{\psi_j}}\cdot\tens{u}}\,,$ associated with the isometry
$\pa_{\psi_j}$\,,
leads to the Lie derivative along this isometry, 
\be\label{PBclcp}
\{\kappa_i,\Psi_j\} 
  = \cv_{\psi_j}\cdot\covd\kappa_i - \pder\kappa_i\cdot\covd\cv_{\psi_j}\cdot\tens{u}
  =\lied_{\partial_{\psi_j\!}}\!\bigl({K}_{(i)}^{ab}\bigr)u_a u_b
\equiv
\lied_{\partial_{\psi_j}}\!\kappa_i=0\;.
\ee
Here, the `generalized' Lie derivative ${\lied_{\partial_{\psi_j}}\!\kappa_i}$ ignores 
the dependence of ${\kappa_i}$ on the momentum ${\tens{u}}$.
It vanishes because ${\cv_{\psi_j}}$ are Killing vectors 
and Killing tensors $\tens{K}^{(i)}$ respect the symmetry of the
spacetime. 

Finally, it remains to evaluate the brackets $\{\kappa_i,\kappa_j\}$.
We shall do it in two steps: first, we prove that  
an equivalent set of observables $\tilde C_j$, given by 
\eqref{Cj_tilde}, Poisson commute and, second, by relating these constants to 
$\kappa_j$ we obtain the desired result. 
So, let us consider the observables $\tilde C_j$.
Using the cyclic property of the trace, 
the derivative of ${\tilde\cp_j}$ in the spacetime direction is
\begin{equation}
  \nabla_a\tilde\cp_j=2j\,{\rm Tr}\bigl[(\nabla_a H) \tilde{P} (H \tilde{P})^{2j-1}\bigr]\;.
\end{equation}
Here, ${\nabla_a H}$ is the matrix of components ${\nabla_a\cKY^b{}_c}$
of the covariant derivative ${\covd\tens{\cKY}}$. Substituting for ${\nabla_a\cKY^b{}_c}$ 
from Eq.~\eqref{PCKY_coords} and using the antisymmetry of $\tens{h}$, we obtain
\ba\label{covdcp}
{\textstyle\frac{D-1}{2j}}\nabla_e \tilde\cp_j
\!\!&=&\!\!\xi_{a_0}\tilde{P}^{a_0}_{\,b_1}\cKY^{b_1}{}_{\!a_1}\tilde{P}^{a_1}_{\,b_2}
\dots\cKY^{b_{2j-2}}{}_{\!\!\!a_{2j-1}}\tilde{P}^{a_{2j-1}}_{\,e}
\!-\!g_{ea_{2j}}\tilde{P}^{a_{2j}}_{\,b_{2j-1}}\cKY^{b_{2j-1}}{}_{\!\!\!a_{2j-1}}
\dots\cKY^{b_1}{}_{\!a_1}\tilde{P}^{a_1}_{b_0}\xi^{b_0}\nonumber\\
\!\!&=&\!\!2\,\xi_{a_0}\tilde{P}^{a_0}_{\,b_1}\cKY^{b_1}{}_{\!a_1}
\tilde{P}^{a_1}_{\,b_2} 
\dots\cKY^{b_{2j-2}}{}_{\!\!\!a_{2j-1}}\tilde{P}^{a_{2j-1}}_{\,e}\;.
\ea
For the derivative with respect to the momentum ${\tens{u}}$ we get
\be\label{pdercp}
  {\textstyle\frac{1}{4j}}\partial^e \tilde\cp_j
=u^e\bigl(\cKY^{d_1}{}_{\!c_1}\tilde{P}^{c_1}_{\,d_2}\cKY^{d_2}{}_{\!c_2}\tilde{P}^{c_2}_{\,d_3}\dots
\tilde{P}^{c_{2j\!-\!1}}_{\,d_{2j}}\cKY^{d_{2j}}{}_{\!d_1}\bigr)
\!+\!\cKY^{e}{}_{\!c_1}\tilde{P}^{c_1}_{\,d_2}\cKY^{d_2}{}_{\!c_2}\tilde{P}^{c_2}_{\,d_3}\dots
\tilde{P}^{c_{2j\!-\!1}}_{\,d_{2j}}\cKY^{d_{2j}}{}_{\!\!c_{2j}}u^{c_{2j}}\;.
\ee
Substituting \eqref{covdcp} and \eqref{pdercp} into \eqref{PBdef} for ${\{\tilde\cp_i,\tilde\cp_j\}}$
and using ${\tilde{P}^a_{\,b}u^b=0}\,,$ we find
\ba\label{PBcpcp}
{\textstyle\frac{D-1}{16ij}}\{\tilde\cp_i,\tilde\cp_j\}
\!\!\!&=&\!\!\xi_{a_0}\tilde{P}^{a_0}_{\,b_1}\cKY^{b_1}{}_{\!a_1}
\dots\tilde{P}^{a_{2i-1}}_{\,b_{2i-1}} 
\cKY^{b_{2i-1}}{}_{\!c_1}\dots
\tilde{P}^{c_{2j-1}}_{\,d_{2j}}\cKY^{d_{2j}}{}_{\!\!c_{2j}}u^{c_{2j}}\nonumber\\
 \!\!&&\!\!\!-\xi_{a_0}\tilde{P}^{a_0}_{\,b_1}\cKY^{b_1}{}_{\!a_1}
\dots\tilde{P}^{a_{2j-1}}_{\,b_{2j-1}} 
\cKY^{b_{2j-1}}{}_{\!c_1}\dots
\tilde{P}^{c_{2i-1}}_{\,d_{2i}}\cKY^{d_{2i}}{}_{\!\!c_{2i}}u^{c_{2i}}=0\,.\quad
\ea
We thus proved that constants $\tilde C_j$ mutually Poisson commute. 
The same is, of course, true for constants $C_j$, \eqref{Cj2}, 
which differ from $\tilde C_j$ only by rescaling \eqref{Cj_tilde}.
The generating function ${Z(\beta)}$, \eqref{Z}, 
is given by power series in ${\beta}$ with coefficients
given (up to constant factors) by constants ${\cp_j}\,$, cf. Eq. \eqref{Zbeta}.
Therefore this function, and similarly ${W(\beta)=\exp Z(\beta)}$,
Poisson commute with $\kappa_0$ and ${\Psi_j}$, as well as with itself
for different choices of $\beta$:
\begin{equation}\label{PBZZ}
  \bigl\{Z(\beta_1),Z(\beta_2)\bigr\}=0\;,\quad\bigl\{W(\beta_1),W(\beta_2)\bigr\}=0\;.
\end{equation}
This means, that also quantities ${\kappa_j}$ generated from ${W(\beta)}$
mutually Poisson commute.
Therefore all the constants of motion are in involution and 
the geodesic motion is completely integrable.

\section{Lie algebra of Killing tensors}
\label{Lie}
Let us use the opportunity to remind here that Killing tensors, as proper symmetry objects, form an appropriate Lie algebra. 
This will give us another point of view on the above calculations.

We start with an observation that the Poisson commutativity of 
constants corresponding to the isometries is equivalent to 
the vanishing of the Lie brackets of these isometries [cf. Eq. \eqref{PBclcl}].
Similarly, the Poisson bracket of a quantity corresponding to 
the Killing tensor and a quantity associated with the isometry leads
to the Lie bracket of the Killing tensor along the isometry [cf. Eq. \eqref{PBclcp}].
More generally, it is well known that Killing tensors form a Lie subalgebra of a Lie
algebra of all totally symmetric contravariant tensor fields on the manifold  with respect to the {\em symmetric Schouten--Nijenhuis} (SN) brackets \cite{Schouten:1940}, \cite{Schouten:1954}, \cite{Nijenhuis:1955}. 
The vanishing of these brackets is equivalent to the Poisson commutativity of 
the corresponding constants of geodesic motion (see, e.g., \cite{BenentiFrancaviglia:1980}).
In particular, we have the following equations: 
\ba
\bigl[K_{(i)},K_{(j)}\bigr]^{abc}_{\rm SN}\!&\equiv&\!  
K_{(i)}^{e(a}\, \nabla_{e} K_{(j)}^{bc)}- 
K_{(j)}^{e(a}\, \nabla_{e} K_{(i)}^{bc)}=0\;,\label{KKSN}\\
\bigl[\partial_{\psi_j},K_{(i)}\bigr]^{ab}_{\rm SN}\!&\equiv&\!  
\mathcal{L}_{\partial_{\psi_j}} K_{(i)}^{ab}=0\,.\label{XKSN}
\ea
Taking the metric $\tens{g}$ as one of the Killing tensors in \eqref{KKSN}, we obtain the Killing tensor equation \eqref{KT_def}.
[Such an equation simply states that an observable corresponding to
the Killing tensor commutes with the Hamiltonian \eqref{Hamiltonian}, and therefore 
constitutes a constant of geodesic motion.]
Using the Schouten--Nijenhuis brackets and the method of generating functions, an independent proof of the Poisson commutativity of constants $\kappa_j$ and $\Psi_k$ generated by a PCKY tensor obeying \eqref{Liexihg} was recently demonstrated \cite{HouriEtal:2008a}.

Finally, we would like to mention that an interesting question whether also Killing--Yano tensors form a closed Lie algebra was 
recently addressed in \cite{KastorEtal:2007}. It is well known that forms on the manifold
form a (graded) Lie algebra with respect to the {\em antisymmetric Schouten--Nijenhuis}
(aSN) brackets. For a $p$-form $\tens{\alpha}$ and a $q$-form $\tens{\beta}$ these are defined as
\be
[\alpha, \beta]_{\rm \,aSN}^{a_1\dots\,a_{p+q-1}}\!\equiv
p\,\alpha^{b[a_1\dots\,a_{p-1}}\nabla_b\beta^{a_p\dots\,a_{p+q-1}]}+
(-1)^{pq}q\,\beta^{b[a_1\dots\,a_{q-1}}\nabla_b\alpha^{a_q\dots\,a_{p+q-1}]}\,.
\ee
The definition is connection independent; covariant derivatives may be replaced 
with partial derivatives. When one of the forms is a vector, the bracket reduces
to the Lie derivative.

One might expect that if Killing--Yano tensors are associated with symmetries in some `appropriately generalized' sense they would form a closed subalgebra with respect to these brackets. 
The (aSN) bracket of a Killing vector and a rank-2 Killing-Yano tensor is indeed a rank-2 Killing--Yano tensor \cite{KastorEtal:2007}.
Unfortunately, for two Killing--Yano tensors this is not, except the special case of a constant curvature spacetime, generally true \cite{KastorEtal:2007}.
The KY tensor (as well as the PCKY tensor) 
in the Kerr spacetime are the counter examples. The geometrical meaning of 
the Killing--Yano symmetry therefore still remains veiled. 


\chapter[Separation of variables]
{Separation of variables}
\label{ch7}
\chaptermark{Separation of variables}

In this chapter, based on \cite{FrolovEtal:2007}, we demonstrate the separability of the Hamilton--Jacobi and Klein--Gordon 
equations in the canonical spacetime \eqref{metric_coordinates}. Such a separability provides an independent proof of complete integrability of geodesic motion.
We also review some related results and briefly discuss an open problem of separability of equations with spin. 

\section{Hamilton--Jacobi equation}
The Hamilton--Jacobi equation for geodesic motion  
on a manifold with metric $\tens{g}$ has the form
\begin{equation}\label{HJ}
\frac{\partial S}{\partial\lambda}+g^{ab}\, \partial_a S\;\partial_b S=0\;.
\end{equation}
Here ${\lambda}$ denotes an `external' time which turns out to be 
an affine parameter of the corresponding geodesic motion. 
We want to demonstrate that in the background (\ref{metric_coordinates}) 
the classical action $S$ allows an additive separation of variables 
\begin{equation}\label{sepS}
S=-w\lambda + \sum_{\mu=1}^n S_{\mu}(x_{\mu})+ \sum_{k=0}^{m} \psic_k\psi_k\,,
\end{equation}
with functions ${S_\mu(x_\mu)}$ of a single argument ${x_\mu}\,$.

Substituting \eqref{sepS} into the Hamilton--Jacobi equation~\eqref{HJ} 
and using the form \eqref{inv_coords} of the inverse metric, we obtain
\begin{equation}\label{HJsep}
\sum_{\mu=1}^n\Bigl[ \frac{X_\mu {S_\mu'}^2}{U_\mu}
  +\frac{1}{X_\mu U_\mu}\Bigl(\sum_{j=0}^m(-x_\mu^2)^{n\!-\!1\!-\!k}\Psi_j\Bigr)^{\!2}\Bigr]-\eps\frac{\Psi_n^2}{cA^{(n)}}-w=0 \;.
\end{equation}
Here, $S_\mu'$ denotes the derivative of function ${S_\mu}$ with respect to its single argument ${x_\mu}$. Using identities (proved in Appendix C.4) 
\be\label{pom_ide1}
1=\sum_{\mu=1}^n \frac{(-x_\mu^2)^{n-1}}{U_\mu}\,,\quad 
\frac{1}{A^{(n)}}=\sum_{\mu=1}^n \frac{1}{x_\mu^2 U_\mu}\,,
\ee
and the definition of $W_\mu$, \eqref{VW}, we can rewrite the last equation in the form
\be\label{ss}
\sum_{\mu=1}^n\frac{F_\mu}{U_\mu}=0\,,
\ee
where $F_\mu$ are functions of $x_\mu$ only:
\be\label{F_HJ}
F_\mu=\frac{W_\mu^2}{X_\mu}+X_\mu{S_\mu'}^2-w(-x_\mu^2)^{n-1}-\eps \frac{\Psi_n^2}{c x_\mu^2}\,.
\ee
Applying Lemma 2 of Appendix C.4, we realize that the general solution of \eqref{ss} is
\be\label{Ft}
F_\mu=\sum_{j=1}^{n-1} \kappa_j (-x_\mu^2)^{n-1-j}\,,
\ee 
where $\kappa_j$ are arbitrary constants. Denoting by 
\be
\kappa_0\equiv w\,,\quad \kappa_n\equiv-\frac{\Psi_n^2}{c}\,, 
\ee
and using the definition of $V_\mu$, \eqref{VW}, we combine \eqref{F_HJ} and 
\eqref{Ft} to obtain equations for $S_\mu'$\,,
\begin{equation}\label{Scond}
  S_\mu'^2=-\frac{W_\mu^2}{X_\mu^2}+\frac{V_\mu}{X_\mu}\,,
\end{equation}
which can be solved by quadratures.
Thus we have shown that the Hamilton--Jacobi equation~\eqref{HJ} in the off-shell gravitational background \eqref{metric_coordinates}
can be solved by the classical action $S$ in the separated form \eqref{sepS}, 
with ${S_\mu}$ satisfying \eqref{Scond}. The separated solution contains $D$ arbitrary constants. Namely, it contains $m+1=D-n$ constants $\Psi_j$ ($j=0,\dots, m$) and $n$ constants $\kappa_k$ ($k=0,\dots, n-1$). 

The gradient of $S$ gives the momentum $p_a=\partial_a S$. Substituting 
our expression for ${S}$ we obtain $p_a$ in terms of the constants $\kappa_k$ and $\Psi_j$:
\be\label{pj_sep}
p_j=\Psi_j\,,\quad p_\mu^2=-\frac{W_\mu^2}{X_\mu^2}+\frac{V_\mu}{X_\mu}\,.
\ee
These relations can be inverted. 
Clearly, ${\Psi_j=p_j}$ are constants linear in the momentum generated
by Killing vectors $\pa_{\psi_j}$.
The constants $\kappa_k$ are quadratic in momenta. They are connected with 
$n$ (irreducible) Killing tensors $\tens{K}^{(k)}$\,,
$(k=0, \dots, n-1)$, 
\begin{equation}\label{kappak_sep}
\kappa_k=K^{(k)}_{ab}p^ap^b\,, \quad \nabla_{\!(c}K^{(k)}_{ab)}=0\,.
\end{equation}
One can easily find the explicit form of $K^{ab}_{(k)}$ by
inverting \eqref{F_HJ}. Let us multiply it by 
$A_{\mu}^{(k)}/U_\mu$, sum over $\mu$, and use identities
(see Appendix C.4)
\begin{equation}\label{pom_ide2}
\sum_{\mu=1}^{n} \frac{(-x_\mu^2)^{n\!-\!1\!-\!j}}{U_\mu}A_\mu^{(k)} 
= \delta_k^j\,,\quad
\sum_{\mu=1}^n \frac{A_{\mu}^{(k)}}{x_\mu^2 U_\mu}
=\frac{A^{(k)}}{A^{(n)}}\,,
\end{equation}
which are valid for $j,k=0,\dots,n-1$. Then we obtain
\be\label{KTk}
\tens{K}_{(k)}=
\sum_{\mu=1}^n \biggl[\frac{A_{\mu}^{(k)}}{X_\mu U_\mu}\Bigl(\sum_{j=0}^m
(-x_\mu^2)^{n-1-j}\pa_{\psi_j}\Bigr)^{\!2}
\!+\!A_{\mu}^{(k)}Q_\mu(\pa_{x_\mu})^2\biggr]\!
-\!\frac{\eps A^{(k)}}{cA^{(n)}}(\pa_{\psi_n})^2\,,
\ee
which are Killing tensors \eqref{KT_metric}, written in the coordinate basis.

\section{Klein--Gordon equation}
The behavior of a massive scalar field $\Phi$ in the gravitational background $\tens{g}$ is governed by the Klein--Gordon equation
\begin{equation}\label{KG}
\Box\Phi=\frac{1}{\sqrt{|g|}}\,\partial_{a}(\sqrt{|g|}g^{ab}\partial_{b}\Phi)=\mu^2\Phi.
\end{equation}
In an Einstein space, this equation remains valid for the non-minimal coupling case as well. (The term $\xi R$ is constant and can be included into the definition of $\mu^2$.)

Now, we demonstrate that the Klein--Gordon equation (\ref{KG}) in the canonical 
background (\ref{metric_coordinates}) allows a multiplicative separation of variables
\begin{equation}\label{multsep}
\Phi=\prod_{\mu=1}^nR_{\mu}(x_{\mu})\prod_{j=0}^{m}e^{i\Psi_j\psi_j}.
\end{equation}
This equation has the following explicit form:
\be\label{KGm}
\sum_{\mu=1}^n\biggl[\partial_{x_\mu}\Bigl(\frac{\sqrt{|g|}}{U_{\mu}}X_{\mu}\partial_{x_\mu}\Phi\Bigr)
+\frac{\sqrt{|g|}}{U_{\mu}X_{\mu}}\Bigl(\sum_{j=1}^{m}(-x_{\mu}^2)^{n-1-j}\partial_{\psi_j}\Bigr)^{\!2}\Phi\biggr]
-\eps\frac{\sqrt{|g|}}{cA^{(n)}}\,\partial^2_{\psi_n}\!\!\Phi=\sqrt{|g|}\mu^2\Phi\;.
\ee 
Here, we have used the quasi-diagonal property of the inverse metric $g^{ab}$, \eqref{inv_coords}, and the fact that $\pa_{\psi_j}$ are Killing vectors. 
We further notice that [cf. Eq. \eqref{detg}]
\begin{equation}
\sqrt{|g|}\propto U P^{\eps}\,,\quad P\equiv \prod_{\mu=1}^n x_{\mu}\,,
\end{equation}
where ``$\propto$'' means equality up to a constant factor [which can be ignored in Eq.~\eqref{KGm}].
Using identities \eqref{pom_ide1} and the obvious fact that $\partial_{x_\mu}(U/U_\mu)=0$, we realize that (\ref{KGm}) is equivalent to 
\be\label{22}
\sum_{\mu=1}^n \!\frac{1}{U_\mu}\! 
\biggl[
\frac{\partial_{x_\mu}\! (P^{\eps}X_{\mu}\partial_{x_\mu}\!\Phi ) }{P^{\eps}}
+\!\frac{1}{X_\mu}\Bigl(\sum_{j=1}^{m}(-x_{\mu}^2)^{n-1-j} \partial_{\psi_k}\Bigr)^{\!2}\!\Phi-\eps\frac{\partial_{\psi_n}^2\!\!\Phi}{cx_{\mu}^2} -\!\mu^2 (-x_{\mu}^2)^{n-1}\Phi\biggr]\!=0\;.
\ee 
Employing the ansatz (\ref{multsep}), we have
\begin{equation}
\partial_{\psi_j}\!\Phi=i\Psi_j\Phi\,,\quad \partial_{x_\mu}\!\Phi=\frac{R_{\mu}'}{R_{\mu}}\Phi\,,\quad  \partial^2_{x_\mu}\!\Phi=\frac{R_{\mu}''}{R_{\mu}}\Phi\,,
\end{equation}
and the Klein--Gordon equation (\ref{22}) takes the form
\begin{equation}\label{KGsep}
\sum_{\mu=1}^n\frac{G_{\mu}}{U_\mu}\,\Phi=0\,,
\end{equation}
where $G_{\mu}$ is function of $x_{\mu}$ only,
\be
G_{\mu}=X_{\mu}\frac{R_{\mu}''}{R_{\mu}}+\frac{R'_{\mu}}{R_{\mu}}\Bigl(X_{\mu}'+\eps\frac{X_{\mu}}{x_{\mu}}\Bigr)-\frac{W_\mu^2}{X_{\mu}}+\frac{\eps\psic_n^2}{cx_{\mu}^2}-\mu^2 (-x_{\mu}^2)^{n-1}.
\ee 
As earlier, the prime means the derivative of functions 
${R_\mu}$ and ${X_\mu}$ with respect to their single argument ${x_\mu}$, and we have used the definition \eqref{VW} for $W_\mu$.
Employing again Lemma 2 of Appendix C.4, we realize that the general solution of (\ref{KGsep}) is 
\begin{equation}
G_{\mu}=-\sum_{j=1}^{n-1} \kappa_{j}(-x_\mu^2)^{n-1-j}\;,
\end{equation}
where $\kappa_j$ are arbitrary constants.

Therefore, we have demonstrated that the Klein--Gordon equation (\ref{KG}) in the 
background (\ref{metric_coordinates}) allows a multiplicative separation
of variables (\ref{multsep}), where functions $R_{\mu}(x_{\mu})$ satisfy the ordinary 
second order differential equations
\be\label{ODE}
\bigl(X_{\mu}R_{\mu}'\bigr)'+\eps\frac{X_{\mu}}{x_{\mu}}R_{\mu}'+
\Bigl(V_\mu-\frac{W_\mu^2}{X_\mu}\Bigr)R_\mu =0\;.
\ee
Here, functions $V_\mu$ and $W_\mu$ are defined in \eqref{VW}. 
They contain 
\be
\kappa_0=-\mu^2\,,\quad \kappa_n=-\frac{\Psi_n^2}{c}\,,
\ee
and arbitrary separation constants $\Psi_j$ ($j=0,\dots, m$) and $\kappa_k$ ($k=1,\dots,n-1$). 
These constants are related to the constants obtained by the separation 
of the Hamilton--Jacobi equation by the {\em geometric optics approximation}.
This connection is briefly discussed in the next section.
 
It should be emphasized that 
in the symmetric form of the metric \eq{metric_coordinates}  all equations \eqref{ODE} `look the same'. 
However, in order to use the proved  separability for concrete calculations in the physical Kerr-NUT-(A)dS spacetime one needs to specify metric functions $X_\mu$ to have the form \eqref{X} and perform a Wick rotation to the `physical space' (see Footnote 1 in Chapter 4, and also Section 9.4.1).
Such a transformation `spoils' the symmetry between essential coordinates but the separability property remains.
The equation for $R_n$ then plays the role of an equation for propagating radial modes, whereas the other equations (with imposed
regularity conditions) represent the eigenvalue problems. For a discussion of special sub-cases of these equations see, e.g, \cite{BertiEtal:2006} and reference therein.

\section{Understanding connections}
To obtain a more complete picture about the above described separability, 
let us in this section review two closely related 
results. Namely, we review the theory of separability structures, 
and briefly describe a recent result \cite{SergyeyevKrtous:2008} on symmetry operators for the Klein--Gordon equation in the canonical background.    

\subsection{Separability structures} 
\label{separability structures} 

The separation of variables for the Hamilton--Jacobi equation in any number of spacetime dimensions allows 
a geometric characterization described by the theory of 
{\em separability structures}, see, e.g., \cite{BenentiFrancaviglia:1979}, \cite{BenentiFrancaviglia:1980}, \cite{DemianskiFrancaviglia:1980}, \cite{KalninsMiller:1981}. Let us briefly recall the main results of this theory.

{\em Separability structures} are classes of separable charts for which the Hamilton--Jacobi equation allows an additive separation of variables.
For each separability structure there exists such a family of separable coordinates
which admits a maximal number of, let us say $r$, ignorable coordinates.
Each system in this family is called a {\em normal separable system} of coordinates. 
The corresponding separability structure is denoted by $\delta_r$.
Its existence is governed by the following central theorem:

{\bf Theorem.}{ A manifold $(M^D, \tens{g})$ admits a $\delta_r$-separability structure if and only if it admits $r$ commuting Killing vectors $\tens{X}_{(k)}$ $(k=0,\dots,r-1)$ and $D-r$ Killing tensors $\tens{K}_{(\alpha)}$ $(\alpha=0,\dots,D-r-1)$, all of them independent, which satisfy:\\
(i)\ in the Lie algebra of Killing tensors with Schouten--Nijenhuis brackets the commutation relations
\ba\label{PoissonK}
\bigl[K_{(\alpha)},K_{(\beta)}\bigr]^{abc}_{\rm SN}\!&\equiv&\!  
K_{(\alpha)}^{e(a}\, \nabla_{e} K_{(\beta)}^{bc)}- 
K_{(\beta)}^{e(a}\, \nabla_{e} K_{(\alpha)}^{bc)}=0\;,\label{KKSN2}\\
\bigl[X_{(k)},K_{(\beta)}\bigr]^{ab}_{\rm SN}\!&\equiv&\!  
\mathcal{L}_{X_{(k)}} K_{(\alpha)}^{ab}=0\,,\label{XKSN2}
\ea
(ii)\ the Killing tensors $\tens{K}_{(\alpha)}$ have in common $D-r$ eigenvectors $\tens{X}_{(\alpha)}$
such that  
\begin{equation}
[\tens{X}_{(\alpha)},\tens{X}_{(\beta)}]=0\,,\quad 
[\tens{X}_{(\alpha)},\tens{X}_{(k)}]=0\,,\quad 
\tens{g}(\tens{X}_{(\alpha)},\tens{X}_{(k)})=0\,.
\end{equation}
}
Let us mention two implications of this theory (we refer to the original publications for more details). 

(1) {\em The existence of 
separability structure implies complete integrability of geodesic motion.}
Indeed, the requirement of independence means
that $r$ linear in momenta constants of motion $c_{(k)}$ associated with Killing vectors $\tens{X}_{(k)}$ and $(D-r)$ quadratic in momenta constants of motion $c_{(\alpha)}$ corresponding to Killing tensors $\tens{K}_{(\alpha)}$ are functionally independent.
Moreover, equations \eqref{KKSN2}, \eqref{XKSN2}, are equivalent to (see also Section 5.5)
\begin{equation}
 \{c_{(\alpha)},c_{(\beta)}\}=0\;,\quad
  \{c_{(k)},c_{(\alpha)}\}=0\;,
\end{equation}
which, together with  $\{c_{(k)},c_{(l)}\}=0$ (following from the commutativity of Killing vectors), implies that all these $D$ constants are in involution and hence the motion is completely integrable. In particular, this means that the proved separability of the Hamilton--Jacobi equation establishes an independent proof of complete integrability of geodesic motion, demonstrated in Chapter 5.
  
(2) $D$ vectors $\{\tens{X}_{(\alpha)},\tens{X}_{(k)}\}$ form  a natural basis $\{\tens{\partial}_x\}$ associated with normal separable coordinates $x^a$.
This allows to write down the {\em canonical metric element} of the `separable' spacetime, e.g., \cite{BenentiFrancaviglia:1980}. The authors of \cite{HouriEtal:2008a}, \cite{HouriEtal:2007} 
used this fact to prove that the existence of a PCKY tensor, obeying the assumptions  \eqref{Liexihg},
restricts the metric of the spacetime to the canonical form \eqref{metric_coordinates}. Their proof consists of showing that the tower of Killing tensors $\tens{K}^{(j)}$ together with the tower of Killing vectors $\tens{\xi}^{(k)}$ generated by a PCKY tensor (see Chapter 3) obey all the requirements of the above theorem. Hence, the Hamilton--Jacobi equation in a spacetime admitting this tensor is separable. The corresponding canonical element turns out to be the off-shell spacetime \eqref{metric_coordinates}. 
In Chapter 7, we show that \eqref{metric_coordinates}
follows from the existence of a PCKY tensor directly, without referring to the theory of separability structures, and even without imposing conditions  \eqref{Liexihg}.

Let us finally mention another theorem which relates the (additive) separability of the 
Hamilton--Jacobi equation with the (multiplicative) separability of the Klein--Gordon equation (see, e.g., \cite{BenentiFrancaviglia:1979}). \\
{\bf Theorem.} The Klein--Gordon equation allows a multiplicative separation of variables if and only if the manifold
$(M^D, \tens{g})$ possesses a separability structure in which the vectors $\tens{X}_{(\alpha)}$ are eigenvectors of the Ricci tensor.\\
{\bf Corollary.} If the manifold is an Einstein space, the Hamilton--Jacobi equation is separable if and only if the same holds for the Klein--Gordon equation.

This corollary explains why, after separating the Hamilton--Jacobi equation, we were able to separate also the Klein--Gordon equation. Another explanation is in the following subsection.

\subsection{Symmetry operators}
After the Hamilton--Jacobi and Klein--Gordon equations in the canonical background were separated \cite{FrolovEtal:2007}, it turned out that 
one can construct symmetry operators for these equations \cite{SergyeyevKrtous:2008}, which invariantly characterize such a separability. Following closely 
the latter paper, let us recapitulate this connection.
 
Following the `first quantization rule',  
$p_a\to -i\alpha \nabla_a$, where $\alpha$ is some scaling constant,
one can consider the operator counterparts of the conserved quantities \eqref{psikappa}.
So, one can introduce the operators 
[cf. Eqs. \eqref{opK}]
\ba
{\hat \xi}_{(k)}&=&-i \alpha \xi^{(k)a}\partial_{a}\,,\quad k=0,\ldots,m\, ,\n{LLL}\\
{\hat K}_{(j)}&=&-\frac{\alpha^2}{\sqrt{|g|}}\,
\partial_{a}\Bigl(\sqrt{|g|}K_{(j)}^{a b}\partial_{b}\Bigr)\,,\quad j=0,\ldots,n-1\, .
\n{KKK}
\ea
 It was proved in \cite{SergyeyevKrtous:2008} that all these operators 
mutually commute in the canonical background \eqref{metric_coordinates}. 
This means that there exist joint eigenfunctions, modes $\Phi$, 
specified by the eigenvalues $\Psi_k$ and $\kappa_j\,$, so that
\be\label{eigen_problem_constants}
{\hat \xi}_{(k)}\Phi=\Psi_k\Phi\,,\quad 
{\hat K}_{(j)}\Phi=\kappa_j \Phi\, .
\ee
In the background \eqref{metric_coordinates} these equations 
allow a separated solution 
\be\label{Phi_alpha}
\Phi=\prod_{\mu=1}^nR_{\mu}(x_{\mu})\prod_{k=0}^{m}e^{\frac{i}{\alpha}\Psi_k\psi_k},
\ee
provided that functions $R_\mu(x_\mu)$ obey the ordinary differential equations
\cite{SergyeyevKrtous:2008}
\be\label{Rmu_alpha}
\alpha^2\bigl(X_{\mu}R_{\mu}'\bigr)'+\eps\alpha^2\frac{X_{\mu}}{x_{\mu}}R_{\mu}'+
\Bigl(V_\mu-\frac{W_\mu^2}{X_\mu}\Bigr)R_\mu =0\;.
\ee

In particular, since $\hat {K}_{(0)}=-\alpha^2 \Box\,,$ 
modes $\Phi$ \eqref{Phi_alpha} are solutions of 
the Klein--Gordon equation
\be
\bigl(\alpha^2\Box +\kappa_0\bigr)\Phi=0\,,
\ee 
and $\{ {\hat \xi}_{(k)}, {\hat K}_{(j)}\}$ form a complete 
set of commuting {\em symmetry operators} for this equation (see, e.g.,  \cite{Miller:book1977}, \cite{FushchichNikitin:book} and references therein).

The constants of separation for the Klein--Gordon equation are related to the constants obtained for the Hamilton--Jacobi equation.
Writing the solution of Eqs. \eqref{eigen_problem_constants} in the form 
\be
\Phi=A\exp\Bigl(\frac{i}{\alpha}S\Bigr)\,,
\ee
one obtains in the {\em geometric optics approximation}, $\alpha\to 0$, a new set of equations
\be\label{problem_approx}
{\xi}_{(k)}^a \partial_a S=\Psi_k\,,\quad 
{K}_{(j)}^{ab}\partial_a S\partial_b S=\kappa_j\, .
\ee
One can easily recognize the Hamilton--Jacobi equation \eqref{HJ} and, upon identifying 
$\partial_a S$ with $p_a$, the separation constants \eqref{pj_sep} and \eqref{kappak_sep}. Moreover, substituting $R_\mu=\exp\bigl(\frac{i}{\alpha}S_\mu\bigr)$ into 
\eqref{Phi_alpha} yields an additive separation ansatz \eqref{sepS}, and 
in geometric optics approximation Eq. \eqref{Rmu_alpha}  gives directly Eq. \eqref{Scond}.  

\section{Discussion}

In this chapter we have demonstrated the separability of the Hamilton--Jacobi and the scalar field equations in the general canonical spacetime \eqref{metric_coordinates}. 
This allows one to study the particle and light propagation in completely general 
higher-dimensional rotating black hole spacetimes, or to calculate the
contribution of a scalar field to the bulk Hawking radiation of these black holes.

These results are very promising and might suggest that also the equations with spin can
be in this background decoupled and separated. 
In fact, the separability of the massive Dirac equation was already demonstrated \cite{OotaYasui:2008}.
We expect that, similar to the 4-dimensional case \cite{CarterMcLenaghan:1979}, \cite{KamranMcLenaghan:1984},
in higher dimensions also the separability of the Dirac equation can be characterized by the corresponding symmetry operators. These operators are well known \cite{BennCharlton:1997}, \cite{Cariglia:2004}.

An important open question is a separability problem for the electromagnetic 
and the gravitational perturbations in higher-dimensional
black hole spacetimes.  A certain progress in this direction was achieved
recently (see, e.g., \cite{KodamaIshibashi:2003}, \cite{KunduriEtal:2006}, \cite{MurataSoda:2008}).
These results are very important for the study of the stability of such
black holes and different aspect of the Hawking radiation produced by
them. Another important direction of research is to study the
quasinormal modes in higher dimensions.
The results obtained in these directions so far 
restricts mainly to non-rotating (or not generally rotating) black holes
(see, e.g., \cite{IdaEtal:2003}, \cite{CardosoEtal:2003a}, \cite{CardosoEtal:2003b}, \cite{CardosoEtal:2003c}, \cite{CardosoEtal:2004}, \cite{Konoplya:2003a}, \cite{Konoplya:2003b}, \cite{CardosoEtal:2004}, \cite{CardosoEtal:2006prl}, 
\cite{Zhidenko:2006}, \cite{KantiEtal:2006}, \cite{LopezOrtega:2006a}, \cite{LopezOrtega:2006b}, \cite{LopezOrtega:2007a}, \cite{KonoplyaZhidenko:2007}, 
\cite{LopezOrtega:2007b}, \cite{Kodama:2007}, \cite{Kodama:2007b}, \cite{CasalsEtal:2008},
\cite{MurataSoda:2008b}, and references therein).

At a first glance it seems that to attack these problems in full generality, for example 
in the way of Teukolsky \cite{Teukolsky:1972}, \cite{Teukolsky:1973}, may not be possible.
On the other hand, it might be useful to study 
the invariant structures determined by the PCKY tensor.
For example, in 4D the method of the Debye potentials \cite{BennEtal:1997} allows one to decouple the electromagnetic perturbations. Unfortunately, this method seems to lie heavily on the self duality property of electromagnetic fields in four dimensions.
Another starting point could be to search for the analogues of the 4-dimensional symmetry operators (see, e.g., \cite{Kamran:1985}, \cite{KalninsEtal:1986}, \cite{KalninsMiller:1989}, \cite{KalninsWilliams:1990}, \cite{KalninsEtal:1992}, \cite{KalninsEtal:1996}) which invariantly characterize the separability of field equations
with spin in the Kerr background.  

It is an important open question to ask whether the existing symmetry connected
with the towers of hidden symmetries generated by the PCKY tensor
is enough to enable the decoupling and separation of the higher spin fields equations.

\chapter[Canonical metric and Kerr-NUT-(A)dS uniqueness]
{Canonical metric and Kerr-NUT-(A)dS uniqueness}
\label{ch8}
\chaptermark{Kerr-NUT-(A)dS uniqueness}

In previous chapters we have seen that the general off-shell metric element
\eqref{metric_coordinates} admits the PCKY tensor $\tens{h}$, \eqref{PCKY_metric}, from which complete integrability of geodesic motion and separability of the Hamilton--Jacobi and Klein--Gordon
equations can be derived.
In this chapter we want to address the question of uniqueness and generality of these results. 
This leads us to the study of 
metric elements admitting a PCKY tensor. 
In particular, we demonstrate the following two important results:
First, we establish that the Kerr-NUT-AdS spacetime \eqref{metric_coordinates}--\eqref{X} is the most general solution of the vacuum Einstein equations with the cosmological constant which possesses the  PCKY tensor.
Second, without imposing the Einstein equations, we explicitly derive the canonical form of the metric admitting such a tensor and show that it coincides with the off-shell metric \eqref{metric_coordinates}.
These results naturally generalize 
the results obtained earlier in four dimensions. This chapter is based on \cite{KrtousEtal:2008}.

\section{Uniqueness of the Kerr-NUT-(A)dS spacetime}

In this section we prove that the most general solution of the Einstein equations
with the cosmological constant which admits a PCKY tensor is the Kerr-NUT-(A)dS spacetime \eqref{metric_coordinates}--\eqref{X}. 
Instead of giving a formally organized proof we shall deduce this statement 
from filling two missing pieces in the mosaic of already known facts.
Namely, it was demonstrated in  \cite{HouriEtal:2008a}, \cite{HouriEtal:2007} that  
the Kerr-NUT-(A)dS spacetime is the only Einstein space admitting the PCKY tensor
obeying the additional restrictions 
\be\label{Liexihg2}
  \lied_\xi \tens{h}=0\,,\quad   \lied_\xi \tens{g}=0\,.
\ee
Here we prove that both these conditions already follow 
from the existence of the PCKY tensor, together with the restriction 
on a vacuum solution of the Einstein equations with the cosmological constant.  

\subsection{Condition on the PCKY tensor}
In this subsection we concentrate on the first condition in \eqref{Liexihg2}.
At first we consider the case of an even dimension, $D=2n$, and then briefly discuss what happens in the odd-dimensional case. 
Besides the canonical basis (Section 3.1.2), it is convenient to introduce also a basis of complex null eigenvectors, $\{\tens{m}_{\hat \mu},\tens{\bar m}_{\hat \mu}\}$, defined by the relations\footnote{%
The eigenvectors of the PCKY tensor play a special role.
One can prove that they coincide with the {\em principal null directions} 
(see Appendix C.1).
}  
\be\label{eveqs}
\PKY \cdot \muv{\mu}\equiv -ix_{\mu} \muv{\mu}\,,\quad 
\PKY \cdot \mbv{\mu}\equiv ix_{\mu} \mbv{\mu}\,.
\ee
Here, $\PKY$ is the operator with components $h^{a}{}_{b}\,$,
bar denotes the complex conjugation, and 
$x_\mu$'s describe the eigenvalues of $\PKY$ [cf. Eq. \eqref{hab}].\footnote{%
In this chapter we do not have a fixed background; we are constructing  the 
metric.
It is therefore necessary to distinguish various positions of indices. 
In particular, we denote by $\tens{h}$ a PCKY 2-form, whereas by $\PKY$ a PCKY operator $h^a{}_b$.
}     
The complex null vectors satisfy the normalization
\be\label{mnorm}
\muv{\mu}\cdot\muv{\nu}=\mbv{\mu}\cdot\mbv{\nu}=0\,,\quad 
\muv{\mu}\cdot\mbv{\nu}=\delta_{\mu\nu}\,.
\ee
They are connected with the canonical vectors 
$\{\tens{e}_{\hat \mu}, \tens{\tilde e}_{\hat \mu}\}$ as follows:
\begin{equation}\label{mdef}
  \muv{\mu}=\frac{1}{\sqrt{2}}(\tens{\tilde e}_{\hat \mu}+i\tens{e}_{\hat \mu})\,,\quad 
  \mbv{\mu}=\frac{1}{\sqrt{2}}(\tens{\tilde e}_{\hat \mu}-i\tens{e}_{\hat \mu})\,.
\end{equation}
Let us further denote $\Du{\mu}\equiv \nabla_{m_\mu}$ and
$\Db{\mu}\equiv \nabla_{{\bar m}_{\hat \mu}}$. Using the PCKY Eq. \eqref{PCKY} one has
\begin{equation}\label{Dh}
(\Du{\mu}\PKY)\cdot \muv{\nu}=(\muv{\nu}\cdot \PKV)\,\muv{\mu}\,.
\end{equation}
Applying ${\Du{\mu}}$ to \eqref{eveqs} and using \eqref{Dh} one obtains
\begin{equation}\label{eqn}
(\PKY+i x_{\nu}\tens{\delta})\cdot\Du{\mu}\muv{\nu}
+ i (\Du{\mu} x_{\nu})\, \muv{\nu}
+(\muv{\nu}\cdot\tens{\xi}) \, \muv{\mu}=0\,.
\end{equation}
By taking a scalar product of \eqref{eqn} with $\mbv{\nu}$,
using antisymmetry of $\tens{h}$ and Eq.~\eqref{eveqs} again, 
the first term cancels out. 
Considering two cases when $\nu=\mu$ and when $\nu\ne \mu$ one gets
\begin{equation}\label{Dmx}
\Du{\mu}x_{\nu}=0\quad\text{for}\; \nu\ne\mu\,,\quad 
\Du{\mu}x_{\mu}=i \muv{\mu}\cdot \tens{\xi}\,.
\end{equation}
Let us define functions ${Q_\mu}$ in terms of magnitudes of complex quantities ${\Du{\mu}x_\mu}\,$:
\begin{equation}\label{Qdef2}
Q_{\mu}\equiv 2|\Du{\mu}x_{\mu}|^2\,,\quad 
\Du{\mu}x_{\mu}=\frac{1}{\sqrt{2}}\sqrt{Q_{\mu}}\,e^{i\alpha}\,.
\end{equation}
As mentioned in Section 3.1.2, the canonical basis is not fixed
by Eqs. \eqref{gab} and \eqref{hab} uniquely. There remains
a freedom of a rotation in each KY 2-plane, $\tens{\omega}^\mu\wedge \,\tens{\tilde \omega}^\mu$, which in terms
of the null basis  \eqref{mdef} reads
${\muv{\mu}\to \exp(i\varphi_\mu)\muv{\mu}}$. We uniquely 
fix the canonical basis by setting the phase factor
${\alpha=\pi/2}$. Then, we have
\begin{equation}\label{eqn1}
\Du{\mu}x_{\mu}=\frac{i}{\sqrt{2}}\,\sqrt{Q_{\mu}}\,.
\end{equation}
Using \eqref{Dmx} and \eqref{eqn1} we find
\begin{equation}\label{xiform}
\tens{\xi} = \frac{1}{\sqrt{2}}\sum_\mu \sqrt{Q_\mu}\,(\muv{\mu}+\mbv{\mu})=
  \sum_\mu\sqrt{Q_\mu}\,\tens{\tilde e}_{\hat \mu}.
\end{equation}
Eqs.~\eqref{Dmx} and \eqref{eqn1} also give us that
the gradient ${\grad x_\mu}$ of the eigenvalue ${x_\mu}$ is proportional
to $\tens{\omega}^{\hat \mu}$,
\begin{equation}\label{dxenf}
  \grad x_\mu = \sqrt{Q_\mu}\,\tens{\omega}^{\hat \mu}\,.
\end{equation}
A simple calculation employing Eqs.~\eqref{gab}, \eqref{xiform} 
and \eqref{dxenf} shows that
\begin{equation}\label{xih}
  \tens{\xi}\cdot\tens{h}= -\sum_\mu x_\mu\sqrt{Q_\mu}\,\tens{\omega}^{\hat \mu}
    =\grad\Bigl(-\frac12\sum_\mu x_\mu^2\Bigr)\,.
\end{equation}
With the help of the fact that this 1-form is exact and using 
the closeness of $\tens{h}$ we immediately obtain the desired relation
\begin{equation}\label{Lixihqed}
\lied_\xi \tens{h} = \tens{\xi}\cdot \tens{d h}
+ \grad (\tens{\xi}\cdot\tens{h}) = 0\,.
\end{equation}

In an odd-dimensional case, equipped with the extra direction $\tens{e}_{\hat 0}$,
we have besides \eqref{eveqs} also an additional equation 
\be\label{he0}
\PKY \cdot \tens{e}_{\hat 0}=0\,.
\ee 
Let us denote $D_{\hat 0}\equiv \nabla_{e_{\hat 0}}$, and apply this operator on  
\eqref{eveqs}. Proceeding analogously to the derivation of \eqref{Dmx} we obtain
\be
D_{\hat 0}x_\nu=0\,,
\ee 
and therefore Eq. \eqref{dxenf} still holds. Moreover, denoting by 
\be
\tens{e}_{\hat 0}\cdot \PKV\equiv \sqrt{-\frac{c}{A^{(n)}}}\,,
\ee
we obtain the expression for $\PKV$, valid in any dimension $D$,
\be
\tens{\xi} = \sum_\mu\sqrt{Q_\mu}\,\tens{\tilde e}_{\hat \mu}+\eps
\sqrt{-\frac{c}{A^{(n)}}}\tens{e}_{\hat 0}\,.
\ee
Using \eqref{he0}, we finally find that Eqs. \eqref{xih} and \eqref{Lixihqed} remain unchanged.

\subsection{Killing vector condition}
The second condition in Eq.~\eqref{Liexihg2}, which 
states that $\tens{\xi}$ is a Killing vector, is automatically satisfied in any
Einstein space. Indeed, it was demonstrated in \cite{Tachibana:1969} (see also Appendix C.2) that 
\begin{equation}\label{Snablaxi}
 \nabla_{(a} \xi_{b)} = \frac{1}{D-2}\,R_{n(a} h_{b)}{}^{n}\,.
\end{equation}
For spaces obeying the vacuum Einstein equations with the cosmological constant
we have the Ricci tensor proportional to the metric  and thanks to the antisymmetry of $\tens{h}$ we immediately get ${\nabla_{(a}\xi_{b)} = 0}$, that is 
${\lied_{\xi} \tens{g}=0}$.

Thus, when the vacuum Einstein equations with the cosmological constant are imposed both conditions \eqref{Liexihg2} are valid and using
the results of \cite{HouriEtal:2008a}, \cite{HouriEtal:2007} one can derive that the metric has to be the Kerr-NUT-(A)dS spacetime \eqref{metric_coordinates}--\eqref{X}.

\section{Canonical metric element}
In this section we explicitly construct the {\em canonical metric} admitting the PCKY tensor. Namely, we show that the most general metric element admitting this tensor is the off-shell metric \eqref{metric_coordinates}.
Our demonstration extends the result of \cite{HouriEtal:2008a}, \cite{HouriEtal:2007} where it was proved provided the additional conditions \eqref{Liexihg2} and with the help of the theory 
of separability structures (see Section 6.3.1).
Let us emphasize that, contrary to the previous section, we work off-shell, 
that is without imposing the Einstein equations. 

It might seem that an obvious path to follow is to prove that 
(yet off-shell) both conditions \eqref{Liexihg2} can still be derived from the very existence of the PCKY tensor and then use the result of \cite{HouriEtal:2008a}, \cite{HouriEtal:2007}.
In fact, going through the proof of the previous section, we realize that the first 
condition is indeed satisfied off-shell. However,
it is not a straightforward task to prove 
the second condition \eqref{Liexihg2} without imposing the Einstein equations.
Therefore, instead we proceed in a different way.
We explicitly demonstrate that besides $n$ natural coordinates\footnote{%
Let us remind that it is a part of the definition of the PCKY tensor
that its eigenvalues ${x_\mu}$ are functionally independent in some spacetime domain.
This means that ${x_\mu}$'s are non-constant, independent, scalar functions
with different values at a generic point of the manifold and one can use them as natural coordinates.
}
$x_\mu$, associated with the eigenvalues of $\tens{h}$,
it is possible to introduce $n+\eps$ additional coordinates
$\psi_{k}$, associated with the tower of vectors generated from $\tens{h}$, \eqref{Primary}--\eqref{xin}, so that the metric and the PCKY tensor take the form \eqref{metric_coordinates} and \eqref{PCKY_metric}, respectively.
Here we sketch only main steps of the derivation and 
for simplicity restrict to an even dimension ${D=2n}$. 
Technical details of this construction, including the odd-dimensional case, 
are in preparation \cite{KrtousEtal:2008b}.

First, taking all projections of equation \eqref{eqn}, 
we collect a partial information about the Ricci coefficients. For example, we obtain
that only those Ricci coefficients with at least two indices equal are nonvanishing.
Next, using ${\PKV\cdot\grad x_\mu=0}$ we can calculate the Lie derivative of $\tens{e}_{\hat \mu}$
in terms of function $q_\mu$\,,
\be
{q}_\mu \equiv \PKV\cdot \grad\bigl[\ln(\sqrt{Q_\mu})\bigr]\,. 
\ee
Using duality relations
and action of the PCKY tensor we find
\begin{equation}\label{Lieev}
  \lied_\xi \env{\mu} = {q}_\mu \env{\mu} + \sum_\nu E^{\nu}_{\mu}\,\tens{\tilde e}_{\hat \nu}\,,\quad  \lied_\xi \tens{\tilde e}_{\hat \mu} = -{q}_\mu \tens{\tilde e}_{\hat \mu}\,,
\end{equation}
where ${E^{\nu}_{\mu}}$ are yet unspecified.
Expressing these Lie derivatives using covariant derivatives gives
an additional information about the Ricci coefficients
and determines ${E^{\nu}_{\mu}}$ in terms of 
the Ricci coefficients and derivatives of ${Q_\mu}$. 
It also guarantees that ${\tens{\tilde e}_{\hat \nu}\cdot\grad Q_\mu=0}$ for ${\mu\ne\nu}$
and ${{q}_\mu=\tens{\tilde e}_{\hat \mu}\cdot\grad\sqrt{Q_\mu}}$. These facts allow
us to calculate the Lie brackets among all the vectors ${\env{\mu},\ehv{\mu}}$ of the canonical basis.
They do not commute, with the exception: ${[\ehv{\mu},\ehv{\nu}]=0}$.

Now, we introduce a new basis ${\{\cxv{\mu},\chv{k}\}}$, ${\mu=1,\dots,n}$, ${k=0,\dots,n-1}$,
\begin{equation}\label{cvframe}
  \cxv{\mu} = \frac1{\sqrt{Q_\mu}}\,\env{\mu}\,,\quad 
  \chv{k} = \sum_\mu A^{(k)}_\mu\sqrt{Q_\mu}\,\ehv{\mu}\,,
\end{equation}
with ${A^{(k)}_\mu}$ given by \eqref{Aj}.
The meaning of the basis vectors ${\chv{k}}$ is elucidated by observing that they coincide with the vector fields $\tens{\xi}_{(k)}$, \eqref{Primary}--\eqref{etaj},
generated from the PCKY tensor.

Using the known Ricci coefficients and the Jacobi identity 
we can prove that vectors of this frame do commute,
\begin{equation}\label{LieBrcframe}
  [\cxv{\mu},\cxv{\nu}]=[\cxv{\mu},\chv{k}]=[\chv{k},\chv{l}]=0\,.
\end{equation}
Moreover, for the dual frame 
\begin{equation}\label{cfframe}
  \cxf{\mu} = \sqrt{Q_\mu}\, \tens{\omega}^{\hat \mu} = \grad x_\mu\,,\quad 
  \chf{k} = \sum_\mu \frac{(-x_\mu^2)^{n{-}1{-}k}}{U_\mu\sqrt{Q_\mu}}\,\tens{\tilde \omega}^{\hat \mu}
\end{equation}
we show
\begin{equation}\label{gradcframe}
  \grad\cxf{\mu}=0 \,,\quad  \grad \chf{k} = 0\,.
\end{equation}
Both conditions \eqref{LieBrcframe} and \eqref{gradcframe}
ensure that additionally to ${x_\mu}$, ${\mu=1,\dots,n}$,
it is possible to introduce coordinates ${\psi_k}$, ${k=0,\dots,n-1}$,
such that
\begin{equation}\label{coors}
\begin{gathered}
  \cxv{\mu}=\pa_{x_\mu}\,,\quad \chv{k} = \pa_{\psi_k}\,,\qquad
  \cxf{\mu}=\grad x_\mu\,,\quad \chf{k} = \grad \psi_k\,.
\end{gathered}
\end{equation}
Taking into account the inverse of Eqs.~\eqref{cfframe} we get
\begin{equation}\label{efcoor}
  \tens{\omega}^{\hat \mu}=\frac{1}{\sqrt{Q_\mu}}\,\grad x_\mu\,,\quad 
  \tens{\tilde \omega}^{\hat \mu}=\sqrt{Q_\mu}\,\sum_{k=0}^{n-1}A^{(k)}_\mu\, \grad\psi_k\,,
\end{equation}
which coincides with the basis 1-forms of the orthonormal form of the 
metric \eqref{omega}, with unspecified metric functions ${Q_\mu}$.
However, in the process, we also learn that metric functions ${Q_\mu}$
must take the form \eqref{Q}, particularly that ${{q}_\mu=0}$ and 
${E^{\nu}_{\mu}=0}$. 
This finishes the proof of our main result: we have constructed a coordinate system 
in which the canonical metric element admitting the PCKY tensor takes the
off-shell form \eqref{metric_coordinates}.
Let us emphasize that this result was achieved without imposing the Einstein equations, starting only from the quantities determined by the PCKY tensor.
As a corollary of this construction, we have established that
all the vectors $\tens{\xi}_{(k)}$ are Killing vectors. 

Let us finally mention, that very recently the authors of \cite{HouriEtal:2008b}, \cite{HouriEtal:2008c} were able to construct 
the most general metric element admitting a closed CKY 2-form. Such a 2-form, besides the functionally independent eigenvalues, may also admit the constant eigenvalues.\footnote{This, in particular, incorporates the case of a covariantly constant 
PCKY tensor, that is a PCKY tensor for which the primary vector $\tens{\xi}$ vanishes. Such a tensor possesses only the constant eigenvalues.
} 
The key observation for the construction is the fact that 
Eq. \eqref{PCKY} for a closed CKY 2-form forbids the possibility of degenerate non-constant eigenvalues, that is, the Darboux subspaces corresponding to the non-constant eigenvalues are always 2-dimensional.
This means, that with respect to the functionally independent eigenvalues the metric 
`behaves' as the canonical spacetime for the PCKY tensor, and one has to find 
only the `trivial' part, corresponding to the constant eigenvalues.
The resulting canonical element turns out to be the `generalized Kerr-NUT-(A)dS spacetime' \cite{HouriEtal:2008b}, \cite{HouriEtal:2008c}, or more precisely, the Kaluza--Klein metric 
on the bundle over K\"ahler manifold whose fibres are canonical metric elements described above. These results complete the classification of all spacetimes admitting a closed CKY 2-form.


\part[Further Developments]{Further Developments}

\chapter{Stationary strings and branes}

In this chapter we demonstrate complete integrability of the Nambu--Goto equations
for a stationary string in the canonical spacetime \eqref{metric_coordinates}.  The stationary string in $D$ dimensions is generated by
a 1-parameter family of Killing trajectories and the problem of finding
a string configuration reduces to a problem of finding a
geodesic line in an effective $(D-1)$-dimensional space. Resulting integrability of 
this geodesic problem is connected with the existence of hidden symmetries 
which are inherited from the black hole background.
More generally, in a spacetime with $p$ mutually commuting Killing vectors it is possible to introduce a concept of a $\xi$-brane, that is a $p$-brane with the worldvolume generated by these fields and a 1-dimensional curve. We
discuss conditions of the integrability of such $\xi$-branes in the Kerr-NUT-(A)dS spacetime
\eqref{metric_coordinates}--\eqref{X}. This chapter is based on \cite{KubiznakFrolov:2008}.

\section{Introduction} 
There are several reasons why the problem of interaction of 
strings and branes with black holes attracted interest recently.
Fundamental strings and branes are basic objects in string theory
\cite{Polchinski:1998}, and black holes (as well as other black objects) form an
important class of solutions of the low-energy effective action in
this theory (see, e.g., \cite{Ortin:2004}). On the other hand, cosmic strings
and domain walls are topological defects which can be naturally
created during phase transitions in the early Universe (see, e.g.,
\cite{VilenkinShellard:1994}, \cite{Polchinski:2004}, \cite{DavisKibble:2005}). Their interaction with astrophysical black holes may
result in interesting observational effects. In both cases we are
dealing with a  problem when the interacting objects are
non-local and relativistic. An important example is an interaction of
a bulk black hole with a brane representing our world in the brane
world models (see, e.g., \cite{EmparanEtal:2000a}, \cite{FrolovEtal:2003}, \cite{FrolovEtal:2004a}, \cite{FrolovEtal:2004b}, \cite{Rodrigo:2006}, \cite{MajumdarMukherjee:2005}). A stationary test brane interacting
with a bulk black hole can be used as a toy model for the study of 
(Euclidean) topology change transitions \cite{Frolov:2006}. This model
demonstrates interesting scaling and self-similarity properties
during such phase transitions, similar to the Choptuik critical
collapse  \cite{Choptuik:1993} and merger black hole transitions \cite{Kol:2006}, \cite{AsninEtal:2006}. These
models may also have far going interesting consequences for the study of
phase transitions in quantum chromodynamics (see, e.g., \cite{MateosEtal:2006}, \cite{KobayashiEtal:2007}, \cite{AlbashEtal:2008}, \cite{HoyosEtal:2007}).  

Even in an idealized case, when one neglects the effects connected
with the thickness of the strings and branes and their tension, this
problem is quite complicated. The reason is evident: the
Dirac--Nambu--Goto action for these objects in an external gravitational
field is very nonlinear. In a general case numerical calculations
are required (see, e.g., \cite{SnajdrEtal:2002}, \cite{SnajdrFrolov:2003}, \cite{DubathEtal:2007}). When the effects of
thickness and tension are taken into account these numerical
calculations become even more involved (see, e.g., \cite{MorisawaEtal:2000}, \cite{MorisawaEtal:2003}, \cite{FlachiTanaka:2005}, \cite{FlachiEtal:2006}, \cite{FlachiTanaka:2007}).

Study of {\em stationary} configurations of strings and branes in a
background of a stationary black hole is simpler problem which
in several cases allows complete solution. One of the examples
is a stationary string in the Kerr spacetime. It was shown
\cite{FrolovEtal:1989} that the Hamilton--Jacobi equation for such a string
allows a complete separation of variables. It was also demonstrated 
\cite{CarterFrolov:1989}, \cite{CarterEtal:1991} that this property is directly connected with the
hidden symmetry of the Kerr metric generated by the Killing tensor
\cite{WalkerPenrose:1970} discovered by Carter in 1968 \cite{Carter:1968cmp}. More recently,
Carters's method was
applied to 5-dimensional rotating black holes and the Killing tensor
was found in these spacetimes \cite{FrolovStojkovic:2003a}, \cite{FrolovStojkovic:2003b}. This result was used
to show that the equations for a stationary string in
the 5-dimensional Myers--Perry metric 
are completely integrable \cite{FrolovStevens:2004}.

Here we demonstrate that this result allows a
generalization to higher dimensional rotating black holes
in an arbitrary number of spacetime dimensions.
Namely, we show that a stationary string configuration is
completely integrable in the canonical spacetime
\eqref{metric_coordinates}. We use the fact that after performing 
a dimensional reduction along the Killing trajectories, the stationary string equation in a $D$-dimensional stationary spacetime can be reduced to a geodesic
equation in a $(D-1)$-dimensional space with a metric conformal to the
reduced metric. The separability of the Hamilton--Jacobi equation in
this effective metric follows from the separability of the
Hamilton--Jacobi equation in the original $D$-dimensional canonical spacetime 
proved in Chapter 6 and a special
property of the primary (timelike) Killing vector. 

There is a natural generalization of the concept of a stationary
string in the case when there exist several mutually commuting
Killing vectors. If $p$ is a number of these fields  one  may consider
a $(p+1)$-hypersurface generated by the Killing vectors passing
through a 1-dimensional line. We call a {\em $\xi$-brane} an extended
object, a $p$-brane, with the worldvolume associated with this
hypersurface. We discuss integrability conditions for $\xi$-branes in
the Kerr-NUT-(A)dS spacetime \eqref{metric_coordinates}--\eqref{X} and give some examples of integrable systems.

\section{Stationary strings}
Consider a string in a stationary $D$-dimensional spacetime $M^D$. Let
$x^a$ ($a=0,\ldots,D-1$) be coordinates
in it and
\be\n{1_ch8}
ds^2=g_{ab}dx^{a}dx^{b}
\ee
be its metric. We denote by $\xi^{a}$ the corresponding Killing vector
which is timelike at least in some domain of $M^D$. We call the string
{\em stationary} if $\xi^a$ is tangent to the 2-dimensional
worldsheet $\Sigma_{\xi}$ of the string in this domain. In other words, the surface
$\Sigma_{\xi}$ is generated by a 1-parameter family of the Killing
trajectories (the integral lines of $\xi^a$). 

A general formalism for studying a stationary spacetime based on its
foliation by Killing trajectories was developed by Geroch
\cite{Geroch:1971}. In this approach, one considers a set $S$ of the
Killing trajectories as a quotient space and introduce the structure of the differential
Riemannian manifold on it. The projector $h_{ab}$ onto $S$ is related
to the metric $g_{ab}$ as follows:
\be\n{2_ch8}
g_{ab}=h_{ab}+\xi_{a}\xi_b/\xi^2\, .
\ee
In this formalism, a stationary string is uniquely determined by a
curve in $S$. 

The equation for this curve follows from the Nambu--Goto
action 
\begin{equation}
I=-\mu\int d^2\zeta\left|\gamma\right|^{1/2}\,.
\end{equation}
Here $\mu$ is the string tension. As it enters the Nambu--Goto action 
as a common factor, its value is not important and one
can always put $\mu=1$. The string worldsheet can be parametrized by
$x^a=x^a(\zeta^A)$, where $\zeta^A$ 
are coordinates on $\Sigma_{\xi}$, ($A=0,1$). We
denote by $\gamma_{AB}$ the induced metric on $\Sigma_{\xi}$
\be
\gamma_{AB}={\partial x^a\over \partial{\zeta^A}}{\partial x^a\over
\partial{\zeta^B}}\,g_{ab}\, ,
\ee
and by $\gamma$ its determinant.

Let Killing time parameter be $t$, so that
$\xi^a\partial_a=\partial_t$, and let $y^i$ be coordinates which are
constant along the Killing trajectories (coordinates in $S$). Then, 
the non-vanishing components of the projection
operator $h_{ab}$ are $h_{ij}$ (reduced metric) and the metric \eq{1_ch8}-\eq{2_ch8} takes the form
\ba
ds^2\!\!&=&\!\!-F(dt+A_i dy^i)^2+h_{ij}dy^i dy^j\, ,\\
F\!\!&=&\!\!g_{tt}=-\xi_a\xi^a\hh A_i=g_{ti}/g_{tt}\, .
\ea
From \eq{2_ch8} it also follows that in these coordinates $h^{ij}=g^{ij}$.

We choose $\zeta^0=t$ and denote $\zeta^1=\sigma$. Then the string
configuration is determined by $y^i=y^i(\sigma)$. The induced metric is 
\ba
d\gamma^2=\gamma_{AB}d\zeta^Ad\zeta^B=-F(dt+A d\sigma)^2+dl^2\,,
\ea
where 
\be
dl^2=h\, d\sigma^2\,,\quad A=A_i \frac{dy^i}{d\sigma}\,,\quad 
h=h_{ij}{dy^i\over d\sigma}{dy^j\over d\sigma}\,, 
\ee
and it has the following determinant
\be
\gamma=\det(\gamma_{AB})=-F h\, .
\ee
So, the Nambu--Goto action is 
\ba
I&=&-\Delta t E\,,\\
\label{E}
E&=&\!\mu\int\!\sqrt{F}dl=
\mu\int d\sigma
\sqrt{Fh_{ij}\frac{dy^i}{d\sigma}\frac{dy^j}{d\sigma}}\,\,.
\ea
In a static spacetime the equation \eq{E} has a very simple meaning: The energy density of a
string is proportional to its proper length $dl$ multiplied
by the red-shift factor $\sqrt{F}$.

The problem of a stationary string configuration therefore reduces to that of a geodesic in the  $(D-1)$-dimensional 
effective background
\begin{equation}\label{H}
dH^2=H_{ij}dy^i dy^j=F h_{ij} dy^i dy^j\,.
\end{equation}

In order to solve this geodesic problem we shall use the Hamilton-Jacobi method. That is, we 
shall attempt for the additive separation of the Hamilton-Jacobi equation 
\begin{equation}\label{HJ_ch8}
\frac{\partial S}{\partial\sigma}+H^{ij}\, \partial_i S\;\partial_j
S=0\, ,
\end{equation}
where ${H}^{ij}$ is the inverse of the effective metric \eqref{H} with 
the components given by
\begin{equation}\label{hij}
FH^{ij}=h^{ij}=g^{ij}.
\end{equation} 
If the Hamilton--Jacobi equation can be separated, the effective geodesic motion 
and hence also the stationary string configuration are completely integrable,
(see Section 6.3.1).

\section{Stationary  strings in canonical spacetime}
In this section we prove complete integrability of a stationary string configuration
in the canonical spacetime \eqref{metric_coordinates}.
As explained earlier, in such a spacetime the primary Killing vector
$\PKV=\pa_{\psi_0}$ plays a special role. This vector is (after the analytical continuation to the physical domain) timelike in the black hole exterior. It is also the one most `directly connected' with the PCKY tensor.\footnote{%
Let us remark here that the asymmetry among the primary 
Killing vector $\pa_{\psi_0}$ and the secondary Killing vectors  $\pa_{\psi_j}$
can be also viewed as arising from the 
requirement that, in addition to the Hamilton--Jacobi, also the 
Klein--Gordon  equation is separable (see, e.g., \cite{CarterFrolov:1989} and references therein). 
}
We call a string stationary if it is tangent to the primary Killing vector. 
For such string one has the following form of the effective metric:
\ba
F{H}^{ij}\pa_i\pa_j\!\!&=&\!\!\sum_{\mu=1}^n\Bigl[ Q_\mu(\pa_{x_\mu})^2
  +\frac{1}{Q_\mu U_\mu^2}\Bigl(\sum_{k=1}^m(-x_\mu^2)^{n\!-\!1\!-\!k}\pa_{\psi_k}\Bigr)^{\!2}\Bigr]-\frac{\eps}{cA^{(n)}}(\pa_{\psi_n})^2 \;,\nonumber \label{invH}\\
F\!\!&=&\!\!\sum_{\mu=1}^n Q_\mu -\frac{\eps c}{A^{(n)}}\,. \label{F_ch8}
\ea
The expression is similar to \eq{inv_coords}, with the only difference 
that in the sum over $k$ the term $k=0$ is omitted. This corresponds to the natural projection given by \eqref{hij}.

In the background of the metric $H_{ij}$ the Hamilton--Jacobi equation
\eq{HJ_ch8} allows the additive  separation of variables 
\begin{equation}\label{sep_ch8}
S=w\sigma + \sum_{\mu=1}^n S_{\mu}(x_{\mu})+ \sum_{k=1}^{m} L_k\psi_k
\end{equation}
with functions ${S_\mu(x_\mu)}$ of a single argument ${x_\mu}$. 
Substituting \eq{sep_ch8} into \eq{HJ_ch8} we obtain
\be\label{HJsep_ch8}
Fw+\sum_{\mu=1}^{n}\Bigl[Q_{\mu}S_\mu'^2+\frac{1}{X_\mu U_\mu }
\Bigl(\sum_{k=1}^m(-x_\mu^2)^{n-1-k}L_k\Bigr)^{\!2}
\Bigr] -\frac{\eps L_n^2}{cA^{(n)}}=0\,,
\ee
where ${S_\mu}'^\;$ denotes the derivative of function ${S_\mu}$
with respect to its single argument ${x_\mu}$. Using the explicit
form of $F$ and algebraic identity 
\begin{equation}\label{Uids}
\frac1{A^{(n)}}=\sum_{\mu=1}^n \frac1{x_\mu^2U_\mu}\,,
\end{equation}
we can rewrite the last equation in the form
\begin{equation}\label{KGsep_ch8}
\sum_{\mu=1}^n\frac{G_{\mu}}{U_{\mu}}=0\,,
\end{equation}
where $G_{\mu}$ are functions of $x_{\mu}$ only:
\be\label{Gmu}
G_{\mu}=X_{\mu}\left(S_{\mu}'^2+w\right)+\frac{1}{X_{\mu}}
\Bigl(\sum_{k=1}^m(-x_\mu^2)^{n\!-\!1\!-\!k}L_k\Bigr)^{\!2}-\eps\,\frac{L_n^2/c+wc}{x_{\mu}^2}\,.
\ee
Applying Lemma 2 of Appendix C.4, we write the general solution 
of \eqref{KGsep_ch8} as  
\begin{equation}
G_{\mu}=\sum_{k=1}^{n-1} c_{k}(-x_\mu^2)^{n\!-\!1\!-\!k}\;,
\end{equation}
where $c_k$ are arbitrary constants. 
So, we have obtained the equations for $S_{\mu}'$,
\be\label{Scond_ch8}
S_\mu'^2=\frac{1}{X_\mu}\Bigl[\sum_{k=1}^{n-1} c_k\, (-x_\mu^2)^{n-1-k}+\eps\,
\frac{L_n^2/c+wc}{x_{\mu}^2}\Bigr]-
\frac1{X_\mu^2}\Bigl(\sum_{k=1}^m\bigl(-x_{\mu}^2\bigr)^{n-1-k}
L_k\Bigr)^{\!2}-w\, ,
\ee 
which can be solved by quadratures.

This completes the demonstration that in the canonical background 
\eqref{metric_coordinates} the reduced $(D-1)$-dimensional geodesic  problem \eqref{E} allows the separation of the Hamilton--Jacobi equation \eq{HJ_ch8} and therefore
the stationary string configuration is completely integrable.

\section{Inherited hidden symmetries}
The resulting complete integrability of the stationary string configuration
in the canonical spacetime \eqref{metric_coordinates} is connected with 
the existence of hidden symmetries of the 
$(D-1)$-dimensional effective metric $H_{ij}$.
Namely, there exist $(n-1)$ irreducible Killing tensors $C^{ij}_{(k)}\,,$
$(k=1, \dots, n-1)$, which 
give the constants of motion 
\begin{equation}
c_k=C_{(k)}^{ij}p_ip_j\,,
\quad D_{\!(m}C^{(k)}_{ij)}=0\,,
\end{equation}
and allow the separation of the Hamilton--Jacobi equation \eq{HJ_ch8}
in the background $H_{ij}$. In the last formula $p_i=\partial_i S$
are the `momenta of geodesic motion'
and $D_i$ denotes the covariant derivative with respect to  
$H_{ij}$.

Similar to Chapter 6, one can easily find the explicit form of $C^{ij}_{(k)}$ by
inverting \eqref{Gmu}. Let us multiply it by 
$A_{\mu}^{(l)}/U_\mu$, sum over $\mu$, and use identities \eqref{pom_ide2}.
Then we obtain
\be
C^{ij}_{(k)}={}^{\tiny (\psi_0)}\!K^{ij}_{(k)}-F_{(k)}H^{ij}\,,\quad 
F_{(k)}\equiv\sum_{\mu=1}^n Q_\mu A_\mu^{(k)}-\eps\,\frac{c
A^{(k)}}{A^{(n)}}\, .
\ee
Here ${}^{\tiny (\psi_0)}\!K^{ij}_{(k)}$ are natural projections of the Killing tensors  \eqref{KTk} for the $D$-dimensional canonical spacetime,
\be\label{KT_ch8}
{}^{\tiny (\psi_0)}\!K^{ij}_{(k)}\pa_i\pa_j=\sum_{\mu=1}^n\Bigl[ A_{\mu}^{(k)}Q_\mu(\pa_{x_\mu})^2
+\frac{A_{\mu}^{(k)}}{Q_\mu U_\mu^2}\Bigl(\sum_{l=1}^m(-x_\mu^2)^{n\!-\!1\!-\!l}\pa_{\psi_l}\Bigr)^{\!2}\Bigr]-\frac{\eps A^{(k)}}{cA^{(n)}}(\pa_{\psi_n})^2 \;.
\ee
Similar to \eqref{invH},  the direction 
$\pa_{\psi_0}$ is projected out.
Therefore, one can say that the hidden symmetries of the 
$(D-1)$-dimensional effective metric $H_{ij}$
are `inherited' from the hidden symmetries  of $g_{ab}$.

A nontrivial property which follows from the separability of the Hamilton--Jacobi equation  
(see Chapter 6) is that the constants $c_k$ mutually Poisson commute,
or equivalently, the Schouten--Nijenhuis brackets, in the background $H_{ij}$, of the corresponding Killing tensors vanish: 
\begin{equation}
\bigl[{C}_{(k)},{C}_{(l)}\bigr]_{H}^{\,{ijm}}\!\!=
C_{(k)}^{n(i}D_n C_{(l)}^{jm)}-
C_{(l)}^{n(i}D_n C_{(k)}^{jm)}=0\,.
\end{equation}
Let us also mention that the objects ${}^{\tiny (\psi_0)}\!K^{ij}_{(k)}$
are the Killing tensors for the reduced metric $h_{ij}$ 
and obey
\begin{equation}
\bigl[{}^{\tiny (\psi_0)}\!{K}_{(k)},{}^{\tiny (\psi_0)}\!{K}_{(l)}\bigr]_{h}^{\,{ijm}}=0\,.
\end{equation}
These results can be easily obtained by separating 
the Hamilton--Jacobi equation in the background of the 
reduced metric $h_{ij}$. We expect them to be more general.
(For a discussion and necessary conditions regarding
the projection of a single Killing tensor see \cite{CarterFrolov:1989}.)

We have seen that the existence of the Killing tensors $C^{ij}_{(k)}$ for the metric 
$H_{ij}$ is the property inherited from the canonical metric $g_{ab}$.
As we have learned in Part II, the latter possesses even more 
fundamental symmetry---connected with the PCKY 
tensor from which all the Killing tensors \eqref{KTk}
are derivable. A natural question arises whether
$H_{ij}$ also `inherits' any Killing--Yano tensor.
In a general case the answer is negative. The necessary conditions for a   
Killing tensor in 4D to be the `square' of a Killing--Yano tensor
were  given by Collinson \cite{Collinson:1976} (see also \cite{FerrandoSaez:2002}). One can
easily check  that they are not satisfied and hence the 4D metric $H_{ij}$ does not admit any Killing--Yano tensor.
In higher dimensions we can exclude the existence of a 
PCKY tensor. 
Indeed, as demonstrated in Chapter 7, the higher-dimensional metric element admitting the PCKY tensor is the canonical spacetime \eqref{metric_coordinates}, i.e., the spacetime different from $H_{ij}$.

\section{$\xi$-branes}
In the above consideration we have focused on stationary strings, that is
strings generated by a 1-parameter family of timelike Killing
trajectories. There are two natural ways how one may try to generalize
this construction. First, one may consider other Killing vector
fields, and/or  second, in the case when there exist more than one Killing
vector, one may consider hypersurfaces formed by the set of Killing
trajectories passing through the same 1-dimensional curve.
Let us discuss these generalizations in more
detail.

For simplicity we assume that the spacetime $M^D$ allows $p$ mutually
commuting Killing  vectors which we denote by $\xi_{(M)}^a\,,$
($M,N=1,\ldots,p$). The Frobenius theorem implies that for each
point of the spacetime $M^D$ there exists (at least locally) a submanifold
of dimension $p$ generated by the Killing vectors $\xi_{(M)}^a$ passing
through this point. In other words, the set $\xi=\{\tens{\xi}_{(M)}\}$ defines
a foliation of $M^D$. Similar to what was done in the Geroch
formalism for one Killing vector field, one can define a quotient
space $S$ of $M^D$ determined by the action of the isometry group
generated by $\xi$. This generalization of the Geroch's
formalism was developed in \cite{MansouriWitten:1984}. The metric $g_{ab}$ of the
spacetime $M^D$ can be written as
\be
g_{ab}=h_{ab}+\Xi_{ab}\,,\quad 
h_{ab}\xi^a_{(M)}=0\, ,\quad 
\Xi_{ab}=\sum_{M,N=1}^p a^{MN}\xi_{(M)a}\xi_{(N)b}\, .
\ee
Here $a^{MN}$ is the $(p\times p)$ matrix which is inverse to the
$(p\times p)$ matrix $a_{MN}=\xi_{(M)a}\xi^a_{(N)}$:
$a^{MN}a_{NK}=\delta^M_K$. A tensor $h_{ab}$ is a projection operator
onto $S$. 

Let us denote by $y^i$ $(D-p)$ coordinates which are constant along
the Killing surfaces generated by the set $\xi$, and by $\psi^M$ the
Killing parameters defined by the conditions
\be
\xi^a_{(M)}\pa_a=\pa_{\psi^M}\, .
\ee
The metric $g_{ab}$ in these coordinates $(x^a)=(y^i,\psi^M)$ takes the form
\be
ds^2=h_{ij}dy^i dy^j+\!\!\sum_{M,N=1}^p\!\!
a^{MN}(\xi_{(M)a}dx^a)(\xi_{(N)b}dx^b)\, .
\ee
In these coordinates we also have
\be\label{NM}
a_{MN}=\xi_{(M)a}\xi^a_{(N)}=\xi_{(N)M}=\xi_{(M)N}\, .
\ee

A natural generalization of stationary strings $\Sigma_{\xi}$ are
$(p+1)$-dimensional objects  $\Sigma_{\xi}^p$ which are formed by a
1-parameter family of Killing surfaces.  We call them {\em $\xi$-branes}. 
In $(y^i,\psi^M)$-coordinates the equation of $\Sigma_{\xi}^p$ is
$y^i=y^i(\sigma)$. For this parametrization coordinates  on $\Sigma_{\xi}^p$ are 
$(\zeta^A)=(\psi^M,\sigma)$ ($A,B=1,\ldots,p+1$). The induced metric on
the $\xi$-brane takes the form
\be\n{gamma}
d\gamma^2=\gamma_{AB}d\zeta^A d\zeta^B=(h+u)d\sigma^2
+2\,d\sigma\!\sum_{M=1}^p \xi_{(M)\sigma}d\psi^M+
\!\!\sum_{M,N=1}^p a_{MN}d\psi^M d\psi^N.
\ee
Here we have defined
\be
h\equiv h_{ij}\frac{dy^i}{d\sigma}\frac{dy^j}{d\sigma}\,,\quad
\xi_{(M) \sigma}\equiv\xi_{(M) i}\frac{dy^i}{d\sigma}\,,\quad
u\equiv\sum_{M,N=1}^p a^{MN}\xi_{(M)\sigma}\xi_{(N)\sigma}\, .
\ee
In order to derive \eq{gamma} we used \eq{NM}. 

The metric $\gamma_{AB}$ can be considered as a block matrix of the
form
\be
\left(   
\begin{array}{cc}
A & B\\
C & D
\end{array}
\right)
\ee
where $A$ is a 1-dimensional matrix and $D$ is a matrix $(p\times p)$.
If $|Z|$ is a determinant of a matrix $Z$, then one has the following
relation for the determinant of a block matrix (see, e.g.,
\cite{Gantmacher:1959})
\be
\left|   
\begin{array}{cc}
A & B\\
C & D
\end{array}
\right|=|D||A-BD^{-1}C|\, .
\ee
Using this equation one obtains
\be
\gamma=\det(\gamma_{AB})=\left|   
\begin{array}{cc}
h+u & \xi_{(M)\sigma}\\
\xi_{(N)\sigma} & a_{MN}
\end{array}
\right| = h\, {\cal F}_{\xi}\,,
\ee
where 
\begin{equation}
{\cal F}_\xi=\det(a_{MN})=\det(\xi_{(M)}^a\xi_{(N) a})
\end{equation}
is the Gram determinant for the set $\xi=\{\tens{\xi}_{(M)}\}$ of the Killing vectors.

The Dirac--Nambu--Goto action for a $(p+1)$-dimensional brane is
\be
I=-\mu\int d^{p+1}\zeta \sqrt{|\gamma|}\, ,
\ee
where $\gamma$ is the determinant of the induced metric on the brane $\gamma_{AB}$.
For a $\xi$-brane this action reduces to the following expression\footnote{
In our derivation we have focused on a $1$-dimensional line in S generating $\xi$-branes.
The same construction remains valid for, let us say, 
$q$-dimensional hyperspace in 
S in the case of a $(p+q)$-dimensional brane. Then, denoting coordinates 
on the worldvolume of such brane by $(\zeta^A)=(\psi^M,\sigma^\alpha)$,
$(\alpha,\beta=1,\dots,q)$, and repeating the same steps one would obtain 
\ba
\gamma\!\!&=&\!\!\det(h_{\alpha\beta})F_\xi=hF_\xi\,,\quad 
h_{\alpha\beta}=h_{ij}\frac{dy^i}{d\sigma^{\alpha}}\frac{dy^j}{d\sigma^{\beta}}\,,\\
I\!\!&=&\!\!-\mu V {\cal E},\quad 
{\cal E}=\int\!\!\!\sqrt{{\cal F}_\xi} dv,\quad 
dv=\sqrt{h}\,d^q\!\sigma\,. 
\ea
}
\be
I=-\mu V {\cal E}\,,\quad dl^2=h \,d\sigma^2\,,\quad 
V=\int d^p \psi^N\,,\quad {\cal E}=\int \sqrt{{\cal F}_\xi} dl\, .
\ee
Thus after the dimensional reduction the problem of finding a
configuration of a $\xi$-brane  reduces to a problem of solving a
geodesic equation in the  reduced $(D-p)$-dimensional space
with the metric
\begin{equation}\n{HH}
dH^2=H_{\,ij}dy^idy^j={\cal F}_\xi h_{\,ij} dy^idy^j\, .
\end{equation}

If the original metric $g_{ab}$ admits a Killing tensor $K^{ab}$
then, since $h^{ij}=g^{ij}$, the natural projection ${}^{\tiny \{\xi\}}\!K^{ij}$ is
a Killing tensor for the metric $h_{ij}$.  However, the full
effective metric $H_{ij}$ does not inherit this symmetry unless the
`red-shift' factor ${\cal F}_{\xi}$ is of the special `separable
form'.   Only then, the Hamilton--Jacobi equation \eq{HJ_ch8} for the geodesic
motion in the metric \eq{HH} allows complete separation of
variables.

\section{$\xi$-branes in Kerr-NUT-AdS spacetime}

\subsection{Separability condition}

Let us discuss now the problem of integrability of $\xi$-branes in
the Kerr-NUT-(A)dS background \eqref{metric_coordinates}--\eqref{X}.  There we have $m+1$
Killing fields $\pa_{\psi_k}\,,$ $k=0,\dots,m$, and we may choose
any arbitrary subset of them as the set $\xi$.  In general, however,
the corresponding red-shift factor ${\cal F}_\xi$ will not be of the
separable form.

More specifically, one requires that the red-shift factor can be written as 
\begin{equation}\label{rs}
{\cal F}_\xi=\sum_{\mu=1}^n \frac{f_\mu(x_\mu)}{U_\mu}\,,
\end{equation}    
with $f_\mu$ functions of $x_\mu$ only, 
in order to allow the separation of variables for the Hamilton--Jacobi equation in 
the effective background ${H}_{ij}$.
The corresponding Killing tensors $(k=1,\dots,n-1)$ would be then
\begin{equation}\label{notriv}
{}^{\tiny \{\xi\}}{C}_{(k)}^{ij}={}^{\tiny \{\xi\}}\!{K}_{(k)}^{ij}-f_{(k)}H^{ij}\,,
\end{equation}
where ${}^{\tiny \{\xi\}}\!{K}_{(k)}^{ij}$ are due natural projections of the 
`primordial' Killing tensors \eqref{KTk}, with directions
from the set $\xi$ projected out, and
\begin{equation}
f_{(k)}\equiv \sum_{\mu=1}^n\frac{f_\mu A_{\mu}^{(k)}}{U_\mu}\,.
\end{equation}

In the case of a stationary string, i.e., for $\xi=\{\pa_{\psi_0}\}$, 
the red-shift factor \eq{F_ch8}, the norm of the 
primary Killing field $\pa_{\psi_0}$, possesses the property
\eqref{rs}, with
\begin{equation}
f_{\mu}=X_\mu-\eps\frac{c}{x_{\mu}^2}\,,
\end{equation}
and the integrability proved in Section 8.3 is justified.

\subsection{$\xi$-branes in 4D}
In 4D a stationary string is the only nontrivial example  of a
$\xi$-brane for which (in these coordinates) integrability can be
proved. Indeed, as discussed in \cite{CarterFrolov:1989} only in the
exceptionally  symmetric case of the de Sitter space itself one can
obtain the integrability of the axially symmetric $\xi$-string with
$\xi=\{\pa_{\psi_1}\}$.

The last possibility of a $\xi$-brane in 4D 
Kerr-NUT-(A)dS spacetime is 
the axially symmetric stationary domain wall, 
$\xi=\{\pa_{\psi_0}, \pa_{\psi_1} \}$.
Let us consider this important example in more detail.
The action takes the form
\begin{equation}
I=-\mu \Delta \psi_0 \Delta \psi_1 
{\cal E}\,,\quad  
{\cal E}=\int\!d\sigma\sqrt{H_{ij}\frac{dy^i}{d\sigma}\frac{dy^j}{d\sigma}}\,,\quad 
\end{equation}
where the effective $2$-dimensional metric is
\begin{equation}
dH^2=H_{ij}{dy^i}{dy^j}=
{\cal F}_\xi\left(\frac{dx_1^2}{Q_1}+\frac{dx_2^2}{Q_2}\right).
\end{equation}
The red-shift factor reads
\begin{equation}
\begin{split}
{\cal F}_\xi=\left|   
\begin{array}{cc}
g_{\psi_0\psi_0} & g_{\psi_0\psi_1}\\
g_{\psi_0\psi_1} & g_{\psi_1\psi_1}
\end{array}
\right|=
\sum_{\mu=1}^2 \frac{f_{\mu}}{U_{\mu}}\,,
\end{split}
\end{equation}
where
\begin{equation}
f_\mu=x_\mu^2X_\mu(X_1+X_2).
\end{equation}
Evidently, $f_\mu$ becomes function of $x_\mu$ only in the case
when all parameters in metric functions $X_\mu$, \eqref{X}, but $c_0$, vanish. Only in that trivial case 
the Hamilton--Jacobi equation for the axially symmetric stationary 
domain wall in 4D can be separated.

\subsection{$\xi$-branes in 5D} 
In 5D the situation is more interesting.
There we can prove integrability 
of the axisymmetric $\xi$-string, $\xi=\{\pa_{\psi_1}\}$, under the condition that parameter $c_1=0$.
Indeed, then the red-shift factor takes the separable form \eq{rs} with
\begin{equation}
f_1(x_1)=2b_2x_1^4+cx_1^2\,,\quad f_2(x_2)=2b_1x_2^4+cx_2^2\,.
\end{equation}

Also, the axially symmetric stationary  
$\xi$-brane, $\xi=\{\pa_{\psi_0}, \pa_{\psi_1}\}$ is completely 
integrable in the case of a vacuum ($c_2=0$) 
5D spacetime \eqref{metric_coordinates}--\eqref{X} with $c_1=0$.
In that case,
\begin{equation}
f_1(x_1)=4b_1b_2x_1^2+2cb_1\,,\quad f_2(x_2)=4b_1b_2x_2^2+2cb_2\,.
\end{equation}
In both cases the nontrivial Killing tensor responsible for the integrability 
is given by  \eqref{notriv}.

However restrictive and  unlikely
to be generally satisfied the condition \eqref{rs} seems,  the above examples
illustrate the special cases where
complete integrability of $\xi$-branes
can be analytically proved. We postpone the discussion of the
existence of other nontrivial examples elsewhere.

\section{Summary}
We have studied integrability of the Nambu--Goto
equations for a stationary string configuration near
a higher-dimensional rotating black hole.  In
a general stationary spacetime this problem reduces to finding a
geodesic in the effective  $(D-1)$-dimensional background ${H}_{ij}$.
In the canonical spacetime \eqref{metric_coordinates}
the geodesic equation can be integrated by a separation of variables of  the corresponding
Hamilton--Jacobi equation. This separability  is a consequence of the
fact that ${H}_{ij}$ inherits some of the hidden symmetries of the
black hole.  Namely, it inherits $(n-1)$ irreducible mutually
commuting Killing tensors which correspond to natural projections of
the Killing tensors present in ${g}_{ab}$. In a general
case there are no Killing--Yano tensors generating
these Killing tensors.

The problem of integrating the equations for $\xi$-branes is more
complicated. We have given some examples where these equations are completely
integrable, but in the general case complete integrability is not
possible. It would be interesting to find other, physically
interesting, examples of completely integrable $\xi$-branes in 
higher dimensional black hole spacetimes. It is also interesting to
study cases where there exist non-complete but non-trivial sets of
(quadratic in momenta) integrals of motion for $\xi$-branes related to
the hidden symmetries of the black hole background.

\chapter{Parallel transport of frames}

In this chapter, based on \cite{ConnellEtal:2008}, we obtain and study the equations describing the parallel
transport of  orthonormal frames along timelike geodesics in a spacetime
admitting the PCKY tensor $\bs{h}$.  
We demonstrate that  the operator $\bs{F}$, obtained by a
projection of $\bs{h}$ to a subspace orthogonal to the velocity, has
in a  generic case eigenspaces of dimension not
greater than 2. Each of these eigenspaces is independently
parallel-propagated.  This allows one to reduce the parallel
transport equations to a set of the first order ordinary differential
equations  for the angles of rotation in the 2D eigenspaces. 
Examples of $D=3, 4, 5$ canonical spacetimes, \eqref{metric_coordinates},
are considered and it is shown that the obtained  first order equations  can be solved by a separation of variables. This chapter is based on \cite{ConnellEtal:2008}.

\section{Introduction} 
One of the remarkable properties of the 4D Kerr metric, 
discovered by Marck in 1983, is that  the equations of
parallel transport can be  integrated \cite{Marck:1983kerr}, \cite{Marck:1983null}. 
Even more generally, a parallel-propagated frame along a geodesic can be
constructed explicitly in any 4D spacetime admitting the
rank-2 Killing--Yano tensor \cite{KamranMarck:1986}.
The purpose of the present chapter is to extend these results
to the case of a spacetime with an arbitrary number of 
dimensions admitting the PCKY tensor $\tens{h}$. 
It was demonstrated in Chapter 7 
that such a spacetime is necessary described by the canonical metric  
\eqref{metric_coordinates}, and in Chapter 5 that the 
 particle geodesic motion is there completely integrable.

Solving the parallel transport equations in curved spacetime 
is useful for many physical problems. In the case of timelike 
geodesics it can be used for studying the behavior of extended 
objects moving in the Kerr and more general geometries. 
In particular, it facilitated the study of tidal
forces acting on a moving body, for example a star, in the background
of a massive black hole (see, e.g., \cite{LuminetMarck:1985}, \cite{LagunaEtal:1993},
\cite{FrolovEtal:1994}, \cite{DienerEtal:1997}, \cite{Shibata:1996}, \cite{IshiiEtal:2005}).
Even more useful is to solve the parallel transport along 
null geodesics.
For example, in geometric optics approximation 
linearly polarized photons and gravitons propagate along null geodesics while the corresponding polarization vectors are parallel-transported along the worldline \cite{MTW}. This
property was used to study the scattering of a polarized radiation by black
holes (see, e.g., \cite{StarkConnors:1977}, \cite{ConnorsStark:1977}, \cite{ConnorsEtal:1980} and references therein). The parallel-propagated frames are very convenient for
investigating the form and shape of a thin `pencil of light' propagating in
an external gravitational field. In the derivation of the equations
for optical scalars such parallel propagating frames play an
important technical role (see, e.g., \cite{Pirani:1965}, \cite{Frolov:1977}).
Another problem where such frames
are useful is the, so called, peeling-off property of the gravitational
radiation in an asymptotically flat spacetime (see, e.g., \cite{KrtousPodolsky:2004b} and references therein). 
In quantum physics the parallel
transport of frames is an important technical element of the point
splitting method which is used for calculation of renormalized values
of local observables in a curved spacetime (such as vacuum expectation values of currents,
stress-energy tensor, etc.). Solving of the
parallel transport equations is especially useful when  fields with
spin are considered (see, e.g., \cite{Christensen:1978}).

Here, we describe how to construct a parallel-propagated frame along 
timelike (spacelike) geodesics. The case of null geodesics 
requires an additional consideration and is under preparation \cite{ConnellEtal:2008null}.
Let us outline the main idea of our construction.
Any 2-form  determines what is called a {\em Darboux basis},
that is a basis in which it has a simple standard form. 
We have already encountered the Darboux basis of $\bs{h}$ 
which we called a {\em canonical} basis (see Section 3.1.2).
Since $\tens{h}$ is non-degenerate its Darboux subspaces
are two-dimensional.\footnote{In an odd number of spacetime dimensions 
there exists an additional one-dimensional 
zero-eigenvalue Darboux subspace of $\tens{h}$.} This means that the `local' 
Darboux basis, defined in the tangent space of any spacetime point, is determined
up to 2D rotations in the Darboux subspaces. The union of local Darboux bases
of $\bs{h}$ forms a global canonical 
basis in the tangent bundle of the spacetime  manifold. 
In the case of the canonical metric \eqref{metric_coordinates},
there exists a special global canonical basis in which
the Ricci rotation coefficients are simplified; the 
{\em principal canonical} basis.
This basis is completely determined by the PCKY tensor (see Chapter 7).

Consider now a timelike geodesic describing the motion of a particle 
with velocity $\bs{u}$. We focus our attention on the 
2-form  $\bs{F}$, \eqref{formF}, obtained as a projection of the PCKY
tensor $\bs{h}$  to a subspace orthogonal to the velocity $\bs{u}$. 
$\tens{F}$ has its own Darboux basis, which we call {\em comoving}. 
For any chosen geodesic the comoving basis is determined along its
trajectory. We have seen in Section 3.3.1 that $\tens{F}$  is
parallel-transported along the geodesic. In particular, this means that
its eigenvalues and its Darboux subspaces, which we call the {\em
eigenspaces} of  $\tens{F}$, are parallel-transported.   We shall
show that for {\em generic} geodesics the eigenspaces of $\bs{F}$ are
at most 2-dimensional. In fact, the eigenspaces with non-zero
eigenvalues are 2-dimensional,  and the zero-value eigenspace is
1-dimensional for an odd number  of spacetime dimensions and
2-dimensional for even. So, the comoving basis is defined up to
rotations  in each of the 2D eigenspaces. The 
{\em parallel-propagated} basis is a special comoving basis.  It can be found by
solving a set of the first order ordinary differential equations for
the angles of rotation in the 2D eigenspaces. 

For special geodesic trajectories the 2-form  $\bs{F}$ may become
{\em degenerate}, that is at least one of its eigenspaces will have
more than 2 dimensions. We shall demonstrate that the eigenspaces
with non-vanishing eigenvalues in such a degenerate case may be
4-dimensional.  In the odd number of spacetime dimensions one may
also have a 3-dimensional eigenspace with a zero eigenvalue. Nevertheless,
in these degenerate cases one can also obtain the parallel-transported basis 
by (now rather more complicated) time dependent rotations of the
comoving basis.

\section{Comoving basis}
In this section, we shall construct a comoving basis, that is a 
Darboux basis of the operator $\tens{F}$, and briefly describe its properties.

\subsection{Operator $\tens{F}$ for timelike geodesics}
Let $\gamma$ be a timelike geodesic affine parameterized by $\tau$, and $u^a=dx^a/d\tau$
be its unit tangent vector (velocity), with the norm $w=\tens{u}\cdot \tens{u}=-1$.
Then the parallel-transported 2-form $\tens{F}$ can be written as [cf. Eq. \eqref{formF}]
\be\n{Fop}
\tens{F}=\tens{h}+\tens{u}^{\flat}\wedge \tens{s}\,,\quad
\tens{s}=\bs{u}\hook \bs{h}\,.
\ee
This form is obtained by a projection of the PCKY
tensor $\bs{h}$  to a subspace orthogonal to the velocity $\bs{u}$ [cf. Eq. \eqref{Fdef}].
Consequently, $\tens{u}$ is an eigenvector of the operator $\tens{F}$
with a zero eigenvalue. 

At a chosen point of the spacetime the tangent space $T$ 
splits into a 1-dimensional space $U$
generated by $\bs{u}$, and a $(D-1)$-dimensional subspace  $V$ 
orthogonal to ${\bs{u}}$;
\be
T=U\oplus V\, .
\ee
$F_{ab}$ and $F^a{}_b$ can be considered  as a 2-form and an
operator, respectively,  either in the subspace $V$ or in the
complete tangent space $T$.

\subsection{Comoving basis}
We demonstrate now that there exists such an orthonormal basis
 in $V$ in which the operator $\tens{F}$ has the (matrix) form (see, e.g.,
\cite{Prasolov:1994})
\be\n{F_ch9}
\mbox{diag}(0,\ldots,0,{\Lambda}_1,\ldots,{\Lambda}_p)\, ,
\ee
where ${\Lambda}_{\mu}$ are matrices of the form
\be
\Lambda_{\mu}=\left(   
\begin{array}{cc}
0 & \lambda_{\mu} {I}_{\mu}\\
-\lambda_{\mu} {I}_{\mu}& 0
\end{array}
\right)\, ,
\ee
and ${I}_{\mu}$ are unit matrices.

The operator $\tens{F}$ maps a linear space $V$ into itself. If
$(\tens{v},\tens{w})=P_{ab}v^a w^b$ is a scalar product in $V$, then an
adjoint operator $\tens{F}^+$ defined by the relation
\be
(\bs{v},\tens{F}\bs{w})=(\tens{F}^+\bs{v},\bs{w})
\ee
obeys the relation $\tens{F}^+=-\tens{F}$, and $\tens{F}^+ \tens{F}=-\tens{F}^2$
is a positive self-adjoint operator. Its spectrum is
\be\n{ll}
\mbox{Spec}(-\tens{F}^2)=\{ 0,\lambda_1^2,\ldots ,\lambda_p^2\}\, .
\ee
We  choose $\lambda_{\mu}$ to be non-negative and order them so that
\be
0=\lambda_0<\lambda_1<\ldots<\lambda_p\, .
\ee
(If $-\tens{F}^2$ does not have a zero eigenvalue, the first term
$\lambda_0$  in \eq{ll} is omitted.) 
The spectrum of $\tens{F}$ is
\be
\mbox{Spec}(\tens{F})=\{
0,i\lambda_1,-i\lambda_1,\ldots,i\lambda_p,-i\lambda_p\}\, .
\ee
Consider a non-zero $\lambda_{\mu}$. We denote 
\be
V_{\mu}^{\pm}=\mbox{Ker}(\tens{F}\pm i\lambda_{\mu}\tens{I})\,,\quad 
q_{\mu}=\mbox{dim}(V_{\mu}^{\pm})\, .
\ee
Thus the eigenvalues and the eigenspaces of $\tens{F}$ are well defined but
they are not real. In order to obtain Darboux form \eqref{F_ch9} it is
sufficient to consider a full space $V_{\mu}$, which is a pair of
eigenspaces for complex conjugate eigenvalues
\be
V_{\mu}=V_{\mu}^{+} + V_{\mu}^{-}\,,\quad 
\mbox{dim}(V_{\mu})=2q_{\mu}\, .
\ee
Using a modified version of the Gram--Schmidt process one can
construct a real orthonormal basis in $V_{\mu}$
\be\n{bas_ch9}
\{ \bsi{{1}}{n}_{\hat{\mu}},\bsi{{1}}{\tilde{n}}_{\hat{\mu}}\ldots, 
\bsi{q_{\mu}}{n}_{\hat{\mu}},
\bsi{{q}_{\mu}}{\tilde{n}}_{\hat{\mu}}\}\, , 
\ee
which has the property (see, e.g., \cite{Prasolov:1994})
\be
\tens{F}\,\bsi{{j}}{n}_{\hat{\mu}}=
-\lambda_{\mu}\,\bsi{{j}}{\tilde{n}}_{\hat{\mu}}\,,\quad 
\tens{F}\,\bsi{{j}}{\tilde{n}}_{\hat{\mu}}=\lambda_{\mu}\bsi{{j}}{n}_{\hat{\mu}}\,.
\ee
Obviously, the space $V_{\mu}$ is the eigenspace of $\tens{F}^2$ and the vectors 
\eqref{bas_ch9} form the complete set of orthonormal eigenvectors\footnote{
One can also introduce the complex eigenvectors of $\tens{F}$: 
\be
\bsi{{j}}{n}_{\hat{\mu}}^{\pm}
=\frac{1}{\sqrt{2}}(\bsi{{j}}{n}_{\hat{\mu}}\pm i\bsi{{j}}{\tilde{n}}_{\hat{\mu}})\,,\quad
\tens{F}\, \bsi{{j}}{n}_{\hat{\mu}}^{\pm}
=\pm i \lambda_{\mu}\bsi{{j}}{n}_{\hat{\mu}}^{\pm}\,,
\ee
which form the bases in $V^{\pm}_{\mu}$ [cf. Eqs. \eqref{eveqs}, \eqref{mdef}]. However, we shall not do so here and consider the real basis of $V_\mu$, \eqref{bas_ch9}, instead.  
}
of $\tens{F}^2$ corresponding to $-\lambda_\mu^2$:
\be\label{Sv_ch9}
\tens{F}^2\tens{v}=-\lambda_\mu^2\tens{v}\,,\quad \tens{v}\in V_\mu\,.
\ee
If $\lambda=0$ and the
corresponding subspace $V_0$ has $q_0$ dimensions, we denote an 
orthonormal basis in $V_0$ by
\be\label{V0}
\{\bsi{1}{n}_{\hat 0},\ldots,\bsi{q_0}{n}_{\hat 0}\}\, .
\ee
The subspaces $V_\mu$ are mutually orthogonal  and
their direct sum forms the space $V$: 
\be
V=V_0\oplus V_{1} \oplus \ldots \oplus V_{p}\, . 
\ee
We further denote by
\be\n{bas*}
\{\fb{1}{\varsigma}{{0}},\ldots,
\fb{{q_{0}}}{\varsigma}{0}\}\,,\quad 
\{\fb{1}{\varsigma}{{\mu}},\fb{{1}}{\tilde{\varsigma}}{{\mu}}
\ldots, \fb{{q_{\mu}}}{\varsigma}{\mu},
\fb{{q}_{\mu}}{\tilde{\varsigma}}{{\mu}}\}\,,
\ee
bases of forms dual to the constructed orthonormal vector bases \eqref{V0}, 
\eqref{bas_ch9}.  These forms give bases in the cotangent spaces
$V_0^*$ and $V_\mu^*$.
We combine the bases \eqref{V0}, \eqref{bas_ch9}, and \eq{bas*} with ${\mu}=0,\ldots,p$ to obtain a complete orthonormal basis of vectors (forms) in the space
$V\,(V^*)$. The duality conditions read
\be\n{bf_ch9}
\fb{s}{\varsigma}{{\mu}}(\bsi{s'}{n}_{\hat{\mu}'})=
\fb{s}{\tilde{\varsigma}}{{\mu}}(\bsi{s'}{\tilde{n}}_{\hat{\mu}'})=
\delta^{\mu}_{\mu'}\delta_{s}^{s'}\,,\quad
\fb{s}{\varsigma}{{\mu}}(\bsi{s'}{\tilde{n}}_{\hat{\mu}'})=
\fb{s}{\tilde{\varsigma}}{{\mu}}(\bsi{s'}{{n}}_{\hat{\mu}'})=0\, .
\ee
Here, for a given $\mu=0,\dots,p$ index $s$ takes the values $s=1,\dots,q_\mu$.
It is evident from the orthonormality of the constructed basis
that we also have
\be\label{bstar_ch9}
(\bsi{s}{n}_{\hat{\mu}})^\flat=\fb{s}{\varsigma}{{\mu}}\,,\quad
(\bsi{s}{\tilde n}_{\hat{\mu}})^\flat=\fb{s}{\tilde \varsigma}{{\mu}}\,,\qquad
(\fb{s}{\varsigma}{{\mu}})^\sharp=\bsi{s}{n}_{\hat{\mu}}\,,\quad
(\fb{s}{\tilde \varsigma}{{\mu}})^\sharp=\bsi{s}{\tilde n}_{\hat{\mu}}\,.
\ee
In this basis  the antisymmetric operator $\bs{F}$, \eqref{Fop}, takes the
form \eqref{F_ch9}. 

For briefness in what follows we shall use the following terminology.
We call $V_{\mu}$ an {\em eigenspace} of $\tens{F}$ corresponding to its
{\em eigenvalue} $\lambda_{\mu}$. 
We call the basis $\{\tens{n}\}$ ($\{\tens{\varsigma}\}$), in which the operator $\tens{F}$ takes the Darboux form \eqref{F_ch9}, an {\em orthonormal Darboux basis}, or
simply the {\em  Darboux basis}.\footnote{
In a symplectic vector space with a non-degenerate 2-form
$\bs{\omega}$ the Darboux basis is defined as a basis in which
$\bs{\omega}$
takes the (matrix) form 
\be\n{form_ch9}
\left(   
\begin{array}{cc}
0 &  {{I}}\\
- {I}& 0
\end{array}
\right)\, ,
\ee
where ${I}$ is the unit matrix. 
When the symplectic space possesses also a positive definite scalar
product, in general it is impossible to find a basis in which
the metric takes the standard diagonal form 
and simultaneously transform $\bs{\omega}$ into \eq{form_ch9}. However, one can put $\bs{\omega}$ into the form similar to \eqref{F_ch9}. This is why we call the above described
modification of the Darboux basis an {\em orthonormal Darboux basis}.}

If we consider $\tens{F}$ as an operator in the complete tangent space
$T$, the corresponding orthonormal Darboux basis  is enlarged by
adding the vector $\bs{u}$ to it. In this enlarged basis the operator
$\tens{F}$ has the same form \eqref{F_ch9}, with the only difference that now
the total number of zeros is not $q_0$, but $q_0+1$.  To remind that
the constructed basis depends on the velocity $\bs{u}$ of a
particle and $\bs{u}$ is one of its elements we call this basis  {\em
comoving}. The characteristic property of the comoving frame is that
all spatial components of the velocity vanish.

Although so far our construction was local (we considered a chosen 
spacetime point), one can naturally extend the
comoving basis along the whole geodesic trajectory. In a general
case, however, the constructed comoving frame is not
parallel-propagated. The parallel-propagated frame can be obtained by
performing additional rotations in each of the  parallel-propagated
eigenspaces of $\tens{F}$. The equations for the corresponding rotation
angles will be derived in the next section. Before we do that we
demonstrate that  due to the fact that the PCKY tensor
$\tens{h}$ is non-degenerate the structure of the eigenspaces of $\tens{F}$, and hence the comoving basis, significantly simplifies.

\subsection{Eigenspaces of $\tens{F}$}

In the comoving frame constructed above the 2-form $\tens{F}$ reads
\be\label{FF_ch9}
\bs{F}=\sum_{{\mu}=1}^{p} \lambda_{\mu} (\sum_{j=1}^{q_{\mu}} 
\fb{j}{\varsigma}{{\mu}}\wedge\fb{{j}}{\tilde{\varsigma}}{{\mu}})\,.
\ee
We shall also use the following
notation
\be\label{spectrum_ch9}
S(\bs{F})=\{ 0^{(q_0+1)},\lambda_1^{(q_1)},\ldots,\lambda_p^{(q_p)}\}
\ee
to encode the complete information about the eigenvalues of $\bs{F}$
and the dimensionality of the corresponding subspaces. 
The extra zero eigenvalue corresponds to the 1-dimensional subspace $U$ spanned by
$\tens{u}$. One also has 
\begin{equation}\label{r_ch9}
D=2n+\eps=1+q_0+2k\,,\quad k = \sum_{\mu=1}^pq_\mu\,.
\end{equation}

\subsubsection{Structure of $V_0$}

Let us now exploit the condition that $\tens{h}$ is {\em non-degenerate}, that is, its (matrix) rank is $2n$. Then one has 
\be\label{q0_ch9}
q_0=\left\{ \begin{array}{cc}
1\, ,  &\mbox{   for  }\varepsilon=0\, ,\\
0 \mbox{  or } 2\, , & \mbox{   for  }\varepsilon=1\, .
\end{array}
\right.
\ee
Let us prove this assertion.
From the definition  \eqref{Fop} of $\tens{F}$ we find
\be\n{hhh_ch9}
\bs{h}^{\wedge m}=\bs{F}^{\wedge m}-m\bs{F}^{\wedge (m-1)}\wedge
\bs{u}^{\flat}\wedge\bs{s}\, ,
\ee
where we have used the property of the exterior product \eqref{dd}.
It is obvious from \eqref{FF_ch9} that the (matrix) rank of $\tens{F}$ is $2k$, that 
is $\tens{F}^{\wedge (k+1)}=0$. So, using \eqref{hhh_ch9} we have $\tens{h}^{\wedge (k+2)}=0$. It means that  
for a non-degenerate (matrix rank $2n$) $\tens{h}$ we have $k+2\ge n+1$.
Employing \eqref{r_ch9} this is equivalent to $q_0\le 1+\varepsilon$ which,  
together with the fact that $q_0$ has to be even for $D$ odd and vice versa,
proves \eqref{q0_ch9}. 

Let us now consider a nontrivial $V_0$, that is $V_0$ with
$q_0=1+\varepsilon$, $n-1=k$.
The vectors spanning it can be found as the eigenvectors of 
the operator $\tens{F}^2$ with zero eigenvalue, not belonging to $U$.
There is, however, a more direct way which was already used by Marck in 4D. Let us consider a Killing--Yano 
$(2+\varepsilon)$-form [cf. Eq. \eqref{fj}] 
\be
\bs{f}=*\bs{h}^{\wedge k}\,,
\ee
and use it to define a $(1+\varepsilon)$-form
\be\label{z_ch9}
\bs{z}\equiv \bs{u}\hook \bs{f}\, .
\ee
Using relation \eqref{prop} and Eq. \eqref{hhh_ch9} one obtains
\be\n{zzz}
\bs{z}=\bs{u}\hook *\bs{h}^{\wedge k}=*(\bs{h}^{\wedge
k}\wedge \bs{u}^{\flat})=
*(\bs{F}^{\wedge k}\wedge \bs{u}^{\flat})\, .
\ee
Employing \eqref{FF_ch9} we have
\be\n{FFF}
\bs{F}^{\wedge k}\!=B\,
\fb{1}{\varsigma}{1}\wedge\fb{1}{\tilde{\varsigma}}{{1}}\wedge \ldots
\wedge\,\fb{{q}_p}{\tilde{\varsigma}}{p}\,,\quad
B\equiv k! \,\prod_{\mu=1}^{p}\lambda_\mu^{q_\mu}\,.
\ee
This means that $\tens{z}$ spans $V_0^*$.
In an even number of spacetime dimensions the space 
$V_0^*$ is 1-dimensional and $\,\fb{\!}{\varsigma}{0}=\tens{z}/|\tens{z}|$. 
Hence, using \eqref{bstar_ch9}, $\,\vb{\!}{n}{0}=\tens{z}^\sharp/|\tens{z}|$ spans $V_0$. 
In the odd number of spacetime dimensions
\be
\bs{z}=\mbox{const} \ \,\fb{1}{\varsigma}{0}\wedge \fb{2}{\varsigma}{0}\, .
\ee
Hence, the 2-form $\bs{z}$ determines the orthonormal
basis $\{\vb{1}{n}{0} , \vb{2}{n}{0}\}$ in $V_0$ up to a 2D rotation.

Let us finally consider the odd-dimensional case in more detail. Expanding the
characteristic equation for the operator $\tens{F}$ one has 
\be
0=\mbox{det}(\tens{F}-\lambda \tens{I})=
a(\bs{u})+b(\bs{u})\lambda^2+\ldots 
\ee
The condition that $q_0=2$ implies that
$a(\bs{u})=\det(\tens{F})=0$. This imposes a constraint on $\bs{u}$.
It means that $q_0=2$ is a {\em degenerate} case which happens only 
for special trajectories $\tens{u}$. For a {\em generic} 
(not special) $\bs{u}$ one has trivial $V_0$ with $q_0=0$.

\subsubsection{Eigenspaces $V_\mu$}
Using the requirement that the eigenvalues of a PCKY tensor $\tens{h}$ are 
functionally independent, or in other words, that in a generic point of the manifold the 
Darboux subspaces of $\tens{h}$ have no more than 2 dimensions, it is
possible to show (see Appendix C.7) that the dimensionalities 
of the eigenspaces of $\tens{F}$ with
non-zero eigenvalues obey the inequalities $q_\mu\le 2$. The case of
$q_\mu= 2$ is possible only in a degenerate case when the vector
$\bs{u}$ obeys a special condition.

\section{Equations of parallel transport}
In this section we describe how to obtain the parallel-transported basis 
from the comoving basis constructed above.
The crucial fact for the construction is that the 2-form 
$\tens{F}$ is parallel-transported along $\tens{u}$ (see Section 3.3.1)
\be
\tens{\dot F}=\nabla_{u}\tens{F}=0\,. 
\ee
This means that any object constructed from $\bs{F}$ and the metric $\bs{g}$ is also
parallel-transported. In particular, this is true for the operator $\tens{F}^2$ and its eigenvalues $-\lambda_{\mu}^2$. We have used this property in Section \ref{other_method} to construct the tower of Killing tensors which, in their turn, imply complete integrability of particle geodesic motion (see Chapter 5).  
Here, we go a little bit further. Namely, we prove that Darboux subspaces of $\tens{F}$, the eigenspaces $V_{\mu}$, are independently parallel-transported, that is
\begin{equation}\label{PTV}
\bs{\dot v}\in V_{\mu} \quad {\rm for} \ 
\forall\, \bs{v}\in V_{\mu}.
\end{equation}  
Indeed, using \eqref{Sv_ch9}, we find
\begin{equation}
\tens{F}^2 \tens{\dot v}=\nabla_u\!\bigl(\tens{F}^2\tens{v}\bigr)=
\nabla_u\!\left(-\lambda_\mu^2\tens{v}\right)=
-\lambda_\mu^2 \tens{\dot v}\,,
\end{equation}
which proves \eqref{PTV}.\footnote{%
Let us emphasize that the dimension of an eigenspace of $\tens{F}$
is also constant along $\gamma$. 
For generic geodesics the eigenspaces of $\tens{F}$ with non-zero eigenvalues are always
2-dimensional, while the subspace with zero eigenvalue ($U\oplus V_0$) has $2-\varepsilon$ dimensions.
There might also exist a zero measure set of special geodesics for which either
an eigenspace of $\bs{F}$ with non-zero eigenvalue has not 2 but 4
dimensions or (in the odd dimensional case) the eigenspace of $\tens{F}$
with zero eigenvalue has 3 dimensions. (See previous section.)
}

This means, that the parallel-propagated basis can be obtained from the comoving basis by time dependent rotations in the eigenspaces of $\tens{F}$. We
denote the corresponding matrix of rotations by $\nbs{O}(\tau)$.
Similar to $\tens{F}$ it has the following structure
\be
\nbs{O}=\mbox{diag}(\nbs{O}_{\hat{0}},\nbs{O}_{\hat{1}},\ldots,\nbs{O}_{\hat{p}})\,.
\ee

For $\lambda_{\mu}>0\,,$  $\nbs{O}_{\hat{\mu}}$ are $2q_{\mu}\times
2q_{\mu}$ orthogonal matrices.  Let $\{ \vb{s}{p}{\mu},
\vb{s}{\tilde{p}}{\mu}\}$ be a parallel-propagated basis in the
eigenspace $V_{\mu}$ and  $\{ \vb{s}{n}{\mu}, \vb{s}{\tilde{n}}{\mu}\}$
be the `original' comoving basis. Then
\be\n{tran}
\left( \begin{array}{c}
\vb{s}{p}{\mu}\\
\vb{s}{\tilde{p}}{\mu}
\end{array}\right)
=\sum_{s'=1}^{q_\mu}\nbs{O}_{\hat{\mu}\ s'}^{\ \ s}
\left(
\begin{array}{c}
\vb{s'}{n}{\mu}\\
\vb{s'}{\tilde{n}}{\mu}
\end{array}\right)\, .
\ee
Here, for fixed values $\{\hat{\mu}, s, s'\}$,  $\nbs{O}_{\hat{\mu}\ s'}^{\ \ s}$ are $2\times 2$ matrices.
Differentiating \eq{tran} along the geodesic and using the fact that
$\{ \vb{s}{p}{\mu}, \vb{s}{\tilde{p}}{\mu}\}$ are parallel-propagated
one gets
\be
\sum_{s'=1}^{q_\mu}{\nbs{{\dot O}}}_{\hat{\mu}\ s'}^{\ \ s}
\left(\!
\begin{array}{c}
\vb{s'}{n}{\mu}\\
\vb{s'}{\tilde{n}}{\mu}
\end{array}\!\right)=-\!\sum_{s'=1}^{q_\mu}{\nbs{{O}}}_{\hat{\mu}\ s'}^{\ \ s}
\left(\!
\begin{array}{c}
\vb{s'}{\dot{n}}{\mu}\\
\vb{s'}{\dot{\tilde{n}}}{\mu}
\end{array}\!\right)\, .
\ee
This gives the following set of the first order differential equations for
$\nbs{O}_{\hat{\mu}\ s'}^{\ \ s}$ 
\be\n{dott}
{\nbs{{\dot O}}}_{\hat{\mu}\ s'}^{\ \ s}=-
\sum_{s''=1}^{q_\mu}{\nbs{{O}}}_{\hat{\mu}\ s''}^{\ \ s}
\nbs{N}_{\hat{\mu}\ s'}^{\ \ s''}
\, ,
\ee
where
\be\label{NN_ch9}
\nbs{N}_{\hat{\mu}\ s'}^{\ \ s''}=
\left(
\begin{array}{c c}
(\vb{s'}{\dot{n}}{\mu},\vb{s''}{n}{\mu}) &
(\vb{s'}{\dot{n}}{\mu},\vb{s''}{\tilde{n}}{\mu})\\
(\vb{s'}{\dot{\tilde{n}}}{\mu},\vb{s''}{n}{\mu}) &
(\vb{s'}{\dot{\tilde{n}}}{\mu},\vb{s''}{\tilde{n}}{\mu})
\end{array}
\right)\, .
\ee
For generic geodesics the parallel transport equations are
greatly simplified. In this case each of the eigenspaces $V_{\mu}$ 
is two-dimensional. 
The equations \eq{tran} take the form
\be\label{pp}
\bs{p}_{\hat{\mu}}=\cos \beta_{\mu} \bs{n}_{\hat{\mu}}-
\sin \beta_{\mu} \bs{\tilde{n}}_{\hat{\mu}}\, ,\quad
\bs{\tilde{p}}_{\hat{\mu}}=\sin \beta_{\mu} \bs{n}_{\hat{\mu}}+
\cos \beta_{\mu} \bs{\tilde{n}}_{\hat{\mu}}\, .
\ee
It is easy to check that Eqs. \eq{dott} reduce to the
following first order equations
\be\n{dot_ch9}
\dot{\beta}_{\mu}=(\bs{\dot {\tilde{n}}}_{\hat{\mu}}, \bs{n}_{\hat{\mu}})
=-(\bs{n}_{\hat{\mu}},\bs{\dot {\tilde{n}}}_{\hat{\mu}})\,.
\ee
If at the initial point $\tau=0$ bases $\{ \tens{p} \}$ and $\{ \tens{n}\}$ coincide,  
the initial conditions for Eqs. \eqref{dot_ch9} are
\begin{equation}\label{ic_ch9}
\beta_{\mu}(\tau=0)=0\,.
\end{equation}

For $\lambda_0=0\,,$ $\nbs{O}_{\hat{0}}$
is a $q_0\times q_0$ matrix. In even number of spacetime dimensions,
$q_0=1$, and $V_0$ is spanned by $\tens{n}_{\hat 0}$ which is already
parallel-propagated. Therefore we have 
$\nbs{O}_{\hat{0}}=1$. For odd number of spacetime dimensions,
$\nbs{O}_{\hat{0}}$ is present only in the degenerate case, 
$q_0=2$, that is when $V_0$ is spanned by $\{\vb{1}{n}{0} , \vb{2}{n}{0}\}$.
The parallel-propagated vectors $\{ \vb{1}{p}{0}, \vb{2}{p}{0}\}$
are then given by the analogue of the equations \eqref{pp}--\eqref{ic_ch9}.

\section{Parallel transport in Kerr-NUT-(A)dS spacetimes}
In this section we shall concretize the above procedure for the 
particular form of the canonical spacetime \eqref{metric_coordinates}.
As it is somewhat unnatural to construct a parallel-propagated frame in the 
unphysical (Wick rotated) space, we use the opportunity to recast 
these metrics into the physical signature.

\subsection{Kerr-NUT-(A)dS spacetimes}
In the physical signature, the Kerr-NUT-(A)dS spacetime \eqref{metric_coordinates}--\eqref{X} can be written as
\begin{equation}\label{metrics_ch9}
\tens{g}=\!\sum_{\mu=1}^{n-1} (\tens{\omega}^{\hat \mu} 
\tens{\omega}^{\hat \mu}\! + \tens{\tilde \omega}^{\hat \mu} 
\tens{\tilde \omega}^{\hat \mu}) + \tens{\omega}^{\hat n} 
\tens{\omega}^{\hat n} -\tens{\tilde \omega}^{\hat n} 
\tens{\tilde \omega}^{\hat n}\!
+ \eps\, \tens{\omega}^{\hat \epsilon} \tens{\omega}^{\hat \epsilon},
\end{equation}
where the basis 1-forms are ($\mu=1,\dots,n-1$)
\ba\label{one-forms_ch9}
\tens{\omega}^{\hat n} \!\!&=&\!\!\frac{\tens{d}r}{\sqrt{Q_n}}\,,\quad 
\tens{\omega}^{\hat \mu} = \frac{\tens{d}x_{\mu}}{\sqrt{Q_{\mu}}}\,,
\quad 
\tens{\omega}^{\hat \epsilon} =\sqrt{\frac{-c}{A^{(n)}}}
\sum_{j=0}^nA^{(j)}\tens{d}\psi_j\;,
\nonumber\\
\tens{\tilde \omega}^{\hat n} \!\!&=&\!\! \sqrt{Q_{n}}
 \sum_{j=0}^{n-1}A_{n}^{(j)}\tens{d}\psi_j\,,\quad 
 \tens{\tilde \omega}^{\hat \mu} = \sqrt{Q_{\mu}}
 \sum_{j=0}^{n-1}A_{\mu}^{(j)}\tens{d}\psi_j\;.
\ea
Notice that we enumerate the basis $\{\tens{\omega}\}$ so that 
$\tens{\tilde \omega}^{\hat n}$ is (the only one) timelike 1-form. 
Here, quantities $A_{\mu}^{(j)},A^{(j)}, Q_{\mu}, U_{\mu}$ are defined exactly 
as before with the only difference that we now understand $x_n^2=-r^2$, and  
\be\label{X_ch9}
X_{n}=-\!\sum\limits_{k=\varepsilon}^{n}c_k(-r^2)^k\!-\!2M r^{1-\varepsilon}
\!+\!\frac{\varepsilon c}{r^2}\,,\quad
X_{\mu}=\sum\limits_{k=\varepsilon}^{n}
c_kx_{\mu}^{2k}\!-\!2b_{\mu}x_{\mu}^{1-\varepsilon}
\!+\!\frac{\varepsilon c}{x_{\mu}^2}\,.
\ee
Time is denoted by $\psi_0$, azimuthal coordinates by $\psi_j$,
$j=1,\dots,m\,,$ $r$ is the Boyer-Lindquist type radial
coordinate, and ${x_\mu}$, ${\mu=1,\dots, n-1}$, stand for latitude
coordinates. 
Again, it is possible to consider a broader class of the {\em off-shell} metrics 
\eqref{metrics_ch9} where $X_n(r), X_\mu(x_\mu)$, are arbitrary functions.

The PCKY tensor reads [cf. Eq. \eqref{PCKY_metric}]
\begin{equation}\label{KYKNA}
\tens{h}=\sum_{\mu=1}^{n-1} x_\mu \tens{\omega}^{\hat \mu}
\wedge \tens{\tilde \omega}^{\hat \mu}
-r \tens{\omega}^{\hat n}\wedge \tens{\tilde \omega}^{\hat n}\,.
\end{equation}
Obviously, the basis $\{\tens{\omega}\}$ remains a principal canonical basis. 
The second-rank irreducible Killing tensors are [cf. Eq. \eqref{KT_metric}] 
\be
\tens{K}^{(j)}=\sum_{\mu=1}^{n-1} A_{\mu}^{(j)} 
(\tens{\omega}^{\hat \mu} \tens{\omega}^{\hat \mu}+
\tens{\tilde \omega}^{\hat \mu} \tens{\tilde \omega}^{\hat \mu})
+A_{n}^{(j)} (\tens{\omega}^{\hat n} \tens{\omega}^{\hat n} 
-\tens{\tilde \omega}^{\hat n} \tens{\tilde \omega}^{\hat n})\!
+ \eps  A^{(j)}  \tens{\omega}^{\hat \epsilon} \tens{\omega}^{\hat \epsilon} \, .
\ee
The velocity of a (timelike) geodesic reads 
\begin{equation}
\tens{u}^\flat=\sum_{\mu=1}^n \bigl( u_{\hat \mu} \tens{\omega}^{\hat \mu} + {\tilde u}_{\hat \mu} \tens{\tilde \omega}^{\hat \mu}\bigr)
+\eps\,u_{\hat \epsilon} \tens{\omega}^{\hat \epsilon}\,,
\end{equation}
where [cf. Eqs. \eqref{basisvectors}, \eqref{VW}]
\ba\label{basisvectors_ch9}
u_{\hat n} \!\!&=&\!\!\frac{\sigma_n}{(X_n U_n)^{1/2}}\,\bigl(
W_n^2 -X_n V_n\bigr)^{1/2}\,,\ \ 
u_{\hat \mu} =\frac{\sigma_\mu}{(X_\mu U_\mu)^{1/2}}\,\bigl(
X_\mu V_\mu-W_\mu^2 \bigr)^{1/2}\,,\quad\nonumber\\
{\tilde u}_{\hat n}\!\!&=&\!\!\frac{W_n}{(X_n U_n)^{1/2}}\,,\quad 
{\tilde u}_{\hat\mu}=\frac{{\rm sign}{(U_\mu)} W_\mu}{(X_\mu U_\mu)^{1/2}}\,,\quad
u_{\hat \epsilon} =-\frac{\Psi_n}{\sqrt{-cA^{(n)}}}\,. 
\ea 
The constants $\sigma_\mu=\pm 1$ ($\mu=1,\dots, n$) are independent 
of one another, and 
\begin{equation}\label{VW_ch9}
\begin{split}
V_n\equiv &-\!\!\sum_{j=0}^{m}r^{2(n-1-j)}\kappa_j\,,\quad 
V_\mu\!\equiv\!\sum_{j=0}^{m}(-x_{\mu}^2)^{n-1-j}\kappa_j\,,\quad \kappa_n=-\frac{\Psi_n^2}{c}\,,\\  
W_n\equiv&\!\sum_{j=0}^{m}r^{2(n-1-j)}\Psi_{\!j}\,,\quad 
W_\mu\!\equiv\!\sum_{j=0}^{m}(-x_{\mu}^2)^{n-1-j}\Psi_{\!j}\,.
\end{split}
\end{equation}
The constant $\kappa_0$ denotes the normalization of the velocity, which for a timelike geodesic is $\kappa_0=-1$.

We shall construct a parallel-propagated frame for geodesic motion in three
steps. At first we use the freedom of local rotations in the 2D Darboux spaces of $\tens{h}$
to introduce the {\em velocity adapted} canonical basis in which 
$n$ components of the velocity vanish.
As the second step, by studying the eigenvalue problem for the
operator $\tens{F}^2$ we find a transformation connecting the velocity
adapted basis to a comoving basis. And finally, we derive the
equations for the rotation angles in the eigenspaces of $\tens{F}$
which transform the obtained comoving basis into the 
parallel-propagated one.

\subsection{Velocity  adapted canonical basis}

To construct the velocity adapted canonical basis we perform  the
boost transformation in the $\{\tens{\tilde \omega}^{\hat n},
\tens{\omega}^{\hat n}\}$ 2-plane and the rotation transformations in
each of the $\{\tens{\tilde \omega}^{\hat \mu}, \tens{\omega}^{\hat
\mu}\}$,  $\mu<n$, 2-planes:
\begin{equation}\label{newframe_ch9}
\begin{split}
\tens{\tilde o}^{\hat n}\equiv&\,\cosh\alpha_n 
\tens{\tilde \omega}^{\hat n}+\sinh\alpha_n\tens{\omega}^{\hat n}\,,\quad
\tens{o}^{\hat n}\equiv\sinh\alpha_n \tens{\tilde \omega}^{\hat n}
+\cosh\alpha_n\tens{\omega}^{\hat n}\,,\\
\tens{\tilde o}^{\hat \mu}\equiv&\,\cos\alpha_\mu
\tens{\tilde \omega}^{\hat \mu}+\sin\alpha_\mu \tens{\omega}^{\hat \mu}\,,\quad 
\tens{o}^{\hat \mu}\equiv-\sin\alpha_\mu
\tens{\tilde \omega}^{\hat \mu}+\cos\alpha_\mu \tens{\omega}^{\hat \mu}\,,\quad 
\tens{o}^{\hat \epsilon}\equiv\tens{\omega}^{\hat \epsilon}\,.
\end{split}
\end{equation}
For arbitrary angles $\alpha_\mu$ ($\mu=1,\dots, n$) this transformation preserves the
form of the metric and of the PCKY tensor:
\be\label{o_basis}
\begin{split}
\tens{g}=&\,\sum_{\mu=1}^{n-1}(\tens{o}^{\hat \mu} \tens{o}^{\hat \mu}\! 
+\!\tens{\tilde o}^{\hat \mu} \tens{\tilde o}^{\hat \mu})\!
+\! \tens{o}^{\hat n} \tens{o}^{\hat n} \! -
\!\tens{\tilde o}^{\hat n} \tens{\tilde o}^{\hat n}\!+ 
\eps\, \tens{o}^{\hat \epsilon} \tens{o}^{\hat \epsilon}\,,\\
\tens{h}=&\,\sum_{\mu=1}^{n-1} x_\mu 
\tens{o}^{\hat \mu}\wedge \tens{\tilde o}^{\hat \mu}
-r \tens{o}^{\hat n}\wedge \tens{\tilde o}^{\hat n}\,.
\end{split}
\ee
Let us define
\be
{\tilde v}_{\hat n}\equiv-\sqrt{{\tilde u}_{\hat n}^2-u_{\hat n}^2}
=-\sqrt{\frac{V_n}{U_n}}\,,\quad 
{\tilde v}_{\hat \mu}\equiv\sqrt{{\tilde u}_{\hat\mu}^2+u_{\hat \mu}^2}
=\sqrt{\frac{V_\mu}{U_\mu}}\,.
\ee
Then, specifying the values of $\alpha_\mu$ to be 
\be\label{br_ch9}
\cosh\alpha_n=\frac{{\tilde u}_{\hat n}}{{\tilde v}_{\hat n}}\,,\ \ 
\sinh\alpha_n=\frac{u_{\hat n}}{{\tilde v}_{\hat n}}\,,\quad 
\cos\alpha_\mu=\frac{{\tilde u}_{\hat \mu}}{{\tilde v}_{\hat \mu}}\,,\ \ 
\sin\alpha_\mu=\frac{u_{\hat \mu}}{{\tilde v}_{\hat \mu}}\,,
\ee
one obtains the following form of the velocity:
\begin{equation}\label{u_simplified}
\tens{u}^\flat=\sum_{\mu=1}^n {\tilde v}_{\hat \mu}\tens{\tilde o}^{\hat \mu}+
\eps u_{\hat \epsilon}\tens{o}^{\hat \epsilon}\,.
\end{equation} 
It means that after this transformation the velocity vector $\tens{u}$ has only 
$(n+\varepsilon)$ non-vanishing components. This simplifies considerably 
the construction of the comoving and the parallel-propagated bases.
Notice also that the boost in the $\{\tens{\tilde \omega}^{\hat n},
\tens{\omega}^{\hat n}\}$ 2-plane is function  of $r$ only and the
rotation in each $\{\tens{\tilde \omega}^{\hat \mu}, \tens{\omega}^{\hat
\mu}\}$ 2-plane is  function of $x_\mu$ only.  The components of
the velocity in the adapted basis $\{\tens{o}\}$ 
depend on constants $\kappa_j$ only; 
constants $\Psi_j$ and $\sigma_\mu$ are
absorbed in the definition of the new frame.

\subsection{Parallel-propagated frame}
At this point we have constructed the velocity adapted basis $\{\tens{o}\}$. Such a basis is still canonical [$\tens{h}$ and $\tens{g}$ take the form \eqref{o_basis}]; in this basis the 
particle's velocity takes the significantly simplified form \eqref{u_simplified}.
The next step is to solve the eigenvalue problem for $\tens{F}^2$ and find the comoving basis.
We further consider only the case of generic geodesics.\footnote{%
A degenerate case which requires a special consideration arises when initially different
elements of $S(\tens{F})$, \eqref{spectrum_ch9}, coincide of one another. It happens for  special values of  integrals of motion characterizing the geodesic trajectories. The 
larger is the number of spacetime dimensions the larger is the number of different degenerate cases. Some of them will be discussed in the next section.
}
For such geodesics  the operator 
$\tens{F}^2$ possesses twice degenerate non-zero eigenvalues. The 
nontrivial eigenspace $V_0$ is present only in even
dimensions, where it is spanned by a properly normalized $\tens{z}^\sharp$.
Therefore the problem 
of finding the parallel-propagated frame in the off-shell spacetimes 
\eqref{metrics_ch9}--\eqref{one-forms_ch9} 
reduces to  finding the eigenvectors $\{\tens{n}_{\hat \mu},
\tens{\tilde n}_{\hat \mu}\}$ spanning  the 2-plane eigenspaces
$V_{\mu}$ and subsequent 2D rotations \eqref{pp}--\eqref{ic_ch9} in these spaces.

In our setup it is somewhat more natural to construct, 
instead of the vector basis $\{\tens{p}\}$, the parallel-propagated basis of forms $\{\tens{\pi}\}$. In the generic case it consists of  
\begin{equation}
\{\tens{u}^\flat, \tens{z}, \tens{\pi}^{\hat 1}, \tens{\tilde \pi}^{\hat 1},\dots,
\tens{\pi}^{\hat n}, \tens{\tilde \pi}^{\hat n}\}\,.
\end{equation}
(The element $\tens{z}$ is present only in even dimensions.)
If $\{ \bs{\varsigma}^{\hat \mu},\bs{\tilde \varsigma}^{\hat
\mu}\}$ are comoving basis forms spanning $V_\mu^*$, then [cf. Eqs. \eqref{pp}--\eqref{ic_ch9}]
\be\label{PTvectors_ch9}
\tens{\pi}^{\hat \mu}=\tens{\varsigma}^{\hat \mu}
\cos\beta_\mu-\tens{\tilde \varsigma}^{\hat \mu}\sin\beta_\mu\,,\quad
\tens{\tilde \pi}^{\hat \mu}=\tens{\varsigma}^{\hat \mu}
\sin\beta_\mu+\tens{\tilde \varsigma}^{\hat \mu}\cos\beta_\mu\,,
\ee
where
\begin{equation}\label{beta_dot_ch9}
{\dot \beta_\mu}=(\tens{\tilde \varsigma}^{\hat \mu},
\tens{\dot \varsigma}^{\hat \mu})=
-(\tens{\varsigma}^{\hat \mu}, \tens{\dot{\tilde \varsigma}}^{\,\hat \mu} )\,,\quad 
\beta_\mu(\tau=0)=0\,.
\end{equation}

The rotation angles $\dot \beta_\mu$\,, as given by \eqref{beta_dot_ch9}, are functions of 
$r$ and $x_\mu$. In the case when $\dot \beta_\mu$ can be 
brought  into the form
\begin{equation}\label{f_ch9}
\dot\beta_\mu=\frac{f^{(\mu)}_n(r)}{U_n}+
\sum_{\nu=1}^{n-1} \frac{f^{(\mu)}_\nu(x_\nu)}{U_\nu}\,,
\end{equation}
the problem \eqref{beta_dot_ch9} is {\em separable} and the particular solution 
is given by (see Appendix C.5)
\begin{equation}\label{separ_ch9}
\beta_\mu=\!\int\!\!\frac{\sigma_n f_n^{(\mu)}dr}{\sqrt{W_n^2-X_n V_n}}
+\sum_{\nu=1}^{n-1}\int\!\!\frac{\sigma_\nu {\rm sign}(U_\nu) f_\nu^{(\mu)}dx_\nu}{\sqrt{X_\nu V_\nu-W_\nu^2}}\,.
\end{equation}

\section{Examples}
We shall now illustrate the above described formalism  by considering  
$D=3, 4, 5$ off-shell spacetimes \eqref{metrics_ch9}.
We take the normalization of the velocity $-1$
and normalize  other vectors of the
parallel-transported frame to $+1$.
In the derivation of the equations for 
$\dot \beta_{\mu}$ we used the Maple program.
  
\subsection{3D spacetime: BTZ black holes}
\subsubsection{Generic case}

As the first example we consider the case when $D=3$, that is when the metric
\eqref{metrics_ch9} describes a BTZ black hole \cite{BanadosEtal:1992}.  We 
first discuss the generic case, $q_0=0$, and then briefly
mention what happens for the degenerate geodesics with $q_0=2$. Since in three dimensions $n=1$ we
drop everywhere index $\mu$.

So, we have the metric
\begin{equation}
\tens{g}=-\tens{\tilde \omega}\tens{\tilde \omega}+\tens{\omega}\tens{\omega}
+\tens{\omega}^{\hat \epsilon}\tens{\omega}^{\hat \epsilon}\,,
\end{equation}
where  
\be\label{triad_ch9}
\tens{\tilde \omega}=\sqrt{X} \tens{d}\psi_0\,,\ \ \tens{\omega}
\!=\!\frac{\tens{d}r}{\sqrt{X}}\,,\ \  
\tens{\omega}^{\hat \epsilon}\!=
\!\frac{\sqrt{c}}{r}(\tens{d}\psi_0\!-r^2 \tens{d}\psi_1 )\,,\ \ 
X=c_1 r^2-2M +\frac{c}{r^2}\,.
\ee
The parameter $c_1$ is proportional to the cosmological constant and parameters 
$M$ and $c>0$ are related to mass and rotation parameter.
    
The PCKY tensor and the Killing tensor are
\begin{equation}\label{KY3D_ch9}
\tens{h}=-r\tens{\omega}\wedge\tens{\tilde \omega}\,,
\quad \tens{K}=-r^2\tens{\omega}^{\hat \epsilon}\tens{\omega}^{\hat \epsilon}\,.
\end{equation}
The geodesic velocity reads
\ba
\tens{u}^\flat\!\!&=&\!\!{\tilde u} \tens{\tilde \omega}+u \tens{\omega}+
u_{\hat \epsilon} \tens{\omega}^{\hat \epsilon}\,,\\
\tilde u \!\!&=&\!\!\frac{W}{\sqrt{X}}\,,\quad  
u=\sigma\sqrt{\frac{W^2}{X}-V}\,,\quad  
u_{\hat \epsilon}=-\frac{\Psi_1}{\sqrt{c}r}\,,\label{u}\\
W\!\!&\equiv&\!\!\Psi_0+\frac{\Psi_1}{r^2}\,,\quad
V\equiv 1+\frac{\Psi_1^2}{cr^2}\,.
\ea
In the velocity adapted frame $\{\tens{\tilde o},\tens{o}, \tens{o}^{\hat \epsilon}\}$, given by \eqref{newframe_ch9}, we have
\ba
\tens{u}^\flat\!&=&\!{\tilde v}\tens{\tilde o}+u_{\hat \epsilon}\tens{o}^{\hat \epsilon}\,,\quad \tilde v =-\sqrt{V}\,,\\
\tens{F}\!&=&\!ru_{\hat 0}\tens{o}\wedge
(u_{\hat 0}\tens{\tilde o}+\tilde v\tens{o}^{\hat \epsilon})\,.
\ea
We find  
\begin{equation}
S(\tens{F})=\{0^{(1)}, \lambda^{(2)}\}\,,\quad
\lambda=\frac{|\Psi_1|}{\sqrt{c}}\,.
\end{equation}
The zero eigenvalue corresponds to the space $U^*$ spanned by $\tens{u}^\flat$.
In the non-degenerate case, that is when $\Psi_1\neq 0$, 
the eigenspace $V_0^*$ is trivial.
The orthonormal forms spanning $V_{\lambda}^*$ are
\begin{equation}
\tens{\varsigma}=\tens{o}\,,\quad 
\tens{\tilde \varsigma}=u_{\hat \epsilon} \tens{\tilde o}+\tilde v \tens{o}^{\hat \epsilon}\,.
\end{equation}
Using \eqref{beta_dot_ch9} one finds 
\begin{equation}
\dot \beta=\frac{C}{r^2+\lambda^2}\,,\quad C\equiv \frac{c-\Psi_0\Psi_1}{\sqrt{c}}\,.
\end{equation}
The parallel-transported forms $\{\tens{\pi}, \tens{\tilde \pi}\}$ are given by 
\eqref{PTvectors_ch9}, where
\begin{equation}\label{beta3d}
\beta=\!\! \int\!\! \frac{\sigma C dr}{(r^2+\lambda^2)\sqrt{W^2-XV}}\,. 
\end{equation}

\subsubsection{Degenerate case}
Let us now consider special geodesic trajectories with 
$\Psi_1=0$ for which $q_0=2$. For such trajectories one has
\be\label{u3d_ch9}
\tilde u =\frac{\Psi_0}{\sqrt{X}}\,,\quad  
u=\sigma\sqrt{\frac{\Psi_0^2}{X}-1}\,,\quad  
u_{\hat \epsilon}=0\,.
\ee
In the adapted basis the velocity is $\tens{u}^\flat=-\tens{\tilde o}\,.$ 
Operators $\tens{F}$ and $\tens{F}^2$ become trivial.  
The space $V_0^*$ is spanned by $\{\fb{{1}}{\varsigma}{0}, \fb{{2}}{\varsigma}{0}\}$, where
\begin{equation}
\fb{{1}}{\varsigma}{0}=\tens{o},\quad 
\fb{{2}}{\varsigma}{0}=-\tens{o}^{\hat \epsilon}\,.
\end{equation}
Similar to \eqref{PTvectors_ch9} and \eqref{beta_dot_ch9} parallel-transported 
forms can be written as follows:
\ba\label{PTvectorsDeg3D}
\tens{\pi}\!&=&\!\fb{{1}}{\varsigma}{0}\cos\beta
-\fb{{2}}{\varsigma}{0}\sin\beta\,,\quad
\tens{\tilde \pi}=\fb{{1}}{\varsigma}{0}\sin\beta
+\fb{{2}}{\varsigma}{0}\cos\beta\,,\nonumber\\
{\dot \beta}\!&=&\!(\fb{{2}}{\varsigma}{0}, \fb{{1}}{\dot \varsigma}{0})=
-(\fb{{1}}{\varsigma}{0}, \fb{{2}}{\dot \varsigma}{0})\,,\quad
\beta(\tau=0)=0\,.
\ea
Using these equations we find $\dot \beta=\sqrt{c}/r^2$ and hence
\begin{equation}
\beta=\!\! \int\!\! \frac{\sigma \sqrt{c} dr}{r^2\sqrt{\Psi_0^2-X}}\,.
\end{equation}
Notice that this relation can be obtained from \eqref{beta3d} by taking the 
limit $\Psi_1\to 0$. 

To conclude, the parallel-propagated orthonormal frame around a BTZ black hole
is $\{\tens{u}^\flat, \tens{\pi}, \tens{\tilde \pi}\}$. This frame remains parallel-propagated 
also off-shell, when $X$ given by \eqref{triad_ch9} becomes an arbitrary function 
of $r$.

\subsection{4D spacetime: Carter's family of solutions}
Let us now consider the case of $D=4$. We have
\begin{equation}\label{g4d}
\tens{g}=-\tens{\tilde \omega}^{\hat 2}\tens{\tilde \omega}^{\hat 2}+\tens{\omega}^{\hat 2}\tens{\omega}^{\hat 2}+
\tens{\tilde \omega}^{\hat 1}\tens{\tilde \omega}^{\hat 1}+\tens{\omega}^{\hat 1}\tens{\omega}^{\hat 1}\,,
\end{equation}
where  
\begin{equation}\label{tetrad_ch9}
\begin{split}
\tens{\tilde \omega}^{\hat 2}=&\,\sqrt{\frac{X_2}{U_2}}(\tens{d}\psi_0+x_1^2\tens{d}\psi_1 )\,,\quad
\tens{\omega}^{\hat 2}=\,\sqrt{\frac{U_2}{X_2}}\,\tens{d}r\,,\\ 
\tens{\tilde \omega}^{\hat 1}=&\,\sqrt{\frac{X_1}{U_1}}(\tens{d}\psi_0-r^2\tens{d}\psi_1 )\,,\quad 
\tens{\omega}^{\hat 1}=\sqrt{\frac{U_1}{X_1}}\, \tens{d}x_1\,. 
\end{split}
\end{equation}
Here, $U_2=-U_1=x_1^2+r^2$,
and we shall not be specifying functions $X_1(x_1), X_2(r)$ at this point. 

The PCKY tensor and the Killing tensor are:
\ba
\tens{h}\!\!&=&\!\!x_1\tens{\omega}^{\hat 1}\wedge\tens{\tilde \omega}^{\hat 1}-r\tens{\omega}^{\hat 2}\wedge\tens{\tilde \omega}^{\hat 2}\,,\label{KY4_ch9}\\
\tens{K}\!\!&=&\!\!x_1^2(\tens{\omega}^{\hat 2}\tens{\omega}^{\hat 2}\!\!-\!\tens{\tilde \omega}^{\hat 2}\tens{\tilde \omega}^{\hat 2})
\!-\!r^2(\tens{\tilde \omega}^{\hat 1}\tens{\tilde \omega}^{\hat 1}\!\!+\!\tens{\omega}^{\hat 1}\tens{\omega}^{\hat 1}).\label{KT4_ch9}
\ea
The components of the velocity are 
\be\label{u4_ch9}
\begin{split}
{\tilde u}_{\hat 2}=&\,\frac{W_2}{\sqrt{X_2U_2}}\,,\quad  
u_{\hat 2}=\frac{\sigma_2}{\sqrt{X_2U_2}}\sqrt{W_2^2-X_2V_2}\,,\\  
{\tilde u}_{\hat 1}=&\frac{-W_1}{\sqrt{X_1U_1}}\,,\quad  
u_{\hat 1}=\frac{\sigma_1}{\sqrt{X_1U_1}}\sqrt{X_1V_1-W_1^2}\,, 
\end{split}
\ee
where
\begin{equation}\label{VW_4D}
\begin{split}
W_2=&\,r^2\Psi_0+\Psi_1\,,\quad
V_2=r^2-\kappa_1\,,\\
W_1=&\,-x_1^2\Psi_0+\Psi_1\,,\quad
V_1=x_1^2+\kappa_1\,.
\end{split}
\end{equation}
The constants of geodesic motion $\Psi_0$ and $\Psi_1$ are associated with isometries and $\kappa_1<0$ corresponds to the Killing tensor \eqref{KT4_ch9}.
In the velocity adapted frame $\{\tens{\tilde o}^{\hat 2},\tens{o}^{\hat 2},\tens{\tilde o}^{\hat 1},\tens{o}^{\hat 1}\}$ given by \eqref{newframe_ch9} we have 
\ba
\tens{u}^\flat\!\!\!&=&\!\!{\tilde v}_{\hat 2}\tens{\tilde o}^{\hat 2}\!\!+{\tilde v}_{\hat 1}\tens{\tilde o}^{\hat 1}\,,\quad
{\tilde v}_{\hat 2}\!=\!-\sqrt{\frac{V_2}{U_2}}\,,\quad
{\tilde v}_{\hat 1}\!=\!\sqrt{\frac{V_1}{U_1}}\,,\label{u4d_adapt}\\
\tens{F}\!\!\!&=&\!\!(r{\tilde v}_{\hat 1}\tens{o}^{\hat 2}+x_1{\tilde v}_{\hat 2}\tens{o}^{\hat 1})\wedge ({\tilde v}_{\hat 1}\tens{\tilde o}^{\hat 2}+{\tilde v}_{\hat 2}\tens{\tilde o}^{\hat 1})\,.\label{F4d_adapt}
\ea
We find 
\begin{equation}\label{spect_4d}
S(\tens{F})=\{0^{(2)}, \lambda^{(2)}\}\,,\quad
\lambda=\sqrt{-\kappa_1}\,.
\end{equation}
The first zero eigenvalue corresponds to $U^*$, while the second one corresponds 
to the eigenspace $V_0^*$, spanned by $1$-form $\tens{z}$ \eqref{z_ch9}. 
When normalized $\tens{z}$ reads
\begin{equation}\label{z4D}
\tens{z}=\lambda^{-1}(-x_1{\tilde v}_{\hat 2}\tens{o}^{\hat 2}+r{\tilde v}_{\hat 1}\tens{o}^{\hat 1})\,.
\end{equation}
The orthonormal forms spanning $V_{\lambda}^*$ are:
\begin{equation}
\tens{\varsigma}={\tilde v}_{\hat 1} \tens{\tilde o}^{\hat 2}+{\tilde v}_{\hat 2}\tens{\tilde o}^{\hat 1}\,,\quad 
\tens{\tilde \varsigma}=\lambda^{-1}(r{\tilde v}_{\hat 1}\tens{o}^{\hat 2}+x_1{\tilde v}_{\hat 2}\tens{o}^{\hat 1})\,.
\end{equation}
Using \eqref{beta_dot_ch9} one finds
\be\label{f1f24D}
\dot \beta=\frac{C}{(x_1^2-\lambda^2)(r^2+\lambda^2)}=
\frac{f_1}{U_1}+\frac{f_2}{U_2}\,,\ \ 
f_1\equiv-\frac{C}{x_1^2-\lambda^2}\,,\ \  
f_2\equiv\frac{C}{r^2+\lambda^2}\,,
\ee
where $C\equiv \lambda\bigl(\lambda^2\Psi_0-\Psi_1\bigr)$.
Therefore, $\beta$ allows a separation of variables and can be written as
\begin{equation}\label{beta4D}
\beta=\!\!\int\!\!\frac{\sigma_2 f_2 dr}{\sqrt{W_2^2-X_2V_2}}-
\!\!\int\!\!\frac{\sigma_1 f_1 dx_1}{\sqrt{X_1V_1-W_1^2}}\ ,
\end{equation}
where functions $f_1$, $f_2$, are defined in \eqref{f1f24D}.
The parallel-transported forms $\{\tens{\pi}, \tens{\tilde \pi}\}$
are given by \eqref{PTvectors_ch9}. 

To summarize, the parallel-propagated orthonormal frame in the spacetime
\eqref{g4d}--\eqref{tetrad_ch9} is $\{\tens{u}^\flat, \tens{z}, \tens{\pi}, \tens{\tilde \pi}\}$. 
This parallel-propagated basis is constructed for arbitrary functions $X_1(x_1), X_2(r)$, and in particular 
for the Carter's class of solutions \cite{Carter:1968pl}, \cite{Carter:1968cmp}---describing among others a rotating charged 
black hole in the cosmological background (see Appendix A). So, we have re-derived the results obtained earlier in \cite{Marck:1983kerr}, \cite{KamranMarck:1986}.

\subsection{5D Kerr-NUT-(A)dS spacetime}
\subsubsection{Generic case}
As the last example we consider the 5D canonical spacetime.
Similar to the 3D case we shall first obtain the parallel-propagated frame 
for generic geodesics and then briefly discuss what happens for the special trajectories characterized by $q_0=2$, or $q_1=2$.
The metric reads
\begin{equation}\label{g5d}
\tens{g}=-\tens{\tilde \omega}^{\hat 2}\tens{\tilde \omega}^{\hat 2}+\tens{\omega}^{\hat 2}\tens{\omega}^{\hat 2}+
\tens{\tilde \omega}^{\hat 1}\tens{\tilde \omega}^{\hat 1}+\tens{\omega}^{\hat 1}
\tens{\omega}^{\hat 1}
+\tens{\omega}^{\hat \epsilon}\tens{\omega}^{\hat \epsilon}\,,
\end{equation}
where  
\begin{equation}\label{quintad}
\begin{split}
\tens{\tilde \omega}^{\hat 2}=&\,\sqrt{\frac{X_2}{U_2}}(\tens{d}\psi_0+x_1^2\tens{d}\psi_1 )\,,\  
\tens{\omega}^{\hat 2}=\,\sqrt{\frac{U_2}{X_2}}\,\tens{d}r\,,\\ 
\tens{\tilde \omega}^{\hat 1}=&\,\sqrt{\frac{X_1}{U_1}}(\tens{d}\psi_0-r^2\tens{d}\psi_1 )\,,\  
\tens{\omega}^{\hat 1}=\sqrt{\frac{U_1}{X_1}}\, \tens{d}x_1\,,\\
\tens{\omega}^{\hat \epsilon}=&\,\frac{\sqrt{c}}{rx_1}\,\left[\tens{d}\psi_0+
(x_1^2-r^2)\tens{d}\psi_1-x_1^2r^2\tens{d}\psi_2\right]\,,
\end{split}
\end{equation}
and $U_2=-U_1=x_1^2+r^2$.
The PCKY tensor and the Killing tensor for this metric are
\ba
\tens{h}\!&=&\!x_1\tens{\omega}^{\hat 1}\wedge\tens{\tilde \omega}^{\hat 1}-r\tens{\omega}^{\hat 2}\wedge\tens{\tilde \omega}^{\hat 2}\,,\\
\tens{K}\!&=&\!x_1^2(-\tens{\tilde \omega}^{\hat 2}\tens{\tilde \omega}^{\hat 2}+\tens{\omega}^{\hat 2}\tens{\omega}^{\hat 2}
+\tens{\omega}^{\hat \epsilon}\tens{\omega}^{\hat \epsilon})
-r^2(\tens{\tilde \omega}^{\hat 1}\tens{\tilde \omega}^{\hat 1}+\tens{\omega}^{\hat 1}\tens{\omega}^{\hat 1}
+\tens{\omega}^{\hat \epsilon}\tens{\omega}^{\hat \epsilon})\,.
\ea
The components of the velocity are 
\ba\label{u5}
{\tilde u}_{\hat 2}\!&=&\!\frac{W_2}{\sqrt{X_2U_2}}\,,\quad  
u_{\hat 2}=\frac{\sigma_2}{\sqrt{X_2U_2}}\sqrt{W_2^2-X_2V_2}\,,\nonumber\\  
{\tilde u}_{\hat 1}\!&=&\!\frac{-W_1}{\sqrt{X_1U_1}}\,,\quad  
u_{\hat 1}=\frac{\sigma_1}{\sqrt{X_1U_1}}\sqrt{X_1V_1-W_1^2}\,,\quad
u_{\hat \epsilon}=-\frac{\Psi_2}{r\sqrt{c x_1^2}}\,, 
\ea
where
\begin{equation}\label{W5}
\begin{split}
W_1=&\,-x_1^2\Psi_0+\Psi_1-\frac{\Psi_2}{x_1^2}\,,\quad 
V_1=x_1^2+\kappa_1+\frac{\Psi_2^2}{cx_1^2}\,,\\
W_2=&\,r^2\Psi_0+\Psi_1+\frac{\Psi_2}{r^2}\,,\quad 
V_2=r^2-\kappa_1+\frac{\Psi_2^2}{cr^2}\,.
\end{split}
\end{equation}
In the velocity adapted frame $\{\tens{o}\}$ given by \eqref{newframe_ch9}
we have 
\be\label{u5c}
\tens{u}^\flat={\tilde v}_{\hat 2}\tens{\tilde o}^{\hat 2}\!+{\tilde v}_{\hat 1}\tens{\tilde o}^{\hat 1}\!+u_{\hat \epsilon}\tens{o}^{\hat \epsilon}\,,\quad
{\tilde v}_{\hat 2}=-\sqrt{\frac{V_2}{U_2}}\,,\quad 
{\tilde v}_{\hat 1}=\sqrt{\frac{V_1}{U_1}}\,.
\ee
The 2-form $\tens{F}$ and the 2-form $\tens{z}$ are
\ba
\tens{F}\!\!&=&\!\!(r{\tilde v}_{\hat 1}\tens{o}^{\hat 2}+x_1{\tilde v}_{\hat 2}\tens{o}^{\hat 1})\wedge ({\tilde v}_{\hat 1}\tens{\tilde o}^{\hat 2}+{\tilde v}_{\hat 2}\tens{\tilde o}^{\hat 1})\nonumber\\
\!\! &\ &\!\!+ru_{\hat \epsilon}\tens{o}^{\hat 2}\wedge(\tilde{v}_{\hat 2}\tens{o}^{\hat \epsilon}+u_{\hat \epsilon}\tens{\tilde o}^{\hat 2})
+x_1u_{\hat \epsilon}\tens{o}^{\hat 1}\wedge({\tilde v}_{\hat 1}\tens{o}^{\hat \epsilon}-u_{\hat \epsilon}\tens{\tilde o}^{\hat 1})\,,\label{F5D}\\
\tens{z}\!\!&=&\!\!\tens{o}^{\hat \epsilon}\wedge (r{\tilde v}_{\hat 1}\tens{o}^{\hat 1}- x_1{\tilde v}_{\hat 2}\tens{o}^{\hat 2})+u_{\hat \epsilon}(x_1\tens{o}^{\hat 2}\wedge
\tens{\tilde o}^{\hat 2}+r\tens{o}^{\hat 1}\wedge\tens{\tilde o}^{\hat 1})\,.\label{z5D}
\ea
One has 
\begin{equation}
S(\tens{F})=\{0^{(1)}, \lambda_1^{(2)}, \lambda_2^{(2)}\}\,,
\end{equation}
where 
\be
\lambda_1=\sqrt{-\frac{\kappa_1+d}{2}}\,,\quad   
\lambda_2=\sqrt{-\frac{\kappa_1-d}{2}}\,,\quad
d\equiv \sqrt{\kappa_1^2-4\frac{\Psi_2^2}{c}}\,.
\ee
The zero eigenvalue corresponds to $U^*$.
The eigenspace $V_{1}^*$ is spanned by 
\ba
\tens{\varsigma}^{\hat 1}\!&=&\!N_1\Bigl(-\frac{x_1 F_r(\lambda_2)}{\sqrt{U_2}}\,\tens{\tilde o}^{\hat 2}+\frac{rF_{x_1}(\lambda_2)}{\sqrt{U_2}}\,\tens{\tilde o}^{\hat 1}+\tens{o}^{\hat \epsilon}\Bigr)\,,\nonumber\\
\tens{\tilde \varsigma}^{\hat 1}\!&=&\!{\tilde N}_{1}\Bigl(F_r(\lambda_2)F_{x_1}(\lambda_1)\tens{o}^{\hat 2}+\tens{o}^{\hat 1}\Bigr)\,.
\ea
Here $N_1$ and ${\tilde N}_{1}$ stand for normalizations,
\ba
N_1\!&=&\!\sqrt{U_2}/\sqrt{U_2+r^2F_{x_1}(\lambda_2)^2-x_1^2F_r(\lambda_2)^2}\,,\nonumber\\
{\tilde N}_{1}\!&=&\!1/\sqrt{1+F_{x_1}(\lambda_1)^2F_r(\lambda_2)^2}\,,
\ea
and we have introduced
\begin{equation}
F_r(\lambda)\equiv\frac{x_1u_{\hat \epsilon}\sqrt{V_2}}{x_1^2u_{\hat \epsilon}^2+\lambda^2}\,,\quad 
F_{x_1}(\lambda)\equiv\frac{ru_{\hat \epsilon}\sqrt{-V_1}}{r^2u_{\hat \epsilon}^2-\lambda^2}\,, 
\end{equation}
which are functions of $r$, $x_1$, respectively.
Using \eqref{beta_dot_ch9} we find
\ba
\dot \beta_{1}\!\!&=&\!\!\frac{C^{(1)}}{(x_1^2-\lambda_1^2)(r^2+\lambda_1^2)}=\frac{f_1^{(1)}}{U_1}+\frac{f_2^{(1)}}{U_2}\,,\nonumber\\
f_1^{(1)}\!\!&\equiv&\!\!-\frac{C^{(1)}}{x_1^2-\lambda_1^2}\,,\ \  
f_2^{(1)}\equiv\frac{C^{(1)}}{r^2+\lambda_1^2}\,,\ \ 
C^{(1)}\equiv-\frac{\Psi_2\Psi_1-\lambda_1^2\Psi_2\Psi_0-c\lambda_2^2}{\lambda_2\sqrt{c}}\,.
\quad\quad
\ea
This means that $\beta_1$ can be separated as follows:
\begin{equation}\label{beta1}
\beta_1=\!\int\!\!\frac{\sigma_2 f_2^{(1)} dr}{\sqrt{W_2^2-X_2 V_2}}
-\!\int\!\!\frac{\sigma_{1} f_1^{(1)} dx_1}{\sqrt{X_1 V_1-W_1^2}}\,.
\end{equation} 
The parallel-propagated forms $\{\tens{\pi}^{\hat 1}, \tens{\tilde \pi}^{\hat 1}\}$ are given by rotation \eqref{PTvectors_ch9}.
Similarly one finds that
\ba
\tens{\varsigma}^{\hat 2}\!&=&\!N_2\Bigl(-\frac{x_1 F_r(\lambda_1)}{\sqrt{U_2}}\,\tens{\tilde o}^{\hat 2}+\frac{rF_{x_1}(\lambda_1)}{\sqrt{U_2}}\,\tens{\tilde o}^{\hat 1}+\tens{o}^{\hat \epsilon}\Bigr)\,,\nonumber\\
N_2\!&=&\!\sqrt{U_2}/\sqrt{U_2+r^2F_{x_1}(\lambda_1)^2-x_1^2F_r(\lambda_1)^2}\,,\nonumber\\
\tens{\tilde \varsigma}^{\hat 2}\!&=&\!{\tilde N}_{2}\Bigl(F_r(\lambda_1)F_{x_1}(\lambda_2)\tens{o}^{\hat 2}+\tens{o}^{\hat 1}\Bigr)\,,\nonumber\\
{\tilde N}_{2}\!&=&\!1/\sqrt{1+F_{x_1}(\lambda_2)^2F_r(\lambda_1)^2}\,,
\ea
span the eigenspace $V_{2}^*$.
Using \eqref{beta_dot_ch9} we have
\ba
\dot \beta_{2}\!\!&=&\!\!\frac{C^{(2)}}{(x_1^2-\lambda_2^2)(r^2+\lambda_2^2)}=\frac{f_1^{(2)}}{U_1}+\frac{f_2^{(2)}}{U_2}\,,\nonumber\\
f_1^{(2)}\!\!&\equiv&\!\!-\frac{C^{(2)}}{x_1^2-\lambda_2^2}\,,\ \ 
f_2^{(2)}\equiv\frac{C^{(2)}}{r^2+\lambda_2^2}\,,\ \ 
C^{(2)}\equiv\frac{\Psi_2\Psi_1-\lambda_2^2\Psi_2\Psi_0-c\lambda_1^2}{\lambda_1\sqrt{c}}\,.\qquad
\ea
The parallel-propagated forms $\{\tens{\pi}^{\hat 2}, \tens{\tilde \pi}^{\hat 2}\}$ 
are given by rotation \eqref{PTvectors_ch9}, where
\begin{equation}\label{beta2}
\beta_2=\!\int\!\!\frac{\sigma_2 f_2^{(2)} dr}{\sqrt{W_2^2-X_2 V_2}}
-\!\int\!\!\frac{\sigma_{1} f_1^{(2)} dx_1}{\sqrt{X_1 V_1-W_1^2}}\,.
\end{equation}

\subsubsection{Degenerate case}
In $D=5$ two different degenerate cases are possible.
One can have either a 2-dimensional $V_0$ ($q_0=2$) which happens for 
the special geodesics characterized by 
$\Psi_2=0$, or a 4-dimensional $V_\lambda$ ($q_1=2$)
which happens when $\kappa_1^2=4\Psi_2^2/c\,.$
The latter case is more complicated and the general formulas \eqref{tran}--\eqref{NN_ch9}
have to be used. We shall not do this here and rather concentrate 
on the first degeneracy which has an interesting consequence.

So, we consider the special geodesics characterized by $\Psi_2=0$.
It can be checked by direct calculations that 
in this case the results can be obtained by taking the 
limit $\Psi_2\to 0$ of previous formulas. 
In particular, one has $u_{\hat \epsilon}=0$, and functions
$W_1, W_2, V_1, V_2,$ are the same as in \eqref{VW_4D}. The velocity $\tens{u}$
and the 2-form $\tens{F}$ become effectively 4-dimensional---equal to \eqref{u4d_adapt} and \eqref{F4d_adapt}, respectively. 
The 2-form $\tens{z}$ \eqref{z5D} reduces to 
\begin{equation}
\tens{z}\!=\!\tens{o}^{\hat \epsilon}\!\wedge (r{\tilde v}_{\hat 1}\tens{o}^{\hat 1}-
x_1{\tilde v}_{\hat 2}\tens{o}^{\hat 2})\,,
\end{equation}
and
\begin{equation}
S(\tens{F})=\{ 0^{(3)}, \lambda^{(2)}\}\,,\quad 
\lambda=\sqrt{-\kappa_1}\,.
\end{equation}
The eigenspace $V_0^*$ is spread by 
$\{\fb{{1}}{\varsigma}{0}, \fb{{2}}{\varsigma}{0}\}$, where
\begin{equation}
\fb{{1}}{\varsigma}{0}=\tens{o}^{\hat \epsilon},\quad  
\fb{{2}}{\varsigma}{0}=\lambda^{-1}(r{\tilde v}_{\hat 1}\tens{o}^{\hat 1}-
x_1{\tilde v}_{\hat 2}\tens{o}^{\hat 2})\,.
\end{equation}
[Notice that $\fb{{2}}{\varsigma}{0}$ is identical to the normalized 4-dimensional 1-form $\tens{z}$ given by \eqref{z4D}.]
The angle of rotation in the $\{\fb{{1}}{\varsigma}{0}, \fb{{2}}{\varsigma}{0}\}$ 2-plane obeys
the equation
\be\label{beta1Deg}
\dot \beta_1=\frac{C^{(1)}}{x_1^2r^2}
=\frac{f_1^{(1)}}{U_1}+\frac{f_2^{(1)}}{U_2}\,,\quad
f_1^{(1)}\equiv-\frac{C^{(1)}}{x_1^2}\,,\ \   
f_2^{(1)}\equiv\frac{C^{(1)}}{r^2}\,,\ \ 
C^{(1)}\equiv\lambda\sqrt{c}\,.
\ee
Thus $\beta_1$ is given by \eqref{beta1}
with functions $f_1^{(1)}, f_2^{(1)},$ defined in \eqref{beta1Deg}.
The eigenspace $V_{\lambda}^*$ is spread by
\begin{equation}
\tens{\varsigma}={\tilde v}_{\hat 1}\tens{\tilde o}^{\hat 2}+{\tilde v}_{\hat 2}\tens{\tilde o}^{\hat 1}\,,\quad 
\tens{\tilde \varsigma}=\lambda^{-1}(r{\tilde v}_{\hat 1}\tens{o}^{\hat 2}+x_1{\tilde v}_{\hat 2}\tens{o}^{\hat 1})\,, 
\end{equation}
which is identical to the 4D case. Thus the rotation angle $\beta_2$ coincides with   
$\beta$ given by \eqref{beta4D}.

\subsubsection{Summary of 5D}

To summarize, we have demonstrated that also in $D=5$ Kerr-NUT-(A)dS
spacetime the rotation angles in 2D eigenspaces can be separated and the
parallel-transported frame $\{\tens{\pi}\}$ explicitly constructed.  This result is
again valid off-shell, that is for arbitrary functions $X_2(r)$,
$X_1(x_1)$.

The special degenerate case characterized by $\Psi_2=0$ 
has the following interesting feature.
The zero-value eigenspace is spanned by the
4-dimensional $\tens{u}^\flat$, by the 4-dimensional  1-form 
$\tens{z}$, and $\tens{o}^{\hat \epsilon}$. The structure of
$V_{\lambda}^*$ is identical to the 4D case and the equation of
parallel transport in this plane is identical to the equation in 4D.
Therefore this 5D degenerate problem effectively reduces to
the generic 4D problem plus the rotation in the 2D 
$\{\tens{z}, \tens{o}^{\hat \epsilon}\}$ plane.
   
This indicates that a similar reduction might be valid also in higher dimensions.
Namely, one may expect that 
the degenerate odd-dimensional problem, $\Psi_n=0$,  effectively reduces to the
problem in a spacetime of one dimension lower plus the rotation  in the 2D
$\{\tens{z}, \tens{o}^{\hat \epsilon}\}$ plane.  
If this is so, one can use this odd-dimensional degenerate case to generate 
the solution for the generic (one dimension lower) even-dimensional problem.

\section{Conclusions} 
 
In this chapter we have described the construction of a
parallel-transported frame in a spacetime admitting the PCKY 
tensor $\bs{h}$. This tensor determines a canonical (Darboux)
basis at each point of the spacetime. The geodesic motion of a particle
in such a space can be characterized by the components of its velocity
$\bs{u}$ with respect to this basis. For a moving particle it is 
also natural to introduce a comoving basis, which is just a
Darboux basis of $\bs{F}$, where $\bs{F}$ is a projection of $\bs{h}$
along the velocity $\bs{u}$. Since $\bs{F}$ is parallel-propagated
along $\bs{u}$, its eigenvalues are constant along the
geodesic and its eigenspaces are parallel-propagated. We have
demonstrated that for a generic motion the parallel-propagated
basis can be obtained from the comoving basis by simple
two-dimensional rotations in the 2D eigenspaces of $\bs{F}$. 

To illustrate the general theory we have considered the parallel transport
in the Kerr-NUT-(A)dS spacetimes. Namely, we have newly constructed the
parallel-propagated frames in three and five dimensions and
re-derived the results \cite{Marck:1983kerr}, \cite{KamranMarck:1986} in 4D.  One of the interesting
features of the 4D construction, observed already by Marck, is that
the equation for the rotation angle allows a separation of variables.
Remarkably, we have shown that also in five dimensions equations
for the rotation angles can be solved by a separation of variables.
Moreover, the 4D result  can be understood as a special degenerate case 
of the 5D construction. 
Is this a general property valid in the  Kerr-NUT-(A)dS spacetime
in any number of dimensions? What underlines the
separability of the rotation angles? These are interesting open questions. 

The present analysis was restricted to the problem of 
parallel transport along timelike geodesics. The 
generalization to the case of spatial geodesics is straightforward.
The case of null geodesics requires additional
consideration and is under preparation \cite{ConnellEtal:2008null}. 
To conclude this chapter we would like to mention that the above described 
possibility of solving the parallel transport equations in
the Kerr-NUT-(A)dS spacetime is one more evidence of the miraculous
properties of these metrics connected with their hidden symmetries.

\chapter{Summary of results}

In this thesis we have described recent developments of the theory of
higher-dimensional black holes. We focused mainly on the problem of
hidden symmetries and separation of variables. (For more general
discussion of the modern status of the theory of higher-dimensional
black holes see, e.g., a recent review \cite{EmparanReall:2008}.)
By studying the hidden symmetries we have demonstrated that
higher-dimensional black holes are in many aspects similar to their
four-dimensional counterparts.

Explicit spacetime symmetries are represented by Killing vectors. Hidden
symmetries are related to generalizations of this concept.
One of the most important of these generalizations is the hidden 
symmetry encoded in the principal conformal Killing--Yano tensor.
We have demonstrated that the Myers--Perry metric, describing the higher-dimensional general rotating black hole, as well as its generalization, the Kerr-NUT-(A)dS metric
which includes the NUT parameters and the
cosmological constant, admit such a tensor. 

The PCKY tensor generates towers of hidden and explicit symmetries.
The tower of Killing tensors is responsible for the existence of irreducible, quadratic in momenta, conserved integrals of geodesic motion. 
These integrals, together with the integrals corresponding to the tower of explicit symmetries, make\break geodesic equations in the Kerr-NUT-(A)dS spacetime 
completely integrable. 
We have further demonstrated that 
the Hamilton--Jacobi and Klein--Gordon
equations allow complete separation of variables in this spacetime.
The separability of the Dirac equation was proved in \cite{OotaYasui:2008}. 
We have also shown that the Nambu--Goto equations 
for a stationary test string in the  Kerr-NUT-(A)dS background can be completely separated, and that 
the problem of finding parallel-propagated frames in these backgrounds reduces to the set of the first order ordinary differential equations.   
It was also demonstrated \cite{ChenLu:2008} that 
the Kerr-NUT-(A)dS solution can be presented in  
the generalized Kerr--Schild form and that it belongs 
to the class of spacetimes of the 
special algebraic type D \cite{HamamotoEtal:2007}, \cite{PravdaEtal:2007}
of the higher-dimensional algebraic classification. 
All these remarkable properties make higher-dimensional black hole solutions very
similar to the 4D black holes.

To complete this analogy we have addressed the question of generality and uniqueness of these developments. Namely, we have studied 
the most general metric elements admitting the 
PCKY tensor---with and without imposing the Einstein equations.
The result can be summarized by the following theorem:\\
{\bf Theorem.} {\em The most general spacetime admitting the PCKY tensor
is the canonical metric \eqref{metric_coordinates}. It possesses the following properties:
\begin{enumerate}
\item
 It is of the algebraic type~D.
\item
 It allows a separation of variables for
the Hamilton--Jacobi, Klein--Gordon, Dirac, and stationary string equations.
\item 
The geodesic motion in such a spacetime is completely integrable. The problem 
of finding parallel-propagated frames reduces to the set of the first order ordinary differential equations.
\item
When the Einstein equations with the cosmological constant are imposed
the canonical metric becomes the Kerr-NUT-(A)dS spacetime 
\eqref{metric_coordinates}--\eqref{X}.
\end{enumerate}
}
This theorem naturally generalizes the results obtained earlier in four dimensions (see Appendix A).

Our work motivates further developments in this field.
For example, the proved separability of the scalar field equations 
has led to the construction of the corresponding symmetry operators \cite{SergyeyevKrtous:2008}.
It also opens the possibility that the equations with spin can be decoupled and separated. As described above, the separation of the Dirac equation was already demonstrated \cite{OotaYasui:2008}. 
An important open question is a separability problem for
the electromagnetic and gravitational perturbations in higher-dimensional
black hole spacetimes.  
Such a separability would provide us the important tools for studying the stability, quasinormal modes,  
and different aspects of the Hawking radiation  of higher-dimensional black holes. 
A certain progress in this direction was already achieved
(see, e.g., \cite{KodamaIshibashi:2003}, \cite{KunduriEtal:2006}, \cite{MurataSoda:2008}).
However, most of the results obtained in these directions so far (see also references in Chapter 6) assumed some additional restrictions on the parameters characterizing black hole solutions. This reminds a situation for the Klein--Gordon and
Dirac equations before the general results on their separability were
proved.

The similarity of higher-dimensional black holes with their four-dimensional cousins also inspires the search for new higher-dimensional solutions. For example,
by relaxing some of the conditions on the PCKY tensor the authors of \cite{HouriEtal:2008b}, \cite{HouriEtal:2008c} were able to find `generalized Kerr-NUT-(A)dS spacetimes'.
Similarly, by generalizing our (unsuccessful) procedure of rescaling the Kerr-NUT-(A)dS metric \cite{KubiznakKrtous:2007} (see also Appendix C.6), new five-dimensional black hole solutions were recently obtained \cite{LuEtal:2008a}, \cite{LuEtal:2008b}.

The curvature of the Kerr-NUT-(A)dS spacetime and its algebraic type were 
studied in \cite{HamamotoEtal:2007}.
The relationship between the existence of the PCKY tensor and the uniqueness of this  spacetime was first addressed in \cite{HouriEtal:2007}, \cite{HouriEtal:2008a}.
Other recent papers which deal with hidden symmetries or the subjects addressed in this thesis are, for example, \cite{AcikEtal:2008a}, \cite{AcikEtal:2008b}, \cite{AcikEtal:2008c}, \cite{AhmedovAliev:2008}, \cite{ConnellFrolov:2008}, \cite{GoodingFrolov:2008}, \cite{HackmannLammerzahl:2008}, \cite{HiokiMiyamoto:2008},  \cite{KagramanovaEtal:2008}, \cite{Papadopoulos:2008},  \cite{Wu:2008}, \cite{Wu:2008b}.

The results presented in this thesis can be used for
studying the particle and light propagation in higher-dimensional
rotating black hole spacetimes. They allow us to calculate the
contribution of the scalar and Dirac fields to the bulk Hawking
radiation, without any restrictions on black hole parameters.
The important open questions are:  Is it possible to decouple the higher spin massless field
equations in the background of the general Kerr-NUT-(A)dS metric?
And do they allow separation of variables? 
Recent result on the separability of the massive Dirac equation is quite promising.
Separability of the higher spin equations, and especially the
equations for the gravitational perturbations, would provide one with
powerful tools, important, for example, for studying the stability
of higher-dimensional black hole solutions. One might  hope that it
will not take too long before this and other interesting open
questions connected with the existence of hidden symmetries in
higher-dimensional black hole spacetimes will find their answers.

%
\appendix

\addcontentsline{toc}{part}{Appendices}
\chapter{On hidden symmetries in 4D}
In this appendix we discuss some aspects of hidden symmetries in 4D. 
The first section plays the role of an introduction
for newcomers to the field where, on the well known 4D case, we
illustrate the main ideas of the more complicated higher-dimensional
theory developed in Parts I and II of this thesis.
In the second section, based on \cite{KubiznakKrtous:2007}, we discuss hidden symmetries for the Pleba\'nski--Demia\'nski family of type D solutions.

\section{Introduction for newcomers}

The key object of the theory in higher dimensions is  a {\em principal conformal Killing--Yano} (PCKY) tensor. 
We start discussing this object and its properties in a $4$D
flat spacetime and demonstrate how it generates other objects
(Killing--Yano and Killing tensors) responsible for hidden
symmetries.  Then we show how this PCKY tensor allows one
easily to `generate' the $4$D Kerr-NUT-(A)dS metric starting from the
flat one---written in the {\em canonical coordinates} determined by
this tensor. Finally, we discuss the separation
of variables in the $4$D Kerr-NUT-(A)dS spacetime in the canonical coordinates.

\subsection{Principal conformal Killing--Yano tensor}

Consider a 4D flat spacetime with the metric
\begin{equation}\label{flat}
dS^2=\eta_{ab}dX^adX^b=-dT^2+dX^2+dY^2+dZ^2\,.
\end{equation}
The PCKY tensor $\tens{h}$ is a (non-degenerate) rank-$2$ closed conformal Killing--Yano tensor. Therefore, there exists a $1$-form potential $\tens{b}$, so
that $\tens{h}=\tens{db}$. Let us consider the following ansatz:
\be\label{bb_A}
\bss{b}={1\over 2}\,\bigl[-R^2\tens{d}T+a(Y\tens{d}X-X\tens{d}Y)\bigr]\,,\quad 
R^2=X^2+Y^2+Z^2\, .
\ee
Our choice of this special form for the
potential $\tens{b}$ will become clear later, when it will be shown that this is
a flat spacetime limit of the potential for the PCKY
tensor in the Kerr-NUT-(A)dS spacetime. For a moment we just mention
that the form \eqref{bb_A} of the potential $\bss{b}$ singles out time coordinate $T$, 
a two-dimensional $(X,Y)$ plane in space, and contains an arbitrary constant $a$.

It can be easily shown that 
\be\label{hh_A}
\bss{h}=\bss{db}=\tens{d}T\wedge(X \tens{d}X+Y \tens{d}Y+Z\tens{d}Z)+a \tens{d}Y
\wedge \tens{d}X\, 
\ee
is a closed conformal Killing--Yano tensor 
[cf. also \eqref{PCKY_KS}, for the higher-dimensional case].
It means that its dual 2-form $\bss{f}=*\bss{h}$ is the Killing--Yano (KY) tensor.\footnote{
In $D$ dimensions the maximum number of (linear independent) Killing--Yano tensors of a given rank-$p$ is  
\be
N_p=\left(\!\!   
\begin{array}{c}
D \\
p\end{array}
\!\!\right)\,+ 
\left(\!\!   
\begin{array}{c}
D \\
p+1\end{array}
\!\!\right)\,=\frac{(D+1)!}{(D-p)!(p+1)!}\,.
\ee
This reflects the fact that, similar to Killing vectors,  
Killing--Yano tensors are completely determined by the values of
their components and the values of their (completely antisymmetric) 
first derivatives at a given point. Flat space has the maximum number of independent 
Killing--Yano tensors of each rank. Any KY tensor there can be written as 
a linear combination of  `translational' KY tensors (which are a simple wedge
product of
translational Killing vectors) and `rotational'  KY tensors (which
are a wedge product of translations with a spacetime rotation,
completely antisymmetrized) \cite{KastorTraschen:2004}. In particular case of $D=4$
we have 10 rank-$2$ KY tensors (6 translational and 4 rotational).
}
\begin{equation}\label{ff_A}
\tens{f}=X \tens{d}Z\wedge \tens{d}Y+Z \tens{d}Y\wedge \tens{d}X+
Y \tens{d}X\wedge \tens{d}Z+a \tens{d}Z\wedge \tens{d}T\,.
\end{equation}

Let us put, for a moment, $a=0$. Then the KY tensor $f_{ab}$ has only
spatial components $f_{ik}\,,$ and the Killing tensor $K_{ab}=f_{ac}f_b{}^{c}$, \eqref{square}, reads
\begin{equation}
K_{ij}=R^2\delta_{ij}-X^iX^j=\!\!\sum_{k=X,Y,Z} \xi_{(k)\,i}\xi_{(k)\,j}\,,
\quad \xi_{(k)\,i}=\epsilon_{kji}X^j\,.
\end{equation}
Here ${\xi}_{(k)\,i}$ are the spatial rotational Killing vectors.
Therefore, the Killing tensor $\tens{K}$ can be written as a sum of
products of  Killing vectors, and thus it is {\em reducible}.
Parallel-propagated vector $L_a=f_{ab}p^b$, \eqref{Lc}, with the nontrivial components
\begin{equation}
L_{i}=f_{ik}p^k=\epsilon_{ijk}X^jp^k={\xi}_{(k)\,i}p^k\,,
\end{equation}
has the meaning of the conserved angular momentum.\footnote{%
In general, for a simple spacelike ($f_{ab}f^{ab}>0$) Killing--Yano
tensor  $\tens{f}$, there exists a close analogy between the angular
momentum of classical mechanics and the vector $L^a=f^{ab}p_{b}$
\cite{DietzRudiger:1981}.}  The conserved quantity $K=K_{ab}p^ap^b$,
\begin{equation}
K(a=0)=\!\!\sum_{k=X,Y,Z} L^2_{k}={\vec L}^2\,,
\end{equation}
is the square of the total angular momentum.

For $a\neq 0$ the conserved quantity
\begin{equation}\label{K_A}
K={\vec L}^2+2aEL_{Z}+a^2(E^2-p_{Z}^2)
\end{equation}
is also reducible. Here $E=-p_T$ and
$p_{Z}$ are the conserved energy and the momentum
in the $Z$-direction, respectively. 

\subsection{`Derivation' of the 4D Kerr-NUT-(A)dS metric} 

Consider a general case with $a\ne 0$. We first introduce the ellipsoidal
coordinates\footnote{In this step, we associate  constant $a$ with
`rotation' parameter.} 
\begin{equation}\n{eq_3.15}
X=\sqrt{r^2+a^2}\sin\theta\cos\phi,\quad 
Y=\sqrt{r^2+a^2}\sin\theta\sin\phi,\quad 
Z=r\cos\theta\,, 
\end{equation}
and rewrite the metric, the potential, the PCKY tensor, and the KY tensor as 
\begin{equation}
\begin{split}
dS^2=&\,-dT^2+(r^2+a^2)\sin^2\!\theta d\phi^2+(r^2+a^2\cos^2\theta)
\bigl(\frac{dr^2}{r^2+a^2}+d\theta^2\bigr)\,,\\
\tens{b}=&\,\frac{1}{2}\bigl[-(r^2+a^2\sin^2\!\theta)\tens{d}T-a\sin^2\!\theta(r^2+a^2)\tens{d}\phi\bigr]\,,\\
\tens{h}=&\,-r \tens{d}r\wedge(\tens{d}T+a\sin^2\!\theta \tens{d}\phi)-a\sin\theta \cos\theta \tens{d}\theta\wedge
\bigl[a\tens{d}T+(r^2+a^2)\tens{d}\phi\bigr]\,,\\
\tens{f}=&\,a\cos\theta \tens{d}r\wedge(\tens{d}T+a\sin^2\!\theta \tens{d}\phi)-r\sin\theta \tens{d}\theta\wedge
\bigl[a\tens{d}T+(r^2+a^2)\tens{d}\phi\bigr]\,.
\end{split}
\end{equation}
Second, we introduce the new coordinates  
\begin{equation}\n{eq_3.18}
y=a\cos\theta,\quad t=T+a\phi,\quad \psi=-\phi/a\,,
\end{equation}
in which the metric takes the `algebraic' form 
\begin{equation}\label{CDP}
dS^2=-\frac{\Delta_r (dt+y^2d\psi)^2}{r^2+y^2}+\frac{\Delta_y (dt-r^2d\psi)^2}{r^2+y^2}
+\frac{(r^2+y^2)dr^2}{\Delta_r}+\frac{(r^2+y^2)dy^2}{\Delta_y}\,,
\end{equation}
where 
\begin{equation}\label{ry_A}
\Delta_r=r^2+a^2,\quad 
\Delta_y=a^2-y^2\,.
\end{equation}
The hidden symmetries are
\begin{equation}\label{form_AA}
\begin{split}
\tens{b}=&\,\frac{1}{2}\bigl[(y^2-r^2-a^2)\tens{d}t-r^2 y^2 \tens{d}\psi\bigr]\,,\\
\tens{h}=&\,y\tens{d}y\wedge(\tens{d}t-r^2\tens{d}\psi)-r\tens{d}r\wedge(\tens{d}t+
y^2\tens{d}\psi)\,,\\
\tens{f}=&\,r\tens{d}y\wedge(\tens{d}t-r^2\tens{d}\psi)+y\tens{d}r\wedge(\tens{d}t+y^2\tens{d}\psi)\,.
\end{split}
\end{equation}
In the potential $\tens{b}$ the term proportional to $a^2$ is constant and may be omitted.
We remind that \eqref{CDP}--\eqref{ry_A} is just a metric of the flat
space written in special coordinates. 

Let us consider now the metric \eq{CDP} without imposing 
conditions \eqref{ry_A} on functions $\Delta_r$ and
$\Delta_y\,,$ but assuming that they are functions of $r$ and $y$,
respectively. 
Remarkably, we realize that the objects $\bss{b}$, $\tens{h}$, and $\bss{f}$ 
\eqref{form_AA} are again  the potential, the PCKY tensor, and the KY tensor. 
We call \eq{CDP} 
with arbitrary $\Delta_r(r)$ and $\Delta_y(y)\,,$ a 
{\em canonical (off-shell) metric}. 
It possesses the hidden symmetries \eqref{form_AA}. 

In particular, let us impose the vacuum Einstein equations 
with the cosmological constant
\be\n{eqq_A}
R_{ab}=-3\lambda g_{ab}\, .
\ee
These equations are satisfied provided that 
\be
{d^2{\Delta}_r\over dr^2}+ {d^2{\Delta}_y\over dy^2}=12\lambda(r^2+y^2)\, .
\ee
The most general solution of this equation is 
\be\label{functions_A}
\Delta_r=(r^2+a^2)(1+\lambda r^2)-2Mr\hhh
\Delta_y=(a^2-y^2)(1-\lambda y^2)+2Ny\,.
\ee 
In other words, a simple
replacement of functions 
\eqref{ry_A} by more general polynomials \eqref{functions_A} generates a
non-trivial solution of the Einstein equations from a flat one. This
solution is  the Kerr-NUT-(A)dS  metric 
written in the {\em canonical form} (see, e.g., \cite{ChenEtal:2006cqg}).
It obeys the Einstein equations
with the cosmological constant, cf. Eq. \eqref{eqq_A}. $M$ stands for mass, and
parameters $a$ and $N$ are connected with rotation and NUT
parameter \cite{GriffithsPodolsky:2006b}.  

Let us remark here that {\em the canonical metric \eqref{CDP} is the most general 
metric element admitting the (non-degenerate) PCKY tensor} \cite{DietzRudiger:1981}, \cite{Taxiarchis:1985}.
The derivation above therefore establishes that {\em the most general Einstein space 
admitting the PCKY tensor is the Kerr-NUT-(A)dS spacetime} 
\eqref{CDP}, \eqref{functions_A}. We also remark that 
even a more general family of solutions--the Carter's spacetime (described in the next section) can be written in the form \eqref{CDP}.

In the Kerr-NUT-(A)dS spacetime neither
the square of the total angular momentum, ${\vec L}^2$, nor the projection of the
momentum on the $Z$-axis, $p_Z$, which enter \eqref{K_A} have well defined
meaning. However, the quadratic in momentum quantity, $K^{ab}p_ap_b$\,,
is well defined and conserved. In the
absence of the cosmological constant and NUT parameter, that is for
the Kerr black hole, this quantity
can be presented in the form \eqref{K_A} in the asymptotic region, where the spacetime is practically
flat. The angular momentum
and other quantities which enter \eqref{K_A} must be then understood as the
corresponding asymptotically conserved quantities. Since the energy
$E$ and the angular momentum along the axis of symmetry $L_Z$ are conserved exactly in any stationary axisymmetric
spacetime they can be excluded from \eqref{K_A} and the asymptotically
conserved quantity can be written as follows \cite{RosquistEtal:2007}:
\be
Q=L_{X}^2+L_{Y}^2-a^2p_Z^2\, .
\ee
For a scattering of particles in the Kerr metric, the presence of an
exact integral of motion connected with the Carter's constant implies
that the quantity $Q$ calculated for the incoming from infinity
particle must be the same as $Q$ calculated at infinity for the
outgoing particle. An interesting question is the following: suppose that 
such a conservation law is established for any scattering of particles by a
localized object, can one conclude that the metric of this object
possesses a hidden symmetry?

\subsection{Symmetric form of the metric}

Let us perform the `Wick' rotation in radial coordinate $r$. This  
transforms the metric \eqref{CDP}, \eqref{functions_A} and its hidden symmetries into 
a {\em symmetric form} \cite{ChenEtal:2006cqg}.
After the transformation
\begin{equation}\n{wr_A}
r=ix\,,\quad M=iN_x\,,\quad N=N_y\,,
\end{equation}
the metric and the KY objects are 
\ba
ds^2\!\!\!&=&\!\!\!\frac{\Delta_x (dt\!+\!y^2d\psi)^2}{x^2-y^2}+
\frac{\Delta_y (dt\!+\!x^2d\psi)^2}{y^2-x^2}
+\frac{(x^2\!-\!y^2)dx^2}{\Delta_x}+\frac{(y^2\!-\!x^2)dy^2}{\Delta_y}\,,\ \ \label{wick}\\
\Delta_x\!\!\!&=&\!\!\!(a^2-x^2)(1-\lambda x^2)+2N_xx\,,\ 
\Delta_y=(a^2-y^2)(1-\lambda y^2)+2N_yy\,,\label{deltawick}\\
\tens{b}\!\!\!&=&\!\!\!\frac{1}{2}\bigl[(x^2+y^2)\tens{d}t+x^2y^2\tens{d}\psi\,\bigr]\,,\n{eq_3.29}\\
\tens{h}\!\!\!&=&\!\!\!y\tens{d}y\wedge(\tens{d}t+x^2\tens{d}\psi)+x\tens{d}x\wedge(\tens{d}t+y^2\tens{d}\psi)\,,\label{hwick}\\
\tens{f}\!\!\!&=&\!\!\!x\tens{d}y\wedge(\tens{d}t+x^2\tens{d}\psi)+y\tens{d}x\wedge(\tens{d}t+y^2\tens{d}\psi)\,.\label{fwick}
\ea
These forms of the Kerr-NUT-(A)dS spacetime and of the KY potential allow the natural
generalizations to higher dimensions \cite{ChenEtal:2006cqg}, \cite{KubiznakFrolov:2007},
which are used throughout the thesis.\footnote{
It is obvious from the derivation that this symmetric form of the metric and of
its hidden symmetries is an analytical continuation of the real physical
quantities \eqref{CDP}, \eqref{form_AA}, \eqref{functions_A}. The signature of the metric for 
this continuation depends on the domain of coordinates $x$ and $y$ and signs of $\Delta_x$ and $\Delta_y$. For example, for $x^2>y^2$ and
$\Delta_x>0$, $\Delta_y<0$ it is of the Euclidean signature. The
transition to the physical space is given by \eqref{wr_A}. 
}

\subsection{PCKY tensor and canonical coordinates}

We demonstrate now that coordinates $(t,x,y,\psi)$ used in 
\eq{wick}--\eq{fwick} have a deep invariant meaning.  We start in a flat spacetime 
\eqref{flat}. Let us define 
\be
{Q}^a_{\ b}=-h^{a c}h_{cb}\hh
\Delta^a_{\  b}={Q}^a_{\ b}-q \delta^a_{\  b}\, ,
\ee
then one has
\be
\Delta^a_{\ b} = \left( 
{\begin{array}{cccc}
-R^2 - q  & a \,Y &  - a \,X & 0 \\
 - a \,Y & a^2-X^{2} - q  & -Y\,X & -Z\,X \\
a \,X & -Y\,X & a^2-Y^{2} - q  & -Z\,Y \\
0 & -Z\,X & -Z\,Y & -Z^{2} - q 
\end{array}}
 \right) \, .
\ee
The condition $\det(\Delta)=0$ which determines the eigenvalues $q$
of the operator $\tens{Q}$
is equivalent to the following equation:\footnote{%
The tensor $\tens{Q}$ is the conformal Killing tensor. 
It is related to $\tens{K}$ as 
\be
K_{ab}=Q_{ab}-\frac{1}{2}g_{ab}Q^c_{\ c}\,,\qquad
Q_{ab}=K_{ab}-\frac{1}{D-2}\,g_{ab}K^c_{\ c}\,.
\ee
}  
\be
q^{2}+(R^2-a^2)q -a ^{2}\,Z^{2}=0\, .
\ee
Hence, the eigenvalues of $\tens{Q}$ are
\be
q_{\pm}={1\over 2}\Bigr[ a^2-R^2\pm\sqrt{(R^2-a^2)^2+4a^2Z^2}\,\Bigr]\, .
\ee
Simple calculations using \eq{eq_3.15}, \eqref{eq_3.18}, and \eqref{wr_A}, give
\be
q_+=a^2\cos^2\theta=y^2\,,\quad
q_-=-r^2=x^2\,.
\ee
Thus the coordinates $x$ and $y$ in \eq{wick} are uniquely
determined as the eigenvalues of the operator $\tens{Q}$ constructed
from the PCKY tensor $\tens{h}$. Let us show
now that the same tensor $\tens{h}$ uniquely determines the coordinates
$t$ and $\psi$. The {\em primary} Killing vector $\tens{\xi}$ and the 
the {\em secondary} Killing vector $\tens{\eta}$ are [cf. Eqs. \eqref{kvv}]
\ba
{\xi}^a\!\!&=&\!\!\frac{1}{3}\nabla_c h^{ca}=(\partial_T)^a\,,\label{vector1_A}\\
{\eta}^{a}\!\!&=&\!\!-K^{ab}{\xi}_{b}=a^2(\partial_{T})^a+
aY(\partial_X)^a-aX(\partial_Y)^a\, \, .\label{vector2_A}
\ea
In coordinates \eq{eq_3.18} one has
\be
\tens{\xi}=\pa_t\hh
\tens{\eta}=\pa_{\psi}\, .
\ee
This means that the coordinates $t$ and $\psi$ are the affine parameters
along the primary and secondary Killing vectors $\tens{\xi}$ and 
$\tens{\eta}$, determined by the tensor $\tens{h}$.

It can be checked that the same is true for (the symmetric form of) the canonical metric \eq{wick} with the PCKY tensor $\tens{h}$ given by \eqref{hwick}. This underlines the exceptional role of the PCKY tensor. 
Remarkably, the existence of a similar object in higher dimensions generates the higher-dimensional 
Kerr-NUT-(A)dS spacetime and determines its canonical coordinates, in a way exactly analogous to four dimensions (see Chapter 7).

\subsection{Separation of variables}

The last subject we would like to discuss in this brief review of
properties of the 4D Kerr-NUT-(A)dS metric is the separation of
variables for the Hamilton--Jacobi and Klein--Gordon equations. 
More generally, we consider these equations in the off-shell spacetime
\eq{wick}, with $\Delta_x(x)$ and $\Delta_y(y)$  
arbitrary functions. 

Let us first discuss the Klein--Gordon equation
\be\n{KG_A}
\Box \Phi-\mu^2\Phi =0\, .
\ee
The separation of variables of equation \eqref{KG_A} in
canonical coordinates $(\tau,x,y,\psi)$ means that $\Phi$ can be
decomposed into modes
\be
\Phi=e^{i\varepsilon \tau+im\psi} X(x) Y(y)\, .
\ee
Indeed, substituting this expression in the Klein--Gordon equation \eqref{KG_A} one
obtains
\be\label{sep1_A}
(\Delta_x X')'+V_x X=0\,,\quad
V_x=\kappa+\mu^2 x^2-{ (\varepsilon x^2-m)^2\over \Delta_x}\, ,
\ee
\be\label{sep2_A}
(\Delta_y Y')'+V_y Y=0\,,\quad
V_y=\kappa+\mu^2 y^2-{ (\varepsilon y^2-m)^2\over \Delta_y}\, .
\ee
Here, the prime stands for the derivative of function with respect to 
its single argument. The separation constants $\varepsilon$ and $m$ are connected
with the symmetries generated by the Killing vectors
$\tens{\xi}=\pa_{t}$ and 
$\tens{\eta}=\pa_{\psi}$. An additional separation constant
$\kappa$ is connected with the hidden symmetry generated by the Killing
tensor $\tens{K}$.
It should be emphasized, that in order to use the proved 
separability for concrete calculations in the physical Kerr-NUT-(A)dS
spacetime \eq{CDP}, \eqref{functions_A}, one needs to specify functions $\Delta_x$
and $\Delta_y$ to have the form \eqref{deltawick} and perform the Wick transformation
inverse to \eqref{wr_A}. This transformation `spoils' the symmetry between
the essential coordinates but the separability property remains.
In coordinates $r$ and $y$ in the `physical' sector equations 
\eqref{sep1_A} and \eqref{sep2_A} play different roles. Eq. \eqref{sep2_A} with imposed
regularity conditions serves as an eigenvalue problem which determines the spectrum
of $\kappa$. Eq. \eqref{sep1_A} is a radial equation for propagating modes.

Similarly, the Hamilton--Jacobi equation for geodesic motion
\be
\partial_{\lambda} S+g^{ab}\partial_{a} S\partial_{b} S=0
\ee
in the background \eq{wick} allows a separation of variables and
$S$ can be written in the form
\be
S=\mu^2\lambda+\varepsilon \tau+m\psi+S_x(x)+S_y(y)\, .
\ee
The functions $S_x$ and $S_y$ obey the equations
\be
({S'}_x)^2={V_x\over \Delta_x}\,,\quad 
({S'}_y)^2={V_y\over \Delta_y}\, .
\ee

The separability of the Hamilton--Jacobi and Klein--Gordon equations demonstrated above 
is directly connected with the existence of the Killing tensor and the corresponding symmetry operator \cite{Carter:1977} (see also Section 6.3.2).
Let us mention that a similar intrinsic characterization of the separation constants (connected with the PCKY tensor) can be found in the case of the Dirac equation \cite{CarterMcLenaghan:1979}, \cite{KamranMcLenaghan:1984}, as well as in the case of the massless equations with spin \cite{Kamran:1985}, \cite{KalninsEtal:1986}, \cite{KalninsMiller:1989}, \cite{KalninsWilliams:1990}, \cite{KalninsEtal:1992}, \cite{KalninsEtal:1996}. For example, one of the very convenient ways how to prove 
that the Maxwell equations in the background \eqref{wick} decouples and separate is the 
method of the Debye potentials, directly based on the existence of the CKY tensor \cite{BennEtal:1997}.

\section{CKY tensors for the Pleba\'nski--Demia\'nski class of solutions}

In this section, based on \cite{KubiznakKrtous:2007},
 we present explicit expressions for the 
conformal Killing--Yano tensors for the
Pleba\'nski--Demia\'nski family of type D solutions.
Some physically important
special cases are discussed in more detail.
In particular, it is demonstrated how the conformal 
Killing--Yano tensor becomes the Killing--Yano tensor 
for the solutions without acceleration.

\subsection{Pleba\'nski--Demia\'nski metric}
\label{sc:intro}
The important family of type D spacetimes in four dimensions, including the black-hole spacetimes 
like the Kerr metric, the metrics describing the accelerating sources as the C-metric, or the
non-expanding Kundt's class type D solutions, 
can be represented by the general seven-parameter 
metric discovered by Pleba\'nski and Demia\'nski  \cite{PlebanskiDemianski:1976}
(cf.~also \cite{Debever:1971}). 
Recently, Griffiths and Podolsk\'y 
\cite{GriffithsPodolsky:2005}, 
\cite{GriffithsPodolsky:2006b}, \cite{GriffithsPodolsky:2006a}, 
\cite{PodolskyGriffiths:2006}, \cite{GriffithsPodolsky:2007},  
put this metric into a new form which enables 
a better physical interpretation of parameters
and simplifies a procedure how to derive all special cases. 
Among subclasses of this solution let us mention  
the six-parameter family of metrics without acceleration derived 
and studied already by Carter \cite{Carter:1968pl}, \cite{Carter:1968cmp} and 
later by Pleba\'nski \cite{Plebanski:1975}.

It turns out that the elegant form of the Pleba\'nski--Demia\'nski metric not only yields 
new solutions in 4D (see, e.g., \cite{KlemmEtal:1997}, \cite{AlonsoEtal:2000}), but also inspires for its generalizations into higher dimensions. For example, 
Chen, L\"u, and Pope \cite{ChenEtal:2006cqg} were able to cast the Einstein space subclass of Carter's non-accelerating solutions into higher dimensions---thus constructing the general 
Kerr--NUT--(A)dS metrics in all dimensions. These are discussed in the main text.
Recently, even more general solutions in 5D
\cite{LuEtal:2008a}, \cite{LuEtal:2008b}, directly inspired by the Pleba\'nski--Demia\'nski metric, were obtained (see also Appendix C.6).
  
One of the most remarkable properties of the Carter's 
subclass of non-\break accelerating solutions,
which is also inherited by its higher dimensional generalization (see Chapter 4),
is the existence of hidden symmetries associated with the Killing--Yano tensor \cite{Penrose:1973}, \cite{Carter:1987}.
In four dimensions the integrability conditions for the existence of a non-degenerate Killing--Yano 
tensor restricts the Petrov type of spacetime to type D (see, e.g., \cite{Collinson:1974}, \cite{DietzRudiger:1981}). 
Demia\'nski and Francaviglia \cite{DemianskiFrancaviglia:1980} demonstrated  that from the known type D solutions only spacetimes without acceleration of sources actually admit this tensor.
The purpose of this section is to show that the general Pleba\'nski--Demia\'nski metric 
admits the conformal generalization of the Killing--Yano tensor.
We also explicitly demonstrate how in the absence of
acceleration this tensor becomes the known Killing--Yano tensor of 
the Carter's metric. The explicit expressions for this tensor for the physically important cases are presented.

The original form of the Pleba\'nski--Demia\'nski metric \cite{PlebanskiDemianski:1976} is given by
\be\label{metric_A}
\tens{g}=\Omega^{2}\Bigl[-\frac{Q(\grad\tau-p^2\grad\sigma)^2}{r^2+p^2}+\frac{P(\grad\tau+r^2\grad\sigma)^2}{r^2+p^2}
+\frac{r^2+p^2}{P}\,\grad p^2+\frac{r^2+p^2}{Q}\,\grad r^2\Bigr]\;.
\ee
This metric obeys the Einstein--Maxwell equations with the electric and magnetic charges $e$ and $g$ and the cosmological 
constant $\Lambda$ provided that functions $P=P(p)$ and $Q=Q(r)$ take the particular form
\begin{equation}\label{functionsQP_A}
\begin{split}
Q&=k\!+\!e^2\!+\!g^2\!-\!2mr\!+\!\epsilon r^2\!-\!2nr^3\!-\!(k\!+\!\Lambda/3)r^4\;,\\
P&=k\!+\!2np\!-\!\epsilon p^2\!+\!2mp^3\!-\!(k\!+\!e^2\!+\!g^2\!+\!\Lambda/3)p^4\,,
\end{split}
\end{equation}
the conformal factor is
\begin{equation}\label{Omega_A}
\Omega^{\!-1}=1-pr\;,
\end{equation}
and the vector potential reads
\begin{equation}\label{A_A}
\tens{A} = -\frac{1}{r^2\!+\!p^2}\,
               \Bigl[ e\,r\,\bigl(\grad \tau \!-\! p^2\,\grad\sigma\bigr)
               +\,g\,p\,\bigl(\grad \tau \!+\! r^2\,\grad\sigma\bigr) \Bigr]\;.
\end{equation}

The general Pleba\'nski--Demia\'nski metric \eqref{metric_A} admits
the CKY tensor \cite{KubiznakKrtous:2007}
\begin{equation}\label{k_A}
\tens{k}=\Omega^{3}\Bigl[p\,\grad r\wedge(\grad\tau-p^2\grad\sigma)+r\,\grad p\wedge(\grad\tau+r^2\grad\sigma)\Bigr]\;.
\end{equation}
Using the Maple program, one can easily verify that Eqs. \eqref{CKY_4rank2},
\begin{equation}\label{CKY_4rank2_A}
\nabla_{a}k_{bc}= \nabla_{[a}k_{bc]}+ {2}\, g_{a[b} \xi_{c]}\;,\quad
\xi_{a}=\frac{1}{3}\,\nabla_{c}k^{c}{}_{a}\;,
\end{equation}
are satisfied. An independent proof is given at the end of this appendix. 
The Hodge dual, ${\tens h}=\tens{*k}$, is also a CKY tensor. It reads 
\begin{equation}\label{bconf_A}
\tens{h}=\Omega^{3}\Bigl[r\,\grad r\wedge(p^2\grad\sigma-\grad\tau)+p\,\grad p\wedge(r^2\grad\sigma+\grad\tau)\Bigr]\;,
\end{equation}
which is equivalent to 
\begin{equation}\label{bform_A}
\tens{h}=\Omega^{3}\,\grad{\tens{b}}\;,
\end{equation}
where 
\begin{equation}\label{b_apA}
\tens{b}=\frac{1}{2}\bigl[(p^2-r^2)\grad\tau+p^2r^2\grad\sigma\bigr]\;.
\end{equation}
It is interesting to mention that $\tens{k}$ and $\tens{h}$ are CKY tensors for the metric \eqref{metric_A}, with 
an arbitrary conformal factor $\Omega$ and arbitrary functions $P(p)$, $Q(r)$, i.e., irrespectively of the fact whether the metric \eqref{metric_A} solves the Einstein equations or not. We shall return to this remark later. We shall also see 
that in the absence of acceleration of sources, $\tens{k}$ becomes the KY tensor and  $\tens{h}$ becomes the closed CKY tensor. 

For ${\Omega}$ given by \eqref{Omega_A} and arbitrary functions
$P(p)$ and $Q(r)$ both isometries of the spacetime follow from the existence 
of $\tens{h}$ and $\tens{k}$ as follows [cf. Eqs.~\eqref{vector1_A}, \eqref{vector2_A}]:
\begin{equation}
\tens{\xi}_{({h})}\equiv-\frac{1}{3}\tens{\delta h}=\tens{\partial}_{\tau}\;,\qquad
\tens{\xi}_{{(k)}}\equiv -\frac{1}{3}\tens{\delta k}=\tens{\partial}_{\sigma}\;.
\end{equation}
The conformal Killing tensor, \eqref{cKT}, associated with $\tens{k}$ reads
\be\label{Qk_A}
\tens{Q}_{(k)}=\Omega^4\Bigl[\frac{Qp^2(\grad\tau\!-\!p^2\grad\sigma)^2}{r^2+p^2}+\frac{Pr^2(\grad\tau\!+\!r^2\grad\sigma)^2}{r^2+p^2}
+\frac{r^2\!+\!p^2}{P}\,r^2\grad p^2-\frac{r^2\!+\!p^2}{Q}\,p^2\grad r^2\Bigr]\,.
\ee
It inherits the `universality' of $\tens{k}$, i.e.,
it is a conformal Killing tensor of the metric \eqref{metric_A} with an arbitrary $\Omega$, and arbitrary $Q(r)$ and $P(p)$. 
In the absence of acceleration $\tens{Q}_{(k)}$ becomes a Killing tensor which generates the Carter's constant for a geodesic motion \cite{Carter:1968pr}.
The conformal Killing tensor associated with ${\tens{h}}$ is 
\be\label{Qh_A}
\tens{Q}_{(h)}=\Omega^4\Bigl[\frac{Qr^2(\grad\tau\!-\!p^2\grad\sigma)^2}{r^2+p^2}+\frac{Pp^2(\grad\tau\!+\!r^2\grad\sigma)^2}{r^2+p^2}
+\frac{r^2\!+\!p^2}{P}\,p^2\grad p^2-\frac{r^2\!+\!p^2}{Q}\,r^2\grad r^2\Bigr]\,.
\ee
Both tensors are related as
\begin{equation}\label{QhQk}
  \tens{Q}_{(h)} = \tens{Q}_{(k)} + \Omega^2(p^2-r^2)\, \tens{g}\;.
\end{equation}

Following \cite{GriffithsPodolsky:2006b} one can easily perform the transformations of coordinates and parameters
to obtain the complete family of type D spacetimes and the corresponding particular forms of CKY tensors.
In the next two sections we consider two special cases.
 First, we deal with the generalized black holes and, second, we demonstrate 
what happens when the acceleration of sources is removed.

\subsection{Generalized black holes}
Following \cite{GriffithsPodolsky:2006b},  let us  
introduce two new continuous parameters $\alpha$ (the acceleration) and $\omega$ (the `twist') by the rescaling
\begin{equation}
p\to \sqrt{\alpha\omega}p\,,\ \ 
r\to \sqrt{\frac{\alpha}{\omega}}r\,,\ \ 
\sigma\to \sqrt{\frac{\omega}{\alpha^3}}\,\sigma\,,\ \ 
\tau\to \sqrt{\frac{\omega}{\alpha}}\,\tau\;,
\end{equation}
and relabel the other parameters as
\be
m\to\left(\frac{\alpha}{\omega}\right)^{\!{3}/{2}}\!\!\!m\;,\ \ 
n\to\left(\frac{\alpha}{\omega}\right)^{\!{3}/{2}}\!\!\!n\;,\ \ 
e\to\frac{\alpha}{\omega}\,e\;,\ \ 
g\to\frac{\alpha}{\omega}\,g\;,\ \ 
\epsilon\to\frac{\alpha}{\omega}\,\epsilon\;,\ \ 
k\to\alpha^2k\;.
\ee
Then the metric and the vector potential take the form
\ba\label{scaledmetric}
\tens{g}\!\!&=&\!\!\Omega^2\Bigl[-\frac{Q(\grad\tau\!-\!\omega p^2\grad\sigma)^2}{r^2+\omega^2p^2}+\frac{P(\omega \grad\tau\!+\!r^2\grad\sigma)^2}{r^2+\omega^2p^2}
+\frac{r^2\!+\!\omega^2p^2}{P}\,\grad p^2+\frac{r^2\!+\!\omega^2p^2}{Q}\,\grad r^2\Bigr]\;,
\nonumber\\
\tens{A} \!\!&=&\!\! -\frac{1}{r^2\!+\!\omega^2p^2}\,
\Bigl[ e\,r\,\bigl(\grad \tau \!-\!\omega p^2\,\grad\sigma\bigr)
+\,g\,p\,\bigl(\omega\grad \tau \!+\! r^2\,\grad\sigma\bigr) \Bigr]\;,
\ea
with 
\ba
\Omega^{\!-1} \!\!&=&\!\!1-\alpha\, pr\;,\label{OmegaA}\\
Q\!\!&=&\!\!\omega^2 k\!+\!e^2\!+\!g^2\!-\!2mr\!+\!\epsilon r^2\!-\!\frac{2\alpha n}{\omega} r^3
   \!-\!\bigl(\alpha^2 k\!+\!\frac\Lambda3\bigr)r^4\;,\\
P\!\!&=&\!\!k\!+\!\frac{2n}{\omega}p\!-\!\epsilon p^2\!+\!2\alpha m p^3
   \!-\!\Bigl[\alpha^2(\omega^2k\!+\!e^2\!+\!g^2)\!+\!\omega^2\frac\Lambda3\Bigr]p^4\;.
\ea
The CKY tensors are (up to trivial constant factors)
\begin{equation}\label{k2_A}
\Omega^{-3}\tens{k}=\omega p\,\grad r\wedge(\grad\tau\!-\!\omega p^2\grad\sigma)+
r\,\grad p\wedge(\omega\grad\tau\!+\!r^2\grad\sigma)\;,
\end{equation}
and, $\tens{h}=\Omega^{3}\,\grad{\tens{b}}$, with $\Omega$ given in \eqref{OmegaA}, and
\begin{equation}\label{b2_A}
\tens{b}=\frac{1}{2}\bigl[(\omega^2p^2-r^2)\grad\tau+\omega p^2r^2\grad\sigma\bigr]\;.
\end{equation}

Let's consider two special cases. First, we relabel ${\omega=a}$, perform an
additional coordinate transformation
\begin{equation}\label{transf1_A}
p\to\cos\theta\,,\quad 
\tau\to \tau-a\phi\,,\quad
\sigma\to-\phi\;,
\end{equation}
and set 
\begin{equation}\label{resc1}
k=1\,,\ \  \epsilon=1-\alpha^2(a^2+e^2+g^2)-\frac{\Lambda}{3}a^2\,,\ \  n=-\alpha a m\,.
\end{equation}
(One parameter---NUT charge---was set to zero and the scaling freedom was used to eliminate the other two.)
We have obtained a six-parameter solution which describes the accelerating 
rotating charged black hole with the cosmological constant:
\be\label{BH_A}
\tens{g}=\Omega^2\Bigl\{-\frac{Q}{\Delta}\bigl[\grad\tau-a\sin^2\!\theta \grad\phi\bigr]^2
\!+\frac{\Delta}{Q}\grad r^2
\!+\frac{P}{\Delta}\bigl[a\grad \tau-(r^2+a^2)\grad\phi\bigr]^2\!+\frac{\Delta}{P} \sin^2\!\theta \grad \theta^2\Bigr\}\;,
\ee
where
\ba\label{cup_A}
\Omega^{\!-1} \!\!&=&\!\!1-\alpha r \cos\theta\;,\quad
\Delta=r^2+a^2\cos^2\theta\;,\non\\
Q \!\!&=&\!\!(a^2\!+\!e^2\!+\!g^2\!-\!2mr\!+\!r^2)(1\!-\!\alpha^2r^2)-\frac{\Lambda}{3}(a^2\!+\!r^2)r^2\;,\non\\
\frac{P}{\sin^2\!\theta} \!\!&=&\!\!1\!-\!2\alpha m \cos\theta
 \!+\!\Bigl[\alpha^2(a^2\!+\!e^2\!+\!g^2)\!+\!\frac{\Lambda a^2}{3}\Bigr]\cos^2\theta\;.
\ea
In the brackets in \eqref{BH_A} we can easily recognize the familiar form of the Kerr solution. The conformal factor and the
modification of metric functions correspond to the acceleration and the cosmological constant.
The CKY tensor $\tens{k}$ takes the form 
\be\label{kgen}
\Omega^{-3}\tens{k}=
a\cos\theta\, \grad r\!\wedge\left[\grad\tau\!-\!a\sin^2\!\theta\, \grad \phi\right]
\!-r\sin\theta\, \grad \theta\wedge\bigl[a\grad\tau\!-\!(r^2\!+\!a^2)\,\grad\phi\bigr]\;,
\ee
where $\Omega$ is given in \eqref{cup_A}. Except the conformal factor we recovered the
Killing--Yano tensor for the Kerr metric derived by Penrose and Floyd \cite{Penrose:1973}, \cite{Floyd:1973}.

The second interesting example is obtained if instead of \eqref{transf1_A} and \eqref{resc1} we perform
\begin{equation}\label{transf2_A}
p\to\frac{l+a\cos\theta}{\omega},\quad 
\tau\to \tau-\frac{(l+a)^2}{a}\,\phi,\quad
\sigma\to-\frac{\omega}{a}\,\phi\;,
\end{equation}
set the acceleration to zero, $\alpha=0$, and adjust 
\be
\epsilon=1-\frac{\Lambda}{3}({a^2}+6l^2)\,,\ \  
n=l+\frac{\Lambda l}{3} (a^2-4l^2)\,,\ \  
\omega^2k=(1-l^2\Lambda)(a^2-l^2)\,.
\ee
Then we have a non-accelerated rotating charged black hole with 
NUT parameter and the cosmological constant:
\begin{equation}\label{NUT_apA}
\begin{split}
\tens{g}=&-\frac{Q}{\Delta}\bigl[\grad\tau-(a\sin^2\theta+4l\sin^2\!\frac{\theta}{2})\,\grad\phi\bigr]^2
+\frac{\Delta}{Q}\,\grad r^2\\
&+\frac{P}{\Delta}\Bigl\{a\grad \tau-\bigl[r^2+(a+l)^2\bigr]\grad\phi\Bigr\}^2+\frac{\Delta}{P}\sin^2\!\theta\, \grad \theta^2\;,\!\!
\end{split}
\end{equation}
where
\ba\label{co_apA}
\frac{P}{\sin^2\!\theta}\!\!&=&\!\!1+\frac{4\Lambda}{3}al\cos\theta +\frac{\Lambda}{3}a^2\cos^2\!\theta\;,\quad \Delta=\,r^2+(l+a\cos\theta)^2\;,\non\\
Q\!\!&=&\!\!a^2\!-\!l^2\!+\!e^2\!+\!g^2\!-\!2mr\!+\!r^2
\!-\!\frac\Lambda3\Bigl[3(a^2\!-\!l^2)\,l^2+({a^2}\!+\!6l^2)r^2\!+\!{r^4}\Bigr]\;.\quad
\ea
The CKY tensor $\tens{k}$ becomes the KY tensor (see also the next subsection) 
and takes the form 
\ba\label{kgen2}
\tens{k}\!\!&=&\!\!
(l+a\cos\theta)\,\grad r\!\wedge \bigl[\grad\tau-(a\sin^2\theta+4l\sin^2\!\frac{\theta}{2})\,\grad\phi\bigr]\nonumber\\
\!\!&&\!\!-r\sin\theta\,\grad \theta\wedge
\Bigl\{a\grad \tau-\bigl[r^2+(a+l)^2\bigr]\grad\phi\Bigr\}\,.
\ea
The dual CKY tensor becomes closed, ${\tens{h}=\grad \tens{b}}$, with 
\be
\tens{b}=\frac{1}{2}\Bigl{\{} \bigl[(l\!+\!a\cos\theta)^2\!-\!r^2\bigr]\bigl[a\grad\tau\!- \!(l\!+\!a)^2\grad\phi\bigr]- r^2(l\!+\!a\cos\theta)^2\grad\phi\Bigr{\}}\;.
\ee
In particular, in vacuum ($e=g=\Lambda=0$) we recover the KY tensor for the Kerr metric ($l=0$), respectively
for the NUT solution ($a=0$) studied recently in \cite{JezierskiLukasik:2006}, respectively \cite{JezierskiLukasik:2007}.

\subsection{Carter's metric}
\label{sc:CartPleb}
Let us take the Pleba\'nski--Demia\'nski metric in the form \eqref{scaledmetric} and 
set the acceleration to zero, $\alpha=0$, and $\omega=1$. Then the conformal factor becomes $\Omega=1$ and 
we recover the Carter's family of non-accelerating solutions \cite{Carter:1968pl}, \cite{Carter:1968cmp}
in the form used in \cite{Plebanski:1975}:
\be\label{Plebanski}
\tens{g}=-\frac{Q(\grad\tau- p^2\grad\sigma)^2}{r^2+p^2}+\frac{P( \grad\tau+r^2\grad\sigma)^2}{r^2+p^2}
+\frac{r^2+p^2}{P}\,\grad p^2+\frac{r^2+p^2}{Q}\,\grad r^2\;,
\ee
where
\begin{equation}
\begin{split}
Q=&\,k+e^2+g^2-2mr+\epsilon r^2-\frac{\Lambda}{3}r^4\;,\\
P=&\,k+{2np}-\epsilon p^2-\frac{\Lambda}{3}p^4\;,
\end{split}
\end{equation}
and the vector potential is given by \eqref{A_A}.
Notice that \eqref{Plebanski} coincides with the canonical form \eqref{CDP} discussed in the previous section.

We also get 
\begin{equation}\label{KY_apA}
\tens{k}=p\,\grad r\wedge(\grad\tau-p^2\grad\sigma)+r\,\grad p\wedge(\grad\tau+r^2\grad\sigma)\;,
\end{equation}
which is the Killing--Yano tensor given by Carter \cite{Carter:1987}. Its dual,
\begin{equation}
\tens{h}=*{\tens{k}}=\grad{\tens{b}}\;,
\end{equation}
with $\tens{b}$ given by \eqref{b_apA}, becomes the closed CKY tensor.
These are the hidden symmetries for the canonical metric [independent of 
the particular form of $P(p)$ and $Q(r)$].
The conformal Killing tensor \eqref{Qk_A} becomes the Killing tensor
\be\label{Qk2_A}
\tens{K}=\frac{Qp^2(\grad\tau-p^2\grad\sigma)^2}{r^2+p^2}+\frac{Pr^2(\grad\tau+r^2\grad\sigma)^2}{r^2+p^2}
+\frac{r^2+p^2}{P}\,r^2\grad p^2-\frac{r^2+p^2}{Q}\,p^2\grad r^2\;.
\ee
Both isometries of spacetime may be derived from 
the existence of $\tens{k}$, but in a different manner than before.
We have $\tens{\xi}_{({h})}=\tens{\partial}_{\tau}$ whereas $\tens{\xi}_{{(k)}}=0$ since $\tens{k}$
is now a KY tensor. Nevertheless, the second isometry is given by [cf. Eq. \eqref{vector2_A}]
\begin{equation}
(\partial_{\sigma})^{a}= K^{a}_{\ b}\,\xi^{b}_{(h)}\;.
\end{equation} 

Let us observe that the full Pleba\'nski--Demia\'nski  metric with acceleration 
is related to the (non-accelerating) Carter's metric 
only by a conformal rescaling
and a modification of metric functions ${P(p)}$ and ${Q(r)}$.\footnote{%
One might hope that such a transition could work also in higher dimensions. For the demonstration that it is not so, see Appendix C.6.
}
 It allows us to use
the theorem (see, e.g., \cite{Tachibana:1969}, \cite{JezierskiLukasik:2006})
which states that whenever $\tens{k}$ is a CKY tensor for the metric $\tens{g}$ then $\Omega^{3}\tens{k}$ is a 
CKY tensor for the conformally rescaled metric $\Omega^{2}\tens{g}$.
This would justify the transition from the known KY tensor \eqref{KY_apA} 
to the CKY tensor \eqref{k_A}, up to the fact, that 
in the transition from \eqref{metric_A} to \eqref{Plebanski}
we also need to change metric functions ${P(p)}$ and ${Q(r)}$.
Fortunately, as mentioned above, the `universality' of $\tens{k}$, i.e., the property that 
\eqref{KY_apA} remains KY tensor for the metric \eqref{Plebanski} with arbitrary function $P(p)$ and $Q(r)$, can be demonstrated.
Indeed, the only nontrivial components of the covariant derivative ${\tens{\nabla}\tens{k}}$, namely
\begin{equation}
\nabla_{\!p}k_{\sigma r}=\nabla_{\!r}k_{p\sigma}=\nabla_{\!\sigma}k_{rp}=r^2+p^2\;,
\end{equation}
are completely independent of the form of $Q(r)$ and $P(p)$.
Therefore one can start with the metric $\tens{g}$ \eqref{Plebanski}, 
with the KY tensor $\tens{k}$ \eqref{KY_apA}, and with arbitrary functions $P(p)$ and
$Q(r)$ so that, after performing  
the conformal scaling $\tens{g}\to \Omega^2\tens{g}$ we obtain the metric \eqref{metric_A}. The theorem ensures that 
$\Omega^3\tens{k}$ is the universal CKY tensor for the new metric, and in particular for 
the Pleba\'nski--Demia\'nski solution, where $\Omega$ is given by \eqref{Omega_A} and functions $P(p)$ and $Q(r)$ by 
\eqref{functionsQP_A}.

\chapter[PCKY tensor in Myers--Perry spacetimes]{PCKY tensor in the Myers--Perry spacetimes}
\chaptermark{PCKY tensor in Myers--Perry spacetimes}

Historically,  the PCKY tensor in higher-dimensional black hole spacetimes was first discovered \cite{FrolovKubiznak:2007} for the Myers--Perry metrics \cite{MyersPerry:1986}, and its existence was verified with the help of the Maple program up to $D\leq 8$.
A little bit later, an (unpublished) analytical calculation, using the Kerr--Schild form of the Myers--Perry metrics, proved its existence in an arbitrary number of spacetime dimensions. Soon after that, the PCKY tensor was discovered
for the Gibbons--L\"u--Page--Pope \cite{GibbonsEtal:2004}, \cite{GibbonsEtal:2005} Kerr-(A)dS metrics (unpublished), and finally \cite{KubiznakFrolov:2007}
for the general Kerr-NUT-(A)dS spacetimes \cite{ChenEtal:2006cqg}. 
In this appendix we give a brief account of these historical developments and present the sketch of the (unpublished) proof justifying the existence of the PCKY tensor in the Myers--Perry spacetimes.

\section{Myers--Perry metrics and their symmetries}
The Myers--Perry (MP) metrics \cite{MyersPerry:1986} are the most general known
vacuum solutions for the higher-dimensional rotating black holes. 
These solutions  allow the Kerr--Schild form \cite{MyersPerry:1986},
they are of the type D of the higher-dimensional algebraic classification 
\cite{MilsonEtal:2005}, \cite{ColeyEtal:2004a}, \cite{Coley:2008}.
The  metrics have slightly different form for the odd and even
number of spacetime dimensions $D$. We can write them compactly as
\be\n{MP}
\tens{g}=-\grad t^2+\frac{U\grad r^2}{V\!-\!2M}\,+\frac{2M}{U} 
\bigl(\,\grad t+
\sum_{i=1}^{m}a_i\mu_i^2 \grad \phi_i\bigr)^{2} 
+\sum_{i=1}^{m}(r^2+a_i^2)(\mu_i^2\grad \phi_i^2
+\grad \mu_i^2)+\epsilon r^2
\grad \nu^2\, ,
\ee 
where
\be
V\equiv r^{\epsilon-2}\prod_{i=1}^{m}(r^2+a_i^2)\,,\quad 
U \equiv V\bigl(1-\sum_{i=1}^m\frac{a_i^2 \mu_i^2}{r^2+a_i^2}\bigr)\, .
\ee
Here $m\equiv [(D-1)/2]$, where $[A]$ means the integer part of $A$. We
define $\epsilon$ to be $1$ for $D$ even and $0$ for odd. (This is different from the  Kerr-NUT-(A)dS $\varepsilon$; $\epsilon=1-\varepsilon$.)
The coordinates $\mu_i$ are not independent. They obey the following
constraint:
\begin{equation}\label{constraint}
\sum_{i=1}^m\mu_{i}^2+\epsilon \nu^2=1\,.
\end{equation} 
The MP metrics possess the PCKY tensor \cite{FrolovKubiznak:2007}, 
and the, derivable from it, towers of hidden symmetries and tower of Killing vectors. The latter is related to the obvious  $m+1$ isometries, $\pa_t$,
$\pa_{\phi_i}\,$, $i=1,\dots,m\,$, present in the spacetime (see, e.g., Eq. \eqref{kv12_MP} below). 
The KY potential $\tens{b}$
for the MP metric \eq{MP} reads \cite{FrolovKubiznak:2007}
\begin{equation}\label{b_MP}
\tens{b}=\frac{1}{2}\Bigl[\bigl(r^2+\sum_{i=1}^ma_i^2\mu_i^2\bigr) \grad t+
\sum_{i=1}^ma_i\mu_i^2(r^2+a_i^2) \grad \phi_i\Bigr]\,.
\end{equation}
The corresponding PCKY tensor $\tens{h}$ is
\be\n{pcky_MP}
\tens{h}=\sum_{i=1}^{m} a_i\mu_i \grad \mu_i\!\wedge\!
\Bigl[a_i\grad t+(r^2+a_i^2) \grad\phi_i\Bigr]+
r\grad r\!\wedge\!  
\bigl(\grad t+\sum_{i=1}^{m}a_i\mu_i^2\grad \phi_i \bigr)\, .
\ee
Here and later on in similar formulas the
summation over $i$ is taken from $1$ to $m$ for both---even and odd number of
spacetime dimensions $D$; the coordinates $\mu_i$  are independent when $D$ is even 
whereas they obey the constraint \eq{constraint} when $D$ is odd.

Historically, the existence of the PCKY tensor was discovered 
with the help of the Maple Program. At the time of discovery it was known that 
the 5D MP metric possesses the Killing
tensor and allows the separation of variables for the Hamilton--Jacobi
and scalar field equations \cite{FrolovStojkovic:2003a}, \cite{FrolovStojkovic:2003b}. It was also known that such a separation
is possible in higher dimensions, provided that the rotation parameters $a_i$ are 
divided into two classes, and within each of the classes they are equal of one another \cite{VasudevanEtal:2005b}. One of the motivations to search for the hidden symmetry associated with Killing--Yano tensors was the task to solve the parallel transport of 
orthonormal frames. (This task was finally accomplished three years later, see Chapter 9.)

We further remark that, by the time of the discovery of the PCKY tensor 
\eqref{pcky_MP} it was completely unknown that such a tensor allows one to generate other hidden symmetries or Killing vectors,
except those of the dual Killing--Yano tensor $\tens{f}=\tens{*h}\,,$
\be\n{fk_MP}
f_{a_1\dots \,a_{D-2}}=(*h)_{a_1\,\dots\, a_{D-2}}
={1\over 2}\,e_{a_1\,\dots\,a_{D-2}}^{\quad \, 
\quad \quad cd}h_{cd}\,,
\ee
the associated Killing tensor
\be\n{Kmn_MP}
K_{ab}={1\over (D-3)!}\, f_{a a_1\,\dots\,a_{D-3}}
f_{b}^{\, \ a_1\,\dots\, a_{D-3}}=
h_{a c}h_{b}^{\, \ c}
-{1\over 2}\, g_{ab}h_{cd}h^{cd}\,,
\ee
and the corresponding two Killing vectors, 
\be\label{kv12_MP}
\xi^{a}={1\over D-1}\,\nabla_c h^{ca}\,,\quad
\eta_{a}=K_{ac} \xi^{c}\,.
\ee
Specifically, the explicit expressions for $\tens{f}$ in
four and five dimensions, and the following (`computer-empirical') formulas:
\be\n{kt_MP}
K^{ab}\pa_a\pa_b=\sum_{i=1}^{m}\Bigl[ a_i^2(\mu_i^2\!-\!1)\,\tens{g}^{-1}
\!+\!a_i^2\mu_i^2\pa_t^2\!+\!
\frac{1}{\mu_i^2}\,\pa_{\phi_i}^2\Bigr]
\!+\!\!\sum_{i=1}^{m-1+\epsilon}\!\!\pa_{\mu_i}^2\!-\!2\tens{Z\!Z}\!-\!\tens{\xi\zeta}\!-\!\tens{\zeta\xi}\,,
\ee
\begin{equation}
\tens{\xi}=\pa_t\,,\quad
\tens{\eta}=\sum_{i=1}^{m}\!a_i^2\,\tens{\xi}-\tens{\zeta}\,,\quad
 \tens{\zeta}\equiv\sum_{i=1}^m a_i\pa_{\phi_i}\,,\quad
\tens{Z}\equiv \sum_{i=1}^{m-1+\epsilon}
\mu_i\pa_{\mu_i}\,,
\end{equation}
were known. 
These expressions were verified with the help of the Maple program 
up to $D=8$. In 4D, the expression for $\tens{f}$ coincided with the KY tensor discovered by Penrose and Floyd \cite{Penrose:1973}, \cite{Floyd:1973}, and the Killing tensor \eqref{kt_MP}
reduced to the Killing tensor obtained in \cite{Carter:1968pr}, \cite{WalkerPenrose:1970}. In $D=5$, the
Killing tensor \eqref{kt_MP} gave the Killing tensor obtained in \cite{FrolovStojkovic:2003a}, \cite{FrolovStojkovic:2003b}, after
the (constant) term  $\tens{\xi\zeta}+\tens{\zeta\xi}$ was omitted. 
All these `computer-empirical' results were put onto the solid ground by the 
(never published) analytical proof proving the existence of the PCKY tensor
in an arbitrary number of spacetime dimensions (see  Appendix B.3).

Soon after this proof was finished, the PCKY tensor for the general Kerr-NUT-(A)dS spacetimes was discovered \cite{KubiznakFrolov:2007}. Let us in the next section briefly mention the pre-stage of this development. The main impulse was the discovery of the PCKY tensor for the Kerr-(A)dS black holes, together with the transformation of this object into its `universal' canonical form.

\section{Kerr-(A)dS black holes and their symmetries}
A generalization of the MP metric which includes the cosmological constant, the 
Kerr-(A)dS solution in all dimensions, was obtained in 2004 by Gibbons, L\"u, Page, and Pope \cite{GibbonsEtal:2004}, \cite{GibbonsEtal:2005}. 
The metric obeys the Einstein equations
\be 
R_{ab}=(D-1)\lambda g_{ab}\,.
\ee
Similar to the MP metric, the solution allows the Kerr--Schild form and
it is of the algebraic type D. In the `generalized' Boyer--Lindquist coordinates it takes the form
\begin{eqnarray}\n{MPC}
\tens{g}\!\!&=&\!\!-W(1-\lambda r^2)\grad t^2+\frac{2M}{U} \Bigl(W\grad t+
\sum_{i=1}^{m}\frac{a_i\mu_i^2\grad \phi_i}{1+\lambda a_i^2}\Bigr)^{2} 
+\sum_{i=1}^{m}\frac{r^2+a_i^2}{1+\lambda a_i^2}\,(\mu_i^2\grad \phi_i^2
+\grad \mu_i^2) \non\\
\!\!&\ &\!\!+\frac{U\grad r^2}{V-2M}+
\frac{\lambda}{W(1-\lambda r^2)}\,
\Bigl(\sum_{i=1}^m\frac{r^2+a_i^2}{1+\lambda a_i^2}\,\mu_i \grad \mu_i+
\epsilon r^2\nu \grad \nu\Bigr)^2
+\epsilon r^2 \grad \nu^2\,,
\end{eqnarray} 
where
\begin{eqnarray}
W\!\!&\equiv&\!\!\sum_{i=1}^m\frac{\mu_i^2}{1+\lambda a_i^2}+\epsilon \nu^2\,,\quad 
V\equiv r^{\epsilon-2}(1-\lambda r^2)\prod_{i=1}^{m}(r^2+a_i^2)\,,\nonumber\\
U\!\!&\equiv &\!\!\frac{V}{1-\lambda r^2}\,\Bigl(1-\sum_{i=1}^m\frac{a_i^2 \mu_i^2}{r^2+a_i^2}\Bigr)\,.
\end{eqnarray}
Here, we use the same notations and constraint \eqref{constraint} as for the MP metrics.
The Kerr-(A)dS spacetime possesses the PCKY tensor, derivable from the KY potential
\be\label{b_MPlambda}
\tens{b}=\frac{1}{2}\biggl\{\Bigl[r^2+\sum_{i=1}^m a_i^2\mu_i^2\bigl(1-\lambda\frac{r^2+a_i^2}{1+\lambda a_i^2}\bigr)\Bigr] \grad t+
\sum_{i=1}^m a_i\mu_i^2\frac{r^2+a_i^2}{1+\lambda a_i^2} \grad \phi_i\biggr\}\,.
\ee
The PCKY tensor, $\tens{h}=\tens{db}$, reads
\be\n{PCKY_MPlambda}
\tens{h}=\!\sum_{i=1}^{m} a_i\mu_i \grad\mu_i\!\wedge\!
\Bigl[a_i\grad t+\frac{r^2+a_i^2}{1+\lambda a_i^2}( \grad \phi_i-\lambda a_i \grad t)\Bigr] 
\!+r\grad r\!\wedge\!  \Bigl[\grad t+\sum_{i=1}^{m}a_i\mu_i^2(\grad \phi_i-\lambda a_i \grad t) \Bigr]\,.
\ee
This formula was explicitly verified, using the Maple program, for $D\leq 7$.
Moreover, when the cosmological constant $\lambda$ is set to zero the KY
potential \eqref{b_MPlambda} reduces to that of the Myers--Perry metric \eqref{b_MP}.

The crucial impulse for the discovery of the PCKY tensor in the general Kerr-NUT-(A)dS spacetimes \eqref{metric_coordinates}, presented in the main text, was the following observation. In $D=4$, the metric \eqref{MPC} describes the Kerr black hole in the (A)dS background. Such a black hole possesses only one rotation parameter which, as usual, we denote by $a$. We also
put $\mu_1=\sin\theta$, $\phi_1=\phi$, and perform the additional linear transformation of $\phi$ and $t$:
\begin{equation}
\grad t\rightarrow \grad t\,,\quad
\grad \phi\rightarrow-a\lambda \grad t-(1+\lambda a^2)\grad \phi\,.
\end{equation} 
Then one recovers the `standard' form of the Kerr-(A)dS metric. Up to a constant factor, one also has  
\be
\tens{f}=r\sin\theta \grad\theta\!\wedge\!\left[a\grad t+(r^2+a^2)\grad \phi\right]
-a\cos\theta \grad r\!\wedge\!\left[\grad t+a\sin\!^2\theta \grad \phi\right]\,,
\ee 
which is the KY tensor discovered by Penrose
and Floyd \cite{Penrose:1973}, \cite{Floyd:1973} for the Kerr metric. In this form, however, it holds also for the nontrivial $\lambda$. Moreover, further transformation
\begin{equation}
t\rightarrow \tau-a^2\sigma\,,\quad 
\phi\rightarrow a \sigma\,,\quad \cos\theta\rightarrow -p/a\,,
\end{equation} 
brings the metric and the KY tensor into their canonical forms \eqref{CDP} and \eqref{form_AA} described in Appendix A. In particular, we have 
\begin{equation}
\tens{f}=r\grad p\!\wedge\!(\grad \tau+r^2\grad \sigma)+p \grad r\!\wedge\!(\grad \tau-p^2\grad \sigma)\,.
\end{equation}
In this form $\tens{f}$ is completely `universal';  it neither depends on $\lambda$ nor on $a$. In fact, it is the KY tensor for the general Carter's solution described in Appendix A.2.3. This inspired the author of this thesis to search for a convenient coordinate transformation which would transform the known higher-dimensional PCKY tensor \eqref{pcky_MP} or \eqref{PCKY_MPlambda} into its `universal' form.
This is how the PCKY tensor \eqref{PCKY_metric} for the Kerr-NUT-(A)dS metric \eqref{metric_coordinates} was discovered.

\section{PCKY tensor in the MP spacetime: the proof}
In this section we prove that the tensor $\tens{h}$, given by \eqref{pcky_MP}, is indeed the PCKY tensor for the MP metric \eqref{MP}. Namely, we prove that it 
obeys the closed CKY equations \eqref{PCKY_coords}.
We proceed in two steps.
First, we transform the metric and $\tens{h}$ into the Kerr--Schild coordinates.
Second, following \cite{MyersPerry:1986}, we introduce
the convenient orthonormal basis in which the verification of \eqref{PCKY_coords} is,
although elaborate, rather straightforward. 

\subsection{Kerr--Schild form}
Let us start with the transformation
\begin{equation}
\grad t=\grad \tau-\frac{2M}{V-2M}\,\grad r\,,\quad
\grad \phi_i=\grad \varphi_i+\frac{V}{V-2M} \frac{a_i}{r^2+a_i^2}\,\grad r\,,
\end{equation}
which transforms the metric element \eqref{MP} into the `Edington-like' form.
We further introduce the Kerr--Schild coordinates 
\be
x_i\equiv\mu_i\sqrt{r^2\!+\!a_i^2}
\cos\bigl(\varphi_i\!-\!\arctan\frac{a_i}{r}\bigr)\,,\    
y_i\equiv\mu_i\sqrt{r^2\!+\!a_i^2}\sin\bigl(\varphi_i\!-\!\arctan\frac{a_i}{r}\bigr)\,,\ 
z\equiv\epsilon \nu r\,.
\ee
Here, index $i=1,\dots,m$, and the last coordinate $z$ is introduced only in an even number of spacetime dimensions. The inverse transformation reads 
\begin{equation}
\mu_i^2=\frac{x_i^2+y_i^2}{r^2+a_i^2}\,,\quad
\varphi_i=\arctan\frac{a_i}{r}+\arctan\frac{y_i}{x_i}\,,\quad 
\epsilon \nu=\frac{z}{r}\,.
\end{equation}
These relations imply 
\begin{eqnarray}
\mu_i \grad \mu_i=\frac{x_i\grad x_i+y_i\grad y_i}{r^2+a_i^2}-\frac{(x_i^2+y_i^2)r\grad r}{(r^2+a_i^2)^2}\,,\quad
\grad \varphi_i=\frac{x_i\grad y_i-y_i\grad x_i}{x_i^2+y^2_i}-\frac{a_i\grad r}{r^2+a_i^2}\,.
\end{eqnarray}
The constraint \eqref{constraint} reads\footnote{%
The `Boyer--Lindquist' form 
\eqref{MP} of the MP metric is in the original paper \cite{MyersPerry:1986} derived from the Kerr--Schild ansatz \eqref{KS_MP}. We are now going backwards. In the original derivation the constraint \eqref{constraint} is understood as a defining equation for the coordinate $r$. It also expresses the fact that
the vector $\tens{k}$, \eqref{k_MP}, is null.
}
\begin{equation}\label{defr_MP}
\sum_{i=1}^m\frac{x_i^2+y_i^2}{r^2+a_i^2}+\epsilon\frac{z^2}{r^2}=1\,.
\end{equation}
Differentiating this expression we find
\ba
\partial_{x_i}r\!\!\!&=&\!\!\frac{rx_i}{F(r^2+a_i^2)}\,,\quad 
\partial_{y_i}r=\frac{ry_i}{F(r^2+a_i^2)}\,,\quad 
\partial_zr=\frac{\epsilon z}{Fr}\,,\label{ru_MP}\\
F\!\!&\equiv&\!\!\frac{U}{V}=1-\sum_{i=1}^m\frac{a_i^2(x_i^2+y_i^2)}{(r^2+a_i^2)^2}=
r^2\sum_{i=1}^m\frac{x_i^2+y_i^2}{(r^2+a_i^2)^2}+\epsilon\frac{z^2}{r^2}\,.
\label{Fff_MP}
\ea
Therefore,
\begin{equation}\label{dr_MP}
\grad r=\frac{r}{F}\sum_{i=1}^m\frac{x_i\grad x_i+y_i\grad y_i}{r^2+a_i^2}+\epsilon
\frac{z\grad z}{Fr}\,.
\end{equation}
Using these relations we find that
the metric takes the `Kerr--Schild' form
\be\label{KS_MP}
\tens{g}=\tens{\eta}+h\tens{kk}\,,
\ee
where
\ba
\tens{\eta}\!\!&=&\!\!-\grad \tau^2+ \sum_{i=1}^m(\grad x_i^2+\grad y_i^2)\,,\quad 
h=\frac{2M}{U}\,,\\
\tens{k}\!\!&=&\!\!\grad \tau+\sum_{i=1}^m\frac{r(x_i\grad x_i+y_i\grad y_i)+a_i(x_i\grad y_i-y_i\grad x_i)}{r^2+a_i^2}+
\epsilon\frac{z\grad z}{r}\,.\label{k_MP}
\ea
The PCKY tensor \eqref{pcky_MP} reads
\begin{equation}\label{PCKY_KS}
\tens{h}=\sum_{i=1}^m\bigl[(x_i\grad x_i+y_i\grad y_i)\!\wedge\!\grad \tau+a_i\grad x_i\!\wedge\!\grad y_i\bigr]+\epsilon z\grad z\!\wedge\!\grad \tau\,.
\end{equation} 
The last expression is particularly 
useful for obtaining the flat space limit of this tensor [cf. Eq. \eqref{hh_A}].

\subsection{Basis forms}
So far, we have covered both cases of the odd(even)-dimensional spacetime 
simultaneously. It is now natural to split the subsequent calculations.  
Here, we concentrate on the even-dimensional case (the proof in an odd number of dimensions is slightly modified but analogous).

Let us introduce the basis of 1-forms $\tens{E}^a=e^a_\mu \grad x^\mu$,
\ba
\tens{E}^u\!\!&\equiv&\!\!\grad u+A^k\grad x^k+\frac{1}{2}A^2\grad v\,,\ \ 
\tens{E}^v\equiv\grad v-H\tens{E}^u\,,\ \ \tens{E}^k\equiv\grad x^k+A^k\grad v\,,\label{forms_MP}\\
\tens{d}v \!\!&=&\!\!\tens{E}^v+H\tens{E}^u\,,\ \ 
\tens{d}u=\tens{E}^u+\frac{1}{2}A^2\grad v-A^k \tens{E}^k\,,\ \ 
\grad x^k=\tens{E}^k-A^k\grad v\,,
\ea
in which the (even-dimensional) metric \eqref{KS_MP} takes the form
\be
\tens{g}=-2\tens{E}^{(u}\tens{E}^{v)}+\tens{E}^k\tens{E}^k\,.
\ee
Here, we have defined $x^k\equiv(x^i, y^{j})\,,$ $A^k\equiv(qB^i, qC^{j})\,,$ $\tens{E}^k\equiv(\tens{E}_x^i, \tens{E}_y^{j})\,,$ 
\begin{equation}
B^i\equiv\frac{rx^i-a_iy^i}{r^2+a_i^2}\,,\quad
C^i\equiv\frac{ry^i+a_ix^i}{r^2+a_i^2}\,,\quad
q\equiv\frac{\sqrt{2}r}{r+z}\,,\quad 
H\equiv\frac{M}{2Uq^2}\,.
\end{equation}
Indices $i, j$ run over $1,\dots, m$, whereas indices $k, l ,o$
through $1,\dots,2m$; due Einstein summation conventions are used. Also, $a_{i_x}= a_{i_y} = a_i$ whenever $i_x=i_y$.
Using the fact that 
\ba
\frac{1}{2}A^2\!\!&\equiv&\!\!\frac{1}{2}A^k\!A^k=\frac{q^2}{2}(B^iB^i+C^iC^i)=\frac{q^2}{2}\frac{x_i^2+y_i^2}{r^2+a_i^2}= \frac{r-z}{r+z}=\sqrt{2}q-1\,,\ \ \label{A2_MP}\\
C^ia_i\!\!&=&\!\!x_i-rB^i\,,\quad B^ia_i=rC^i-y_i\,,\ \ \label{id1_MP}\\
X\!\!&\equiv&\!\!x^kA^k=qr\frac{x_i^2+y_i^2}{r^2+a_i^2}=
\frac{q}{r}(r^2-z^2)=\sqrt{2}(r-z)\,,\label{X_MP}
\ea
we find
\begin{eqnarray}
\grad v\!\wedge\!\grad u\!\!&=&\!\!-\tens{E}^u\!\wedge\!\tens{E}^v-\grad v\wedge
A^k\!\tens{E}^k\,,\nonumber\\
a_i\grad x_i\!\wedge\!\grad y_i\!\!&=&\!\!a_i\tens{E}^i_x\!\wedge\!\tens{E}^i_y+\grad v\wedge(qx^k\!\tens{E}^k-rA^k\!\tens{E}^k)\,,\nonumber\\
x^k\grad x^k\!\wedge\!(\grad u\!+\!\grad v)\!\!&=&\!\!X\tens{E}^u\!\wedge\!\tens{E}^v\!+\!x^k\!\tens{E}^k\!\wedge\!(\tens{E}^u\!-\!A^l\!\tens{E}^l)\!
+\!\grad v\!\wedge\!(XA^l\!\tens{E}^l\!-\!\sqrt{2}qx^l\!\tens{E}^l)\,.\nonumber
\end{eqnarray}
Plugging these expressions into \eqref{PCKY_KS}, we find the following form of the PCKY tensor in the chosen basis:
\be\label{kbasis_MP}
\tens{h}=r\tens{E}^u\!\wedge\!\tens{E}^v+\frac{x^k}{\sqrt{2}}\tens{E}^k\!\wedge\!\tens{E}^u+(a_i\delta^{ki_x}\delta^{li_y}
-\frac{x^k\!A^l}{\sqrt{2}})\tens{E}^k\!\wedge\!\tens{E}^l\,.
\ee

Let us conclude this subsection with introducing the dual basis operators $D_a~=~e^{\mu}_a\partial_{\mu}$\,,
\ba
D\!\!&\equiv&\!\!D_v=\partial_v\!-\!A^k\partial_k\!+\!\frac{1}{2}\,A^2\partial_u\,,\ \ 
\Delta\equiv D_u=\partial_u\!+\! HD\,,\ \ 
\delta^k\equiv D_k=\partial_{k}\!-\!A^k\partial_u\,,\nonumber\\
\partial_u\!\!&=&\!\!\Delta-HD,\quad
\partial_v=D+\frac{1}{2}\,A^2\partial_u+A^k\delta^k,\quad
\partial_k=\delta^k+A^k\partial_u\,.
\ea

\subsection{Connection coefficients for the MP metric}
The connection coefficients $\Gamma_{abc}$ (antisymmetric in the first two indices) 
are obtained from relations
\begin{equation}\label{connections_MP}
\tens{d E}^a=-\frac{1}{2}\,D^a_{\ bc}\,\tens{E}^b\!\wedge\tens{E}^c\,,
\quad \Gamma_{abc}=\frac{1}{2}\,(D_{cab}+D_{bac}-D_{abc})\,.
\end{equation}
When calculating these coefficients, we shall use the fact that 
the exterior derivative $\tens{d}$ can be expressed as 
\begin{equation}
\tens{d}=\tens{E}^u\Delta +\tens{E}^vD+\tens{E}^l\delta^l\,.
\end{equation}
Calculations are aided by the fact (proved in the next subsection) that
\begin{equation}\label{DA_MP}
DA^k=0\,.
\end{equation}
So we find 
\begin{eqnarray}
\tens{dE}^u\!\!&=&\!\!\grad A^k\!\wedge\!\tens{E}^k=
-\Delta A^k\tens{E}^k\!\wedge\!\tens{E}^u+\delta^l\!A^k\tens{E}^l
\!\wedge\!\tens{E}^k\,,\nonumber\\
\tens{dE}^v\!\!&=&\!\!DH \tens{E}^u\!\wedge\!\tens{E}^v+(H\Delta A^k-\delta^k\!H)\tens{E}^k\!\wedge\!\tens{E}^u-
H\delta^l\!A^k\tens{E}^l\!\wedge\!\tens{E}^k\,,\nonumber\\
\tens{dE}^l\!\!&=&\!\!\Delta A^l \tens{E}^u\!\wedge\!\tens{E}^v+\delta^k\!A^l\tens{E}^k\!\wedge\!\tens{E}^v
+H\delta^k\!A^l\tens{E}^k\!\wedge\!\tens{E}^u\,.
\end{eqnarray}
Comparing with \eqref{connections_MP} we identify
\begin{eqnarray}
D^u_{\ ku}\!\!&=&\!\!\Delta A^k,\quad D^u_{\ lk}=-F^{lk},\quad
D^v_{\ uv}=-DH,\quad 
D^v_{\ ku}=\delta^k\!H-H\Delta A^k,\nonumber\\
D^v_{\ lk}\!\!&=&\!\!HF^{lk},\quad
D^l_{\ uv}=-\Delta A^l,\quad
D^l_{\ kv}=-\delta^k\!A^l,\quad D^l_{\ ku}=-H\delta^k\!A^l,
\end{eqnarray}
where we introduced
\begin{equation}
F^{lk}\equiv \delta^l\!A^k-\delta^k\!A^l=-F^{kl}=F_{lk}\,.
\end{equation}
This leads to the following coefficients: 
\begin{eqnarray}
\Gamma^v_{\ vu}\!\!&=&\!\!-DH\,,\ \ 
\Gamma^v_{\ ku}=H\Delta A^k-\delta^k\!H\,,\ \ 
\Gamma^v_{\ kl}=-H\delta^k\!A^l\,,\ \ 
\Gamma^u_{\ ku}=-\Delta A^k\,,\nonumber\\
\Gamma^u_{\ kl}\!\!&=&\!\!-\delta^l\!A^k\,,\ \ 
\Gamma^u_{\ uu}=DH\,,\ \ \Gamma^k_{\ uu}=H\Delta A^k-\delta^k\!H\,,\ \ 
\Gamma^k_{\ ul}=-H\delta^k\!A^l\,,\nonumber\\
\Gamma^k_{\ vu}\!\!&=&\!\!-\Delta A^k\,,\ \ \Gamma^k_{\ vl}=-\delta^l\!A^k\,,\ \ 
\Gamma^l_{\ ku}=-HF_{lk}\,.
\end{eqnarray}

\subsection{Covariant derivatives of the PCKY tensor}
In order to verify the closed CKY equation \eqref{PCKY_coords},
\be\label{PCKY_coords_MP}
\nabla_{c} h_{ab}=2g_{c[a}\xi_{b]},\quad
\xi_b=\frac{1}{D-1}\nabla_dh^{d}_{\ b}\,,
\ee
we need to calculate the 
covariant derivatives of $\tens{h}$, 
\begin{equation}\label{covdiff_MP}
h_{ab;\,c}=D_{\!c}h_{ab}-\Gamma^d_{\ ac}h_{db}-\Gamma^d_{\ bc}h_{ad}\,,
\end{equation}
where [cf. Eq. \eqref{kbasis_MP}]
\begin{equation}
h_{uv}=r\,,\quad h_{ku}=\frac{1}{\sqrt{2}}\,x^k\,,\quad
 h_{kl}=-\frac{2}{\sqrt{2}}\,x^{[k}\!A^{l]}+ 2a_i\delta^{i_x[k}\delta^{l]i_y}\,.
\end{equation}
The most lengthy parts of the calculation (the details of which we moved to the next subsection) are summarized by the following lemmas: \\
{{\bf Lemma 1} (`Orthogonality relations').} 
\begin{equation}\label{OR_MP}
{\rm a)}\quad r\delta^l\!A^k-h_{ok}\delta^l\!A^o=q\delta^{kl}\,,\quad
{\rm b)}\quad r\delta^k\!A^l-h_{ko}\delta^o\!A^l=q\delta^{kl}\,.
\end{equation}
{\bf Lemma 2.}
\begin{equation}\label{l2_MP}
r\Delta A^k-h_{lk}\Delta A^l=-\frac{1}{\sqrt{2}}A^k\,.
\end{equation}
{\bf Lemma 3.} 
\begin{equation}
h_{ku;u}=0\,.
\end{equation}
Summing a) and b) in \eqref{OR_MP}, we immediately get
\begin{equation}\label{col_MP}
h_{ko}F^{lo}=2q\delta^{kl}-2r\delta^{(l}\!A^{k)}\,,\quad
h_{[k|o|}F_{l]o}=0\,.
\end{equation}
We also need the following relations:
\begin{eqnarray}
\Delta h_{kl}+\frac{2}{\sqrt{2}}\,x^{[k}\Delta A^{l]}=0\,,\ \ 
Dh_{kl}=0\,,\ \ \delta^o\! h_{kl}+
\frac{2}{\sqrt{2}}\,x^{[k}\delta^{|o|}\!A^{l]}=\frac{2}{\sqrt{2}}\delta^{o[l}A^{k]}\,,
\end{eqnarray}
which follow from previous identities.

Applying all these lemmas and identities, we find that 
the only nontrivial covariant derivatives of $\tens{h}$ are
\begin{eqnarray}\label{covderupr}
h_{uv;\,u}\!\!&=&\!\!\frac{1}{\sqrt{2}}-qH\,,\ \  h_{uv;\,v}=-q\,,\ \ 
h_{ku;\,v}=-\frac{1}{\sqrt{2}}\,A^k\,,\ \ 
h_{ku;\,l}=\delta^{lk}(\frac{1}{\sqrt{2}}-qH)\,,\nonumber\\
h_{kl;\,o}\!\!&=&\!\!\frac{2}{\sqrt{2}}\,\delta^{o[l}\!A^{k]}\,,\ \ 
h_{vk;\,u}=\frac{1}{\sqrt{2}}\,A^k\,,\ \ 
h_{kv;\,l}=q\delta^{kl}\,.
\end{eqnarray} 
Specifically, we find
\be
\tens{\xi}^\flat=-\frac{1}{D-1}\,\tens{\delta h}= \Bigl(\frac{1}{\sqrt{2}}-qH\Bigr)\tens{E}^u+q\tens{E}^v-\frac{1}{\sqrt{2}}\,A^k\tens{E}^k\,.
\ee
It is now straightforward to verify that Eq. \eqref{PCKY_coords_MP} holds. $\quad\heartsuit$

\subsection{Proofs of lemmas}
In this subsection we gather the proofs of the above statements. 
Let us denote $c_k\equiv 1/(r^2+a_k^2)$.
For example, the constraint \eqref{defr_MP}, and the definition of $F$, \eqref{Fff_MP},
are
\be
c_kx_k^2+\frac{z^2}{r^2}=1\,,\quad
F=1-a_k^2x_k^2c_k^2=r^2x_k^2c_k^2+\frac{z^2}{r^2}\,.
\ee
We also find
\be\label{ckak_MP}
c_kx^k\!A_k= qr x_k^2c_k^2=
\frac{q}{r}\bigl(F-\frac{z^2}{r^2}\bigr)=\frac{r}{q}\,c_k A_k^2\,,
\quad a_k^2x_kA_kc_k^2=qra_k^2x_kc_k^3\,.
\ee

Let us first prove the relation \eqref{DA_MP}.
Using Eqs. \eqref{ru_MP}, \eqref{A2_MP}, \eqref{id1_MP}, and \eqref{ckak_MP}, we find 
\begin{eqnarray}
Dr\!\!&=&\!\!\partial_vr\!-\!A^k\partial_kr\!+\!\frac{1}{2}A^2\partial_ur=-\frac{z}{\sqrt{2}Fr}-\frac{r c_kx^k\!A^k}{F} +\frac{(\sqrt{2}q\!-\!1)z}{\sqrt{2}Fr}=-q\,,\nonumber\\
Dz\!\!&=&\!\!q-\sqrt{2}\,,\quad Dq=0\,,\quad Dx^k=-A^k\,,\quad
Dc_k=2rqc_k^2\,.
\end{eqnarray}
Therefore one has [and similarly for $D(qC^i)=0$]
\begin{equation}
D(qB^i)=qD(B^i)={q^2}\Bigl[-x^ic_i-rc_iB^i+a_ic_iC^i+2rc_i^2(rx^i-a_iy^i)\Bigr]=0\,,
\end{equation}
where \eqref{id1_MP} and the definition of $B^i$ were used.$\qquad  \heartsuit$

{\bf Proof of Lemma 1.} Let us decompose $h_{kl}$ into its constant part $\hat h_{kl}$ and the `rest'
\begin{equation}\label{decomposekkl}
h_{kl}= \hat h_{kl}+ \tilde h_{kl}\,, \quad
\hat h_{kl}\equiv 2a_i\delta^{i_x[k}\delta^{l]i_y}\,,\quad \tilde h_{kl}\equiv-\frac{2}{\sqrt{2}}\,x^{[k}\!A^{l]}\,.
\end{equation}
We first notice that [see \eqref{A2_MP}]
\ba
\frac{1}{2}\,\delta^l\!A^2\!\!&=&\!\!\sqrt{2}\,\delta^lq\,,\quad
\delta^l\!x^k=\delta^{lk}\,,\quad \delta^lz=-\frac{1}{\sqrt{2}}A^l\,, \label{hol_MP}\\
x^k\delta^l\!A^k\!\!&=&\!\!\delta^l\!X-A^l=\sqrt{2}\delta^l\!r\,,\quad
\tilde h_{ok}\delta^l\!A^o=-A^k\delta^lr+x^k\delta^lq\,.\label{sum1_MP}
\ea
Furthermore, using \eqref{id1_MP}, we find
\begin{eqnarray}\label{sum2_MP}
\hat h_{ok}\delta^l\!A^o \!\!&=&\!\!\hat h_{i_xk}\delta^l\!A^{i_x}+\hat h_{i_yk}\delta^l\!A^{i_y}=
\delta^l(qB^ia_i)\delta^{i_yk}-\delta^l(qC^ia_i)\delta^{i_xk}\nonumber\\
\!\!&=&\!\!\delta^l(qrC^i-qy_i)\delta^{i_yk}-\delta^l(qx_i-qrB^i)\delta^{i_xk}\nonumber\\
\!\!&=&\!\!A^k\delta^lr-x^k\delta^lq+r\delta^l\!A^k-q\delta^{lk}.
\end{eqnarray}
Combining \eqref{sum2_MP} with \eqref{sum1_MP} completes the proof of \eqref{OR_MP} a).

To prove \eqref{OR_MP} b), we successively find  
\ba
\delta^lr\!\!&=&\!\!\frac{rx^l\!c_l}{F}-\frac{zA^l}{\sqrt{2}rF}\,,\ \ 
\delta^lq=\frac{q^2A^l}{2r}+\frac{zq^2\delta^lr}{\sqrt{2}r^2}=\frac{q^2zx^l\!c_l}{\sqrt{2}rF}+\frac{q^2A^l}{2r}\bigl(1-\frac{z^2}{Fr^2}\bigr)\,,\quad\ \\
\delta^k\!A^l\!\!&=&\!\!A^k\!A^l\Bigl(\frac{q}{2r}-\frac{qz^2}{2r^3F}+\frac{\sqrt{2}zc_l}{F}\Bigr)
+A^l\!x^k\Bigl(\frac{qzc_k}{\sqrt{2}rF}-\frac{2r^2c_kc_l}{F}\Bigr)
-A^k\!x^l\frac{qzc_l}{\sqrt{2}rF}\quad\nonumber\\
\!\!&&+x^lx^k\frac{qrc_lc_k}{F}+\delta^{kl}{qrc_k}
+2qa_ic_i\delta^{i_x[k}\delta^{l]i_y}\,.\label{deltaA_MP}
\ea
Moreover, using identities \eqref{ckak_MP}, we find
\be\label{pom1_MP}
\begin{split}
rA^k-h_{kl}A^l=A^k(r+z)-\frac{qz}{r}x^k\,,\ \ 
rx^kc_k-h_{kl}x^lc_l=\frac{zA^k}{rq}+\frac{qx^k}{\sqrt{2}r}(F-\frac{z^2}{r^2})\,,\\
a_ic_i(r\delta^{i_x[k}\delta^{l]i_y}-h_{ko}\delta^{i_x[o}\delta^{l]i_y})=-\frac{A^k\!A^l}{\sqrt{2}q}+
\frac{rc_lx^l\!A^k}{\sqrt{2}}+\frac{rc_lx^k\!A^l}{\sqrt{2}}-\frac{qc_lx^l\!x^k}{\sqrt{2}}
+a_l^2c_l\delta^{lk}\,.
\end{split}
\ee
Using these relations one can express,
$r\delta^k\!A^l-h_{ko}\delta^o\!A^l$, in terms of `independent' 
coefficients $A^k\!A^l, A^l\!x^k, A^k\!x^l, x^l\!x^k, \delta^{lk}$. Finally, using
\begin{equation}
\frac{q}{2}+\frac{zq}{2r}=\frac{1}{\sqrt{2}}\,,
\end{equation} 
and the definition of $q$ one can verify that each term, but $q\delta^{lk}$, vanishes
which completes the proof of \eqref{OR_MP} b).$\qquad  \qquad\heartsuit$

{\bf Proof of Lemma 2.}
Let us decompose $h_{kl}$ as in (\ref{decomposekkl}). Then we find
\ba
\frac{1}{2}\Delta A^2\!\!\!&=&\!\!\!\sqrt{2}\Delta q\,,\ \  
\Delta x^k=-A^k\! H\,,\ \ \Delta z=\frac{1}{\sqrt{2}}+H(q-\sqrt{2})\,, \label{61}\\
x^k\Delta A^k\!\!\!&=&\!\!\!\sqrt{2}(\Delta r\! +\!qH)\!-\!1\,,\ \ 
\tilde h_{lk}\Delta A^l=x^k\Delta q\!-\!A^k\Delta r-qHA^k\!+\!\frac{A_k}{\sqrt{2}}\,\,.
\quad\label{sum3_MP}
\ea
With the help of \eqref{id1_MP} we find
\begin{equation}\label{sum4}
\hat h_{lk}\Delta A^l=-x^k\Delta q+A^k\Delta r+qHA^k+r\Delta A^k\,.
\end{equation}
Combination of \eqref{sum3_MP} and \eqref{sum4}, gives \eqref{l2_MP}. $\qquad\heartsuit$

{\bf Proof of Lemma 3.} This proof is the most difficult part of the whole calculation.
We sketch only the main steps.
Using \eqref{l2_MP} and (\ref{61}), we find
\begin{equation}
h_{ku;\,u}=2rH\Delta A^k-r\delta^k\!H-\frac{1}{\sqrt{2}}\,
x^kDH+h_{kl}\delta^l\!H+\frac{1}{\sqrt{2}}Hx^lF^{lk}\,.
\end{equation}
Our task is now to show that this expression is equal to zero.
First of all, using the following
identities:
\begin{eqnarray}
x^l\delta^l r\!\!&=&\!\!\frac{r-z}{F}\,,\ \ 
A^l\!\delta^lr=n-\frac{nz}{rF}\,,\ \ 
x^l\delta^lq=\frac{q^2X}{2r}\Bigl(1+\frac{z}{rF}\Bigr)\,,\nonumber\\
x^l\delta^lA^k\!\!&=&\!\!\frac{r-z}{F}\,qx^kc_k+A^k\Bigl[\sqrt{2}q+\frac{X}{2F}\bigl(\frac{zq}{r^2}-2\sqrt{2}rc_k\bigr)\Bigr]\,,
\end{eqnarray}
we find
\begin{equation}\label{jedna_MP}
\frac{1}{\sqrt{2}}\,x^lF^{lk}=
\frac{x^l\delta^l A^k}{\sqrt{2}}-\delta^kr=
-\frac{\sqrt{2}qzx^kc_k}{F}+A^k\Bigl(q+\frac{qz}{rF}-\frac{Xrc_k}{F}\Bigr)\,.
\end{equation}
Next, from \eqref{ru_MP} it follows that  
\begin{equation}\label{dva_MP}
2r\Delta A^k=A^k\Bigl(-q+\frac{qz^2}{Fr^2}-\frac{2\sqrt{2}zrc_k}{F}\Bigr)+\frac{x^kc_k\sqrt{2}zq}{F}\,.
\end{equation}
Contracting \eqref{deltaA_MP}, we find 
\begin{equation}\label{deltalAl}
\delta^l\!A^l=2qr\Bigl(\sum_{i=1}^mc_i+\frac{N_3}{F}\Bigr)\,,\quad
N_3\equiv a_k^2x_k^2c_k^3=\frac{a_k^2x_kA_kc_k^2}{qr}\,.
\end{equation}
So we have
\begin{eqnarray}
DF\!\!&=&\!\!-2qrN_3\,,\quad
D\lg V=\frac{q}{r}-2qr\!\sum_{i=1}^{m} c_i\,,\quad
\nonumber\\
\frac{DH}{H}\!\!&=&\!\!\frac{1}{H}D\Bigl(\frac{M}{2q^2F V}\Bigr)=-D\lg(FV)
=-\frac{q}{r}+2qr\Bigl(\sum_{i=1}^mc_i+\frac{N_3}{F}\Bigr)\,,
\end{eqnarray}
and therefore, using (\ref{deltalAl}), we get 
\begin{equation}\label{tri_MP}
\frac{x^k}{\sqrt{2}}\frac{DH}{H}=\frac{x^k}{\sqrt{2}}\Bigl(\delta^l\!A^l-\frac{q}{r}\Bigr)\,.
\end{equation}
Similarly, we obtain 
\begin{eqnarray}
\delta^k\!\lg q^2\!\!&=&\!\!\frac{qA^k}{r}+\frac{\sqrt{2}zq\delta^kr}{r^2}\,,\quad 
\delta^k\!\lg V=\frac{\delta^kr}{r}\bigl(2r^2\!\sum_{i=1}^m c_i-1\bigr)\,,\\
\delta^k\!F\!\!&=&\!\!-2a_k^2x^kc_k^2+4r\delta^krN_3\,\,,\nonumber\\
\frac{\delta^k\!H}{H}\!\!&=&\!\!-\delta^k\!\lg(q^2\!FV)
=\frac{2a_k^2x^kc_k^2}{F}-\frac{qA^k}{r}+
\frac{\delta^kr}{r}\Bigl(1-\frac{\sqrt{2}zq}{r}-\frac{2r^2N_3}{F}-
\frac{r\delta^l\!A^l}{q}\Bigr)\,.\nonumber
\end{eqnarray}
Finally, using \eqref{pom1_MP}, and 
\be
r\delta^kr-h_{kl}\delta^lr=\frac{qx^k}{\sqrt{2}}\,,\ \ 
ra_k^2x^kc_k^2-h_{kl}a_l^2x^lc_l^2=A^k\Bigl(\frac{F-1}{\sqrt{2}}+\frac{a_k^2c_k}{q}\Bigr)+
\frac{qrN_3x^k}{\sqrt{2}}\,,
\ee
one has 
\begin{equation}\label{ctyri}
\frac{1}{H}\bigl(r\delta^k\!H-h_{kl}\delta^l\!H\bigr) =
A^k\Bigl(\frac{2a_k^2c_k}{Fq}-\frac{\sqrt{2}}{F}\Bigr)+
x^k\Bigl(\frac{q}{\sqrt{2}r}-\frac{\delta^l\!A^l}{\sqrt{2}}\Bigr)\,.
\end{equation}
Combining the results \eqref{jedna_MP}, \eqref{dva_MP}, \eqref{tri_MP}, and (\ref{ctyri}), we find that 
\begin{equation}
h_{ku;\,u}=\frac{1}{\sqrt{2}}\,x^lF^{lk}+2r\Delta A^k-\frac{x^k}{\sqrt{2}}\frac{DH}{H}-\frac{1}{H}(r\delta^k\!H-h_{kl}\delta^l\!H)
=0\,.\quad \heartsuit
\end{equation}

\chapter{Miscellaneous results}
In this appendix we gather various results concerning the PCKY tensor
and other related topics. Namely, we prove that the eigenvectors of the PCKY tensor coincide with the principal null directions, 
we review the algebraic integrability conditions for
the existence of a CKY 2-form and relate them with 
the algebraic type of the spacetime, 
we review some algebraic identities used throughout the text and comment on the separability of the first order differential equations in Kerr-NUT-(A)dS spacetimes, we 
outline the unsuccessful attempt to generalize these metrics to the Pleba\'nski--Demia\'nski solution in higher dimensions, and finally briefly 
comment on the degeneracy of the 
eigenvalues of the operator $\tens{F}$ used in Chapter 9.  

\section[Principal null directions]{Principal null directions as eigenvectors of the PCKY tensor} 
{\bf Lemma.} {\em Eigenvectors of the PCKY tensor 
$\tens{h}$ coincide with the `principal null directions'. That is,
the solution of the eigenvalue problem
\be\label{ee_C}
\tens{l}\hook\tens{h}=\lambda \tens{l}^\flat\,,
\ee
is a geodesic WAND (Weyl aligned null direction).}

{\em Proof:} Contracting \eqref{ee_C} with $\tens{l}$ immediately implies that $\tens{l}$ is null. To prove that it is geodesic 
let us introduce the complex null Darboux basis for $\tens{h}$,
\be\label{he_C}
\tens{h}=\lambda \,\tens{l}^\flat\! \wedge \tens{n}^\flat +\sum_i \nu_i {\tens{m}_i}^\flat\wedge {\tens{\bar m}_i}^\flat\,,
\ee
with the only non-vanishing scalar products 
\be\label{mm_C}
(\tens{l},\tens{n})=-1\,,\quad (\tens{m}_i,\tens{\bar m}_i)=1\,.
\ee 
Let us denote by $\tens{\dot T}\equiv \nabla_l \tens{T}$\,, and in particular $\tens{z}\equiv \tens{\dot l}$.
Using the PCKY equation \eqref{PCKY} and \eqref{ee_C} we find
\be
\nabla_l (\tens{l}\hook \tens{h})=\tens{z}\hook \tens{h}-\tens{l}^\flat (\tens{l}\hook \tens{\xi})=\lambda \tens{z}^\flat +\dot \lambda \tens{l}^\flat\,.\nonumber
\ee
Re-arranging the last equation we get
\be \label{zh_C}
\tens{z}\hook \tens{h}= \tens{l}^\flat \bigl(\dot \lambda+\tens{l}\hook \tens{\xi}\bigr)
+\lambda \tens{z}^\flat\,.
\ee
On the left-hand-side we plug the expression \eqref{he_C} for $\tens{h}$, and contract both sides with $\tens{n}$\,, to obtain
\be
\tens{n}\hook (\tens{z}\hook \tens{h})=\lambda (\tens{n},\tens{z})=
\lambda (\tens{n},\tens{z})
-\bigl(\dot \lambda+\tens{l}\hook \tens{\xi}\bigr)\,.\nonumber
\ee
From here it follows that $\dot \lambda=-\tens{l}\hook \tens{\xi}$.
Plugging this expression into \eqref{zh_C} we find that $\tens{z}\hook \tens{h}=\lambda \tens{z}^\flat$. Comparing with \eqref{ee_C} we conclude that 
\be
\tens{z}=\nabla_l \tens{l}=\alpha_l \tens{l}\,.
\ee
This means that $\tens{l}$ is a (non-affine parametrized) null geodesic.
One can restore the affine parametrization by performing a proper boost in $\{\tens{l},\tens{n}\}$ 2-plane, so that afterwards 
\be
\nabla_l\tens{l}=0\,.
\ee
Similarly, one can consider null geodesics in other directions, such as 
\be
\tens{m}_i\hook\tens{h}=-\nu_i {\tens{m}_i}^\flat\,\ \ \Longrightarrow\ \ 
\nabla_{\!m_i} \tens{m}_i=\alpha_{i}{\tens{m}_i}\,.
\ee
It remains to prove that $\tens{l}$ is WAND. 
The most simple way to show this, is to use the explicit form of the eigenvector $\tens{l}$ 
in the most general spacetime admitting the PCKY tensor (see Chapter 7), and 
refer to the paper \cite{HamamotoEtal:2007} where it was shown  that such a vector 
is WAND. $\qquad \qquad \heartsuit$

\section{Integrability conditions for a CKY 2-form}
In this section, following closely \cite{Tachibana:1969}, we repeat the derivation of the integrability conditions for the existence of a CKY 2-form $\tens{k}$,
\eqref{CKY_4rank2}, written as the algebraic relations between components of $k_{cd}$ and the curvature tensor. We use the conventions of \cite{Wald:book1984}. For example, we have
\ba
\bigl(\nabla_a\nabla_b\!-\!\nabla_b\nabla_a\bigr)k_{cd}\!\!&=&\!\!R_{abc}^{\ \ \ \,e}k_{ed}\!+\!
R_{abd}^{\ \ \ \,e}k_{ce}\,,\quad\\
R_{ac}=R_{ca}\!\!&=&\!\!R_{abc}^{\ \ \ \,b}=R_{\ abc}^{b}\,,
\ea
and the following definition of the Weyl tensor $C_{abcd}$:
\be\label{Weyl_C}
R_{abcd}=C_{abcd}+\frac{2}{D-2}\bigl(g_{a[c}R_{d]b}-g_{b[c}R_{d]a}\bigr)
- \frac{2}{(D-1)(D-2)}Rg_{a[c}g_{d]b}\,.
\ee

The defining equation for a CKY 2-form reads 
\be\label{2CKY_C}
\nabla_{(c} k_{a)b}=g_{ca}\xi_{b}\!-\!\xi_{(c}g_{a)b}\,,\quad
\xi_b=\frac{1}{D-1}\nabla_dk^{d}_{\ \,b}\,,
\ee 
or equivalently
\be\label{2CKYb_C}
\nabla_b k_{cd}=\nabla_{[b} k_{cd]}+2g_{b[c}\xi_{d]}\,.
\ee 
What conditions follow from these equations?
Differentiating Eq. \eqref{2CKY_C} and shuffling the indices we have
\ba
\nabla_a\bigl(\nabla_{(b}k_{c)d}\bigr)\!&=&\!g_{bc}\rho_{ad}-\rho_{a(b}g_{c)d}\,,
\label{A1_C}\\
\nabla_b\bigl(\nabla_{(a}k_{c)d}\bigr)\!&=&\!g_{ac}\rho_{bd}-\rho_{b(a}g_{c)d}\,,
\label{A2_C}\\
\nabla_c\bigl(\nabla_{(a}k_{b)d}\bigr)\!&=&\!g_{ab}\rho_{cd}-\rho_{c(a}g_{b)d}\,.
\label{A3_C}
\ea
Here and later we use the abbreviations
\be
\rho_{ab}\equiv \nabla_a\xi_b\,,\quad  
S_{ab}\equiv \rho_{ab}+\rho_{ba}\,,\quad 
A_{ab}\equiv \rho_{ab}-\rho_{ba}\,.\nonumber
\ee
Calculating \eqref{A1_C}$+$\eqref{A2_C}$-$\eqref{A3_C}, and using the Bianchi identity 
$R_{[abc]}^{\ \ \ \ \,e}=0\,,$ we obtain
\ba\label{A4_C}
2\nabla_a\nabla_b k_{cd}\!&=&\!k_{eb}R_{acd}^{\ \ \ \,e}\!+\!
k_{ec}R_{bad}^{\ \ \ \,e}\!+\! k_{ea}R_{bcd}^{\ \ \ \,e}
-g_{cd}S_{ab}+g_{bd}A_{ca}+g_{ad}A_{cb}\,\nonumber\\
\!&+&\!\!2(k_{ed}R_{cba}^{\ \ \ \,e}\!+\!g_{bc}\rho_{ad}\!+\!g_{ac}\rho_{bd}\!-\!g_{ab}\rho_{cd})
\ea
Similarly, we have
\ba
2\nabla_a\nabla_c k_{db}\!&=&\!k_{ec}R_{adb}^{\ \ \ \,e}\!+\!
k_{ed}R_{cab}^{\ \ \ \,e}\!+\!k_{ea}R_{cdb}^{\ \ \ \,e}
-g_{db}S_{ac}+g_{cb}A_{da}+g_{ab}A_{dc}\nonumber\\
\!&+&\!\! 2(k_{eb}R_{dca}^{\ \ \ \,e}\!+\!g_{cd}\rho_{ab}\!+\!g_{ad}\rho_{cb}\!-\!g_{ac}\rho_{db})\,, \label{A5_C}\\
2\nabla_a\nabla_d k_{bc}\!&=&\!k_{ed}R_{abc}^{\ \ \ \,e}\!+\!
k_{eb}R_{dac}^{\ \ \ \,e}\!+\!k_{ea}R_{dbc}^{\ \ \ \,e}
-g_{bc}S_{ad}+g_{dc}A_{ba}+g_{ac}A_{bd}\nonumber\\
\!&+&\!\! 2(k_{ec}R_{bda}^{\ \ \ \,e}\!+\!g_{db}\rho_{ac}\!+\!g_{ab}\rho_{dc}\!-\!g_{ad}\rho_{bc})\,.\label{A6_C}
\ea
From here it follows that
\be\label{A7_C}
\nabla_a\bigl(\nabla_{[b}k_{cd]}\bigr)=\nabla_a\nabla_b k_{cd}-2\rho_{a[d}g_{c]b}\,.
\ee
Adding equations \eqref{A4_C}, \eqref{A5_C}, and \eqref{A6_C}, and using equation \eqref{A7_C},
we have
\ba
2\nabla_a\nabla_bk_{cd}\!\!\!&=&\!\!k_{ed}R_{cba}^{\ \ \ e}\!+\!k_{eb}R_{dca}^{\ \ \ e}\!+\!k_{ec}R_{bda}^{\ \ \ e}\!+\!2g_{cb}\rho_{ad}\nonumber\\
\!\!\!&-&\!\!2g_{db}\rho_{ac}\!+\!g_{ac}A_{bd}\!+\!g_{ab}A_{dc}\!
+\!g_{ad}A_{cb}\,.\nonumber
\ea
Subtracting \eqref{A4_C} from this equation we get
\be\label{skoro_C}
k_{ea}R_{bcd}^{\ \ \ e}\!+\!k_{eb}R_{adc}^{\ \ \ e}
\!+\!k_{ec}R_{dab}^{\ \ \ e}\!+\!k_{ed}R_{cba}^{\ \ \ e}
\!+\!g_{db}S_{ac}\!+\!g_{ac}S_{db}\!-\!g_{ab}S_{dc}\!-\!g_{cd}S_{ab}=0\,.
\ee
Contracting indices $a$ and $b$ in this equation and using 
$k^{ea}(R_{adce}+R_{acde})=0$\,, we have
\be
2k_{e(d}R_{c)}^{\ e}+(D-2)S_{dc}+S^a_a g_{cd}=0\,.
\ee
Contracting the last equation 
once more we obtain $S^a_a=0$. So we get\footnote{In an Einstein space, that is when  $R_{ec}\propto g_{ec}$, the last equation implies that $\xi^a$ is a Killing vector.}
\be\label{Sab_C}
\frac{1}{2}S_a^a=\nabla_a\xi^a=0\,,\quad
\frac{1}{2}S_{dc}=\nabla_{(d}\xi_{c)}=\frac{1}{D-2} R_{e(c}k_{d)}^{\ \ e}\,.
\ee 
Denoting by 
\be\label{TT_C}
T_{bca}^{\ \ \ e}\equiv R_{bca}^{\ \ \ e}+\frac{2}{D-2}R^e_{[b} g_{c]a}\,,
\ee
and plugging \eqref{Sab_C} into \eqref{skoro_C} we finally obtain the following algebraic conditions for the existence of a rank-2 CKY tensor $k_{ab}$:
\be\label{algeb_cond_C}
\bigl(T_{bcd}^{\ \ \ e}\delta_a^f+
T_{adc}^{\ \ \ e}\delta_b^f+
T_{dab}^{\ \ \ e}\delta_c^f+
T_{cba}^{\ \ \ e}\delta_d^f\bigr) k_{fe}=0\,.
\ee

\section{On algebraic type and CKY tensors}
Let us briefly comment on the relationship of the algebraic type of the spacetime and
the existence of a CKY 2-form. It was proved by Collinson \cite{Collinson:1974}, that 
the vacuum spacetime admitting a non-degenerate Killing--Yano 2-form is necessary of 
the algebraic type D. Here we outline the proof that the same remains true for the existence a non-degenerate CKY 2-form. This proof can be easily extended to higher dimensions \cite{ColeyEtal:2008}. 

Our starting point are the algebraic relations between the components of $k_{ab}$ and the curvature tensor \eqref{algeb_cond_C}.
In a particular case of vacuum, we have
\be\label{Ck_C}
\bigl(C_{bcd}^{\ \ \ e}\delta_a^f+
C_{adc}^{\ \ \ e}\delta_b^f+
C_{dab}^{\ \ \ e}\delta_c^f+
C_{cba}^{\ \ \ e}\delta_d^f\bigr) k_{fe}=0\,,
\ee  
where $C_{abcd}$ denotes the Weyl tensor---related to the Riemann tensor by \eqref{Weyl_C}. 
In an arbitrary number of dimensions we have the following canonical forms of 2-forms (see, e.g., \cite{Milson:2004}):
\ba
k_{ab}\!\!&=&\!\!\lambda_0 n_{[a}l_{b]}+\!\!\!\sum_{p=1}^{[D/2]-1}\!\!\!\lambda_p m^{2p}_{[a}m^{2p+1}_{b]}\,,\\
k_{ab}\!\!&=&\!\!\lambda_0 n_{[a}m^{D-1}_{b]}+\!\!\!\sum_{p=1}^{[(D-3)/2]}
\!\!\!\lambda_p m^{2p}_{[a}m^{2p+1}_{b]}\,,\\
k_{ab}\!\!&=&\!\!\lambda_0\bigl(n_{[a}m^{D-1}_{b]}\!\!+\!l_{[a}m^{D-1}_{b]}\bigr)\!+\!
\!\!\!\!\!\sum_{p=1}^{(D-3)/2}\!\!\!\!\!\lambda_p m^{2p}_{[a}m^{2p+1}_{b]}\,.
\ea
Here, $\tens{n}$ has boost weight 1, $\tens{l}$ has boost weight -1, and the remaining spacelike forms are of boost weight 0. 

To prove that the existence of CKY 2-form $k_{ab}$ in a vacuum implies that the spacetime is of the algebraic type $D$, it is sufficient to prove that the algebraic relations \eqref{Ck_C} with these canonical forms eliminate all, but the boost weight zero, components of the Weyl tensor.  

\subsection{Non-degenerate CKY in four dimensions}
In 4D, the non-degenerate 2-form takes the canonical form
\be\label{4dk_C}
k_{ab}=\lambda_0 n_{[a}l_{b]}+\lambda_1 m^{2}_{[a}m^{3}_{b]}\,.\\
\ee
Denoting by $F(a,b,c,d)$ the left-hand-side of \eqref{Ck_C}, we successively find
\ba
F(1,1,2,2)&=&2\lambda_0 C_{1212}-2\lambda_1 C_{1213}=0\,,\nonumber\\
F(1,1,2,3)&=&-\lambda_1 C_{1313}=0\,,\nonumber\\
F(0,0,2,2)&=&-2\lambda_0 C_{0202}-2\lambda_1 C_{0203}=0\,,\nonumber\\
F(0,0,2,3)&=&-\lambda_1C_{0303}=0\,,\nonumber\\
F(0,1,3,1)&=&\lambda_0 C_{0113}+\lambda_1C_{0112}=0\,,\nonumber\\
F(1,2,3,2)&=&\lambda_0 C_{1223}-\lambda_1C_{1323}=0\,,\nonumber\\
F(0,0,1,2)&=&-\lambda_0 C_{0102}-\lambda_1C_{0103}=0\,,\nonumber\\
F(0,2,3,2)&=&-\lambda_0 C_{0223}-\lambda_1C_{0323}=0\,,\nonumber\\
C_{0212}&=&C_{0313}\,,\quad 
C_{0312}=-C_{0213}\,.
\ea
This implies that only the following components of the Weyl:
\be
C_{0101}\,,\ C_{0123}\,,\ C_{0313}\,,\ C_{2323}\,, 
\ee
and the components obtainable from them by symmetries of this tensor may be present in 
the spacetime admitting a CKY tensor \eqref{4dk_C}. All of them are of boost zero and hence the spacetime is necessary of the algebraic type D.

\section{Some algebraic identities}
Throughout the thesis we use various algebraic identities. Most of them are 
directly related to the canonical form of the PCKY tensor or the canonical form of the Kerr-NUT-(A)dS spacetime. In this section 
we make a short overview of such identities.

The following functions are used throughout the text:
\begin{gather}
U_{\mu}=\prod_{\substack{\nu=1\\\nu\ne\mu}}^{n}(x_{\nu}^2-x_{\mu}^2)\,,\quad 
A_{\mu}^{(k)}=\!\!\!\!\!\sum_{\substack{\nu_1, \dots,\, \nu_k=1\\ \nu_1<\dots\,<\nu_k,\, \nu_i\ne\mu}}^n\!\!\!\!\!x^2_{\nu_1}\dots x^2_{\nu_k}\,,\quad 
A^{(k)}=\!\!\!\!\!\sum_{\substack{\nu_1, \dots,\, \nu_k=1\\ \nu_1<\dots\,<\nu_k}}^n\!\!\!\!\!x^2_{\nu_1}\dots x^2_{\nu_k}\;.
\end{gather}
These functions appear in the definition of the canonical metric \eqref{metric_coordinates}, they 
appear in the expressions for the PCKY tensor \eqref{PCKY_metric}, or the expressions for the Killing tensors \eqref{KT_metric}.
One can easily generate $A^{(k)}$, $A_{\mu}^{(k)}$ with the help of \cite{KrtousEtal:2007jhep}, \cite{OotaYasui:2008}
\ba\label{eqA_C}
\prod_{\nu=1}^n (t-x_\nu^2)&=&A^{(0)} t^{n}-
A^{(1)} t^{n-1}+\dots + (-1)^{n}A^{(n)}\,,\\
\prod_{\substack{\nu=1\\ \nu \neq \mu}}^n (t-x_\nu^2)&=&A^{(0)}_\mu t^{n-1}-
A^{(1)}_\mu t^{n-2}+\dots + (-1)^{n-1}A^{(n-1)}_\mu\,.
\ea
One might also introduce quantity $U$, \cite{FrolovEtal:2007}, \cite{KubiznakFrolov:2007}
\begin{equation}\label{Udef_C}
  U\equiv\det\left[A_\mu^{(j)}\right]=\prod_{\substack{\mu,\nu=1\\\mu<\nu}}^n(x_\mu^2-x_\nu^2)\;,
\end{equation}
which is simply related to the determinant of the canonical metric [cf. Eq. \eqref{detg}] 
\begin{equation}\label{det_C}
g={\det(g_{ab})}=\bigl(-c A^{(n)}\bigr)^\eps\, U^2\;.
\end{equation}
In the first expression \eqref{Udef_C}, $A_\mu^{(j)}$ $(j=0,\dots, n-1)$, is understood as the $n \times n$ matrix.\\
{{\bf Lemma 1} (\cite{FrolovEtal:2007}).} {\em 
The $n \times n$ matrix 
${B^{\,\mu}_{(k)}\equiv (-x_\mu^2)^{n\!-\!1\!-\!k}/U_\mu}$ is an inverse of ${A_\mu^{(k)}}$.
That is,
\begin{equation}\label{Ainverse_C}
\begin{gathered}
\sum_{k=0}^{n-1}\frac{(-x_\mu^2)^{n\!-\!1\!-\!k}}{U_\mu}A_\nu^{(k)}=\delta_\mu^\nu\;,\quad
\sum_{\mu=1}^{n} \frac{(-x_\mu^2)^{n\!-\!1\!-\!k}}{U_\mu}A_\mu^{(l)} = \delta_k^l\;.
\end{gathered}
\end{equation}
}
In particular, we obtain the following important identities:
\begin{subequations}\label{Uids_C}\allowdisplaybreaks
\begin{align}
&\sum_{\mu=1}^n \frac{(-x_\mu^2)^{n-1}}{U_\mu}= 1\;,\label{Un-1id_C}\\
&\sum_{\mu=1}^n \frac{x_\mu^{2k}}{U_\mu}  = 0    
         \quad\text{for}\quad k=0,\dots,n-2\;,\label{Ukid_C}\\
&\sum_{\mu=1}^n \frac{1}{x_\mu^2 U_\mu} = \frac{1}{A^{(n)}}\;,\label{U-1id_C}\\
&\sum_{\mu=1}^n \frac{A_\mu^{(k)}}{x_\mu^2U_\mu} = \frac{A^{(k)}}{A^{(n)}}
         \quad\text{for}\quad k=0,\dots,n-1\;,\label{UA-1id_C}
\end{align}
\end{subequations}
The first two relations follow immediately from \eqref{Ainverse_C} (set ${l=0}$ in the latter expression).  \eqref{U-1id_C} follows from \eqref{Un-1id_C}
by substituting ${x_\mu\to1/x_\mu}$. \eqref{UA-1id_C} can be verified using 
\eqref{U-1id_C}, \eqref{Ainverse_C},
and the fact that ${A_\mu^{(k)}=A^{(k)}-x_\mu^2 A_\mu^{(k-1)}}$. 

The following lemma plays the central role for the separability  
in the canonical spacetimes:\\
{\bf Lemma 2.} {\em 
The most general solution of the equation 
\be
\sum_{\nu=1}^n \frac{f_\nu(x_\nu)}{U_\nu}=0
\ee
is given by 
\be
f_\nu(x_\nu)=\sum_{j=1}^{n-1}C_j(-x_\nu^2)^{n-1-j}\,,
\ee
where $C_j$ are arbitrary constants.
} \\
This lemma was already used in \cite{HamamotoEtal:2007}, \cite{FrolovEtal:2007}.
Its proper proof can be found in \cite{Krtous:2007}. 
We finally mention the identity 
\begin{equation}\label{upUder_C}
\partial_{x_\mu} \Bigl[\,\frac{U}{U_\mu}\,\Bigr]=0\;,
\end{equation}
(used in the separation of the Klein--Gordon equation) which obviously follows from the definition of $U$ and $U_\mu$. Many more useful relations can be found, for example, in \cite{Krtous:2007}.

\section[Integrability in Kerr-NUT-(A)dS]{Integrability of some functions in Kerr-NUT-(A)dS spacetimes through separation of variables}
In the main text we have encountered several situations where we have to solve 
an ordinary differential equation (or the set of equations, $j=1,\dots,l$) 
\be\label{Fj_CC}
\dot F_j=G_j(x_\mu)\,.
\ee
Here the dot denotes the derivative with respect to an affine parameter and the 
right-hand-side is in general complicated function of $x_\mu$'s (or possibly $r$ in the Lorentzian case).
Such equations were, for example, obtained for the components $\psi_j$, \eqref{xpsi_dot}, of the geodesic velocity in Chapter 5, or for the rotation angles $\beta_\mu$ in Chapter 9.  
It turns out that some of these equations may be `symbolically' integrated as they allow 
an additive separation of variables. 
Let us probe this possibility in more detail. The separability means, that we seek the solution in the form 
\begin{equation}\label{psidot_CC}
F_j=\sum_{\nu=1}^{n} F^{(\nu)}_j(x_\nu)\,.
\end{equation}
Using the first relation \eqref{xpsi_dot}, one finds 
\be\label{beta_dot2_CC}
\dot F_j=\sum_{\nu=1}^{n} \bigl(F^{(\nu)}_j\bigr)'{\dot x_\nu}=\sum_{\nu=1}^{n} 
\frac{\sigma_\nu  {\rm sign}(U_\nu) \bigl( F^{(\nu)}_j \bigr)' \sqrt{X_\nu V_\nu-W_\nu^2}}{U_\nu}\,.
\ee
Prime denotes the derivative with respect to a single argument.
For each $\nu$ the numerator of the last expression is function of $x_\nu$ only.
If $\dot F_j$ given by \eqref{Fj_CC} can be brought  into the form
\begin{equation}\label{f_CC}
\dot F_j=\sum_{\nu=1}^{n} \frac{f^{(\nu)}_j(x_\nu)}{U_\nu}\,,
\end{equation}
the problem is separable. By comparing \eqref{beta_dot2_CC} with \eqref{f_CC}
we arrive at 
\begin{equation}\label{g_CC}
\sum_{\nu=1}^{n} \frac{g^{(\nu)}_j(x_\nu)}{U_\nu}=0\,,\quad
g^{(\nu)}_j=\sigma_\nu  {\rm sign}(U_\nu) \bigl( F^{(\nu)}_j \bigr)' \sqrt{X_\nu V_\nu-W_\nu^2}-f^{(\nu)}_j\,.
\end{equation}
The general solution of \eqref{g_CC} is (see Lemma 2 of the previous section)
\be
g^{(\nu)}_j=\sum_{k=1}^{n-1}C_j^{(k)}(-x_\nu^2)^{n-1-k}\,.
\ee
However, what we need is a particular solution.
For such a solution, we may choose
all the constants $C_j^{(k)}=0$. (In fact, a different choice of $C_j^{(k)}$ leads only to 
a different additive constant for $F_j$.) So we have 
\begin{equation}\label{separ_CC}
F_j=\sum_{\nu=1}^{n}\int \frac{\sigma_\nu {\rm sign}(U_\nu) f_j^{(\nu)}dx_\nu}{\sqrt{X_\nu V_\nu-W_\nu^2}}\,.
\end{equation}

The situation in the Lorentzian case is exactly analogous. If the right-hand-side of 
\eqref{Fj_CC} can be brought into the form
\begin{equation}\label{f2_CC}
\dot F_j=\frac{f^{(r)}_j(r)}{U_n}+\sum_{\nu=1}^{n-1} \frac{f^{(\nu)}_j(x_\nu)}{U_\nu}\,,
\end{equation}
the separated solution reads
\begin{equation}\label{separ2_CC}
F_j=\!\int\!\!\frac{\sigma_n f_j^{(r)}dr}{\sqrt{W_n^2-X_n V_n}}
+\sum_{\nu=1}^{n-1}\int\!\!\frac{\sigma_\nu {\rm sign}(U_\nu) f_j^{(\nu)}dx_\nu}{\sqrt{X_\nu V_\nu-W_\nu^2}}\,.
\end{equation}

As a particular example let us consider the affine parameter itself. Due to the identity
\eqref{Un-1id_C},
\be
1=\sum_{\mu=1}^n \frac{f^{(\mu)}}{U_\mu}\,,\quad f^{(\mu)}=(-x_\mu^2)^{n-1}\,,
\ee
 we find that 
\be\label{tau_CC}
\tau=\sum_{\mu=1}^n\int\frac{\sigma_\mu{\rm sign}(U_\mu)(-x_\mu^2)^{n-1}dx_\mu}{\sqrt{X_\mu V_\mu-W_\mu^2}}\,.
\ee

\section{Higher-dimensional Pleba\'nski--Demia\'nski?}
As we have mentioned in Chapter 4,
the form \eqref{metric_coordinates} of the higher-dimensional Kerr-NUT-(A)dS spacetime 
can be considered as a `natural' higher-dimensional generalization of the 
Carter's canonical form \eqref{Plebanski} of the 4D Kerr-NUT-(A)dS spacetime.
In four dimensions, it is well known how even more general solutions can be `generated' from the canonical form.  The electromagnetic charge is added by a simple change of  metric functions $Q$ and $P$, 
the accelerated class of solutions is obtained by a conformal rescaling of the 
canonical element. All these classes are uniformly described by the Pleba\'nski--Demia\'nski
class of solutions, see Appendix A.2.

This inspires the search for more general higher-dimensional solutions. 
It is natural to ask, whether it is not possible to generalize
the higher-dimensional Kerr-NUT-(A)dS spacetimes in a way exactly 
analogous to the 4D case.
The problem of charging these solutions was addressed in \cite{Krtous:2007}.
It turned out that such a procedure is not sufficiently general.\footnote{The higher-dimensional `Kerr--Newman solution' is still analytically unknown. 
We conjecture that such a solution may not be of the algebraic type D. If this is so,
its form might be quite different from the form of the Myers--Perry metric.
The similarity of these solutions in four dimensions stems from the special properties of electromagnetism in 4D.}
Here we demonstrate that also the attempt to `accelerate' Kerr-NUT-(A)dS 
spacetime in a way analogous to 4D, i.e., by a conformal rescaling (cf. Appendix A.2), generally fails
\cite{KubiznakKrtous:2007}.

We consider the metric 
\begin{equation}\label{scaledHDBH}
\tilde{\tens{g}}=\Omega^2\tens{g}\;,
\end{equation}
where $\tens{g}$ is a canonical metric \eqref{metric_coordinates} and  
${\Omega^2}$ is an unknown conformal factor. 
We ask the question, whether it is possible to 
adjust $\Omega$ and metric functions ${X_\mu(x_\mu)}$ so that the scaled metric $\tilde{\tens{g}}$ satisfy the vacuum Einstein equations. Due to the same argument which we used in four dimensions (see Appendix A.2.3) such a metric would possess a conformal Killing--Yano tensor ${\tilde{\tens{h}}=\Omega^3\tens{h}}$.

The Ricci tensor ${\tilde{\Ric}}$ of the rescaled metric ${\tilde{\tens{g}}}$
is related to the Ricci tensor of the unscaled metric ${\tens{g}}$
by a well known expression (see, e.g., appendix in \cite{Wald:book1984}),
which can be written as
\begin{equation}\label{confRic}
  \tilde{\Ric}=\Ric 
     +(D-2)\Omega\covd\covd\Omega^{\!-1}
     +\tens{g} \Bigl[ \Omega\covd^2\Omega^{\!-1} - (D-1)\Omega^2\bigl(\covd\Omega^{\!-1}\bigr)^2\Bigr]\;.
\end{equation}
Here, the `square' of 1-forms is defined using 
the inverse unscaled metric ${\tens{g}^{-1}}$.
We require ${\tilde{\Ric} = -\lambda\,\tilde{\tens{g}}}$ 
with ${\lambda}$ proportional to the cosmological constant.
The Ricci tensor ${\tilde{\Ric}}$ thus must be diagonal in the frame $\{\tens{\omega}\}$.
The conditions on off-diagonal terms give 
equations for the conformal factor ${\Omega}$. 

In a generic odd dimension these conditions are too strong---they 
admit only a constant conformal factor ${\Omega}$.
In even dimensions the conditions on off-diagonal terms lead to equations
\begin{equation}\label{OmegaCond}
  \Omega^{-1}{}_{\!\!,\mu\nu} = 
    \frac{x_\nu\, \Omega^{-1}{}_{\!\!,\mu}}{x_\nu^2-x_\mu^2}
    +\frac{x_\mu\,\Omega^{-1}{}_{\!\!,\nu}}{x_\mu^2-x_\nu^2}\;,\quad
  0=\frac{x_\mu\,\Omega^{-1}{}_{\!\!,\mu}}{x_\nu^2-x_\mu^2}
  +\frac{x_\nu\,\Omega^{-1}{}_{\!\!,\nu}}{x_\mu^2-x_\nu^2}\;.
\end{equation}
It gives the conformal factor depending on two constants ${c}$ and ${a}$\,,
\begin{equation}\label{OmegaHD}
  \Omega^{\!-1} = c + a\, x_1 \dots x_n\;,
\end{equation}
which is obviously a generalization of the 
four-dimensional factor \eqref{Omega_A} (with ${c=1}$ and ${a=i\alpha}$).

Unfortunately, the conditions for diagonal terms of the
Ricci tensor are in even dimensions ${D>4}$ rather restrictive.
Analyzing first the condition for the scalar curvature and then checking
all diagonal terms one finds
that either\footnote{%
The trivial global scaling was eliminated by setting ${a=1}$
in \eqref{dualsol} and ${c=1}$ in \eqref{trivsol}.}
\begin{equation}\label{dualsol}
  \Omega^{\!-1} = x_1\dots x_2\;,\quad
  X_\mu = \bar b_\mu\, x_\mu^{2n-1} + \sum_{k=0}^{n}\, c_{k}\, x_\mu^{2k}\;,
\end{equation}
with ${\lambda=(D-1)\,c_0}\,$, or
\begin{equation}\label{trivsol}
  \Omega^{\!-1} = 1+a\, x_1\dots x_2\;,\quad
  X_\mu = \sum_{k=0}^{n}\, c_{k}\, x_\mu^{2k}\;,
\end{equation}
with ${\lambda=(D-1)\bigl[(-1)^{n-1}c_n+a^2 c_0\bigr]}\,$.
The first case is not a new solution: the substitution
\begin{equation}\label{dualsoltrans}
  x_\mu = 1/\bar x_\mu\;,\quad
  \psi_j=\bar\psi_{n-1-j}\;,\quad
  X_\mu = \bar x_\mu^{-n+1}\bar X_\mu\,,
\end{equation}
transforms the rescaled metric ${\tilde{\tens{g}}}$ back to
the form \eqref{metric_coordinates} in `barred' coordinates.
In the second case metric functions ${X_\mu}$
depend on a smaller number of parameters and one has 
to expect that the metric describes only a subclass of the `accelerated black hole solutions'. 
It is actually the trivial subclass---%
it was shown in \cite{HamamotoEtal:2007} that
the metric \eqref{metric_coordinates} with ${X_\mu}$ given by \eqref{trivsol}
represents the maximally symmetric spacetime;
therefore the scaled metric ${\tilde{\tens{g}}}$, being the Einstein space 
conformally related to the maximally symmetric spacetime,
must describe also the maximally symmetric spacetime.
In analogy with the four-dimensional case 
we expect that the metric \eqref{scaledHDBH} with 
metric functions \eqref{trivsol}
describes the Minkowski or \mbox{(anti-)de~Sitter} space
in some kind of `rotating accelerated' coordinates.
Such an interpretation, however, 
requires a further detailed study.\footnote{%
We have to conclude that a non-trivial 
generalization of the Pleba{\'n}ski--Demia\'nski metric into a generic 
higher dimension cannot be found by a conformal rescaling \eqref{scaledHDBH} of the canonical element \eqref{metric_coordinates}.
However, recently it was demonstrated that
a nontrivial generalization of the 5D MP metric can be obtained  
with the help of two scaling factors \cite{LuEtal:2008a}.  
The solution contains three independent and one adjustable parameter. It is obtained by gluing the rescaled 4D canonical metric with a (differently rescaled) part 
corresponding to the fifth dimension.
This solution also gives rise to a new charged black hole of 5D minimal supergravity \cite{LuEtal:2008b}.
}

The form of the Pleba\'nski--Demia\'nski metric
motivates 
the construction of new solutions in higher dimensions. 
Its non-accelerated subclass, the Carter's metric,
has been already generalized into higher dimensions by 
Chen, L\"u, and Pope \cite{ChenEtal:2006cqg}. However, it seems that 
further generalizations, although almost obvious 
at a first sight, cannot be easily obtained.
For example, the attempts to `naturally' charge these solutions failed so far (see, e.g., \cite{Krtous:2007}, \cite{ChenLu:2008}).
Here we have demonstrated that also the generalization to accelerated solutions is not straightforward.  
In particular, we have shown that the direct analogue of the Pleba\'nski--Demia\'nski  family of solutions (with acceleration)
cannot be in higher dimensions obtained in a manner similar to the four-dimensional 
case, that is, by a conformal scaling of the Chen--L\"u--Pope metric, possibly with the `natural' change of metric functions.
The question about the existence of the C-metric in higher dimensions therefore still remains 
open.

\section{Remarks on the degeneracy of eigenvalues of $\bs{F}$}
In this section we comment on the degeneracy of the eigenvalues of 
the 2-form $\tens{F}$ used in Chapter 9.\footnote{As this appendix 
directly relates to Chapter 9, we consider the case of the Lorentzian signature 
and $\tens{F}$ for timelike geodesics, that is, $\tens{F}$ given by \eqref{Fop}. Also other notations are the same as in Chapter 9.}
Namely, we prove that due to that fact that the PCKY tensor $\tens{h}$ is by definition non-degenerate, the corresponding 2-form $\tens{F}$, \eqref{Fop}, possesses $(q_0+1)$-times 
degenerate zero eigenvalue, where $q_0=1$ in even number of spacetime dimensions and $q_0=0,2$ for generic, special trajectories in an odd number of spacetime dimensions.   
Moreover, the multiplicity of non-zero eigenvalues is governed by $q_\mu\leq 2$, and the equality occurs only for special geodesic trajectories.

Consider a {\em non-degenerate} $\bs{h}$, then
\be\n{Del_CC}
\Delta(\lambda)=\det(h^a_{\ b}-\lambda \delta^a_b)
=(-\lambda)^{\varepsilon}
\prod_{k=1}^n(\lambda^2+\nu_k^2)\, ,
\ee
where all $\nu_k$ are different.
Let us re-calculate this determinant in terms of $\tens{F}$ and compare the results.
For this calculation we use the  Darboux basis of $\tens{F}$ and 
corresponding matrix form of the objects. In particular, we have 
the expression \eqref{F_ch9} for $\tens{F}$, and 
\be\label{S_CC}
s=(s_{\hat 0}, s_{\hat 1}, \dots s_{\hat p})\,,\quad 
s_{\hat 0}=(\vnb{1}{s}{{0}},\dots, \!\vnb{q_0}{s}{{0}})\,,\ \ 
s_{\hat \mu}=(\vnb{1}{s}{{\mu}},\vnb{1}{\bar s}{{\mu}},\dots,
\vnb{q_\mu}{s}{{\mu}},\vnb{q_\mu}{\bar s}{{\mu}})\,,
\ee
for the 1-form $\tens{s}$ defined in \eqref{Fop}. Using \eq{Fop}, \eqref{F_ch9}, and \eqref{S_CC}, one can rewrite \eqref{Del_CC} as 
\be
\Delta(\lambda)\!=\det(F^a_{\ \,b}\!-\!u^as_b\!+\!s^au_b\!-\!\lambda \delta^a_b)\!=\!\left| 
\begin{array}{cc}
A & B\\
C & E
\end{array}
\right|\, ,
\ee
where $A=-\lambda$, $B=-s$, $C=-s^T$, and $E$ is the $(D-1)$-dimensional  
matrix of the form  
\ba\n{DD_CC}
E=\mbox{diag}(-\lambda I_{q_0}, \nbsi{1}{Z}, 
\ldots, \nbsi{p}{Z})
\,,\quad 
\nbsi{\mu}{Z}=\left(   
\begin{array}{cc}
-\lambda {I}_\mu & \lambda_\mu I_\mu\\
-\lambda_\mu I_\mu & -\lambda I_\mu
\end{array}
\right)
\, .
\ea
Here $I_{q_0}$  is a unit $q_0\times q_0$ matrix, 
and we use $X^T$ to denote a matrix transposed to $X$. It is easy to check that 
\ba\n{DDD_CC}
E^{-1}\!\!&=&\!\mbox{diag}(-\lambda^{-1}I_{q_0},
 \nbsi{1}{Z}^{-1}, \ldots, \nbsi{p}{Z}^{-1}) \,,\nonumber\\
\nbsi{\mu}{Z}^{-1}\!\!&=&\!Q_\mu^{-1}\ \nbsi{\mu}{Z}^T \,,\ \ 
Q_\mu=(\lambda^2+\lambda_\mu^2)\,,\ \ \det(\nbsi{\mu}{Z})
=Q_\mu^{q_\mu}\,.
\ea
One has the following
relation for the determinant of a block matrix (see, e.g.,
\cite{Gantmacher:1959})
\be
\left|   
\begin{array}{cc}
A & B\\
C & E
\end{array}
\right|={\cal A}\,|E|\,,\quad {\cal A}=|A-BE^{-1}C|\, .
\ee
Using \eqref{DD_CC} and \eqref{DDD_CC}, one finds
\be\label{cco}
\det(E)=(-\lambda)^{q_0}\!\! \prod_{\mu=1}^p Q_\mu^{q_\mu}\,,\quad   
{\cal A}=-\lambda-s {E}^{-1}s^T.
\ee
Combining all these relations one obtains
\ba\n{DL_CC}
\Delta(\lambda)\!\!&=&\!\!(-\lambda)^{q_0-1} \prod_{\mu=1}^p Q_\mu^{q_\mu}
\left[\lambda^2 \Bigl(1-\sum_{\mu=1}^p
\frac{s_{\hat \mu}^2}{Q_\mu}\Bigr)- s_{\hat 0}^2 \right]\,
,\\
s_{\hat 0}^2\!\!&=&\!\! \sum_{i=1}^{q_0} \vnb{i}{s}{{0}}^2\,,\quad  
s_{\hat \mu}^2=\sum_{i=1}^{q_\mu} (\vnb{i}{s}{{\mu}}^2+\vnb{i}{\bar s}{{\mu}}^2)\,.\nonumber
\ea 

Let us now compare \eqref{Del_CC} and \eqref{DL_CC}.
First of all, let us compare the powers of $(-\lambda)$. 
For $s_{\hat 0}^2\ne 0$ we have match 
for $q_0-1=\varepsilon$, whereas the case $s_{\hat 0}^2=0$ may happen only in 
odd dimensions and one must have $q_0=0$ [cf. \eqref{q0_ch9}].
Another result of the comparison is that $q_\mu\le 2$. Really,
if $q_\mu>2$, then at least 2 roots of $\Delta(\lambda)$ in \eqref{DL_CC} coincide. This
contradicts the assumption previously stated, since for a non-degenerate
operator $\bs{h}$ the characteristic polynomial has only single roots
$\lambda^2=-\nu_k^2$. The case when $q_\mu=2$ is degenerate. It is valid
only for a special value of the velocity $\bs{u}$. Really, in this case
one of the eigenvalues, say $\nu_k$, of $\bs{h}$ coincides with one of
the eigenvalues of $\bs{F}$ so that one has
$\mbox{det}(\bs{F}-\nu_k\tens{I})=0$. The latter is an equation
restricting the value of $\bs{u}$.

%
%

\end{document}